\pdfoutput=1

\documentclass[a4paper, 11pt, twoside]{ms}  
\graphicspath{{./}}  
\setappendixpath{./}

\usepackage[square, numbers, comma, sort&compress]{natbib}  
\usepackage{verbatim}  
\usepackage{verbatimbox}
\usepackage[usenames]{xcolor}
\usepackage{colortbl}
\usepackage{tabularx}
\usepackage{multirow}
\usepackage{wrapfig}
\usepackage{framed}
\usepackage{amssymb}
\usepackage{amsmath}
\usepackage{amsthm}
\usepackage{stmaryrd}
\usepackage{framed}
\usepackage{todonotes}
\usepackage{fancyhdr}
\reversemarginpar
\presetkeys{todonotes}{inline, size=\tiny, prepend, caption={TODO}}{}
\usepackage{listings}
\usepackage{longtable}
\usepackage{multicol}
\usepackage{makecell}
\usepackage{pifont}
\usepackage{tabu}
\usepackage{fancyvrb}
\usepackage{textcomp}
\usepackage{ascii}
\usepackage{lmodern}
\usepackage{bm}
\usepackage{boldline}
\usepackage{placeins}
\usepackage{contour}
\usepackage{adjustbox}
\usepackage{pbox}

\newcommand\YAMLcolonstyle{\color{red}\bfseries\small\ttfamily}
\newcommand\YAMLkeystyle{\color{black}\bfseries\small\ttfamily}
\newcommand\YAMLvaluestyle{\color{blue}\bfseries\small\ttfamily}
\makeatletter
\newcommand\language@yaml{yaml}
\expandafter\expandafter\expandafter\lstdefinelanguage
\expandafter{\language@yaml}
{
  keywords={true,false,null,y,n},
  keywordstyle=\color{violet}\bfseries\small\ttfamily,
  basicstyle=\YAMLkeystyle,                                 
  sensitive=false,
  comment=[l]{\#},
  morecomment=[s]{/*}{*/},
  commentstyle=\color{purple}\ttfamily,
  stringstyle=\YAMLvaluestyle\ttfamily,
  moredelim=[l][\color{orange}]{\&},
  moredelim=[l][\color{magenta}]{*},
  moredelim=**[il][\YAMLcolonstyle{:}\YAMLvaluestyle]{:},   
  morestring=[b]',
  morestring=[b]",
  literate = {---}{{\ProcessThreeDashes}}3
             {|}{{\textcolor{red}\textbar}}1 
             {\ -\ }{{\mdseries\ -\ }}3,
}
\lst@AddToHook{EveryLine}{\ifx\lst@language\language@yaml\YAMLkeystyle\fi}
\makeatother
\newcommand\ProcessThreeDashes{\llap{\color{cyan}\mdseries-{-}-}}

\definecolor{darkblue}{rgb}{0.0, 0.0, 0.55}
\setboolean  {proposal}{false} 
\university  {Saarland University}{}
\department  {Department of Computer Science}{}
\group       {}{}
\faculty     {Faculty of Mathematics and Computer Science}{}
\supervisor  {Your Supervisor}
\advisor     {Your Advisor}
\examiner    {Prof. Dr. Anja Feldmann}
\secondexaminer {Dr. Keon Jang}
\degree      {Master of Science}
\thesiskind  {Master's Thesis}
\authors     {Evgeniya Khasina}
\thesistitle {The CoCo-Beholder: Enabling Comprehensive Evaluation of Congestion Control Algorithms}
\date        {\today}

\doifproposal{
	\let\thesisnameproposal\thesisname
	\thesiskind{Proposal for a \thesisnameproposal}
}

\usepackage[pdfpagemode={UseOutlines},bookmarks=true,
   bookmarksopenlevel=0,bookmarksnumbered=true,hypertexnames=false,
   colorlinks,linkcolor={black},citecolor={black},urlcolor={black},
   pdfstartview={FitV},unicode,breaklinks=true,bookmarksopen=true]{hyperref}
\usepackage{bookmark}

\begin{document}

\hypersetup{urlcolor={blue!80!black}}

\frontmatter	  

\maketitle

\doifnotproposal{
\begin{declaration}
  \begin{center}
    \bf \large
    Erkl\"{a}rung
  \end{center}
  Ich erkl\"{a}re hiermit,
  dass ich die vorliegende Arbeit selbst\"{a}ndig verfasst und keine
  anderen als die angegebenen Quellen und Hilfsmittel verwendet habe. 

  \begin{center}
    \bf \large
    Statement
  \end{center}
  I hereby confirm that I have written this thesis on my own and that I have not used any other media or materials than the ones referred to in this thesis. \\[2cm]

  \begin{center}
    \bf \large
    Einverst\"{a}ndniserkl\"{a}rung
  \end{center}
  Ich bin damit einverstanden, dass meine (bestandene) Arbeit in beiden 
  Versionen in die Bibliothek der Informatik aufgenommen und damit 
  ver\"{o}ffentlicht wird. 

  \begin{center}
    \bf \large
    Declaration of Consent
  \end{center}
  I agree to make both versions of my thesis (with a passing grade) 
  accessible to the public by having them added to the library of the
  Computer Science Department.\\[2.5cm]
  
      Saarbr\"{u}cken, \noindent\makebox[2in]{\hrulefill} \qquad\makebox[2.0in]{\hrulefill} \\
   \makebox[1.4in][l]{}\makebox[2.07in][l]{(Datum/Date)}\makebox[2.0in][l]{(Unterschrift/Signature)}%

\end{declaration}
  \cleardoublepage  
}

\begin{abstract}

The recent endeavors of the research community to unite efforts on the design and evaluation of congestion control algorithms have created a growing collection of congestion control schemes called Pantheon. However, the virtual network emulator that comes with the collection has very limited capabilities: it can run flows of only one scheme at once, and the flows cannot have individual network settings, as the \mbox{topology} is\nolinebreak[4] \mbox{point-to-point}. This thesis addresses those limitations and presents CoCo-Beholder, a human-friendly \mbox{emulator} \mbox{providing} the popular dumbbell topology of any size, each link of which may have \mbox{individual} rate, delay, and queue size. The central link of the topology may also have a variable delay with optional\nolinebreak[4] \mbox{jitter}. Flows of different schemes may run between the halves of the topology at once, and for each flow, the direction and starting time can be chosen. CoCo-Beholder's reliability is ensured by testing schemes in the dumbbell topology of size one and comparing the results against those of Pantheon emulator. With CoCo-Beholder, the thesis successfully reproduces experiments from a recent paper that evaluated the fairness and RTT-fairness of schemes using a real hardware dumbbell\nolinebreak[4] testbed. \mbox{Finally}, the thesis explores the behavior of schemes under the square-wave delay using CoCo-Beholder's variable delay feature.

\end{abstract}

\doifnotproposal{

\begin{acknowledgements}
I am very grateful to Professor Anja Feldmann for supervising this work.\\

I am very grateful to Dr. Corinna Coupette for help and advice during the whole work.
\end{acknowledgements}
\cleardoublepage
}

\addtotoc{Contents}
\tableofcontents
\clearpage

\doifnotproposal{
  \thispagestyle{empty}  
}

\mainmatter	  


\loadchapter{intro}{Introduction}

The life of the modern human society heavily depends on computer applications that rely on Internet communication service~\cite{meeker2018internet}. Internet is a network of networks interconnected by packet switches referred to usually as routers~\cite{rfc1122, kurose}. In order to benefit from the Internet communication service, a user should make their host computer a part of a network.

According to the end-to-end principle~\cite{saltzer1984end}, more complex communication functions should lie on the network edge~\cite{kurose} represented by hosts, rather than on the network core~\cite{kurose} represented by routers. This is why a host intended to become part of a network typically must implement at least one protocol from each layer of the four layers comprising the Internet protocol suite: Application, Transport, Internet, and Link Layers~\cite{rfc1122}. At the same time, for a router it is enough to operate on the two lower layers: Internet and Link Layers~\cite{kurose}. It must be noted here that there also exists an alternative seven-layered Open Systems Interconnection (OSI) model~\cite{zimmermann1980osi} of the Internet.

Transport Layer protocols implemented at the host define how the application data passed from the upper Application Layer (e.g., via socket API~\cite{rfc147}) should be processed so that the data would finally be pushed further down the network stack and delivered by Internet and Link Layer protocols across the Internet to a destination host~\cite{kurose}.

In the early days of the Internet both the main transport protocols, TCP~\cite{rfc793} and UDP~\cite{rfc768}, used to blast data into the network without any constraint~\cite{jacobson1988congestion, jacobson1995congestion}. This ended up into the first-ever observed congestion collapse, which happened in October 1986 when the throughput of the connection between Lawrence Berkeley Lab and UC Berkeley (around 365 meters and three hops) dropped down from 32 Kbps to 40 bps~\cite{jacobson1988congestion, jacobson1995congestion}.

The problem was mitigated by enhancing TCP with the end-to-end TCP congestion control algorithm by Van Jacobson~\cite{jacobson1988congestion, jacobson1995congestion} that was consequently named TCP Tahoe~\cite{fall1996simulation}. The situation was further improved in the early 1990s when Random early detection (RED) queuing discipline for routers was invented by  Sally Floyd and Van Jacobson~\cite{floyd1993random}.

UDP still does not provide any congestion control, and Application Layer protocols that are built on top of UDP should take care of not overwhelming the network with their traffic themselves~\cite{rfc8085}.

Congestion control algorithms (schemes) can be end-to-end like the above-mentioned TCP Tahoe and network-assisted. End-to-end or implicit congestion control is implemented at hosts and gets no explicit support from the routers of the network core~\cite{kurose}. Network-assisted or explicit congestion control, on the contrary, implies that routers should notify hosts explicitly about the upcoming congestion~\cite{kurose}. 

The most prominent example of the network-assisted approach is Explicit Congestion Notification (ECN)~\cite{rfc3168}, which requires an active queue management like RED being adopted by routers as well as special extensions to IP~\cite{rfc791} and TCP being supported by the network stack. As such schemes make routers keep the congestion-related state, they depart~\cite{rfc6077} from the strict adherence to the end-to-end principle and are slowly adopted~\cite{kuhlewind2013state, trammell2015enabling}. This thesis further considers only end-to-end congestion control as one most widely used.

To protect network links in the path between sender and receiver hosts from overload, the sender is to restrain its TCP congestion window, that is, the number of TCP segments that can be inflight yet unacknowledged, while keeping it close to the optimum~\cite{geist2019overview}. \mbox{The optimal} congestion window size is the Bandwidth Delay Product (BDP) calculated as the product of the total round trip time (RTT) of the path and the bandwidth of the slowest link along the path~\cite{geist2019overview}. There are different approaches to achieving the goal.

\begin{figure}[h]
\centering
\includegraphics[width=7.5cm]{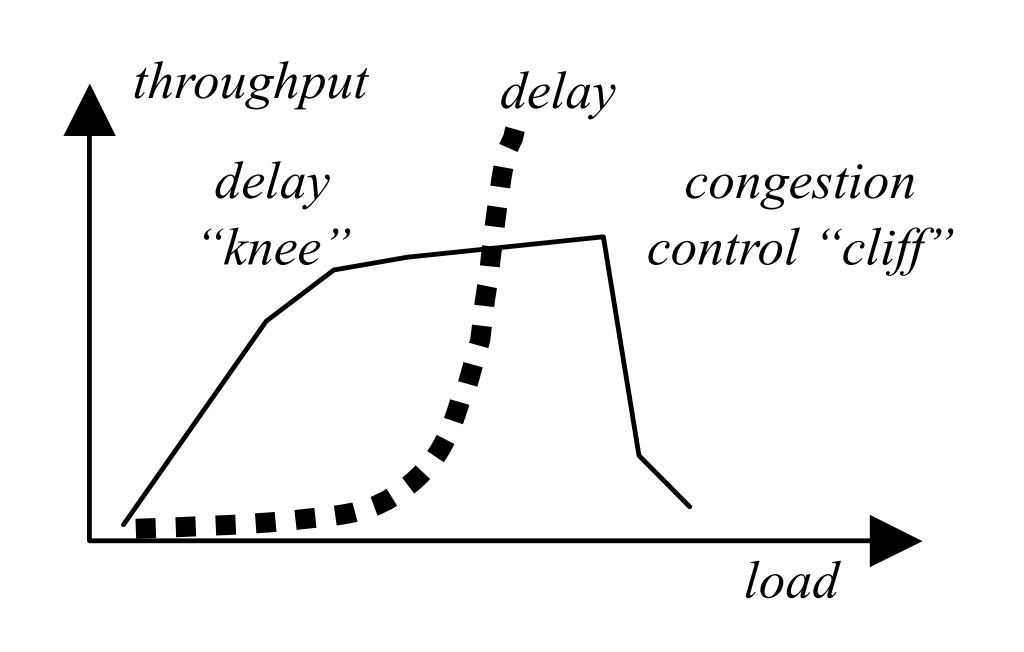}
\caption{Congestion collapse~\cite{ahn2002supporting}.}
\label{fig:collapse}
\end{figure}

The most actively used TCP congestion control schemes are loss-based ones~\cite{al2019survey}. For example, loss-based TCP Cubic~\cite{cubic} is currently used in Linux kernel by default~\cite{geist2019overview}. As inferred from their name, loss-based schemes detect congestion only when the queues of routers in the path are already full and packets start to get dropped, thus, being lost. That is, the schemes try to keep the network to the left of the ``cliff" -- the congestion collapse point, at which the network throughput radically decreases (see Figure~\ref{fig:collapse})~\cite{jain1998congestion}. 

Loss-based schemes are successors of TCP Tahoe, or rather, its improved version TCP Reno~\cite{rfc5681}, which became the first widely deployed congestion control algorithm~\cite{turkovic2019fifty} and the first TCP version that probed the bandwidth with the additive-increase/multiplica-tive-decrease (AIMD) algorithm resulting in the classical TCP sawtooth graph of the congestion window size over time~\cite{kurose, dordal2017introduction}. 

However, the loss-based congestion control has certain disadvantages: increasing the congestion window incessantly till the packet loss occurs means that at least one packet is to be dropped periodically~\cite{geist2019overview} and that queues are to build up at routers, which introduces big delays and can harm low-latency applications~\cite{turkovic2019fifty}. 

These problems are addressed by delay-based congestion control algorithms, the most famous representative of which is TCP Vegas~\cite{vegas}. The delay-based schemes monitor the delays of packet acknowledgments to be able to detect the imminent network overload when the queues at routers only start growing~\cite{turkovic2019fifty}. That is, such algorithms try to keep the network at the ``knee" -- the point after which the throughput grows slowly but the packet delay increases rapidly (see Figure~\ref{fig:collapse}) -- and are also referred to as congestion avoidance algorithms~\cite{jain1998congestion}. 

The disadvantage of delay-based schemes is that they may provide poor link utilization due to their conservative behavior comparing to that of loss-based schemes~\cite{turkovic2019fifty}. Also, delay-based schemes show bad performance in presence of loss-based schemes, as the latter introduce big queuing delays~\cite{geist2019overview}.

Hybrid congestion control algorithms combine the loss-based and delay-based approaches ~\cite{turkovic2019fifty}. The examples are TCP Veno~\cite{veno}, which was the very first hybrid scheme~\cite{turkovic2019fifty}, and TCP BBR developed by Google~\cite{bbr}.

A separate group of congestion control algorithms that are trained to adjust the TCP congestion window using machine learning (e.g., Indigo~\cite{pantheon} and PCC Allegro~\cite{pcc}) or are computer-generated (e.g., Tao 100x~\cite{taova}) are called learned in this thesis.

Only in the paper~\cite{turkovic2019fifty} more than 30 congestion control schemes are mentioned and new and new schemes are created. Comprehensive evaluation of the algorithms is crucial to compare them and decide which of the algorithms should be deployed in the real world and to guide further research.

Several Requests for Comments (RFC) specify the characteristics by which  congestion control algorithms should be evaluated~\cite{rfc2914, rfc5033, rfc5166}. The schemes should be measured in challenging environments with different BDP. Throughput, delay, and loss should be measured when running the schemes in the environments to assess their performance. The reaction of the schemes to the variability of the environment should be considered. 

Intra- and inter-fairness of a scheme should be inspected, which is how fairly the flows of the scheme share the available bandwidth with other flows of this very scheme and with flows of other schemes correspondingly (the competing flows may be of different RTTs). The fairness quality is crucial, as it ensures that the congestion control scheme is indeed capable of preventing the congestion collapse. Deployability and security issues like the robustness of the schemes in the presence of misbehaving hosts are also the aspects that the RFC documents counsel to take into account but are not explored in this thesis.
 
\phantomsection
\label{pr1}

The goal of the thesis is to build an easy-to-use tool enabling comprehensive evaluation of congestion control algorithms individually and against each other using the above-mentioned metrics from~\cite{rfc2914, rfc5033, rfc5166} and providing a highly customizable emulation environment that would allow to have the popular~\cite{turkovic2019fifty, ma2017fairness, zhang2019evaluation, tsao2007taxonomy, copa, vivace, pcc, cdg, cubic, bic, veno} testing setups involving the dumbbell network topology.

The work makes the following contributions:
\begin{itemize}
\item The author of the thesis built this tool (further referred to as CoCo-Beholder).

\item The author compared the results by CoCo-Beholder against those by the tool from~\cite{pantheon} for the simplest point-to-point testing setups possible for~\cite{pantheon}.  

\item The author compared the results by CoCo-Beholder against those by the real testbed from~\cite{turkovic2019fifty} for more complicated dumbbell topology testing setups.

\item The author explored the reaction of different congestion control schemes to a variable delay present at the central link of the dumbbell topology using CoCo-Beholder, which provides the variable delay as one of its features. 
\end{itemize}

\loadchapter{related}{Related Work}

This chapter focuses on the approaches and environments that allow performing computer network research experiments, including those on network congestion control. The reader interested in the experimentation results, rather than in the experimentation methodology, can find the comparative performance evaluation of loss-based, delay-based, and hybrid congestion control schemes in the paper ``Fifty Shades of Congestion Control: A Performance and Interactions Evaluation"~\cite{turkovic2019fifty} (2019) by Belma Turkovic, Fernando Kuipers, and Steve Uhlig.

Section~\ref{sec:first} of this chapter overviews the three prevalent experimentation strategies. Section~\ref{sec:second} takes a closer look at the strategy to which CoCo-Beholder adheres. Section~\ref{sec:third} considers the experimentation tool closest to CoCo-Beholder in its goal to enable research into congestion control schemes.

\section{Simulation, Emulation, and Live Testing}
\label{sec:first}

There are three approaches commonly used to conduct a network experiment: simulation (modeling the network fully in software), live Internet testing, and emulation (the hybrid of the two previous approaches). None of these strategies are perfect, each being a trade-off between such goals as ease of use, control, and realism~\cite{allman1999effective, emulab}.

Table~\ref{tab:tools} shows which experimental environments were used by inventors of different congestion control schemes and referenced in experimental and evaluation sections of their papers presenting the schemes. 

From the table, it can be observed that ns-2~\cite{ns-2} is the leader among simulation tools. This object-oriented, event-driven, packet-level network simulator with graphical visualization was written in C++ ~\cite{arvind2016comparative, torkey2008performance}. Ns-2 is able to simulate different network topologies and protocols, wired and wireless networks,  various router queue managements~\cite{arvind2016comparative, torkey2008performance}. The experiments should be configured by writing scripts in OTcl, which is the object-oriented extension of Tcl programming language and may require some time and effort for a newcomer to get acquainted with~\cite{arvind2016comparative, harju2001network}. Ns-2 is open-source and so provides support of the big research community and unlimited customization -- the opportunities lacked by proprietary simulators like OPNET also appearing in Table~\ref{tab:tools}~\cite{rampfl2013network}.   

Packet-level discrete event simulators are highly controllable and repeatable~\cite{emulab}. However, building the right model can be too hard or time consuming~\cite{nussbaum2009comparative} and its level of abstraction may turn out to be too high to render important low-level details~\cite{emulab}, which can prevent researchers from detecting features or bugs of the actual implementation~\cite{dummynet}. 

Performing experiments in live networks is the whole opposite alternative to simulation. It gives researchers great realism but reduces the level of control they have~\cite{emulab}. It is difficult or impossible to achieve repeatability and to configure or monitor the state of nodes in a live network~\cite{emulab}. Platforms often provide little choice of configuration, if any, which leads to poor generalizability of results~\cite{nussbaum2009comparative}. Table~\ref{tab:tools} shows that researchers prefer to use live experiments more as an addition to simulation or emulation testing.

One of the greatest disadvantages of live networks is also that it can be challenging to get access to their resources. To join PlanetLab spanning its nodes across more than 25 countries, one should belong to an institution that is a member of PlanetLab Consortium~\cite{planetlab}. To lead a project utilizing GENI nodes located in the USA, one should be a faculty or a senior member of an institution (GENI can be found in Table~\ref{tab:tools}).

Emulation is an approach that is intermediate between simulation and live testing and tries to balance pluses and minuses of the two. Real applications, protocols, and operating systems are used but in a synthetic, simulated network environment configured to provide the required delay, bandwidth, loss, and other specific network conditions in place of real network infrastructure~\cite{nussbaum2009comparative, emulab}. Examples of emulation tools are considered in the next section. Most of the examples are present in Table~\ref{tab:tools}.

Concerning Table~\ref{tab:tools}, there are some points that should be commented on. 

TCP BBR and QUIC Cubic, present in the table, were developed by Google, and from their papers, it is unclear which simulation or emulation tools were used for evaluation of the schemes. However, the papers state that Google actively deploys both the inventions at its backbones and Web servers and, thus, has an exclusive opportunity to assess the performance of the schemes with the help of millions of users. 

\begin{table}[!p]
\caption{\mbox{Experimental environments used in the papers of congestion control schemes.}}
\label{tab:tools}
\small
\renewcommand{\arraystretch}{2.15}
\resizebox{\textwidth}{!}{\begin{tabular}{|c||>{\centering\arraybackslash}p{1cm}|>{\centering\arraybackslash}p{1cm}|>{\centering\arraybackslash}p{2cm}|>{\centering\arraybackslash}p{2cm}|>{\centering\arraybackslash}p{2cm}|c|} \hline\hline

\textbf{Scheme} & \textbf{Year} & \textbf{Type} & \textbf{Simulation} & \textbf{Emulation} & \textbf{Live tests} & \textbf{\thead{Dumbbell\\topology}}\\ [0.1cm]\hline

Indigo~\cite{pantheon} & 2018 & learned & \ding{55} & \ding{55} & Pantheon & \ding{55} \\ [0.1cm]\hline

Copa~\cite{copa} & 2018 & delay & ns-2 & tc qdisc & Pantheon & \ding{51} \\ [0.1cm]\hline

\makecell{PCC\\Vivace}~\cite{vivace} & 2018 & learned & \ding{55} & \makecell{Emulab,\\ Mahimahi} & AWS~\cite{miller2010amazon} & \ding{51} \\ [0.1cm]\hline

TCP BBR~\cite{bbr} & 2016 & hybrid & \textbf{?} & \textbf{?} & \makecell{Google\\services} & \textbf{?} \\ [0.1cm]\hline

Verus~\cite{verus} & 2015 & delay & OPNET & tc qdisc & \makecell{Etisalat\\UAE}~\cite{ayish2005virtual} & \ding{51} \\ [0.1cm]\hline

\makecell{PCC\\Allegro}~\cite{pcc} & 2015 & learned & \ding{55} & Emulab & GENI~\cite{berman2014geni} & \ding{51} \\ [0.1cm]\hline

Tao 100x~\cite{taova} & 2014 & learned & ns-2 & \ding{55} & \ding{55} & \ding{51} \\ [0.1cm]\hline

Sprout~\cite{sprout} & 2013 & delay & \ding{55} & Cellsim~\cite{sprout} & \ding{55} & \ding{55} \\[0.1cm] \hline

\makecell{QUIC\\Cubic}~\cite{quic} & 2012 & loss & \textbf{?} & \textbf{?} & \makecell{Google\\services} & \textbf{?} \\ [0.1cm]\hline

TCP CDG~\cite{cdg} & 2011 & hybrid & \ding{55} & DummyNet & \ding{55} & \ding{51}\\ [0.1cm]\hline

TCP Cubic~\cite{cubic} & 2008 & loss & \ding{55} & DummyNet & \ding{55} & \ding{51} \\ [0.1cm]\hline

\makecell{TCP\\Illinois}~\cite{illinois} & 2008 & hybrid & ns-2 & \ding{55} & \ding{55} & \ding{51} \\ [0.1cm]\hline

YeAH-TCP~\cite{yeah} & 2007 & hybrid & \ding{55} & \textbf{?} & \ding{55} & \ding{51} \\ [0.1cm] \hline

\makecell{TCP\\New Vegas}~\cite{nv} & 2005 & delay & ns-2 & \ding{55} & \ding{55} & \ding{55} \\ [0.1cm]\hline

H-TCP~\cite{htcp0, htcp} & 2004 & loss & ns-2 & DummyNet & \ding{55} & \ding{51}\\ [0.1cm]\hline

BIC TCP~\cite{bic} & 2004 & loss & ns-2 & \ding{55} & \ding{55} & \ding{51} \\ [0.1cm]\hline

TCP Hybla~\cite{hybla} & 2004 & loss & ns-2 & \ding{55} & \ding{55} & \ding{51} \\ [0.1cm]\hline

\makecell{HighSpeed\\TCP}~\cite{hstcp} & 2003 & loss & ns-2 & \ding{55} & \ding{55} & \ding{51}\\ [0.1cm]\hline

TCP-LP~\cite{lp} & 2003 & delay & ns-2 & \ding{55} & \ding{55} & \ding{51} \\ [0.1cm]\hline

\makecell{Scalable\\TCP}~\cite{scalable} & 2003 & loss & \ding{55} & \ding{55} & \makecell{DataTAG\\~\cite{datatag}} & \ding{51}\\ [0.1cm]\hline

TCP Veno~\cite{veno}& 2003 & hybrid & \ding{55} & DummyNet & \makecell{Internet\\in China} & \ding{51} \\ [0.1cm]\hline

\makecell{TCP\\Westwood}\cite{westwood} & 2001 & hybrid & ns-2 & \textbf{?} & \ding{55} & \ding{51} \\ [0.1cm]\hline

TCP Vegas~\cite{vegas} & 1994 & delay & $x$-Sim~\cite{x-sim} & \ding{55} & \makecell{Internet\\in the USA} & \ding{55} \\ [0.1cm]
\hline\hline
\end{tabular}}
\end{table}

The inventors of YeAH-TCP and TCP Westwood use experimental testbeds of the same topology: two source hosts sending traffic to one destination host through a single router. 

The link between the router and the destination host is set up to be a bottleneck link using a network link emulator (see the next section) but neither paper gives the name of the emulator so there are question marks in the corresponding rows of the table.   

The last column of the table informs if a scheme in a particular row was tested using a chosen experimental tool in a single-bottleneck bandwidth scenario, which can be modeled with a dumbbell topology. Often, as a principal topology, a standard dumbbell topology is used with one or two central routers connecting the ``halves" of the network: this is true for TCP CDG, TCP Cubic, BIC TCP, and TCP Veno papers and for the modern papers devoted to comparison of performance of different schemes against each other~\cite{turkovic2019fifty, ma2017fairness, zhang2019evaluation, tsao2007taxonomy}. Sometimes a reduced dumbbell topology is used like in the already mentioned cases of YeAH-TCP and TCP Westwood. 

There are also cases like that of TCP-LP paper in which, besides a dumbbell topology, more complicated topologies with multiple bottlenecks are used as well. Regardless of this, it is clear that a dumbbell-like topology is a popular choice among researchers. Apart from providing a shared link with the bottleneck bandwidth not immediately incident to sender nodes, such a topology allows one to choose individual settings for each flow moving through the shared link. In particular, this enables researchers to explore intra- and inter-RTT-fairness of congestion control schemes.

\phantomsection
\label{pr2}

RTT-fairness (also referred to as RTT-unfairness and RTT heterogeneity) shows how fair the flows with different RTTs are towards the flows of the same scheme in case of intra-RTT-fairness and to the flows of other schemes in case of inter-RTT-fairness. Modern papers consider RTT-fairness the important metric by which the congestion control schemes should be compared~\cite{turkovic2019fifty, ma2017fairness, zhang2019evaluation, tsao2007taxonomy}. The papers of the following schemes present in Table~\ref{tab:tools} also evaluate RTT-fairness: Copa, PCC Allegro, TCP Cubic, TCP Illinois, BIC TCP, YeAH-TCP, H-TCP, TCP CDG -- that is, 8 out of 23 schemes.

\section{Network Emulators}
\label{sec:second}

Network emulators can be classified into virtual network emulators, which emulate the entire network adjusting a cluster of nodes according to a topology configuration supplied by the user, and network link emulators, which tune a network interface to have certain parameters like delay, rate, loss, etc., chosen by the user~\cite{nussbaum2009comparative}. Virtual network emulators often reuse network link emulators~\cite{emulab, pantheon, mininet}. 

The history of emulators is quite old. In 1995, in order to assess TCP Vegas, a virtual network emulator, called WAN, consisting of a dozen nodes was used with the interfaces of the nodes being configured using a network link emulator Hitbox~\cite{ahn1995evaluation}.

Table~\ref{tab:tools} shows that Traffic Control (tc) with an appropriate queuing discipline (qdisc)~\cite{qdisc} -- usually NetEm qdisc~\cite{netem} -- and Dummynet~\cite{dummynet} are network link emulators most employed in research testbeds. Dummynet is in FreeBSD kernel, while tc qdisc is in Linux kernel~\cite{nussbaum2009comparative}. Dummynet allows one to control (to shape and schedule) both ingress and egress traffic of an interface, while tc qdisc -- only egress traffic~\cite{nussbaum2009comparative}. There is also a tool, called NISTNet, that allows controlling only ingress traffic but the tool is not maintained anymore and works only for Linux kernels up to version 2.6~\cite{nussbaum2009comparative, NISTNet}. The disadvantage of Dummynet, however, is that, as opposed to tc qdisc and NISTNet, it does not provide delay jitter, packet reordering, packet duplication, and packet corruption~\cite{nussbaum2009comparative, armitage2004some}.

Unlike Dummynet and tc qdisc, Mahimahi~\cite{mahimahi} is a network link emulator operating in user space. It works as a set of nested UNIX shells started by user one inside another. Each shell process is in its individual Linux's network namespace~\cite{linux-ns} and so has a separate network stack. The network namespaces of each pair of the nested shells are connected with a pair of TUN~\cite{tun} virtual network interfaces exchanging raw IP data. The network namespace of the outermost shell is connected to the global namespace of the machine running Mahimahi, i.e. to the outside world, with a pair of TUN devices too. There are several types of shells and they can be nested with or without repetition.  
    
DelayShell introduces link latency by providing two queues in which ingress and egress traffic of the namespace is retained for an amount of time specified by the user. LinkShell also manages two special queues for ingress and egress traffic to emulate a constant-rate or variable-rate link according to a packet-delivery trace file supplied by the user. LossShell drops packets of either ingress or egress traffic with a probability specified by the user. Mahimahi also provides record-and-replay~\cite{mahimahi} feature via RecordShell and ReplayShell. All applications, launched in a shell, run under network conditions emulated by this and outer shells. Mahimahi can be seen in Table~\ref{tab:tools}. 

Network link emulators are single-node and it requires much labor to set them up for a whole testbed of nodes~\cite{emulab}. So there are platforms like Emulab~\cite{emulab} present in Table~\ref{tab:tools} that do it for the user. Emulab is not only a virtual network emulator but also integrates simulation and live testing. The user defines an experiment configuration with ns-2 script and chooses if the experiment is simulated with ns-2, emulated with nodes of one of Emulab testbeds or run on the Internet with geographically-distributed machines of institutions participating in Emulab project~\cite{emulab}. Emulab testbeds use Dummynet and tc qdisc to emulate wide-area network links according to the topology specified by the user~\cite{nussbaum2009comparative}. To spare resources of the testbeds, Emulab implements virtual nodes by extending FreeBSD jails which enables the user to have ten times many nodes as there are physical machines in Emulab~\cite{stoller2008large}. The disadvantage of the platform is the same as of live networks: only faculty and senior staff members can lead a project in Emulab~\cite{emulab}.

As hardware testbeds are shared and expensive, even big projects like Emulab use virtualization. For the same reason, there exist virtual network emulators that allow creating a large network on a single computer. An example of such an emulator is Mininet~\cite{mininet, mininet1} (not present in Table~\ref{tab:tools}), which is also the notable emulator among those very few~\cite{wang2013estinet} supporting OpenFlow protocol and Software-Defined Networking (SDN)~\cite{sdn}. A virtual network topology includes virtual hosts -- UNIX shell processes, each in its own Linux's network namespace -- and software OpenFlow layer 3 switches and controllers. Hosts and switches are interlinked with virtual Ethernet pairs (veth pairs)~\cite{veth}, with each host being connected to only one switch and, thus, having one veth interface. The links may have certain rate, delay, and loss with tc qdisc used under the hood. The user specifies a network topology with a Python script using the Python module \verb+mininet+.     

\section{Pantheon of Congestion Control}
\label{sec:third}

Unlike other tools considered in this chapter, Pantheon (2018)~\cite{pantheon} was created specifically for research on congestion control. Pantheon keeps 17 transport protocols and congestion control schemes. New schemes can be added to Pantheon by other researchers. Pantheon also provides emulation and live testing opportunities through its testing tool.

Pantheon testing tool spans a virtual private network, called Pantheon-tunnel, between sender and receiver nodes. In this point-to-point topology, one or several flows of a single tested scheme are launched all at once or with a fixed time interval and run for a specified runtime of seconds. It can be specified which schemes should be tested (each individually) and how many times they should be tested. For each run of a scheme, sizes and delays of traffic packets are recorded into special log files. Pantheon analysis tool then analyzes the log files and generates individual and aggregate plots and statistics.

Pantheon testing tool has two modes: remote and local. In remote mode, the authors of Pantheon periodically run all the schemes between pairs of real-life nodes located in wired, Wi-Fi, and cellular networks of 9 countries all around the globe. The results of the tests are publicly archived on Pantheon website \url{https://pantheon.stanford.edu}.

In local mode, Pantheon testing tool becomes a virtual network emulator. On a single computer, a virtual point-to-point network is created with the ends of Pantheon-tunnel lying in different Linux's network namespaces and, thus, representing the two virtual nodes. A set of nested Mahimahi shells specified by the user is applied to one end of the raw IP link connecting the virtual nodes, in order to emulate certain network conditions like delay, constant or variable rate, queue size, loss, etc.

The drawback of Pantheon-tunnel is that it attaches a unique identifier (UID) to each packet and encapsulates the result into a UDP datagram. This makes all traffic UDP and gives 36 bytes of per-packet overhead so that three schemes (PCC, Verus, Sprout) had to be patched to reduce the Maximum Transmission Unit (MTU)~\cite{kurose} hardcoded in their source code. However, the UID allows tracking a packet between sender and receiver nodes to measure its one-way delay. Also, in remote mode, Pantheon-tunnel enables to initiate flows between two nodes even if one of the nodes is behind a NAT~\cite{kurose}.   

Researchers often struggle to build a new testbed and older congestion control schemes to evaluate their new scheme against those older schemes. Pantheon is a solid attempt to solve the problem. Therefore, this thesis uses Pantheon as storage of schemes and as a point of reference when building a new emulator resolving such shortcomings of Pantheon as the impossibility to run flows of different schemes together or to have a dumbbell topology. The new emulator, CoCo-Beholder, is presented in the next chapter.

\loadchapter{tool}{CoCo-Beholder}
\vspace*{-0.5cm}
CoCo-Beholder is a virtual network emulator implemented by the author of the thesis. The emulator consists of the three main tools: the testing tool, the analysis tool, and the plotting tool. There is also the auxiliary cleaning tool that allows one to clean output files and directories in a flexible way. The help messages of all the tools are present in Appendix~\ref{appendix:AppendixA}. The repository of CoCo-Beholder can be downloaded by the\nolinebreak[4] URL \url{https://www.mpi-inf.mpg.de/inet/software/coco-beholder}.\footnote[2]{The alternative URL is \url{https://github.com/ZhenyaKh/coco-beholder}.}

CoCo-Beholder is written in Python and is checked to work with Python versions 2.7, 3.5, 3.6, and 3.7. The emulator is developed for Linux operating system (the testing tool uses Linux command-line utilities extensively) and is ensured to work on Ubuntu 16.04 LTS, Ubuntu 18.04 LTS, and Debian 10. CoCo-Beholder repository includes the installation script and the comprehensive README file with step-by-step instructions on how to install CoCo-Beholder and Pantheon collection of congestion control schemes for each of the three Linux distributions. Chapter~\ref{chap:testing} gives more details on the matter.

As discussed further in the chapter, CoCo-Beholder utilizes some parts of Mininet library. The library provides three kinds of installation: the pre-built Virtual Machine image, the Ubuntu package, and the installation from the source code meant for Ubuntu, Debian, and Fedora Linux distributions. The first variant is too cumbersome. The second variant confines to a single Linux distribution. The third variant is intrusive, may damage the operating system of the user, and supplies no deinstallation script -- the developers of Mininet admit the problem themselves~\cite{faq}. Therefore, the author of the thesis included the needed parts of Mininet 2.3.0d5 into CoCo-Beholder repository as a third-party library in accord with the license of Mininet. Not only does this solve the installation problem, but also any possible compatibility issues that could arise with the future versions of Mininet. The included Mininet assets are the license, a dozen Python scripts (slightly modified), and one file in C programming language. The installation script of CoCo-Beholder compiles the C file in advance.

Sections~\ref{sec:testing_tool},~\ref{sec:analysis_tool}, and~\ref{sec:plotting_tool} of this chapter are devoted to the testing, analysis, and plotting tools of CoCo-Beholder correspondingly. Section~\ref{sec:comparison} provides the comparison of CoCo-Beholder to Pantheon.\\

\section{CoCo-Beholder Testing Tool}
\label{sec:testing_tool}

\begin{sidewaysfigure}[p]
\centering
\includegraphics[trim=0 -10mm 0 0, scale=0.7]{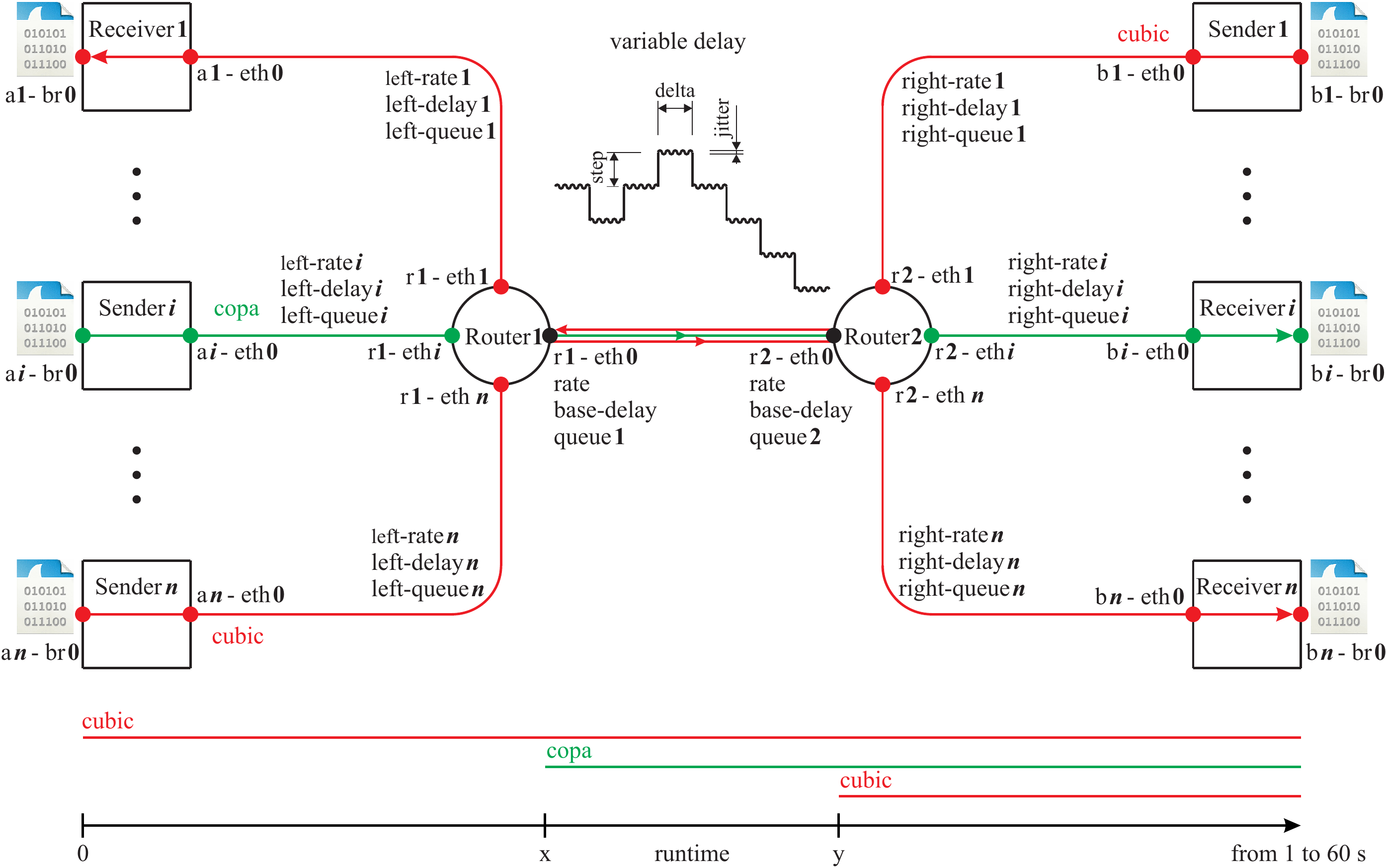}
\caption{The plan of CoCo-Beholder, done by the author of the thesis.}
\label{fig:plan}
\end{sidewaysfigure}

The testing tool enables to run flows of optionally different congestion control schemes in the dumbbell topology during a chosen number of runtime seconds (see the plan in Figure~\ref{fig:plan}). Each flow has a host in the left half and a host in the right half of the topology and the hosts exchange the traffic of a scheme, with one host being the sender and one being the receiver. There is the left router that interconnects all the hosts in the left half and the right router that interconnects all the hosts in the right half of the topology. All the flows share the common central link between the two routers.

The user can specify how many flows of which schemes should be run by defining groups of flows in a layout file having the YAML~\cite{yaml} format, with one entry of the file corresponding to one group. An example of such an entry can be seen in Listing~\ref{list:layout}.
\\

\begin{minipage}{\linewidth}
\begin{lstlisting}[caption=An example entry of the layout file., label=list:layout,language=yaml]
- direction: ->
  flows: 2
  left-delay: 0us
  left-queues: 2000
  left-rate: 100.0
  right-delay: 5ms
  right-queues: 3000
  right-rate: 100
  scheme: vegas
  start: 5
\end{lstlisting}
\end{minipage}

A group of flows is defined by ten properties. Four of the properties are necessary: the scheme name, the number of flows, the second of runtime at which the group of flows should be started, and the direction of the flows (leftward or rightward). The rest of the properties are optional and define the delay, rate, and queue size installed at the interfaces on the ends of the links, belonging to the flows, in the left half and in the right half of the dumbbell topology using tc qdisc NetEm (see Section~\ref{sec:second}) network link emulator. It must be reminded here that tc qdisc is applicable only to egress traffic.

The rate is a float or integer value in Mbit/s. The delay is a float or integer number of nanoseconds, milliseconds or seconds (if no unit is specified, the milliseconds are assumed). The queue size is measured in the number of packets, and queue management is a simple tail-drop~\cite{rfc2309}. If a rate or delay property is lacking or null in the\nolinebreak[4] layout file, it is set to be zero, which, for tc qdisc NetEm, means leaving the parameter unset. \mbox{If a queue size property is lacking or null in the layout file, it is set to 1000.}

Plan~\ref{fig:plan} shows an example setup generated by CoCo-Beholder for $n$ flows. Among the flows, one can observe: a leftward TCP Cubic flow started at the very beginning of the runtime, a rightward Copa flow started at the $x^{\text{th}}$ second of the runtime, and a rightward TCP Cubic flow started at the $y^{\text{th}}$ second of the runtime. It can be noticed that the side links of the dumbbell topology, along which the flows run, all have individual delay, rate, and queue size values. In particular, for any single flow, the values differ for the links in the left and right halves of the topology. For the sake of brevity, the delay, rate, and queue size labels in the plan are placed in the middle points of the side links of the topology, while in reality the parameters are installed at both the ends of each link, so that, in particular, the RTT of a link would be twice the (one-way) delay.

The dumbbell topology is built using the lower-level parts of Mininet (see Section~\ref{sec:second}) library. Using Mininet out-of-the-box, that is, utilizing its higher-level entities like Switch, Controller, Topology, etc. was impossible. First, CoCo-Beholder does not need OpenFlow protocol: as it can be seen in Figure~\ref{fig:plan}, there are no OpenFlow layer 3 switches or controllers in the plan. Secondly, Mininet does not ordinarily provide neither routers nor regular hosts having more than one veth interface (any Linux host can be turned into a Linux router), while Figure~\ref{fig:plan} indicates that each of $2n$ hosts should have one bridge~\cite{bridge} interface and one veth interface attached to the bridge, and each of the  two routers should have $(n + 1)$ veth interfaces. Thirdly, at the time when CoCo-Beholder was being implemented, there was a bug~\cite{bug} in Mininet that prevented the usage of its TCLink class, whose instances are links with an off-the-shelf tc qdisc configuration. 

Because of all these difficulties, it was deemed reasonable to reuse only the lower-level parts of Mininet: those providing a virtual host as a UNIX shell process in a separate Linux's network namespace, a virtual link as a veth pair connecting a couple of the virtual hosts, and the application programming interface that allows launching processes at the virtual hosts conveniently. Taking the advantage of these essentials, the author of the thesis implemented the dumbbell topology of the required configuration with the help of the following Linux command-line utilities: \verb+tc+, \verb+sysctl+, \verb+ip+, \verb+arp+, \verb+route+, \verb+ethtool+, \verb+tcpdump+, \verb+ifconfig+, \verb+lsof+. The installation script of CoCo-Beholder installs all the necessary utilities into the operating system of the user in advance, and the testing\nolinebreak[4] tool checks their presence before running any test. Exit codes of the tasks performed by the utilities are checked to ensure the configuration consistency.

The current solution has IPv6~\cite{ipv6} turned off at all the virtual hosts. Thus, the dumbbell topology consists of $(2n + 1)$ IPv4~\cite{rfc791} subnetworks with the network prefix length 30 that accommodates four network addresses, one pair of which are reserved for the network address and the broadcast address and the other pair is assigned to the interfaces at the ends of the corresponding link of the topology. ECN~\cite{rfc3168} is not supported by the network. TCP segmentation offload~\cite{siemon2013queueing} and UDP fragmentation offload~\cite{siemon2013queueing} are turned off for all the interfaces with \verb+ethtool+ because, otherwise, Ethernet frames of the scheme traffic with the payload size much greater than the MTU~\cite{kurose} (1500 bytes) were observed in the network. The ARP caches and the routing tables of all the nodes in the topology are filled in beforehand in a static way to eliminate service packets inflight.

From Figure~\ref{fig:plan}, it can be observed that the interfaces of the central link of the dumbbell topology also can be configured with tc qdisc NetEm. The parameters should be supplied by the user as command-line arguments. The rate of the central link can be chosen explicitly or, if left unspecified, it is set to 100 Mbit/s. Each queue of the two interfaces on the ends of the central link can have an individual size. The variable delay at the link is set up by specifying the initial (base) delay, the delta defining the time periodicity of the delay change, and the step by which the delay is increased or decreased each delta time. Whether the increase or decrease takes place at a certain point is defined by the pseudo-random number generator, the seed for which either can be provided by the user as a command-line argument or is assigned the current UNIX time~\cite{matthew2008beginning}. To switch off the variability feature, one can choose the delta greater than the runtime command-line argument. The jitter affecting the delay can be added with an optional argument too.

In Figure~\ref{fig:plan}, one can see an icon of a PCAP~\cite{tcpdump} dump file near each host in the dumbbell topology. The icons mean that an individual \verb+tcpdump+~\cite{tcpdump} process captures network traffic at the bridge interface of each host during the testing runtime. The necessity of the bridge interfaces at the hosts is explained in the next section. With a filter expression, the capture is limited to TCP and UDP packets having IP addresses of the sender and receiver hosts.  In case of a flow with a high rate, the Linux kernel can drop a certain number of packets, in which situation CoCo-Beholder gives the user an informative warning message. To solve the issue, the user has the opportunity to increase the operating system capture buffer with a special command-line argument (the default value is 2 MiB), which is then passed to each \verb+tcpdump+ instance started. All the PCAP dumps are written into a chosen output directory, their file names having the format \verb+<flow's starting #>-<scheme>-<sender/receiver>.pcap+.

The tested congestion control schemes are taken from Pantheon (see Section~\ref{sec:third}) Github repository~\cite{pantheongit}. The repository includes the YAML configuration file with the list of the schemes kept in Pantheon, the directory with the submodule references to the implementations of the schemes, and the directory with the wrapper scripts of the schemes. All the wrapper scripts provide the same interface that allows one to get the dependencies of a scheme, to build a scheme, to set up a scheme (after the reboot), to start the sender of a scheme, to start the receiver of a scheme, and to learn if the receiver or the sender of a scheme is the server and so should be started first. The README file of CoCo-Beholder instructs to download the repository and build the schemes \emph{not} applying the patches that reduce the MTU (the MTU problem of Pantheon was described in Section~\ref{sec:third}). Schemes not present in Pantheon can be added to the collection by the user locally or globally, as explained in Section~\ref{sec:tfirst}. When launching CoCo-Beholder testing tool, the path to Pantheon repository is provided by the user via a command-line argument.

The whole operation of CoCo-Beholder testing tool looks as following. The user gives the layout file and the command-line arguments to the executable script of the tool. For the future reproducibility, all the relevant testing parameters are processed and saved to the output directory in a special JSON~\cite{rfc7159} metadata file \verb+metadata.json+. The testing module then reads the parameters and sorts the flows by the start second indicated in the layout file. The schemes demanded in the layout file are set up via their wrapper scripts. The dumbbell topology of the required size is built, and its links are configured using tc qdisc NetEm. The recording of network traffic with \verb+tcpdump+ is started at all the virtual hosts (the PCAP dumps being saved next to the metadata file). The server of each flow is launched using the wrapper script at a host in the left or in the right half of the topology depending on the direction of the flow. At this point, the auxiliary Python thread is launched. The thread is provided by Python \verb+threading+ module and is pseudo-concurrent, that is, it does not benefit from the multiple cores of a CPU~\cite{van1995python}.

The auxiliary thread cyclically sleeps and wakes up each second to launch the clients of those flows that are scheduled to be started at the second, if any, using the wrapper scripts of the corresponding schemes. To improve the precision, the start of a batch of the flows is performed concurrently on the multiple cores of a CPU, if available, by means of Python \verb+multiprocessing+ module~\cite{van1995python}. The auxiliary thread stops its execution after the last batch of the flows is started. The main thread is synchronized with the auxiliary thread and waits asleep until the latter launches the flows, scheduled for the start at the $0^{\text{th}}$ second of the runtime. After that, the main thread cyclically sleeps and wakes up each delta time to change the delay at the interfaces on the ends of the central link of the topology by the step using tc qdisc NetEm until the runtime ends, in order to emulate the variable delay. Finally, the main thread joins the auxiliary thread, stops the \verb+tcpdump+ recordings, kills \emph{all} the descendant processes, and removes zombie processes.

Additional steps adopted to enhance the accuracy of the variable delay (e.g., the pre-computation of the delays for all the delta time intervals) are discussed in \mbox{Section}~\ref{sec:tthird}.

In Listing~\ref{list:testoutput}, there can be seen an example output of CoCo-Beholder testing tool. The runtime is 10 seconds. The rate at the central link is 120 Mbit/s, the base delay is 30 ms, the delta is 500 ms, the step is 10 ms, and the jitter is 5 ms.  The default path to the layout file -- \verb+layout.yml+ -- was used, as no other was provided by the command-line argument. In the example, no file with the path \verb+layout.yml+ existed, so it was generated with the default contents: two TCP Cubic flows starting at the $0^{\text{th}}$ second and two TCP Vegas flows starting in half the runtime, i.e. at the $5^{\text{th}}$ second (other properties of the layout are not considered here). The two benchmarks in Listing~\ref{list:testoutput} show the execution time of the cycles performed by the auxiliary and main threads correspondingly.\\

\begin{minipage}{\linewidth}
\begin{lstlisting}[caption=An example output of CoCo-Beholder testing tool.,label=list:testoutput,mathescape]  
$\text{\textbf{\textcolor{teal}{\$} ./run.py \textcolor{darkblue}{-p} \(\sim\)/pantheon 30ms 0.5s 10 5000us \textcolor{darkblue}{-t} 10 \textcolor{darkblue}{-r} 120}}$
Script not started as root. Running sudo...
Testing:
Total number of flows is 4
Flows have been sorted by their start
Creating the dumbbell topology...
Calling setup_after_reboot on wrappers of the schemes...
Setting rates, delays and queue sizes at the topology's interfaces...
Starting tcpdump recordings at hosts...
Starting servers...
Starting clients and optionally varying delay...
debug benchmark 1: $\text{\textbf{\textcolor{blue}{5.005561}}}$
debug benchmark 2: $\text{\textbf{\textcolor{blue}{10.001110}}}$
Killing descendent processes properly...
SUCCESS
Done.
\end{lstlisting}
\end{minipage}

\section{CoCo-Beholder Analysis Tool}
\label{sec:analysis_tool}

The analysis tool is very straightforward and has only two command-line arguments: the paths to the input and output directories. The input directory is expected to contain the PCAP dumps (two per flow) and the metadata file generated by the testing tool. The analysis tool extracts data from each PCAP dump into a log file in a special format. The log files are written and the metadata file is copied into the output directory. The directory with the PCAP dumps is not needed anymore for the future plotting and statistics generation. 

The plotting tool can be run many times with different arguments over the log data files to get different types of plots and statistics without the necessity to analyze the possibly heavy PCAP dumps again and again. The biggest PCAP dump that the author of the thesis managed to get on the machines, on which CoCo-Beholder was tested, was 12 GB in size. The size of the resulting log data file for such a dump is around 300 MB. This and the optimal log format gives a tremendous speed boost for the forthcoming plotting. 

For packet parsing, the following Python libraries were considered: \verb+dpkt+~\cite{dpkt}, \verb+scapy+~\cite{scapy}, and \verb+pyshark+~\cite{pyshark}. To make the right choice, the author of the thesis tested the variants, and \verb+dpkt+ showed 15 times better performance over \verb+scapy+ and 40 times -- over \verb+pyshark+. Therefore, CoCo-Beholder handles PCAP dumps with \verb+dpkt+. 

Both on the Ubuntu and Debian machines (their detailed characteristics can be found in Chapter~\ref{chap:testing}), the analysis is about 1.3 times faster when executed with Python 2, rather than with Python 3. As an exception, the execution on the Debian machine with Python 3.7 has the speed comparable to that with Python 2.7. Processing of a 12 GB PCAP dump takes around 4 minutes on the Ubuntu machine and around 5 minutes on the Debian machine, regardless of the latter having much more processor cores and memory. This is explained by the fact that the analysis is single-threaded and utilizes the reasonable amount of memory available on both the machines, while the processor frequency of the Ubuntu machine is higher than that of the Debian machine.

Each log data file has a name in the format \verb+data-<flow's starting #>.log+. Each log data file is in pure JSON and contains the data extracted from the corresponding PCAP dumps recorded at the sender and receiver hosts. The log format is present in Listing~\ref{list:contents}.\\

\begin{minipage}{\linewidth}
\begin{lstlisting}[caption=The format of a log data file with $N$ packets.,label=list:contents,mathescape]  
[arrival$_1$, arrival$_N$]          
[bytes_lost, bytes_sent] 
[arrival$_1$, arrival$_2$, ..., arrival$_N$] 
[delay$_1$, delay$_2$, ..., delay$_N$]
[size$_1$, size$_2$, ..., size$_N$]
\end{lstlisting}
\end{minipage}

When parsing both the sender and receiver dumps, the main focus is on the packets directed from the sender to the receiver host. Thus, every following packet of a dump is taken into consideration if its source IP address is that of the sender host. However, in the dumps, two IP addresses are always present. To find out which of the two belongs to the sender host, the flow direction is learned from the metadata file. The left half of the dumbbell topology~\ref{fig:plan} is built before the right one, so the hosts in it have lower IP addresses. Thus, if the flow is rightward, the sender IP address is the lower one.

Any packet in a PCAP dump has a timestamp that is the UNIX time at which the packet was captured. The minimum timestamp of the first packets of all the $2n$ dumps recorded during the testing phase is chosen as a base time ($n$ is the number of the flows). The base time is subtracted from the timestamp of any processed packet in any dump. This ensures that the duration of the entire recorded network traffic starts at zero.

For each flow, the sender dump is processed first. The dictionary with the departure timestamps of all the packets sent by the sender host is created, with the keys being the digests of the packets. The digest of a packet is calculated as SHA-1 hash~\cite{rfc3174} over IP Identification field~\cite{rfc6864} and IP payload of the packet. The digest generation is a simplified version of IPv4 Hash-Based Duplicate Packet Detection technique~\cite{rfc6621}. If a collision happens for a pair of packets, the error is reported to the user. The author of the thesis has never encountered any collisions even for huge dumps for the 29 schemes considered in Section~\ref{sec:tfirst}. However, if a certain scheme generated many packets with identical payload, a collision could happen because, on Linux, the IP Identification values periodically repeat for packets with a given source/destination/protocol tuple~\cite{rfc6864}.

After that, the receiver dump is processed. The digest of each packet received from the sender host is calculated. If the digest is found in the dictionary, the follow\nolinebreak[4]ing\nolinebreak[4] \mbox{information} about the $i^{\text{th}}$ packet will be written to the log data file, as it is shown in Listing~\ref{list:contents}: the arrival timestamp \verb+arrival+$_{\verb+i+}$, the size \verb+size+$_{\verb+i+}$, and the delay \verb+delay+$_{\verb+i+}$ computed as the difference between the arrival and departure timestamps. To spare the memory, the found entry \textlangle{}digest, departure\textrangle{} is removed from the dictionary. The timestamps of the first and the last packet arrivals are additionally present in the first line of any log data file, as appears in the listing, to allow learning the flow duration from the file quickly.

If the digest is not found in the dictionary, then it means that the packet was sent by the sender host but not recorded at it. It happens to quite a small number of packets, but the more the traffic rate is, the more such ``phantom" packets appear. These packets together with the packets, sent by the sender host and found in the sender dump, comprise all the packets sent by the sender host, and their total size in bytes is denoted by \verb+bytes_sent+ in Listing~\ref{list:contents}. The number of lost packets is equal to the final size of the dictionary after all the removals, that is, to the number of packets sent by the sender host and found only in the sender dump. Their total size is denoted by \verb+bytes_lost+.

The loss in percents is computed as follows: \verb+float(bytes_lost) / bytes_sent * 100+. When the receiver dump is empty, the loss is 100\%. When the sender dump is empty, the loss is 0\%. When both the dumps are empty, the loss calculation is skipped, as \verb+bytes_sent+ is zero and one cannot divide by it. The number of the lost and sent bytes is saved to the log data file, rather than the loss in percents itself, to later allow calculating the aggregate loss for several flows together.

Listing~\ref{list:empty} shows how a log data file looks when no packet among the packets sent by the sender host was found in both the sender and receiver dumps at once. This includes the edge cases with one or both the dumps being empty.\\

\begin{minipage}{\linewidth}
\begin{lstlisting}[caption=The format of a log data file with no packets.,label=list:empty,mathescape]  
[null, null]          
[bytes_lost, bytes_sent] 
[ ] 
[ ]
[ ]
\end{lstlisting}
\end{minipage}

As explained above, the one-way delay is computed for each packet that made its way from the sender to the receiver. The possibility of deriving the correct delay value from the pair of dumps is provided by the bridge interfaces present at each virtual host. For each host, the network traffic is sent and received using the bridge interface to which the host IP address is assigned. The network traffic is also recorded with \verb+tcpdump+ at the bridge interface. On each host, the regular veth interface, configured with tc qdisc NetEm according to the layout file, is attached to the bridge interface, so that all the traffic, sent or received by the bridge interface, goes through the veth interface and gets affected by the installed NetEm parameters. The whole setup can be seen in Figure~\ref{fig:plan}. 

If a host had only veth interface $X$ configured with NetEm parameters $Y$ and interface $X$ was used both for the sending and recording, then any latency introduced by parameters $Y$ would not be reflected in the timestamps of the egress packets recorded at $X$. This is because, in a host, \verb+tcpdump+ captures an outgoing packet \emph{after} the queuing discipline is applied to the packet. This can be observed in the lower right corner of the plan~\cite{flow} that shows the general packet flow within the network stack of a Linux host.

The author of the thesis came up with the setup~\ref{fig:plan} utilizing the Linux bridge and has not encountered such a solution to the described issue in any research works.

\begin{figure}[h]
\centering
\includegraphics[width=\textwidth]{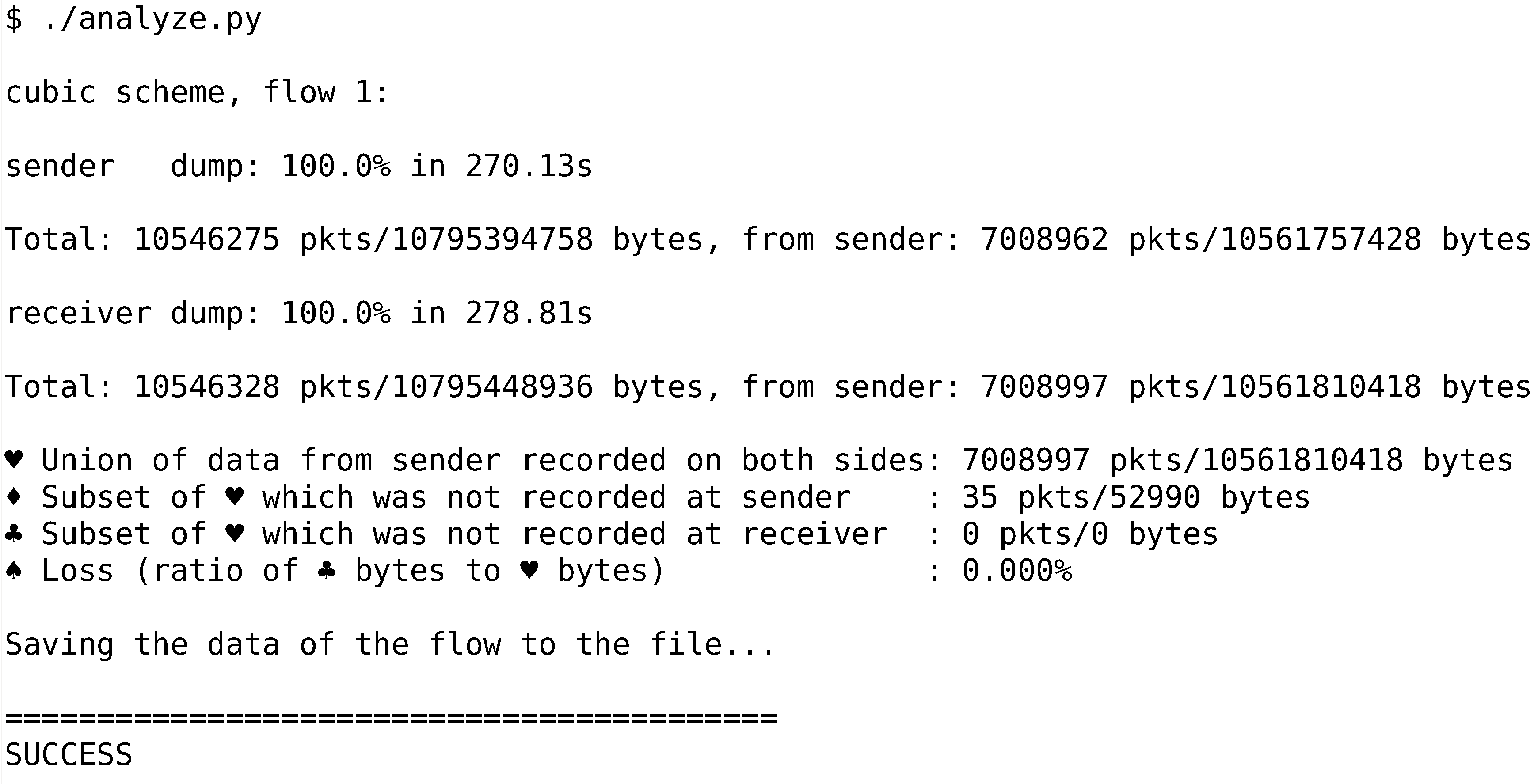}
\caption{An example output of CoCo-Beholder analysis tool.}
\label{fig:analysis_output}
\end{figure}

Figure~\ref{fig:analysis_output} demonstrates the output of the analysis tool for one TCP Cubic flow running for 60 seconds without any delay or rate limitation set up at any links in the topology. For each of the pair of dumps, the total number of packets/bytes and the number of packets/bytes directed from the sender to the receiver host are printed. The numbers of bytes denoted by \ETX and \ENQ are also written to the resulting log data file \verb+data-1.log+ in place of \verb+bytes_sent+ and \verb+bytes_lost+ correspondingly. The number of packets denoted by \EOT is the number of the above-mentioned ``phantom" packets.

The example output is for one flow. If there are more flows, the analogous analysis output for each flow is printed one after another, separated visibly.

The analysis of a dump can take several minutes like in the example output above, where two 11 GB dumps were processed. Hence, the author of the thesis implemented the lightweight progress bar. It is not actually a bar, as it only redraws the completion percentage and the passed time (e.g., \verb+78 % in 5 s+). The redrawing takes place only every second. In Figure~\ref{fig:analysis_output}, each of the two progress ``bars" is in the final state indicating how long it took to process the dump. Real progress bars like one from Python module \verb+progress+ are unsatisfactory because they make the analysis two times slower.

\section{CoCo-Beholder Plotting Tool}
\label{sec:plotting_tool}

CoCo-Beholder plotting tool generates plots and statistics over the log data files of the flows. The type of the plots and statistics can be specified by the user with a command-line argument: per-flow, total, and per-subset. Depending on the chosen type, the flows are divided into what here is referred to as ``curves" and the plots and statistics are actually created for the resulting curves, with the data of the flows in a curve merged. A curve's start and end (i.e., the curve duration) are the minimum first arrival and the maximum last arrival of its constituent flows.

The per-flow type means that for each flow, a line plot has a separate curve, a scatter plot has a separate collection of points, and a statistics file has a separate entry. The total type implies that for all the flows together, a line plot has exactly one curve, a scatter plot has exactly one collection of points, and a statistics file has exactly one entry. The per-subset type means that for each subset, a line plot has a separate curve, a scatter plot has a separate collection of points, and a statistics file has a separate entry.

Flows are in one subset iff they have the same values of the chosen layout file properties, supplied by the user in the format \verb+"property+$_{\verb+1+}$\verb+ property+$_{\verb+2+}$ \verb+... property+$_{\verb+k+}$\verb+"+ with a command-line argument. E.g., for \verb+"direction scheme"+, a curve includes flows having both the same direction and scheme name. Currently, a subset can be defined using the following properties of the layout file: \verb+scheme+, \verb+direction+. But CoCo-Beholder plotting tool is implemented in such a way that the list of the supported properties can be easily extended with minor changes to the source code.

For a chosen plot type, line plots are created for the following measures: average rate (i.e., average throughput), average fairness, average one-way delay. Also, a scatter plot is created for per-packet one-way delay. The line plots are averaged per an aggregation time interval specified by the user with a command-line argument (0.5 seconds by default). The runtime duration, which lasts from zero to the maximum end of all the curves, is split into the intervals, and the packets of each curve, that is, of the curve's flows, get distributed across the slots according to their arrival timestamps. To get the average rate of a certain interval, the sizes of all the packets in the corresponding slot are summed up and the sum is divided by the interval. To get the average delay in the interval, the arithmetic mean of the delays of all the packets in the slot is computed. 

As opposed to the average rate and delay, the average fairness is computed over the curves, rather than over the flows, and, hence, is the only measure, the plot of which has exactly one curve, or rather a ``meta-curve", regardless of the chosen plot type. The value of the ``meta-curve" in a certain interval, is computed over the average rates of the $m$ curves in the interval using Jain's fairness index formula~\ref{eq:jain}~\cite{jain1984quantitative}. A curve is involved in the computation only if its duration encompasses the interval, i.e., the interval lies between the curve's start and end.
\begin{equation}
\label{eq:jain}
\mathcal{J} (x_1, x_2, \dots, x_m) = \frac{( \sum_{i=1}^m x_i )^2}{m \cdot \sum_{i=1}^m {x_i}^2} \in [\frac{1}{m}, 1]
\end{equation}

A scatter per-packet one-way delay plot contains a point \textlangle{}arrival, delay\textrangle{} for each packet in a curve. An average one-way delay plot with a smaller aggregation interval, e.g. 0.01 seconds, can look very close to the corresponding per-packet one-way delay plot.

Exactly one statistics file of a chosen type is generated. The file is independent of the chosen aggregation interval. It has two sections, and each section has an entry per curve. The first section is devoted to average and loss statistics. In the section, the entry of each curve contains the curve's overall average rate over the curve's whole duration in Mbit/s, the average one-way delay for all the curve's packets in milliseconds, and the curve's overall loss in percents. Also, the section contains the overall Jain's index value for all the curves together, computed over the average rate statistics of the curves.

The second section is devoted to per-packet statistics. The entry of each curve contains the median, average, and $95^{\text{th}}$ percentile per-packet one-way delays over all the curve's packets in milliseconds. The overall average one-way delay values in the two sections are actually the same values computed in different ways.

For a chosen type, five files are created in a chosen output directory named in the format: \verb+<type>-avg-rate.png+, \verb+<type>-avg-jain.png+, \verb+<type>-avg-delay.png+, \verb+<type>-ppt-+
\verb+delay.png+, \verb+<type>-stats.log+. Depending on the type, \verb+<type>+ can be \verb+per-flow+, \verb+total+, \verb+per-scheme+, \verb+per-direction+, \verb+per-scheme-direction+, etc.

The average plots and statistics generation does not consume much memory, as, at maximum, it holds the arrivals, delays, and sizes of the packets of only one flow in the memory at once. On the contrary, the per-packet plots and statistics generation is memory consuming because it holds the arrivals and delays of the packets of one whole curve at once, and the curve may consist of all the flows. For that reason, the average plots and statistics generation is performed first to be completed regardless of any further problems with the memory shortage. 

Both for Python 2 and Python 3, the plots and statistics generation over the data, extracted by the analysis tool from the two 12 GB dumps, takes around 20 seconds, 11 of them spent on the per-packet plots and statistics. Python 3, however, releases memory more effectively than Python 2~\cite{pythonmemory}.

Edge cases are processed properly and informatively for the user: when there are no packets in a curve or the curve duration is less than 5 ms, when the loss of a curve cannot be computed, when a curve contains packets with coinciding timestamps, etc.

The plotting tool also allows choosing custom colors for the curves in the plots.

\begin{figure}[h!]
\centering
\includegraphics[width=\textwidth]{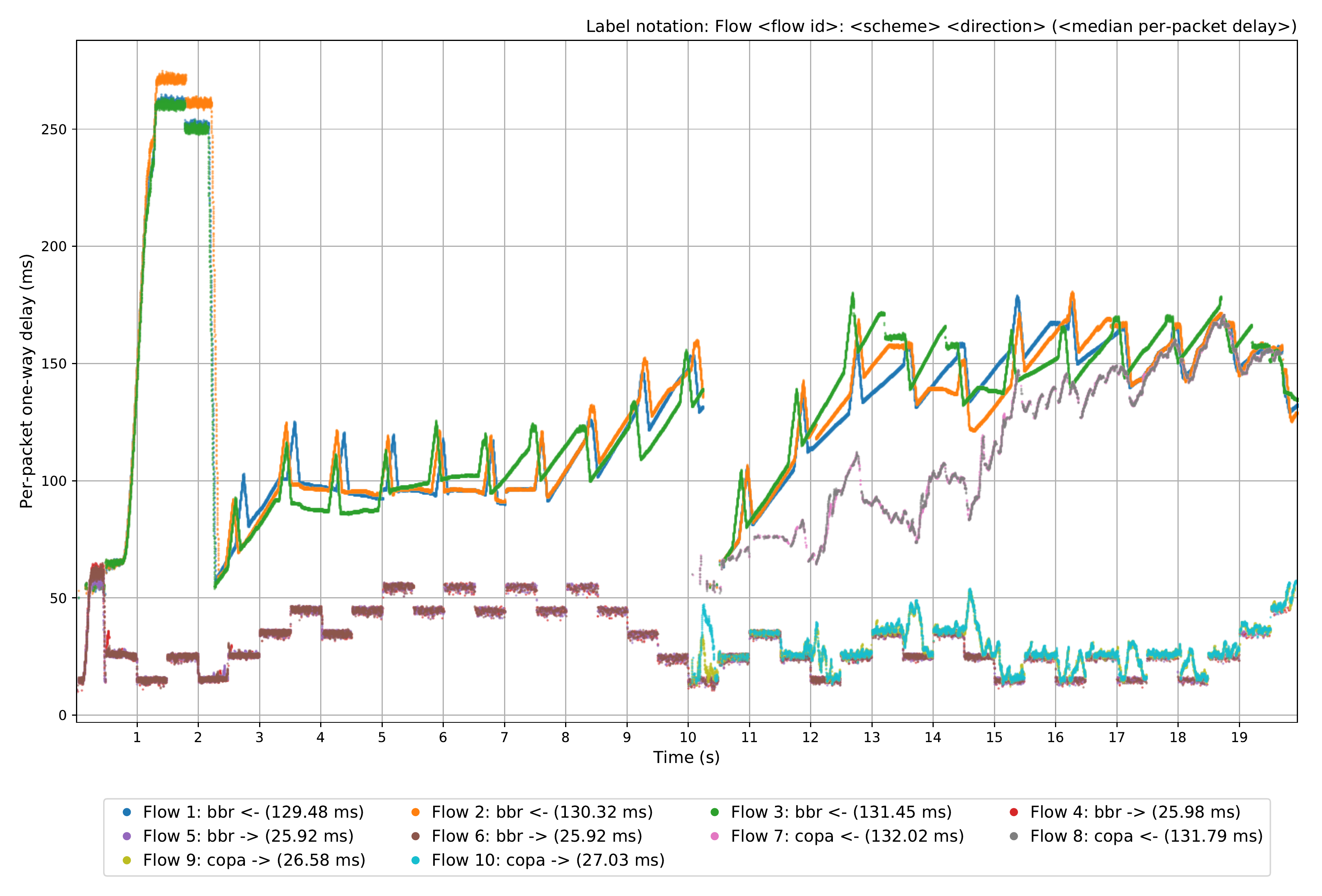}
\caption{Per-flow per-packet one-way delay plot of the example in Appendix~\ref{appendix:AppendixB}.}
\label{fig:pptdelayappendixb}
\end{figure}

\begin{figure}[h!]
\centering
\includegraphics[width=\textwidth]{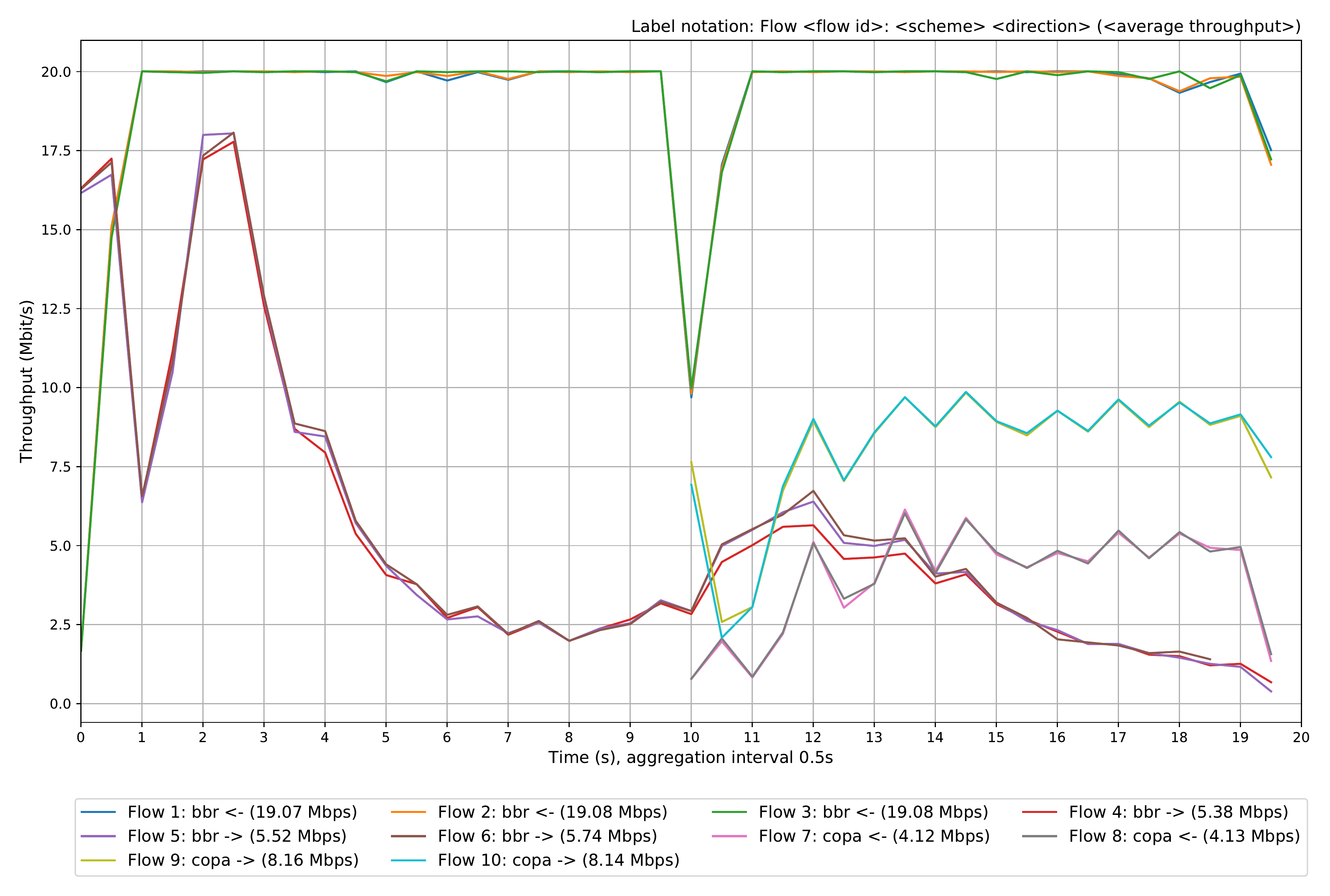}
\caption{Per-flow average throughput plot of the example in Appendix~\ref{appendix:AppendixB}.}
\label{fig:avgrateappendixb}
\end{figure}

In Appendix~\ref{appendix:AppendixB}, one can see all the per-flow, total, per-scheme, per-direction, per-scheme-and-direction plots and statistics generated with various aggregation intervals for the testing setup in Listing~\ref{list:example_appendix_b}. Also, the layout file, the command with which the testing tool was run, and the example output of the plotting tool can be seen there. Here, only the per-flow average rate plot~\ref{fig:avgrateappendixb} and the per-packet one-way delay plot~\ref{fig:pptdelayappendixb} are presented. Though by default the plotting tool outputs the plots in a raster-graphics format, all the plots in this thesis were purposely output in a vector-graphics format.

In the average rate plot~\ref{fig:avgrateappendixb}, one can observe the aggregation interval indicated. In both the plots, there is the label notation at the top-right corner explaining the meaning of the labels of the curves. In particular, each label includes a suitable statistic of the curve in the parentheses like e.g. \verb+Flow 4: bbr -> (5.38 Mbps)+. This statistic is the same that is written to the above-described statistics file.  

In the per-packet delay plot~\ref{fig:pptdelayappendixb}, the variable delay with the delta 500 ms, step 10 ms, and jitter 5 ms is indeed noticeable. As expected, the initial delay is 50 ms for the three leftward BBR flows and 10 ms (5 ms + 5 ms) for the three rightward BBR flows. 

\definecolor{c1}{HTML}{1F77B4}
\definecolor{c2}{HTML}{FF7F0E}
\definecolor{c3}{HTML}{2CA02C}
\definecolor{c4}{HTML}{D62728}
\definecolor{c5}{HTML}{9467BD}
\definecolor{c6}{HTML}{8C564B}
\definecolor{c7}{HTML}{E377C2}
\definecolor{c8}{HTML}{7F7F7F}
\definecolor{c9}{HTML}{BCBD22}
\definecolor{c10}{HTML}{17BECF}

\begin{minipage}{\linewidth}
\begin{lstlisting}[caption=The testing setup of the example in Appendix~\ref{appendix:AppendixB}.,label=list:example_appendix_b,mathescape,basicstyle=\linespread{0.9}\ttfamily\small,mathescape,frame=b] 
                           ----   0-TH SEC:   ----
    <-------$\text{\textbf{20Mbps}}$---------|  |               |  |<----$\text{\textbf{50ms}}$------------$\text{\textcolor{c1}{\textbf{bbr}}}$
    <-------$\text{\textbf{20Mbps}}$---------|  |               |  |<----$\text{\textbf{50ms}}$------------$\text{\textcolor{c2}{\textbf{bbr}}}$
    <-------$\text{\textbf{20Mbps}}$---------|  |               |  |<----$\text{\textbf{50ms}}$------------$\text{\textcolor{c3}{\textbf{bbr}}}$
                           |  |     $\text{\textbf{500ms}}$     |  |
    $\text{\textcolor{c4}{\textbf{bbr}}}$-----$\text{\textbf{20Mbps,5ms}}$---->|  |     ^^^^^$\text{\textbf{5ms}}$  |  |-----$\text{\textbf{5ms}}$--------------->
    $\text{\textcolor{c5}{\textbf{bbr}}}$-----$\text{\textbf{20Mbps,5ms}}$---->|  | $\text{\textbf{10ms}}$|   |     |  |-----$\text{\textbf{5ms}}$--------------->
    $\text{\textcolor{c6}{\textbf{bbr}}}$-----$\text{\textbf{20Mbps,5ms}}$---->|  |---$\text{\textbf{70Mbps,0ms}}$--|  |-----$\text{\textbf{5ms}}$--------------->
                           |  |               |  |
                           |  |   10-TH SEC:  |  |
    <-------$\text{\textbf{50ms}}$-----------|  |               |  |<----$\text{\textbf{10Mbps}}$---------$\text{\textcolor{c7}{\textbf{copa}}}$
    <-------$\text{\textbf{50ms}}$-----------|  |               |  |<----$\text{\textbf{10Mbps}}$---------$\text{\textcolor{c8}{\textbf{copa}}}$
                           |  |               |  |
    $\text{\textcolor{c9}{\textbf{copa}}}$----$\text{\textbf{5ms}}$----------->|  |               |  |-----$\text{\textbf{10Mbps,5ms}}$-------->
    $\text{\textcolor{c10}{\textbf{copa}}}$----$\text{\textbf{5ms}}$----------->|  |               |  |-----$\text{\textbf{10Mbps,5ms}}$-------->
                           ----    20 SECS    ----    
\end{lstlisting}
\end{minipage}

In accord with the testing setup in Listing~\ref{list:example_appendix_b}, Copa flows indeed start at the $10^{\text{th}}$ second in both the plots. The initial delays of the leftward Copa flows expectedly coincide with those of the leftward BBR flows, and the initial delays of the rightward Copa flows -- with those of the rightward BBR flows. The rates of the six BBR flows is expectedly up to 20 Mbit/s, and the rates of the four Copa flows -- up to 10 Mbit/s. The overall rate of the five flows with the same direction is limited to 70 Mbit/s by the central link.

\section{CoCo-Beholder vs Pantheon}
\label{sec:comparison}

\textbf{Applicability.} Pantheon gives uniform access to the collection of congestion control schemes, which is enriched by different researchers. Pantheon performs the periodical live testing of the schemes. Additionally, Pantheon provides a virtual network emulator. CoCo-Beholder is a virtual network emulator and here is compared to Pantheon only in the scope of the emulation capabilities of the two tools.

\textbf{Network Topology.} CoCo-Beholder provides a dumbbell topology of a chosen size, while Pantheon provides a point-to-point topology out-of-the-box. The popularity of the dumbbell topology was explored in Section~\ref{sec:first}. Furthermore, the point-to-point topology is a particular case of the dumbbell one.

\textbf{Layout Properties.} Having the dumbbell topology allows CoCo-Beholder to paramet\-rize flows running in the network in a much more flexible way. CoCo-Beholder enables to run flows of different congestion control schemes in the network at the same time. Each of the flows may have an individually chosen direction. A flow can be started at any second of the runtime. A flow may have an individual delay, rate, and queue size parameters, and, moreover, the parameters can be different for the flow's side link in the left half of the dumbbell topology and that one in the right half. \mbox{On the contrary,} Pantheon allows running flows of only a single congestion control scheme simultaneously. All the flows in Pantheon have the same direction and the same parameters of the single link, and the flows can be started only either all together at the beginning of the runtime or with a fixed time interval (in seconds). Pantheon has an advantage, though, that the single link may have a probabilistic loss and various queue management policies.

\textbf{Environment Variability.} CoCo-Beholder provides the variable delay with optional delay jitter, as shown in Section~\ref{sec:testing_tool}. Pantheon does not have this but, instead, it enables the variable rate via Mahimahi packet-delivery trace files.

\textbf{Changes to Schemes.} As described in Section~\ref{sec:third}, Pantheon uses the UDP tunneling to compute the packet delays and to solve the NAT problem. For that reason, the authors of Pantheon have to provide the patches for some schemes to reduce the MTU hardcoded in their source code. Also, encapsulating all the traffic into UDP datagrams gives 36 bytes of per-packet overhead and makes the traffic look UDP even if it is not. CoCo-Beholder does not use the tunneling and so does not suffer from these problems.

\textbf{Traffic Recording.} Pantheon records only the metadata of every packet\nolinebreak[4] (its \mbox{timestamp}, size, and delay) into a special log file. The log files are difficult to comprehend, especially for a newcomer. Besides, one cannot see the real traffic there, i.e. the payload of the packets. CoCo-Beholder records the real traffic at all the hosts of the topology with \verb+tcpdump+, and the PCAP dumps can be conveniently seen with, e.g., Wireshark~\cite{wireshark}.  

\textbf{Reproducibility.} It is laborious to reproduce a test performed by Pantheon. One should create the proper packet-delivery trace files and reconstruct the command with which Pantheon was launched. The command should include the paths to the new trace files for the Mahimahi LinkShell, the proper Mahimahi shells prepended and appended to the LinkShell, any extra arguments for the LinkShell, etc. A special problem is to guess the right order of the nested shells. CoCo-Beholder, meanwhile, saves \emph{all} the parameters of a test into a special JSON metadata file, as explained in Section~\ref{sec:testing_tool}, which can be fed to CoCo-Beholder testing module to fully reproduce the test. Even the randomization seed is saved to the metadata file, so the delay variableness is also\nolinebreak[4] preserved.

\textbf{Cleaning Up.} Pantheon does not kill the descendant processes started by the wrappers of the tested schemes even during the normal operation and with \verb+--pkill-cleanup+\nolinebreak[4] flag. So after running a test, the user has to kill the descendant (\verb+iperf+~\cite{iperf}) processes manually. If the user commands Pantheon to run a test several times repeatedly, the processes are not killed between the runs, which can affect the results. CoCo-Beholder always kills all the descendant processes even in cases of an interruption or an emergency\nolinebreak[4] exit.

\textbf{Link Utilization.} As appears in Section~\ref{sec:tfirst}, Pantheon shows the poor utilization of high bandwidth, while CoCo-Beholder does have this issue.

\textbf{Test Scheduling.} The user can give a list of schemes to Pantheon to test each of them continuously a chosen number of times and in random order. The statistics for all the runs of each scheme are averaged and written to the final report together with all the plots. CoCo-Beholder cannot run a test several times automatically. To have such a functionality, the user can create a shell script wrapping around CoCo-Beholder. 

\textbf{Runtime.} Both the tools ask the user to choose the testing runtime as an integer number of seconds from 1 to 60. CoCo-Beholder does not allow a longer runtime because it takes the schemes from Pantheon collection, and the wrappers of the schemes send the traffic only for the time required by the authors of Pantheon.

\textbf{Plots and Statistics.} Here, CoCo-Beholder has lots of advantages over Pantheon:
\vspace{-0.4cm}
\begin{itemize}
  \item Pantheon plots only average rate and per-packet one-way delay, while CoCo-Beholder additionally plots average Jain's index and average one-way delay.
  
  \item Pantheon computes the overall average rate, loss, and $95^{\text{th}}$ percentile one-way delay statistics, while CoCo-Beholder also computes the overall median and average one-way delay and the overall Jain's index statistics.
  
 \item Pantheon makes only per-flow plots and statistics, while CoCo-Beholder can also make total, per-direction, per-scheme-and-direction, etc. plots and statistics.
 
 \item The plots made by Pantheon are always ``shifted" 3 seconds right because \mbox{Pantheon} waits 3 seconds after starting the scheme servers to ensure that they listen to \nolinebreak[4]clients. CoCo-Beholder checks the servers' readiness intellectually with \verb+lsof+~\cite{lsof} utility. 
 
 \item Pantheon has a fixed 0.5-second aggregation interval for the average plots, while CoCo-Beholder allows the user to choose any positive float aggregation interval.
 
 \item CoCo-Beholder allows the user to change the colors of the curves in the plots. 
\end{itemize}
\vspace{-0.4cm}
The advantage of Pantheon over CoCo-Beholder is that it can make plots and statistics also for the (acknowledgment) traffic, directed from the scheme receiver to the sender.

\textbf{Python.} Pantheon supports only Python 2, while CoCo-Beholder -- Python 2 and 3.

\bigbreak
\bigbreak
\bigbreak

This section concludes the chapter describing how CoCo-Beholder is built and works. CoCo-Beholder enables the user to have a much greater range of various testing setups comparing to Pantheon emulator. Nevertheless, CoCo-Beholder still has room for improvement: having the variable rate feature, analyzing the acknowledgment traffic, etc. The next chapter is devoted to the results of the testing performed with CoCo-Beholder.

\loadchapter{testing}{The Testing}

Installing CoCo-Beholder is effortless, and its installation script is simple and concise. On the contrary, making the schemes in Pantheon collection work is often problematic. While some schemes may fail to be compiled due to their library dependencies, others have special requirements like, e.g., TCP BBRv1.0, which is available only in Linux kernels 4.9 or higher~\cite{bbr-kernel}. Besides, Linux distributions sometimes just have bugs.

For that reason, the README file of CoCo-Beholder gives the detailed installation instructions for Ubuntu 16.04 LTS, Ubuntu 18.04 LTS, and Debian 10 independently. The instructions were tested by the author of the thesis on virtual machines with fresh releases of the distributions.

The installation process is straightforward for Ubuntu 18.04 LTS and causes no problems. Ubuntu 16.04 LTS comes with Linux kernel 4.15 that has the delay jitter feature of tc qdisc NetEm completely broken due to the bug~\cite{jbug}, and the jitter is used by CoCo-Beholder testing tool, as discussed in Section~\ref{sec:testing_tool}. Thus, the user is guided on how to downgrade the kernel to 4.13 version, though upgrading the kernel is also an option. The installation on Debian 10 is most tiresome because it requires to downgrade some libraries, to make changes to the source code or wrappers of some schemes, etc. To make it easier, the user is given the exact \verb+sed+~\cite{sed} commands changing the files as needed.

CoCo-Beholder was being implemented by the author of the thesis on the machine with a 4-core (8-thread) processor (Intel Core i7-4710MQ, 2.5 GHz) and 8 GB memory. The operating system was 64-bit Ubuntu 16.04.6 LTS with 4.13 kernel. 

The testing, the results of which are present in this chapter, was performed by the author of the thesis on the virtual machine with a 64-core (128-thread) processor (AMD EPYC 7601, 2.2 GHz) and 2 TB memory, running 64-bit Debian 10 with\nolinebreak[4] 4.19\nolinebreak[4] kernel.

Section~\ref{sec:tfirst} compares the results of the testing by CoCo-Beholder against those by Pantheon for the simplest point-to-point testing setups, possible for both the emulators. Section~\ref{sec:tsecond} compares the results of the testing by CoCo-Beholder against those by the real testbed from the paper~\cite{turkovic2019fifty} for more complicated dumbbell topology testing setups. Section~\ref{sec:tthird} explores the reaction of different congestion control schemes to a variable delay present at the central link of the dumbbell topology using CoCo-Beholder.

\section{Testing CoCo-Beholder vs Pantheon}
\label{sec:tfirst}

In this series of experiments, one flow of a scheme was run for 10 seconds in CoCo-Beholder and in Pantheon and the resulting overall rate, $95^{\text{th}}$ percentile one-way delay, and loss statistics generated by the two tools were compared. 

The topology used in Pantheon is the only one supported by this tool: point-to-point. The single link was set up to have the rate 100 Mbit/s, the delay $X$ ms, and the queue size 1000 packets. 

The topology used in CoCo-Beholder is a dumbbell of size one with the three links: the left, the central, and the right one. All the three links were set up to have the rate 100 Mbit/s and the queue size 1000 packets. The left and the right links had 1 ms delay, while the central link had $(X - 2)$ ms delay. 

The experiments for the delay $X \in \{3, 5, 10, 20\}$ ms were performed ten times for both the tools. The mean and the sample standard deviation was computed for the rate, delay, and loss results of the ten runs of a scheme for each of the tools. The difference was computed for the mean rate, delay, and loss values of the two tools.

Listing~\ref{list:pantheontesting10ms} shows an example command for Pantheon for $X=10$ ms. The \mbox{packet-delivery} trace file \verb+100mbps.trace+ contains 25 lines. Each of the first 8 lines contains the digit 1, each of the next 8 lines contains the digit 2, each of the last 9 lines contains the\nolinebreak[4] digit\nolinebreak[4] 3. This implies 8 packets sent during the $1^{\text{st}}$ ms, 8 packets sent during the $2^{\text{d}}$ ms, 9 packets sent during the $3^{\text{d}}$ ms, and the file then wraps around for all the following milliseconds. With a packet delivery opportunity being 1500 bytes~\cite{mahimahi}, this gives the rates 96 Mbit/s, 96 Mbit/s, and 108 Mbit/s, the average of which is the desired 100 Mbit/s.\\

\begin{minipage}{\linewidth}
\begin{lstlisting}[caption=\mbox{An example command for Pantheon for $X=10$ ms.},label=list:pantheontesting10ms,mathescape]  
$\text{\textbf{\textcolor{teal}{\$} ./test.py local \textcolor{darkblue}{--scheme} "cubic" \textcolor{darkblue}{-t} 10 \textcolor{darkblue}{--flows} 1}}$ \
$\text{\textbf{\textcolor{darkblue}{--uplink-trace} 100mbps.trace \textcolor{darkblue}{--downlink-trace} 100mbps.trace}}$ \
$\text{\textbf{\textcolor{darkblue}{--prepend-mm-cmds} "mm-delay 10" }} $\
$\text{\textbf{\textcolor{darkblue}{--append-mm-cmds} "--uplink-queue=droptail --uplink-queue-args=packets=1000}}$ \
$\text{\textbf{--downlink-queue=droptail --downlink-queue-args=packets=1000"}}$
\end{lstlisting}
\end{minipage}

Listings~\ref{list:testingbeh10ms} and~\ref{list:testingbehlay10ms} show an example command for CoCo-Beholder for $X=10$ ms and the contents of the layout file, which is the same for any $X$. As the delay is constant, the delta is set to 100 seconds: a value, equal or greater than the 10 seconds of the\nolinebreak[4] runtime. The \verb+-r+ rate and all the queue sizes could be omitted: they are 100 and 1000 by default.\\

\begin{minipage}{\linewidth}
\begin{lstlisting}[caption=\mbox{An example command for CoCo-Beholder for $X=10$ ms.},label=list:testingbeh10ms,mathescape]  
$\text{\textbf{\textcolor{teal}{\$} ./run.py \textcolor{darkblue}{-p} \(\sim\)/pantheon 8ms 100s 0ms \textcolor{darkblue}{-t} 10 \textcolor{darkblue}{-r} 100 \textcolor{darkblue}{-q} 1000}} $
\end{lstlisting}
\end{minipage}

\begin{minipage}{\linewidth}
\begin{lstlisting}[caption=The layout file for CoCo-Beholder., label=list:testingbehlay10ms,language=yaml]
- direction: <-
  flows: 1
  left-delay: 1ms
  left-queues: 1000
  left-rate: 100
  right-delay: 1ms
  right-queues: 1000
  right-rate: 100
  scheme: cubic
  start: 0
\end{lstlisting}
\end{minipage}

The testing was performed for the schemes in Pantheon collection. Additionally, the schemes not present in Pantheon but available as Linux kernel modules were tested. For that, they should have been added to Pantheon. A new scheme can be added to Pantheon either globally or locally. The global way is the same as the local one but with an additional step: one should make a pull request to Pantheon, so that the scheme would become publicly available and would be run between the real-life nodes of Pantheon, with the results of the live testing archived on Pantheon website.

To test a new scheme with CoCo-Beholder or with Pantheon emulator, it is enough to add the scheme to the collection locally though.  A new entry for the scheme should be added to the YAML configuration file of Pantheon with the list of all the kept schemes. Then the wrapper for the scheme should be created. TCP Vegas scheme kept in Pantheon is available as a Linux kernel module and is launched by the wrapper using \verb+iperf+ utility. Thus, the wrapper for a new scheme can be produced by copying the wrapper file of TCP Vegas and replacing all the entries of the string \verb+vegas+ in the file by the name of the new scheme. The README file of CoCo-Beholder contains exact instructions and commands to add a scheme locally and also to troubleshoot a scheme.

For the delay $X=10$ ms, Table~\ref{tab:tab1} contains the results of the testing  for the schemes present in Pantheon and Table~\ref{tab:tab5} contains the results for the schemes added locally.  

\definecolor{myg}{HTML}{C1FFFF}
\definecolor{myr}{HTML}{FFC1C1}

\begin{table*}[hp!]
\centering
\Large
\caption{Testing results for the delay $X=10$ ms.}
\renewcommand{\arraystretch}{1.65} 
\resizebox*{\textwidth}{!}{\begin{tabu}{|c|cV{5}c|c|cV{5}c|c|cV{5}c|c|cV{5}}
\hline
\multirow{2}*{\parbox[c][2.5cm]{2.1cm}{\centering \bf \large Scheme}} &  \multicolumn{1}{c|}{\multirow{2}*{\parbox[c][2.5cm]{0.0cm}{}}} & \multicolumn{3}{c|}{\parbox[c][1cm]{2.7cm}{\bf \large Rate (Mbps)}} & \multicolumn{3}{c|}{\parbox[c][1cm]{2.3cm}{\bf \large Delay (ms)}} & \multicolumn{3}{c|}{\parbox[c][1cm]{1.8cm}{\bf \large Loss (\%)}}\\ 
\cline{3-11}
 & \multicolumn{1}{c|}{} & \parbox[c][1.5cm]{1.8cm}{\centering \bf \normalsize CoCo-Beholder}  & \parbox[1cm]{1.9cm}{\centering \bf \normalsize Pantheon} & \multicolumn{1}{c|}{\parbox{1.8cm}{\centering \pmb{$d_r$}}} & \parbox{1.8cm}{\centering \bf \normalsize CoCo-Beholder} & \parbox{1.8cm}{\centering \bf \normalsize Pantheon} & \multicolumn{1}{c|}{\parbox{1.8cm}{\centering \pmb{$d_r$}}} & \parbox{1.8cm}{\centering \bf \normalsize CoCo-Beholder} & \parbox{1.8cm}{\centering \bf \normalsize Pantheon} &\multicolumn{1}{c|}{ \parbox{1.8cm}{\centering \pmb{$\Delta$}}}\\
\cline{1-2}\tabucline[2pt]{3-11}
\multirow{2}*{\parbox[c][1cm]{2.1cm}{\centering bbr}} & \large \pmb{$\mu$} & 99.06 & 98.93 & 0.14\% & \cellcolor{myg}18.58 & \cellcolor{myr}36.32 & 64.65\% & 0.17  & 0.33 & 0.16\\
\cline{2-11}
& \large \pmb{$\sigma$} & 0.00 & 0.05 &  & 0.16 & 0.20 &  & 0.04 & 0.03 &\\
\cline{1-2}\tabucline[2pt]{3-11}
\multirow{2}*{\parbox[c][1cm]{2.1cm}{\centering copa}} & \large \pmb{$\mu$} & \cellcolor{myg}82.23 & \cellcolor{myr}70.77 & 14.99\% & \cellcolor{myr}20.46 & \cellcolor{myg}12.81 & 46.01\% & 0.18 & 0.11 & 0.07\\
\cline{2-11}
& \large \pmb{$\sigma$} & 5.13 & 2.44 &  & 2.63 & 0.29 & & 0.13 & 0.02 & \\
\cline{1-2}\tabucline[2pt]{3-11}
\multirow{2}*{\parbox[c][1cm]{2.1cm}{\centering cubic}} & \large \pmb{$\mu$} & 99.10 & 98.90 & 0.20\% & 125.14 & 124.23 & 0.73\% & 1.08 & 1.04 & 0.04\\
\cline{2-11}
& \large \pmb{$\sigma$} & 0.03 & 0.11 &  & 15.61 & 6.49 & & 0.18 & 0.15 & \\
\cline{1-2}\tabucline[2pt]{3-11}
\multirow{2}*{\parbox[c][1cm]{2.1cm}{\centering fillp}} & \large \pmb{$\mu$} & 95.26 & 97.47 & 2.29\% & \cellcolor{myg}29.51 & \cellcolor{myr}41.66 & 34.15\% & 0.79 & 1.08 & 0.29\\
\cline{2-11}
& \large \pmb{$\sigma$} & 0.76 & 0.48 & & 1.94 & 5.19 & & 0.17 & 0.15 & \\
\cline{1-2}\tabucline[2pt]{3-11}
\multirow{2}*{\parbox[c][1cm]{2.1cm}{\centering fillp\_sheep}} & \large \pmb{$\mu$} & 88.27 & 87.07 & 1.36\% & \cellcolor{myg}23.74 & \cellcolor{myr}26.27 & 10.10\% &  0.24 & 0.31 & 0.08\\
\cline{2-11}
& \large \pmb{$\sigma$} & 0.80 & 2.06 & & 1.00 & 1.10 & & 0.05 & 0.16 & \\
\cline{1-2}\tabucline[2pt]{3-11}
\multirow{2}*{\parbox[c][1cm]{2.1cm}{\centering indigo}} & \large \pmb{$\mu$} & 98.01 & 93.66 & 4.54\% & \cellcolor{myr}22.33 & \cellcolor{myg}14.58 & 42.00\% & 0.24 & 0.13 & 0.10\\
\cline{2-11}
& \large \pmb{$\sigma$} & 0.09 & 3.86 & & 0.75 & 0.95 & & 0.06 & 0.02 & \\
\cline{1-2}\tabucline[2pt]{3-11}
\multirow{2}*{\parbox[c][1cm]{2.1cm}{\centering ledbat}} & \large \pmb{$\mu$} & 90.81 & 84.25 & 7.50\% & 46.31 & 42.72 & 8.07\% & 0.58 & 0.49 & 0.09\\
\cline{2-11}
& \large \pmb{$\sigma$} & 0.01 & 9.38 &  & 0.00 & 5.72 & & 0.06 & 0.08 & \\
\cline{1-2}\tabucline[2pt]{3-11}
\multirow{2}*{\parbox[c][1cm]{2.1cm}{\centering pcc}} & \large \pmb{$\mu$} & \cellcolor{myg}91.28 & \cellcolor{myr}67.14 & 30.48\% & 13.93 & 13.43 & 3.60\% &  0.09 &  0.13 & 0.04\\
\cline{2-11}
& \large \pmb{$\sigma$} & 1.47 & 6.47 & & 1.54 & 2.25 & & 0.11 & 0.01 & \\
\cline{1-2}\tabucline[2pt]{3-11}
\multirow{2}*{\parbox[c][1cm]{2.1cm}{\centering pcc\_exp}} & \large \pmb{$\mu$} & 89.48 & 88.39 & 1.22\% & 52.62 & 49.05 & 7.01\% & 0.35 & 0.35 & 0.00\\
\cline{2-11}
& \large \pmb{$\sigma$} & 1.19 & 1.52 & & 34.90 & 17.44 & & 0.60 & 0.37 & \\
\cline{1-2}\tabucline[2pt]{3-11}
\multirow{2}*{\parbox[c][1cm]{2.1cm}{\centering quic}} & \large \pmb{$\mu$} & 71.76 & 70.28 & 2.09\% & 10.41 & 11.43 & 9.32\% & 0.17 & 0.13 & 0.04\\
\cline{2-11}
& \large \pmb{$\sigma$} & 1.25 & 2.80 & & 0.01 & 0.11 & & 0.13 & 0.06 & \\
\cline{1-2}\tabucline[2pt]{3-11}
\multirow{2}*{\parbox[c][1cm]{2.1cm}{\centering scream}} & \large \pmb{$\mu$} & 0.22 & 0.22 & 1.02\% & \cellcolor{myg}10.49 &\cellcolor{myr}11.60 & 10.00\% & 0.24 &  0.11 & 0.14\\
\cline{2-11}
& \large \pmb{$\sigma$} & 0.00 & 0.00 & & 0.02 & 0.08 & & 0.21 & 0.17 & \\
\cline{1-2}\tabucline[2pt]{3-11}
\multirow{2}*{\parbox[c][1cm]{2.1cm}{\centering sprout}} & \large \pmb{$\mu$} & 23.95 & 24.30 & 1.43\% & 16.78 & 18.01 & 7.08\% & 0.24 & 0.21 & 0.03\\
\cline{2-11}
& \large \pmb{$\sigma$} & 0.26 & 0.20 & & 2.20 & 0.96 & & 0.26 & 0.26 & \\
\cline{1-2}\tabucline[2pt]{3-11}
\multirow{2}*{\parbox[c][1cm]{2.1cm}{\centering taova}} & \large \pmb{$\mu$} & \cellcolor{myg}92.61 & \cellcolor{myr}80.00 & 14.61\% & \cellcolor{myg}11.82 & \cellcolor{myr}17.59 & 39.25\% & 0.19 & 0.20 & 0.01\\
\cline{2-11}
& \large \pmb{$\sigma$} & 0.16 & 0.66 & & 0.00 & 0.17 & & 0.01 & 0.12 & \\
\cline{1-2}\tabucline[2pt]{3-11}
\multirow{2}*{\parbox[c][1cm]{2.1cm}{\centering vegas}} & \large \pmb{$\mu$} & 99.09 & 91.23 & 8.26\% & 43.28 & 39.51 & 9.10\% & 0.28 &  0.17 & 0.11\\
\cline{2-11}
& \large \pmb{$\sigma$} & 0.03 & 6.79 &  & 17.95 & 17.88 & & 0.12 & 0.06 & \\
\cline{1-2}\tabucline[2pt]{3-11}
\multirow{2}*{\parbox[c][1cm]{2.1cm}{\centering verus}} & \large \pmb{$\mu$} & 97.08 & 92.37 & 4.97\% & 118.84 & 123.72 & 4.02\% & 4.69 & 4.71 & 0.02\\
\cline{2-11}
& \large \pmb{$\sigma$} & 0.85 & 2.75 & & 1.04 & 1.55 & & 0.21 & 0.60 & \\
\cline{1-2}\tabucline[2pt]{3-11}
\multirow{2}*{\parbox[c][1cm]{2.1cm}{\centering vivace}} & \large \pmb{$\mu$} & \cellcolor{myg}83.23 &\cellcolor{myr}70.42 & 16.67\% & \cellcolor{myg}10.72 & \cellcolor{myr}19.02 & 55.82\% & 0.09 & 0.17 & 0.08\\
\cline{2-11}
& \large \pmb{$\sigma$} & 0.19 & 18.81 & & 0.09 & 9.45 & & 0.07 & 0.10 & \\
\cline{1-2}\tabucline[2pt]{3-11}
\end{tabu}}
\label{tab:tab1}
\end{table*}

\newpage

\begin{table*}[hp!]
\centering
\Large
\caption{Testing results for the delay $X=10$ ms \emph{(continuation)}.}
\renewcommand{\arraystretch}{1.65} 
\resizebox*{\textwidth}{!}{\begin{tabu}{|c|cV{5}c|c|cV{5}c|c|cV{5}c|c|cV{5}}
\hline
\multirow{2}*{\parbox[c][2.5cm]{2.1cm}{\centering \bf \large Scheme}} &  \multicolumn{1}{c|}{\multirow{2}*{\parbox[c][2.5cm]{0.0cm}{}}} & \multicolumn{3}{c|}{\parbox[c][1cm]{2.7cm}{\bf \large Rate (Mbps)}} & \multicolumn{3}{c|}{\parbox[c][1cm]{2.3cm}{\bf \large Delay (ms)}} & \multicolumn{3}{c|}{\parbox[c][1cm]{1.8cm}{\bf \large Loss (\%)}}\\ 
\cline{3-11}
 & \multicolumn{1}{c|}{} & \parbox[c][1.5cm]{1.8cm}{\centering \bf \normalsize CoCo-Beholder}  & \parbox[1cm]{1.9cm}{\centering \bf \normalsize Pantheon} & \multicolumn{1}{c|}{\parbox{1.8cm}{\centering \pmb{$d_r$}}} & \parbox{1.8cm}{\centering \bf \normalsize CoCo-Beholder} & \parbox{1.8cm}{\centering \bf \normalsize Pantheon} & \multicolumn{1}{c|}{\parbox{1.8cm}{\centering \pmb{$d_r$}}} & \parbox{1.8cm}{\centering \bf \normalsize CoCo-Beholder} & \parbox{1.8cm}{\centering \bf \normalsize Pantheon} &\multicolumn{1}{c|}{ \parbox{1.8cm}{\centering \pmb{$\Delta$}}}\\
\cline{1-2}\tabucline[2pt]{3-11}
\multirow{2}*{\parbox[c][1cm]{2.1cm}{\centering bic}} & \large \pmb{$\mu$} & 99.11 & 98.84 & 0.27\%  & 131.03 & 126.77 & 3.31\% & 3.77 & 3.31 & 0.46\\
\cline{2-11}
& \large \pmb{$\sigma$} & 0.01 & 0.15 & & 0.87 & 6.54 & & 0.09 & 0.86 & \\
\cline{1-2}\tabucline[2pt]{3-11}
\multirow{2}*{\parbox[c][1cm]{2.1cm}{\centering cdg}} & \large \pmb{$\mu$} & \cellcolor{myg}90.22 & \cellcolor{myr}62.00 & 37.08\% & \cellcolor{myr}16.88 & \cellcolor{myg}13.49 & 22.30\% & 0.16 & 0.16 & 0.01\\
\cline{2-11}
& \large \pmb{$\sigma$} & 2.58 & 7.52 & & 2.60 & 1.17 & & 0.08 & 0.06 & \\
\cline{1-2}\tabucline[2pt]{3-11}
\multirow{2}*{\parbox[c][1cm]{2.1cm}{\centering highspeed}} & \large \pmb{$\mu$} & 99.10 & 98.91 & 0.19\% & \cellcolor{myr}122.96 & \cellcolor{myg}111.08 & 10.15\% & 2.68 & 2.48 & 0.19\\
\cline{2-11}
& \large \pmb{$\sigma$} & 0.00 & 0.08 & & 1.92 & 9.91 & & 0.12 & 0.14 & \\
\cline{1-2}\tabucline[2pt]{3-11}
\multirow{2}*{\parbox[c][1cm]{2.1cm}{\centering htcp}} & \large \pmb{$\mu$} & 99.10 & 98.9 & 0.21\% &  \cellcolor{myr}130.29 &  \cellcolor{myg}109.47 & 17.36\% & 2.17 & 2.16 & 0.00\\
\cline{2-11}
& \large \pmb{$\sigma$} & 0.00 & 0.06 & & 0.34 & 17.55 & & 0.20 & 0.33 &\\
\cline{1-2}\tabucline[2pt]{3-11}
\multirow{2}*{\parbox[c][1cm]{2.1cm}{\centering hybla}} & \large \pmb{$\mu$} & 98.4 & 98.04 & 0.37\% & 119.21 & 117.86 & 1.14\% & 2.33 & 1.86 & 0.47\\
\cline{2-11}
& \large \pmb{$\sigma$} & 0.01 & 0.21 & & 0.83 & 16.31 & & 0.08 & 0.21 & \\
\cline{1-2}\tabucline[2pt]{3-11}
\multirow{2}*{\parbox[c][1cm]{2.1cm}{\centering illinois}} & \large \pmb{$\mu$} & 99.11 & 98.85 & 0.26\% & 93.56 & 98.57 & 5.21\% & 2.36 & 2.32 & 0.05\\
\cline{2-11}
& \large \pmb{$\sigma$} & 0.00 & 0.15 & & 1.09 & 3.45 & & 0.11 & 0.10 & \\
\cline{1-2}\tabucline[2pt]{3-11}
\multirow{2}*{\parbox[c][1cm]{2.1cm}{\centering lp}} & \large \pmb{$\mu$} & 99.10 & 98.94 & 0.17\% & 89.49 & 95.14 & 6.13\% & 2.14 & 2.02 & 0.11\\
\cline{2-11}
& \large \pmb{$\sigma$} & 0.00 & 0.06 & & 0.81 & 2.22 & & 0.06 & 0.18 & \\
\cline{1-2}\tabucline[2pt]{3-11}
\multirow{2}*{\parbox[c][1cm]{2.1cm}{\centering nv}} & \large \pmb{$\mu$} & 87.45 & 83.02 & 5.20\% & \cellcolor{myg}12.24 & \cellcolor{myr}13.85 & 12.35\% & 0.04 & 0.17 & 0.13\\
\cline{2-11}
& \large \pmb{$\sigma$} & 0.01 & 0.75 & & 0.04 & 0.06 & & 0.08 & 0.02 & \\
\cline{1-2}\tabucline[2pt]{3-11}
\multirow{2}*{\parbox[c][1cm]{2.1cm}{\centering reno}} & \large \pmb{$\mu$} & 99.10 & 98.87 & 0.24\% & 92.10 & 93.85 & 1.88\% & 2.16 & 1.93 & 0.23\\
\cline{2-11}
& \large \pmb{$\sigma$} & 0.01 & 0.12 & & 8.93 & 5.34 & & 0.00 & 0.15 & \\
\cline{1-2}\tabucline[2pt]{3-11}
\multirow{2}*{\parbox[c][1cm]{2.1cm}{\centering scalable}} & \large \pmb{$\mu$} & 99.11 & 98.89 & 0.22\% & 131.38 & 129.93 & 1.11\% & 5.04 & 4.71 & 0.33\\
\cline{2-11}
& \large \pmb{$\sigma$} & 0.01 & 0.10 & & 0.61 & 0.23 & & 0.08 & 1.21 &\\
\cline{1-2}\tabucline[2pt]{3-11}
\multirow{2}*{\parbox[c][1cm]{2.1cm}{\centering veno}} & \large \pmb{$\mu$} & 99.10 & 98.55 & 0.56\% & 89.16 & 94.42 & 5.73\% & 2.14 & 1.86 & 0.28\\
\cline{2-11}
& \large \pmb{$\sigma$} & 0.00 & 0.95 & & 0.31 & 4.11 & & 0.04 & 0.19 & \\
\cline{1-2}\tabucline[2pt]{3-11}
\multirow{2}*{\parbox[c][1cm]{2.1cm}{\centering westwood}} & \large \pmb{$\mu$} & 89.63 & 84.68 & 5.67\% & \cellcolor{myr}97.53 &\cellcolor{myg}88.19 & 10.06\%  & 1.94 & 2.01 & 0.07\\
\cline{2-11}
& \large \pmb{$\sigma$} & 0.02 & 2.25 & & 0.53 & 4.27 & & 0.00 & 0.04 & \\
\cline{1-2}\tabucline[2pt]{3-11}
\multirow{2}*{\parbox[c][1cm]{2.1cm}{\centering yeah}} & \large \pmb{$\mu$} & 99.10 & 98.26 & 0.85\% & \cellcolor{myg}30.03 & \cellcolor{myr}34.33 & 13.38\% & 0.40 & 0.32 & 0.07\\
\cline{2-11}
& \large \pmb{$\sigma$} & 0.01 & 0.91 & & 1.62 & 8.64 & & 0.00 & 0.08 &\\
\cline{1-2}\tabucline[2pt]{3-11}
\end{tabu}}
\label{tab:tab5}
\end{table*}

For the delays $X \in \{3, 5, 20\}$ ms, the six more tables with the results for all the tested schemes  can be found in Appendix~\ref{appendix:AppendixD}. The full name, the type, and the reference to the research paper of each of the tested schemes can be seen in Appendix~\ref{appendix:AppendixC}.

All the schemes present in Pantheon use UDP as a Transport Layer protocol, except for the three TCP schemes: Cubic, Vegas, and BBR. All the locally added schemes\nolinebreak[4] are\nolinebreak[4] TCP.

The schemes in the tables are sorted in alphabetical order by their names. The values in the cells are rounded to two decimal points. The symbols $\mu$ and $\sigma$ stand for the mean and the sample standard deviation computed over the ten runs of a scheme correspondingly. The symbol $d_r$ stands for the relative difference calculated using the formula~\ref{eq:differ}, and its values in the tables are represented in percentage form.
\begin{equation}
\label{eq:differ}
d_r(x, y)=\frac{|x-y|}{\left(\frac{x+y}{2}\right)}\, .
\end{equation}
  
If the mean rate or delay results by CoCo-Beholder and Pantheon differ more than by 10 percent, then the value of the winner is highlighted with blue color and the value of the opponent is highlighted with red color. The symbol $\Delta$ stands for the absolute difference and is computed between the mean loss results of the two tools, instead of the relative difference, because the mean loss values are small, and if they are, for instance, 0.06\% and 0.07\%, the relative difference would already be 15\%. 

The consideration of the tables in Appendix~\ref{appendix:AppendixD} and the two tables in this section shows that, according to what is expected, the greater the delay $X$ configured in the topologies of the two tools is, the smaller the resulting mean rate statistics and the bigger the resulting mean delay and loss statistics, output by each tool, are for a chosen scheme.

The mean loss values output by the two emulators are close. Their mean one-way delays are also comparable. However, e.g., the tested schemes TCP BBR, FillP, and Tao 100x show considerably smaller one-way delays of their data packets when tested with CoCo-Beholder, rather than with Pantheon, with the relative differences being around 50\%, 30\%, and 35\% correspondingly, regardless of the delay $X$ installed in the topologies.

The most reliable and clear conclusions can be drawn from the mean rate statistics in the tables. The results by CoCo-Beholder are substantially better than those by Pantheon. Only in several cases Pantheon showed a better mean rate but the corresponding value shown by CoCo-Beholder was still very close. The worst case of CoCo-Beholder can be seen in Table~\ref{tab:tab4} for $X=20$ ms: the mean rate of Fillp-Sheep was 86.97 Mbit/s for CoCo-Beholder against 90.29 Mbit/s for Pantheon. For Copa, PCC Allegro, PCC Vivace, Tao 100x, and TCP CDG, the mean rate statistics output by CoCo-Beholder are considerably better than those by Pantheon, and the rates are colored in the tables. E.g., for PCC Allegro, the relative differences are around 40, 45, 30, 25 percent for the delays $X=$ 3, 5, 10, 20\nolinebreak[4] ms in favor of CoCo-Beholder. Interestingly, in most cases, when a mean rate by CoCo-Beholder is much bigger than by Pantheon, the corresponding sample standard deviation is, on the contrary, much smaller.\\

\begin{figure}[h!]
\vspace*{-0.3cm}
    \centering
    \subfloat[BIC TCP by CoCo-Beholder]{{\includegraphics[width=0.5\textwidth]{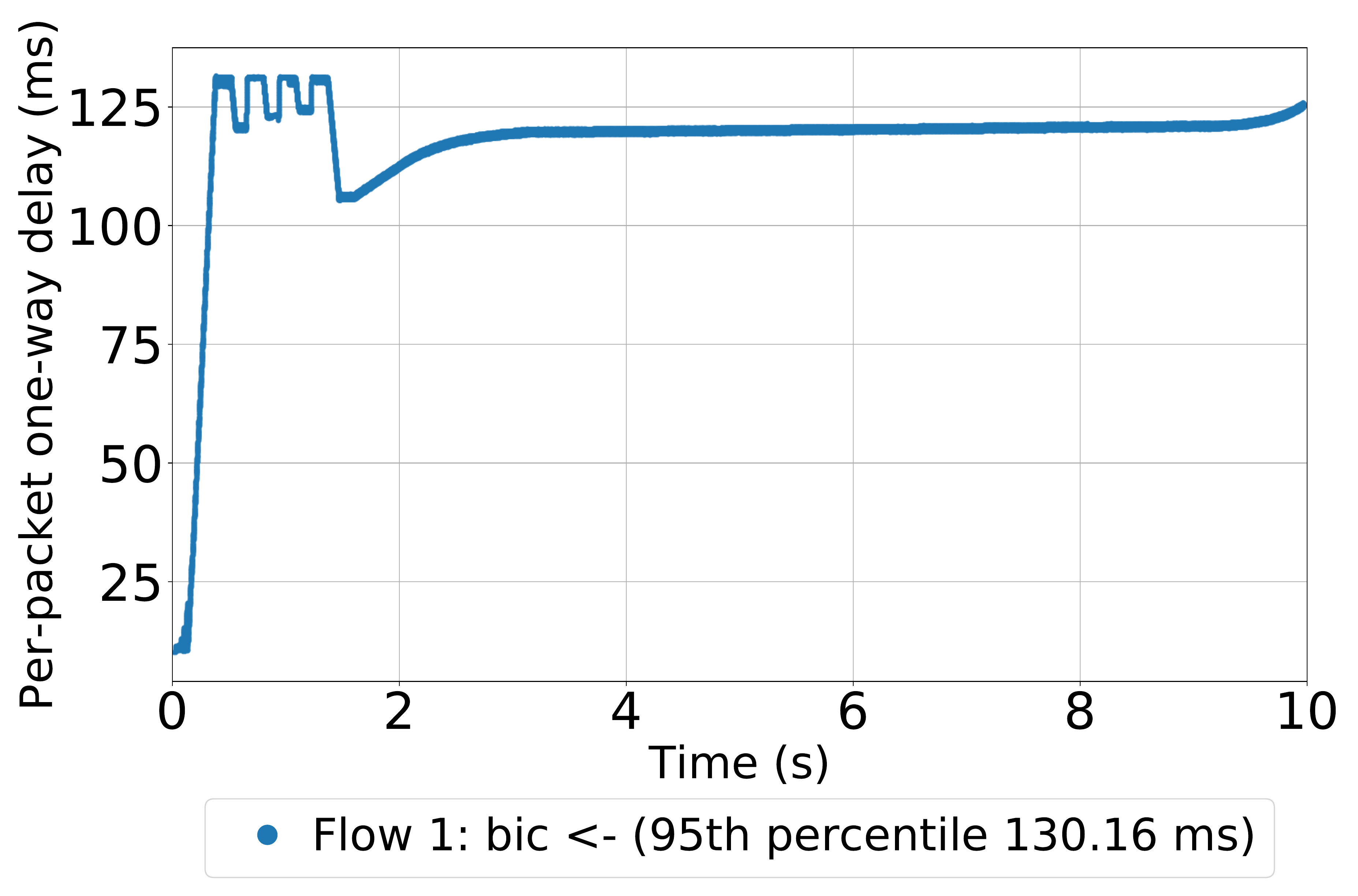} }}%
    \subfloat[BIC TCP by Pantheon]{{\includegraphics[width=0.5\textwidth]{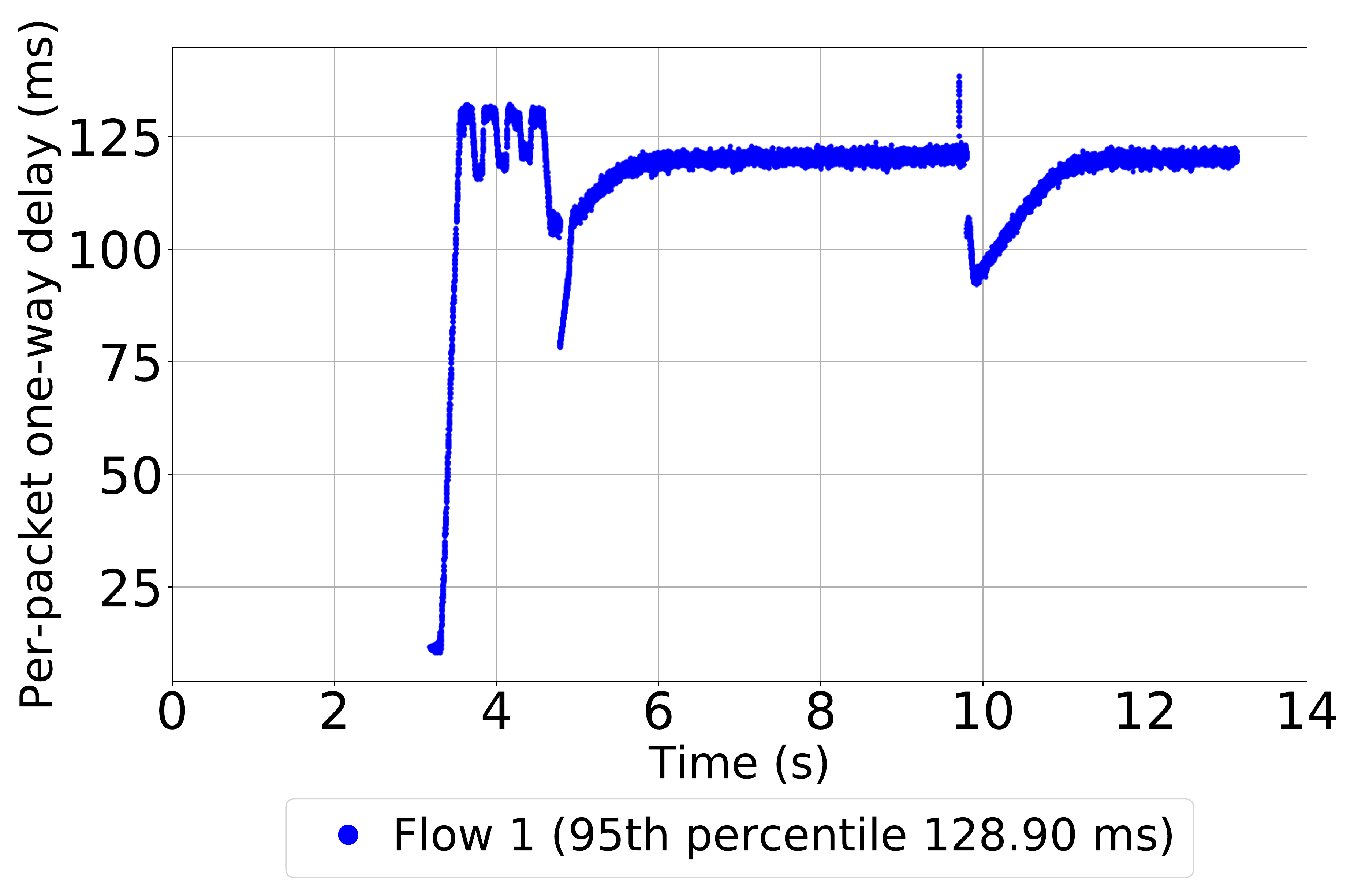} }}\\
    \vspace{-0.3cm}
    \subfloat[Scalable TCP by CoCo-Beholder]{{\includegraphics[width=0.5\textwidth]{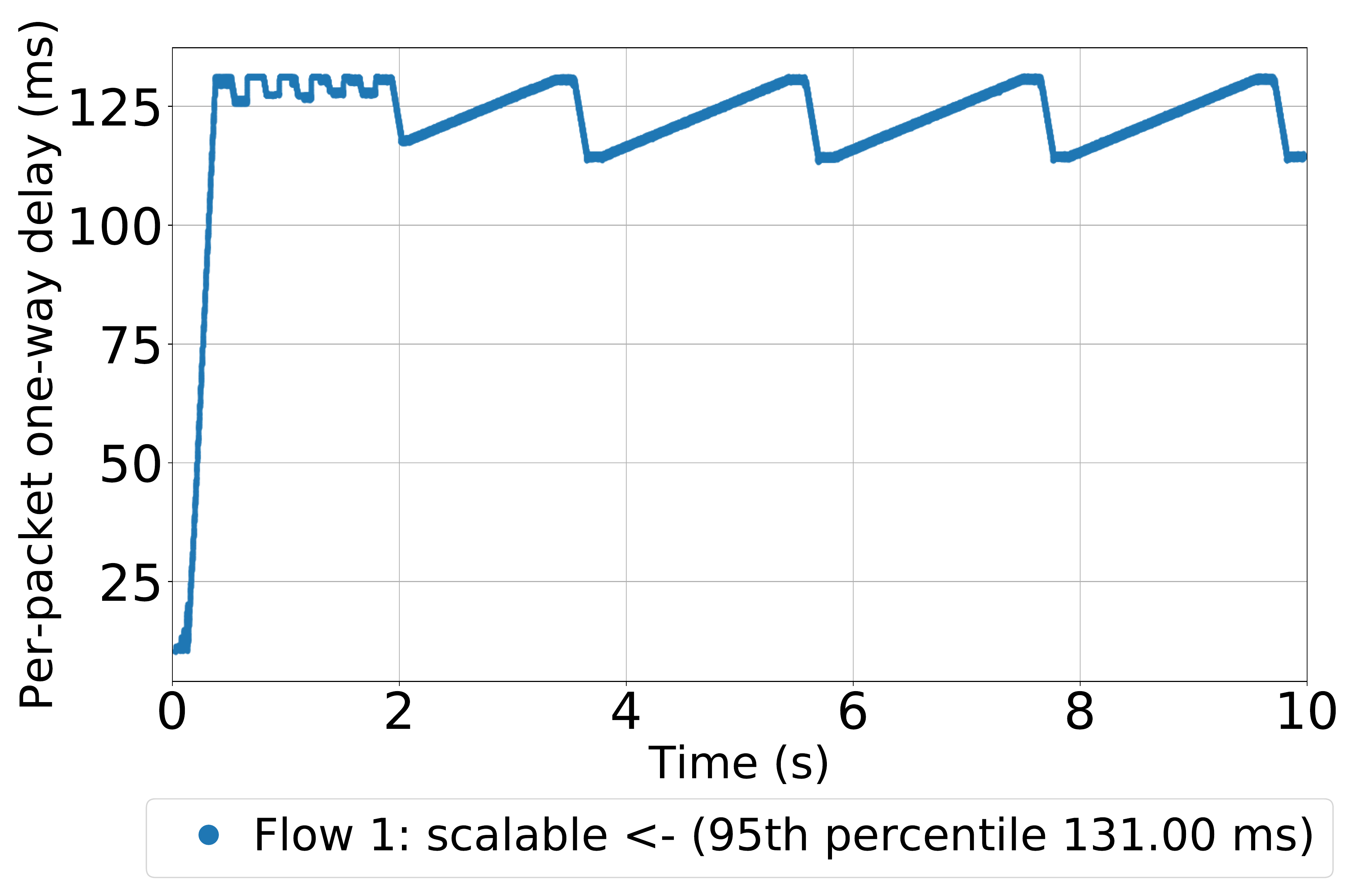} }}%
    \subfloat[Scalable TCP by Pantheon]{{\includegraphics[width=0.5\textwidth]{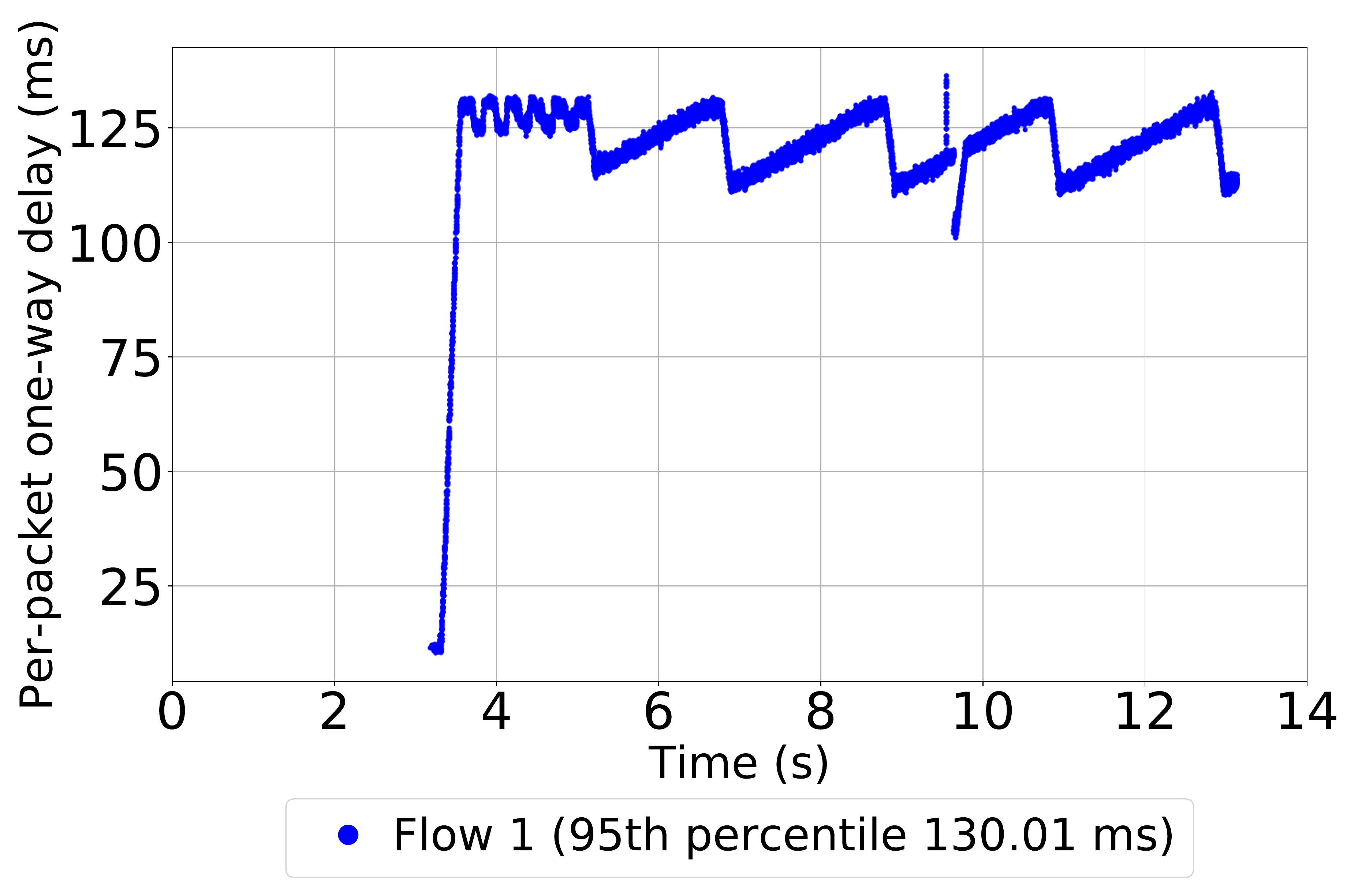} }}\\
    \vspace{-0.3cm}
    \subfloat[Sprout by CoCo-Beholder]{{\includegraphics[width=0.5\textwidth]{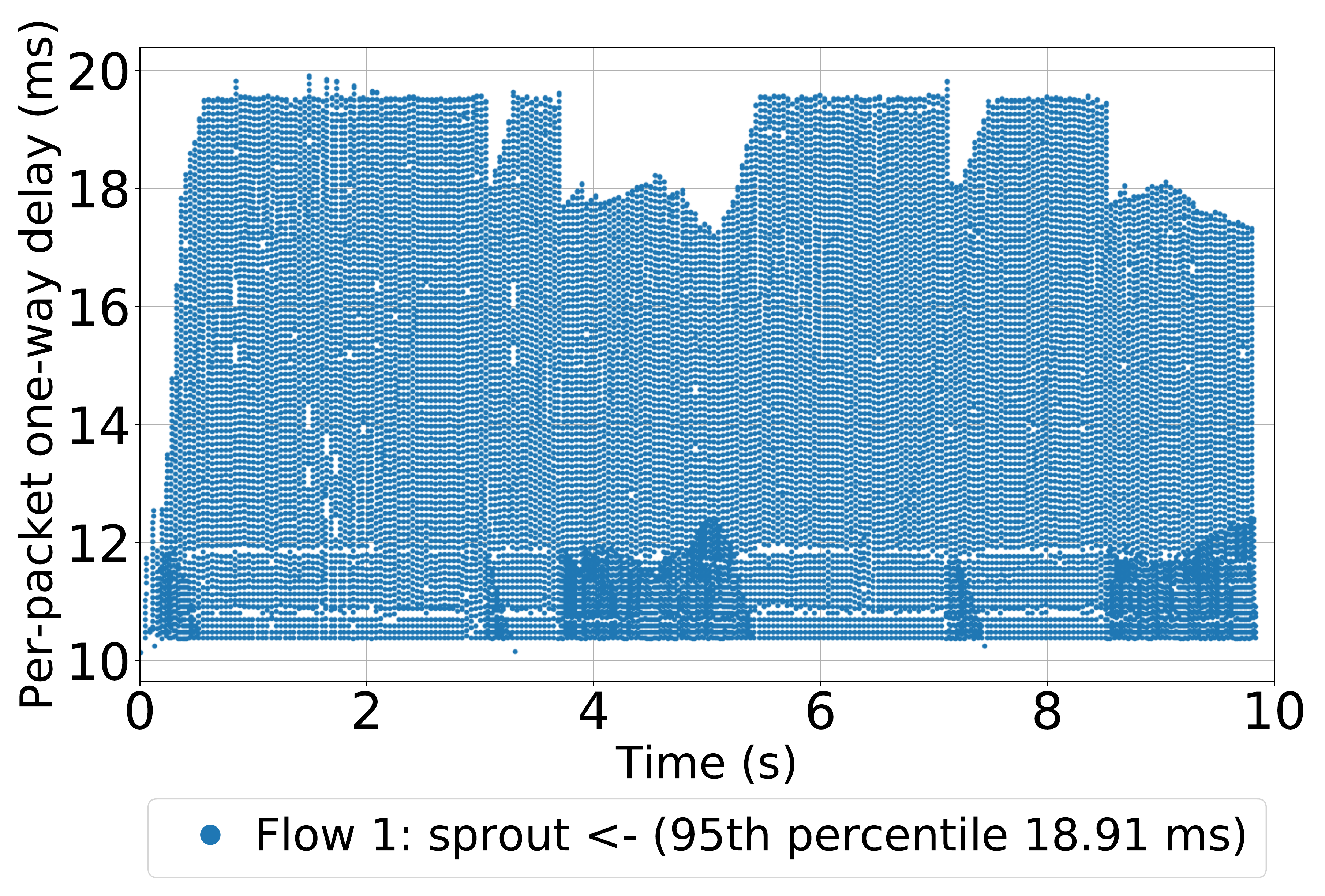} }}%
    \subfloat[Sprout by Pantheon]{{\includegraphics[width=0.5\textwidth]{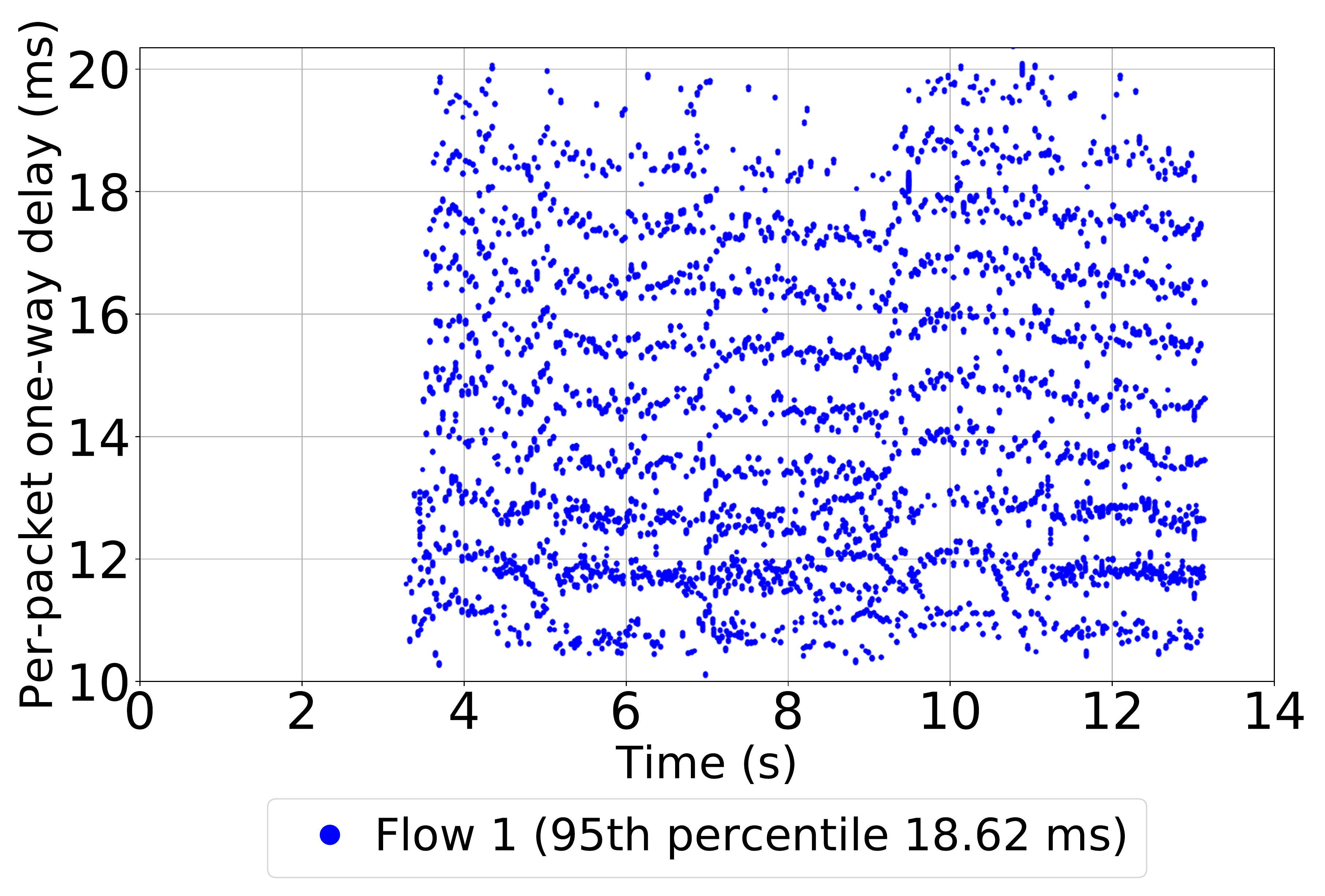} }}%
    \caption{Example per-packet one-way delay plots, the delay $X=10$ ms.}%
    \label{fig:compd}
\end{figure}

\vspace*{-1.1cm}
The reproducibility of both the emulators is acceptable: the sample standard deviation values in the tables are moderate, and each emulator generated very similar delay and rate plots for all the ten runs of a chosen scheme with a chosen delay\nolinebreak[4] $X$.

Comparing the delay and rate plots generated by the two tools, it can be stated that they look very much alike for each scheme, except for the above-mentioned schemes, whose mean statistics differed too much for the tools. In particular, the delay plots of a scheme have a common pattern that is like a signature of the scheme. Figure~\ref{fig:compd} shows the per-packet one-way delay plots by CoCo-Beholder (on the left) and by Pantheon (on the right) for BIC TCP, Scalable
TCP, and Sprout. The average rate plots of the schemes are less distinctive and can be found in Figure~\ref{fig:compr}\nolinebreak[4] in\nolinebreak[4] Appendix~\ref{appendix:AppendixD}.

\begin{table*}[h!]
\vspace*{-0.4cm}
\centering
\Large
\caption{The utilization of high bandwidth, the delay $X=3$ ms.}
\vspace*{-0.2cm}
\renewcommand{\arraystretch}{1.65} 
\resizebox*{\textwidth}{!}{\begin{tabu}{|c|cV{5}c|c|cV{5}c|c|cV{5}c|c|cV{5}}
\hline
\multirow{2}*{\parbox[c][2.5cm]{2.1cm}{\centering \bf \large Scheme}} &  \multicolumn{1}{c|}{\multirow{2}*{\parbox[c][2.5cm]{0.0cm}{}}} & \multicolumn{3}{c|}{\parbox[c][1cm]{4cm}{\bf \large 100 Mbps Capacity}} & \multicolumn{3}{c|}{\parbox[c][1cm]{4cm}{\bf \large 500 Mbps Capacity}} & \multicolumn{3}{c|}{\parbox[c][1cm]{3.4cm}{\bf \large 1 Gbps Capacity}}\\ 
\cline{3-11}
 & \multicolumn{1}{c|}{} & \parbox[c][1.5cm]{1.8cm}{\centering \bf \normalsize CoCo-Beholder}  & \parbox[1cm]{1.9cm}{\centering \bf \normalsize Pantheon} & \multicolumn{1}{c|}{\parbox{1.8cm}{\centering \pmb{$d_r$}}} & \parbox{1.8cm}{\centering \bf \normalsize CoCo-Beholder} & \parbox{1.8cm}{\centering \bf \normalsize Pantheon} & \multicolumn{1}{c|}{\parbox{1.8cm}{\centering \pmb{$d_r$}}} & \parbox{1.8cm}{\centering \bf \normalsize CoCo-Beholder} & \parbox{1.8cm}{\centering \bf \normalsize Pantheon} &\multicolumn{1}{c|}{ \parbox{1.8cm}{\centering \pmb{$d_r$}}}\\
\cline{1-2}\tabucline[2pt]{3-11}
\multirow{3}*{\parbox[c][1cm]{2.1cm}{\centering bbr}} & \large \pmb{$\mu$} & 99.79 & 99.67 & 0.13\% & 497.79 & 494.99 & 0.56\% & 979.93 & 777.14 & 23.08\%\\
\cline{2-11}
& \large \pmb{$\sigma$} & 0.03 & 0.14 &  & 0.21 & 1.32 & & 2.76 & 28.58 & \\
\cline{2-11}
& \large \pmb{$U$} &  99.79\% & 99.67\%  &  & 99.56\% & 99.00\% & &\cellcolor{myg}97.99\% & \cellcolor{myr}77.71\% & \\
\cline{1-2}\tabucline[2pt]{3-11}
\multirow{3}*{\parbox[c][1cm]{2.1cm}{\centering cubic}} & \large \pmb{$\mu$} & 99.81 & 99.53 & 0.28\% & 497.53 & 307.91 & 47.09\% & 901.28 & 263.84 & 109.42\%\\
\cline{2-11}
& \large \pmb{$\sigma$} & 0.00 & 0.25 &  & 0.98 & 18.12 & & 16.14 & 12.63 & \\
\cline{2-11}
& \large \pmb{$U$} &  99.81\% & 99.53\%  & & \cellcolor{myg}99.51\% & \cellcolor{myr}61.58\% & & \cellcolor{myg}90.13\% & \cellcolor{myr}26.38\%& \\
\cline{1-2}\tabucline[2pt]{3-11}
\multirow{3}*{\parbox[c][1cm]{2.1cm}{\centering highspeed}} & \large \pmb{$\mu$} & 99.78 & 99.61 & 0.18\% &  498.16 & 341.33 & 37.36\%  & 923.75 & 291.02 & 104.17\%\\
\cline{2-11}
& \large \pmb{$\sigma$} & 0.05 & 0.08 &  & 0.08 & 16.66 & & 13.79 &12.12 & \\
\cline{2-11}
& \large \pmb{$U$} &  99.78\% & 99.61\%  &  & \cellcolor{myg}99.63\% & \cellcolor{myr}68.27\% & &\cellcolor{myg}92.37\% &\cellcolor{myr}29.10\% & \\
\cline{1-2}\tabucline[2pt]{3-11}
\multirow{3}*{\parbox[c][1cm]{2.1cm}{\centering illinois}} & \large \pmb{$\mu$} & 99.80 & 99.59 & 0.21\% &  498.06 & 411.02 & 19.15\% &926.65 &328.27 & 95.37\%\\
\cline{2-11}
& \large \pmb{$\sigma$} & 0.01 & 0.15 &  & 0.17 & 28.10 & & 8.57&31.50 & \\
\cline{2-11}
& \large \pmb{$U$} &  99.8\% & 99.59\% &  & \cellcolor{myg}99.61\% &\cellcolor{myr}82.20\% & &\cellcolor{myg}92.67\% & \cellcolor{myr}32.83\%& \\
\cline{1-2}\tabucline[2pt]{3-11}
\multirow{3}*{\parbox[c][1cm]{2.1cm}{\centering lp}} & \large \pmb{$\mu$} & 99.79 & 99.55 & 0.24\% & 498.15 & 291.58 & 52.31\% & 900.45& 251.12& 112.77\%\\
\cline{2-11}
& \large \pmb{$\sigma$} & 0.00 & 0.06 &  & 0.09 & 15.52 & &12.87 & 12.88& \\
\cline{2-11}
& \large \pmb{$U$} &  99.79\% & 99.55\%  &  & \cellcolor{myg}99.63\% & \cellcolor{myr}58.32\%  & & \cellcolor{myg}90.04\%& \cellcolor{myr}25.11\%& \\
\cline{1-2}\tabucline[2pt]{3-11}
\multirow{3}*{\parbox[c][1cm]{2.1cm}{\centering vegas}} & \large \pmb{$\mu$} &  99.66 & 65.76 & 40.99\% & 389.50 & 68.70 & 140.02\% & 543.35 & 48.43 & 167.26\%\\
\cline{2-11}
& \large \pmb{$\sigma$} & 0.02 & 8.72 &  & 38.32 & 15.12 & &161.39 & 14.50& \\
\cline{2-11}
& \large \pmb{$U$} & \cellcolor{myg}99.66\% & \cellcolor{myr}65.76\%  &  & \cellcolor{myg}77.90\% & \cellcolor{myr}13.74\% & &\cellcolor{myg}54.34\%&\cellcolor{myr}4.84\% & \\
\cline{1-2}\tabucline[2pt]{3-11}
\end{tabu}}
\label{tab:tab9}
\end{table*}
\vspace*{-1cm}
This section is concluded with Table~\ref{tab:tab9} showing the utilization of high bandwidth by the two tools. The table shows neither delay nor loss. The tested schemes are loss-based TCP Cubic, delay-based TCP Vegas, and hybrid TCP BBR present in Pantheon and loss-based TCP HighSpeed, delay-based TCP-LP, and hybrid TCP Illinois added\nolinebreak[4] locally. 

The testing setup is the same as before for $X=3$ ms: the delay of the single link is 3 ms for Pantheon, and the delays of all the three links are 1 ms for CoCo-Beholder. Each experiment is run ten times, and the mean rate $\mu$ over the ten runs is computed. The column ``100 Mbps Capacity" contains the already-discussed mean rates from Tables~\ref{tab:tab2} and~\ref{tab:tab6} for the tested schemes. The two other columns contain the new results for the schemes: when all the links in the topologies of the tools have high bandwidth (500\nolinebreak[4] Mbit/s or 1 Gbit/s). The symbol $U$ in the table stands for the percentage of the mean rate $\mu$ from the total bandwidth, i.e., from 100 Mbit/s, 500 Mbit/s or\nolinebreak[4] 1000\nolinebreak[4] Mbit/s.

Unlike CoCo-Beholder, Pantheon shows the poor utilization of the available link capacity, which is unsatisfactory because the utilization should be determined by a congestion control algorithm and should not be influenced by an emulator.

\section{Testing CoCo-Beholder vs the Real Dumbbell Testbed}
\label{sec:tsecond}

The paper ``Fifty Shades of Congestion Control: A Performance and Interactions Evaluation"~\cite{turkovic2019fifty} (2019) by Belma Turkovic, Fernando Kuipers, and Steve Uhlig, cited several times in this thesis, contains the comparative performance evaluation of the three notable representatives of the loss-based, delay-based, and hybrid congestion control schemes: TCP Cubic, TCP Vegas, and TCP BBR. The author of the thesis is very grateful to Belma Turkovic, who kindly agreed to answer some questions on the paper~\cite{turkovic2019fifty}.

In the research~\cite{turkovic2019fifty}, a hardware testbed was used, having the standard dumbbell topology of size $n$, i.e. with $2\cdot(n+1)$ nodes. Each of $n$ clients in the left half of the topology sent the traffic of a chosen congestion control scheme generated with \verb+iperf3+~\cite{fffff} utility to the corresponding one of $n$ servers in the right half. The two halves were connected by the central link with the two routers on its ends. Each node had a 4-core (4-thread) processor (Intel Xeon, 3 GHz), 4 GB memory, and six 1 Gbps\nolinebreak[4] NICs. The Linux kernel version was 4.13. The txqueuelen was set to 1000, which means that the queue size of the interfaces at the ends of all the links was 1000 packets. With \verb+ethtool+ utility, the rate of the central bottleneck link was set to 100 Mbit/s. In those experiments, which required RTTs to be set at the links in the right half, that was achieved with\nolinebreak[4] tc\nolinebreak[4] qdisc\nolinebreak[4] NetEm. 

The dumbbell topology together with the configuration of network conditions provided by CoCo-Beholder perfectly satisfies to enable the detailed emulation of the testbed. The author of the thesis repeated the experiments from the paper~\cite{turkovic2019fifty} and compared the results to those present in the paper. 

This section has six subsections, the first four of which are devoted to the exploration of fairness, and the last two -- of intra-RTT-fairness.\hypersetup{pdfborderstyle={/S/U/W 1},linkbordercolor=green} The definitions of the different kinds (intra- and inter-) of fairness and RTT-fairness were given on pages~\pageref{pr1} and~\pageref{pr2}.\hypersetup{pdfborderstyle=} 

Subsections~\ref{subsec:ssec1} and~\ref{subsec:ssec2} explore the intra- and inter-fairness for $n=2$, i.e. for two flows. Subsections~\ref{subsec:ssec3} and~\ref{subsec:ssec4} explore the intra- and inter-fairness for $n=4$, i.e. for four flows. The four subsections use a common testing setup referred to here and in the paper~\cite{turkovic2019fifty} as BW scenario. Listing~\ref{list:bw} demonstrates the BW scenario adjusted for Subsection~\ref{subsec:ssec4}.

\begin{minipage}{\linewidth}
\begin{lstlisting}[caption=BW scenario: inter-fairness of 4 flows.,label=list:bw,mathescape,basicstyle=\linespread{0.9}\ttfamily\small,mathescape,frame=]
                             ----  0-TH SEC:  ----
                    $\text{\contour{black}{\textcolor{green}{\textbf{cubic}}}}$--->|  |             |  |--->
                      $\text{\contour{black}{\textcolor{red}{\textbf{bbr}}}}$--->|  |             |  |--->
                      $\text{\contour{black}{\textcolor{blue}{\textbf{bbr}}}}$--->|  |---100Mbps---|  |--->
                      $\text{\contour{black}{\textcolor{yellow}{\textbf{bbr}}}}$--->|  |             |  |--->
                             |  |             |  |
                             ----   60 SECS   ----
$\text{\textcolor{white}{\textbf{.}}}$
\end{lstlisting}
\end{minipage}

Subsection~\ref{subsec:ssec5} explores the intra-RTT-fairness for $n=2$, i.e. for two flows. Subsection~\ref{subsec:ssec6} explores the intra-RTT-fairness for $n=4$, i.e. for four flows. The two subsections use a common testing setup referred to here and in the paper~\cite{turkovic2019fifty} as RTT scenario. Listings~\ref{list:rtt2} and~\ref{list:rtt4} show the RTT scenarios for the two subsections.\\

\begin{minipage}{\linewidth}
\begin{lstlisting}[caption=RTT scenario: intra-RTT-fairness of 2 flows.,label=list:rtt2,mathescape,basicstyle=\linespread{0.9}\ttfamily\small,mathescape,frame=]
                             ----  0-TH SEC:  ----
                    $\text{\contour{black}{\textcolor{green}{\textbf{cubic}}}}$--->|  |             |  |----0ms-->
                    $\text{\contour{black}{\textcolor{red}{\textbf{cubic}}}}$--->|  |             |  |--100ms-->
                             |  |             |  |
                             ----   60 SECS   ----
$\text{\textcolor{white}{\textbf{.}}}$
\end{lstlisting}
\end{minipage}

\begin{minipage}{\linewidth}
\begin{lstlisting}[caption=RTT scenario: intra-RTT-fairness of 4 flows.,label=list:rtt4,mathescape,basicstyle=\linespread{0.9}\ttfamily\small,mathescape,frame=]
                             ----  0-TH SEC:  ----
                    $\text{\contour{black}{\textcolor{green}{\textbf{cubic}}}}$--->|  |             |  |---50ms-->
                    $\text{\contour{black}{\textcolor{red}{\textbf{cubic}}}}$--->|  |             |  |--100ms-->
                    $\text{\contour{black}{\textcolor{blue}{\textbf{cubic}}}}$--->|  |---100Mbps---|  |--150ms-->
                    $\text{\contour{black}{\textcolor{yellow}{\textbf{cubic}}}}$--->|  |             |  |--200ms-->
                             |  |             |  |
                             ----   60 SECS   ----
$\text{\textcolor{white}{\textbf{.}}}$
\end{lstlisting}
\end{minipage}

In the paper~\cite{turkovic2019fifty}, the per-flow delays are defined using the RTT term, i.e., the four flows have 100 ms, 200 ms, 300 ms, and 400 ms RTTs. Here, the per-flow delays are defined using the one-way delay term, i.e., the flows have 50 ms, 100 ms, 150 ms, and 200 ms one-way delays. CoCo-Beholder was instructed to set a one-way delay of a flow with the \verb+right-delay+ property of the layout file. As CoCo-Beholder installs the one-way delay at both the ends of the link using tc qdisc NetEm, the resulting RTT of the flow is twice the one-way delay, as expected. CoCo-Beholder uses the one-way delay context, rather than the RTT context, because the emulator supports both TCP- and UDP-based schemes, and UDP does not provide an acknowledgment of each packet the way TCP \nolinebreak[4]does.

The wrappers of the schemes TCP Cubic, TCP Vegas, and TCP BBR in Pantheon collection use \verb+iperf+ to generate the traffic of the schemes. As in the paper~\cite{turkovic2019fifty} \verb+iperf3+ was used, the wrappers were locally changed to use \verb+iperf3+ too on the Debian machine, on which the author of the thesis performed the testing.

The authors of the paper~\cite{turkovic2019fifty} did multiple runs of each experiment and observed similar results, so the plots in the paper are for one run and the statistics in the tables correspond to these very plots. The plots are the per-flow average RTT, the per-flow average rate (i.e., per-flow average throughput), and the average Jain's index over all the flows.  The plots were averaged per an aggregation time interval (0.3 or 0.6\nolinebreak[4] seconds). The statistics -- the per-flow average RTT, per-flow average rate, and average Jain's index -- were computed as a mean over all the time intervals (slots) in the corresponding plots.

The plots by CoCo-Beholder in this section are the per-flow average one-way delay (\mbox{unfortunately} for the comparison of the results, not RTT), per-flow average rate, and average Jain's index over all the flows, with the aggregation interval equal to that one chosen in the paper~\cite{turkovic2019fifty}. In general, CoCo-Beholder also generates the following statistics: the per-flow average rate over the whole duration of the flow, the per-flow average one-way delay over all the data packets of the flow, and the average Jain's index computed over the average rate statistics of all the flows. CoCo-Beholder generates the statistics in this way because the resulting values do not depend on the aggregation interval. However, to make the statistics more comparable to those from the paper~\cite{turkovic2019fifty}, the author of the thesis temporarily changed the statistics generation to be like in the paper~\cite{turkovic2019fifty}, i.e., to compute the statistic values as a mean over the time slots. Still, the average one-way delay statistics by CoCo-Beholder again have to be compared against the average RTT statistics from the paper~\cite{turkovic2019fifty}.

Each experiment was run in CoCo-Beholder ten times. The mean $\mu$ and the sample standard deviation $\sigma$ of the rate, one-way delay, and Jain's index statistics were computed over the ten runs. The results for each experiment are placed in a separate table further in this section against the rate, RTT, and Jain's index values from the paper~\cite{turkovic2019fifty}. For the rate and Jain's index, the relative differences $d_r$ between the results by CoCo-Beholder and those from the paper were computed. The table cells with the rates are colored if the difference is more than 10\% according to the principle described in Section~\ref{sec:tfirst}. 

Also, for each experiment, the plots by CoCo-Beholder belonging to one of the ten runs are placed in an individual figure against the corresponding plots from the paper~\cite{turkovic2019fifty}. The plots of a run were chosen if the resulting rates of the flows in this run have the minimum Euclidean distance from the computed mean rate statistics of the flows among all the ten runs. The plots of the paper~\cite{turkovic2019fifty} were extracted from the electronic version of the paper directly and are raster-graphics. The curves in the plots by CoCo-Beholder were colored the same way, as in the plots from the paper, by supplying the needed color cycle to CoCo-Beholder plotting tool as a string command-line argument. The least possible Jain's index is 0.5 for $n=2$ and 0.25 for $n=4$, which is why the minimum y-axis limit for Jain's index plots is either of these values in this section.

\subsection{Intra-Fairness For Two Flows}
\label{subsec:ssec1}

\begin{figure}[p!]
\vspace*{-0.3cm}
\captionsetup[subfigure]{labelformat=empty}
    \centering
    \subfloat[]{{\includegraphics[width=\textwidth]{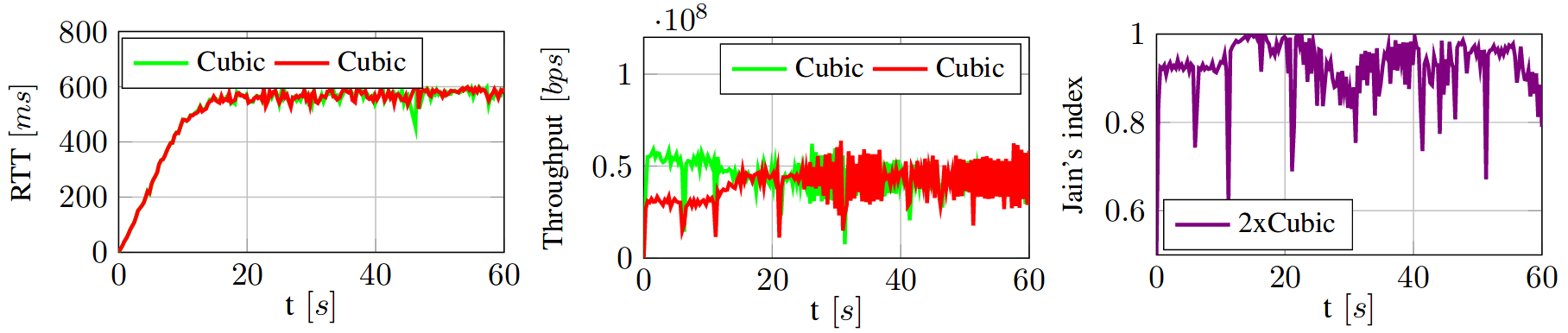} }}\\
    \vspace{-0.7cm}
    \subfloat[]{{\includegraphics[width=\textwidth/3]{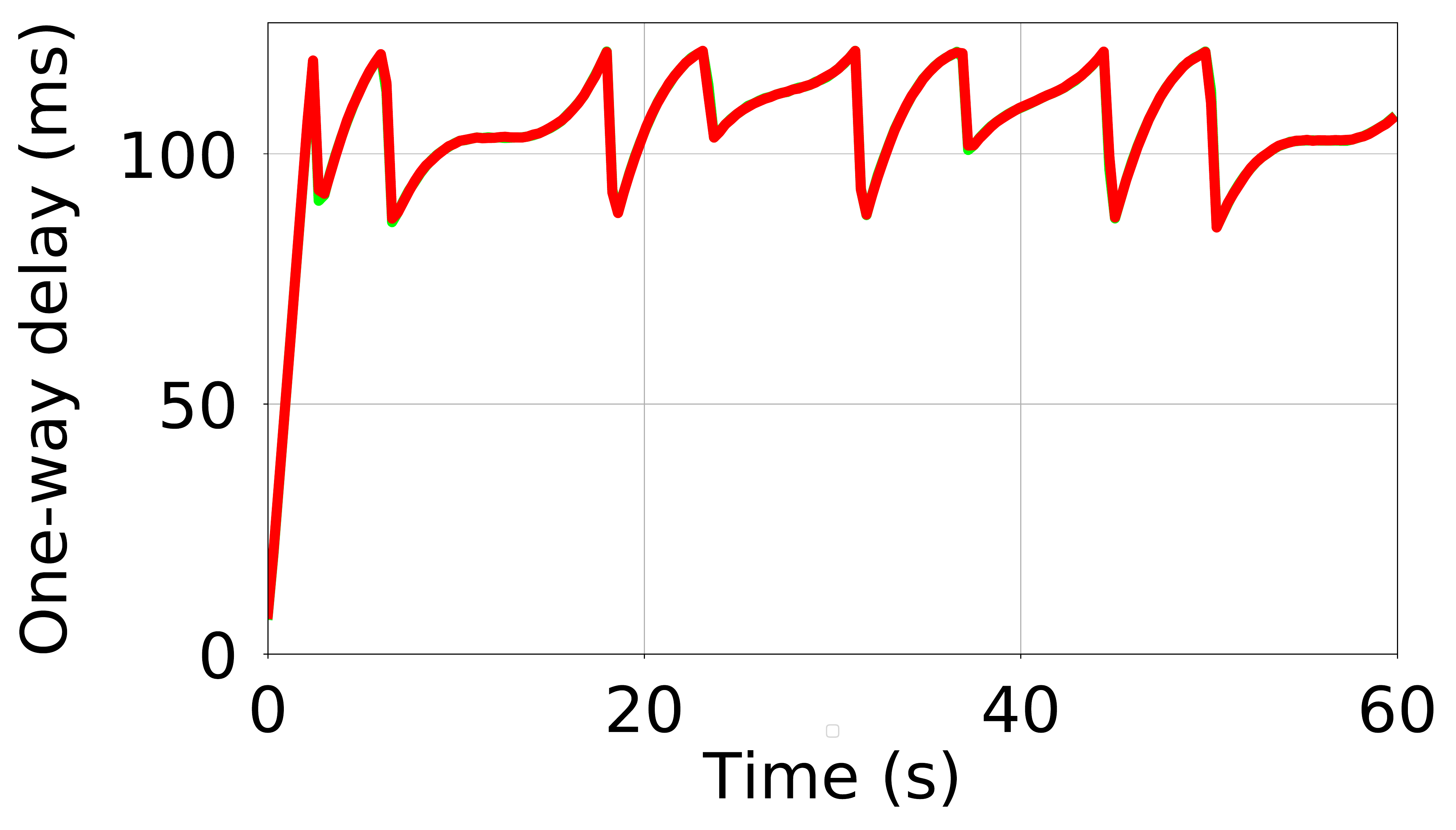} }}%
    \subfloat[]{{\includegraphics[width=\textwidth/3]{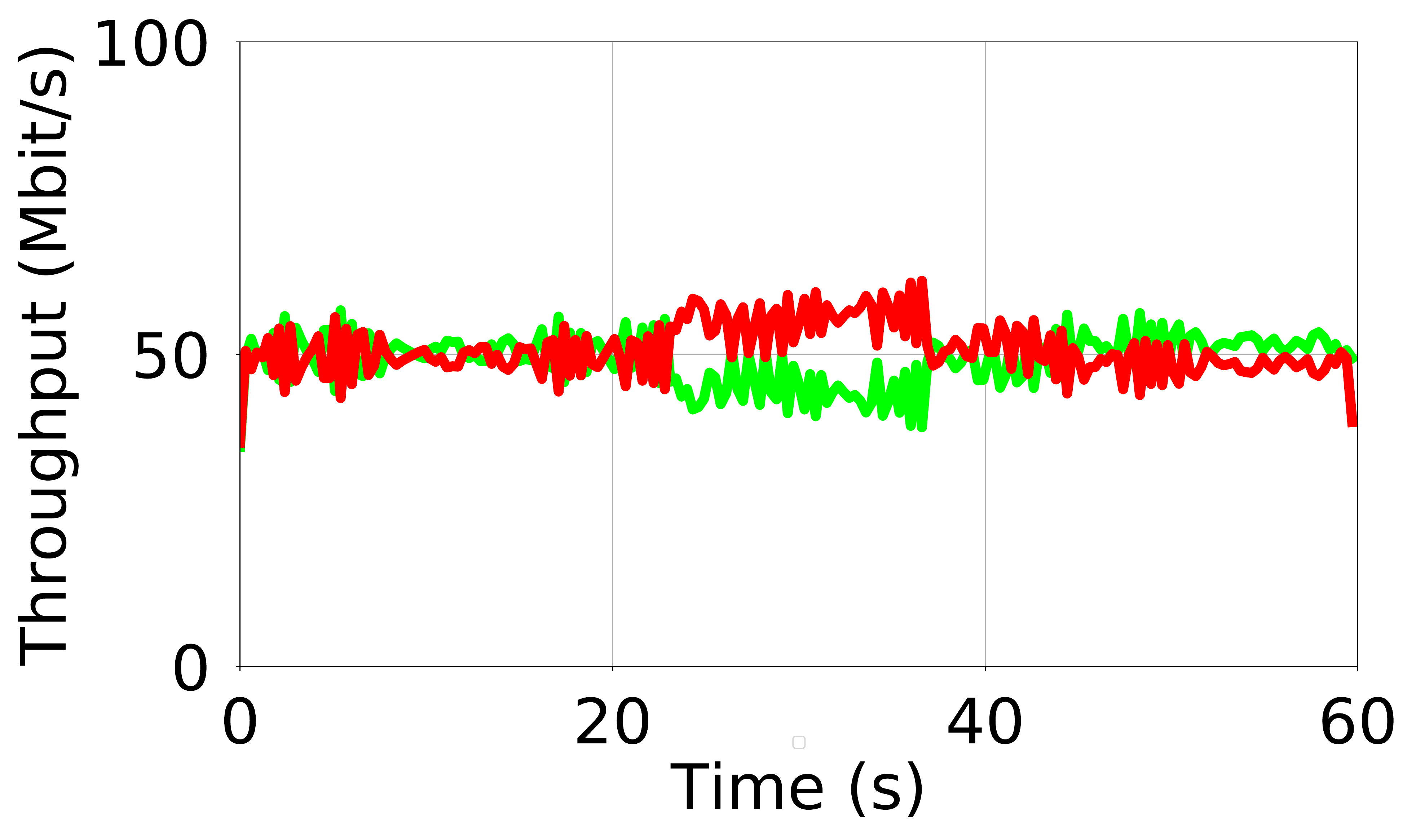} }}%
    \subfloat[]{{\includegraphics[width=\textwidth/3]{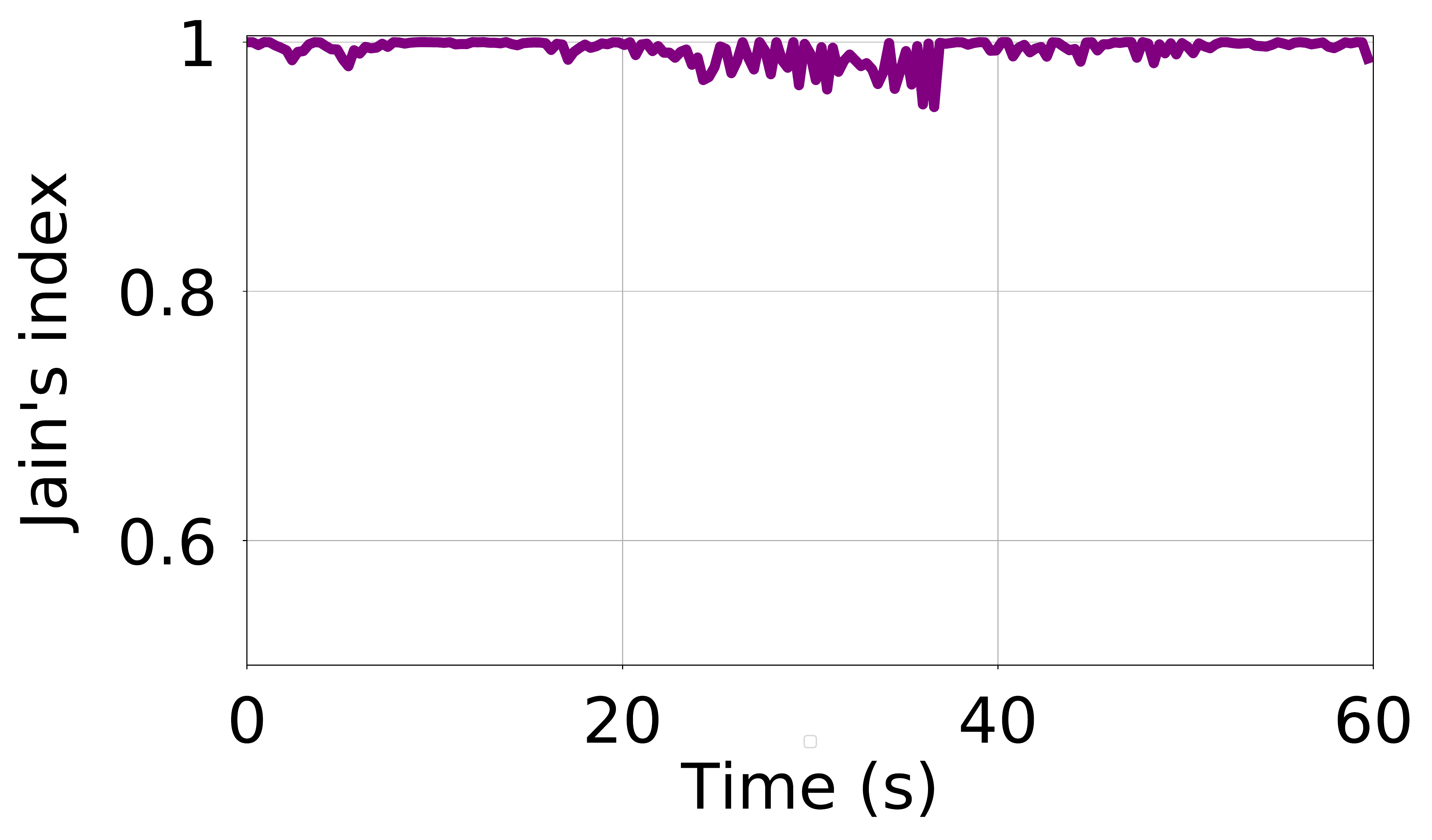} }}\\
    \vspace{-0.5cm}
    \caption{BW scenario: 2 Cubic flows. The aggregation interval is 300 ms.\\The top-row plots are by the testbed, the bottom-row -- by CoCo-Beholder.}%
    \label{fig:fig4211}
\end{figure}

\begin{figure}[h!]
\vspace*{-0.2cm}
\captionsetup[subfigure]{labelformat=empty}
    \centering
    \subfloat[]{{\includegraphics[width=\textwidth]{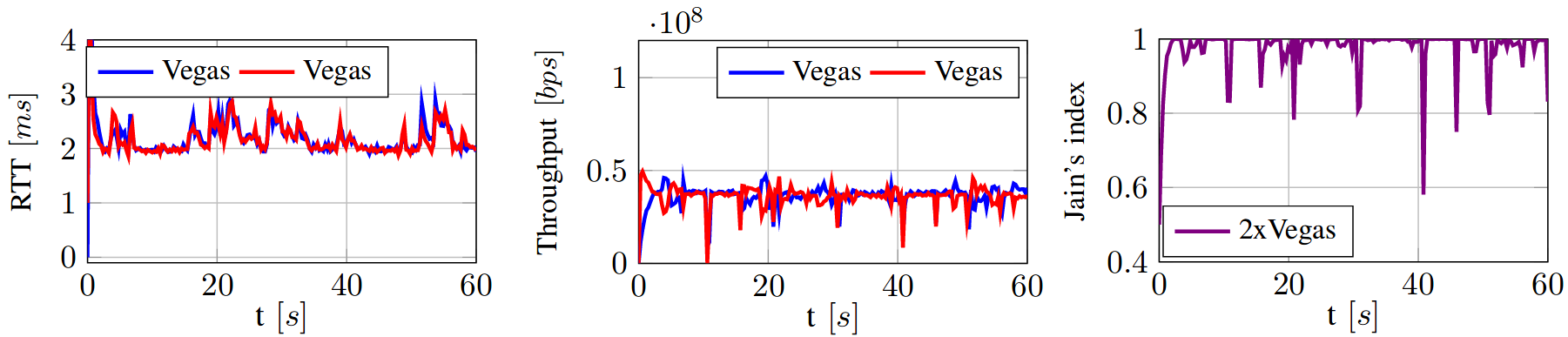} }}\\
    \vspace{-0.7cm}
    \subfloat[]{{\includegraphics[width=\textwidth/3]{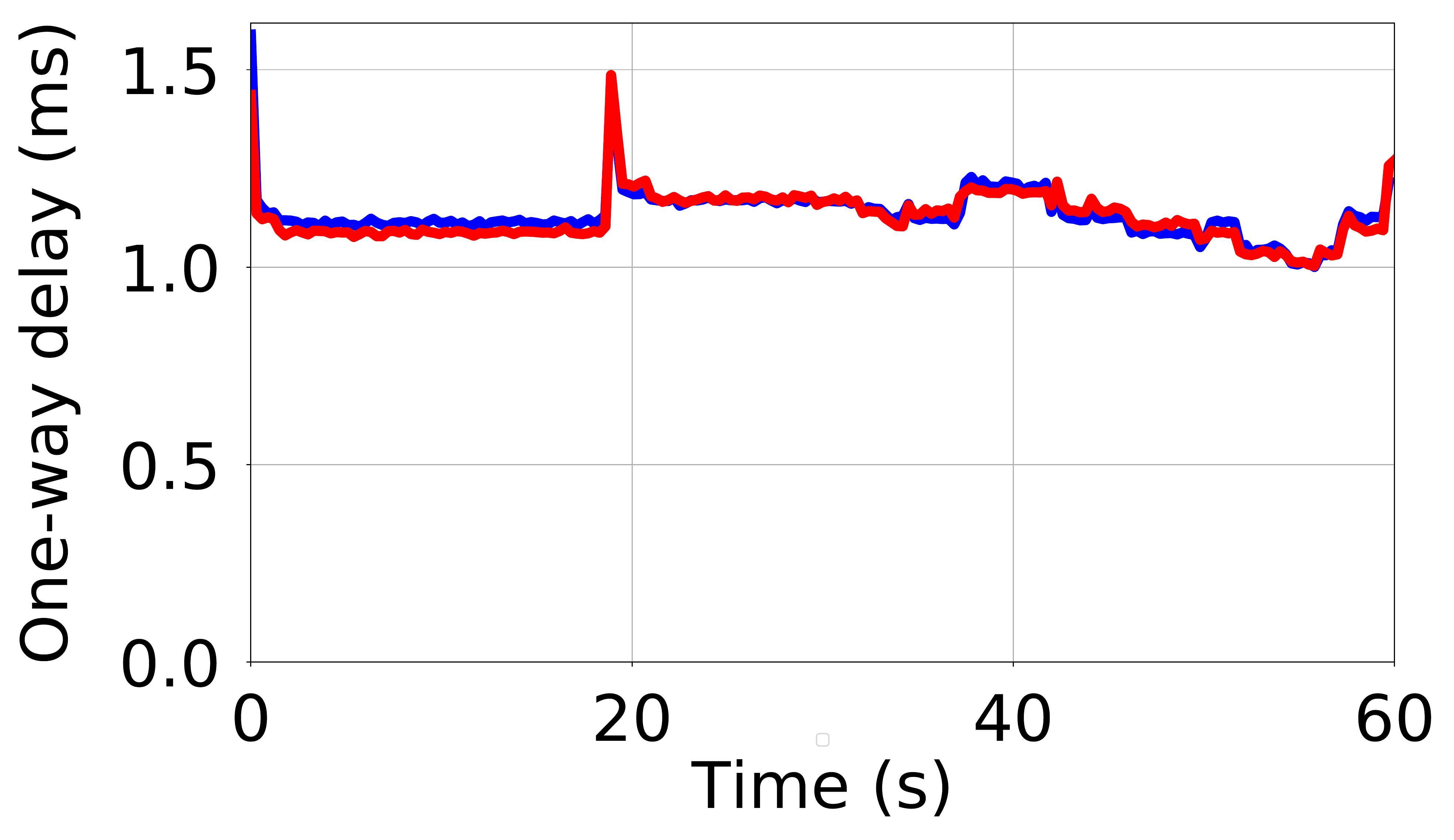} }}%
    \subfloat[]{{\includegraphics[width=\textwidth/3]{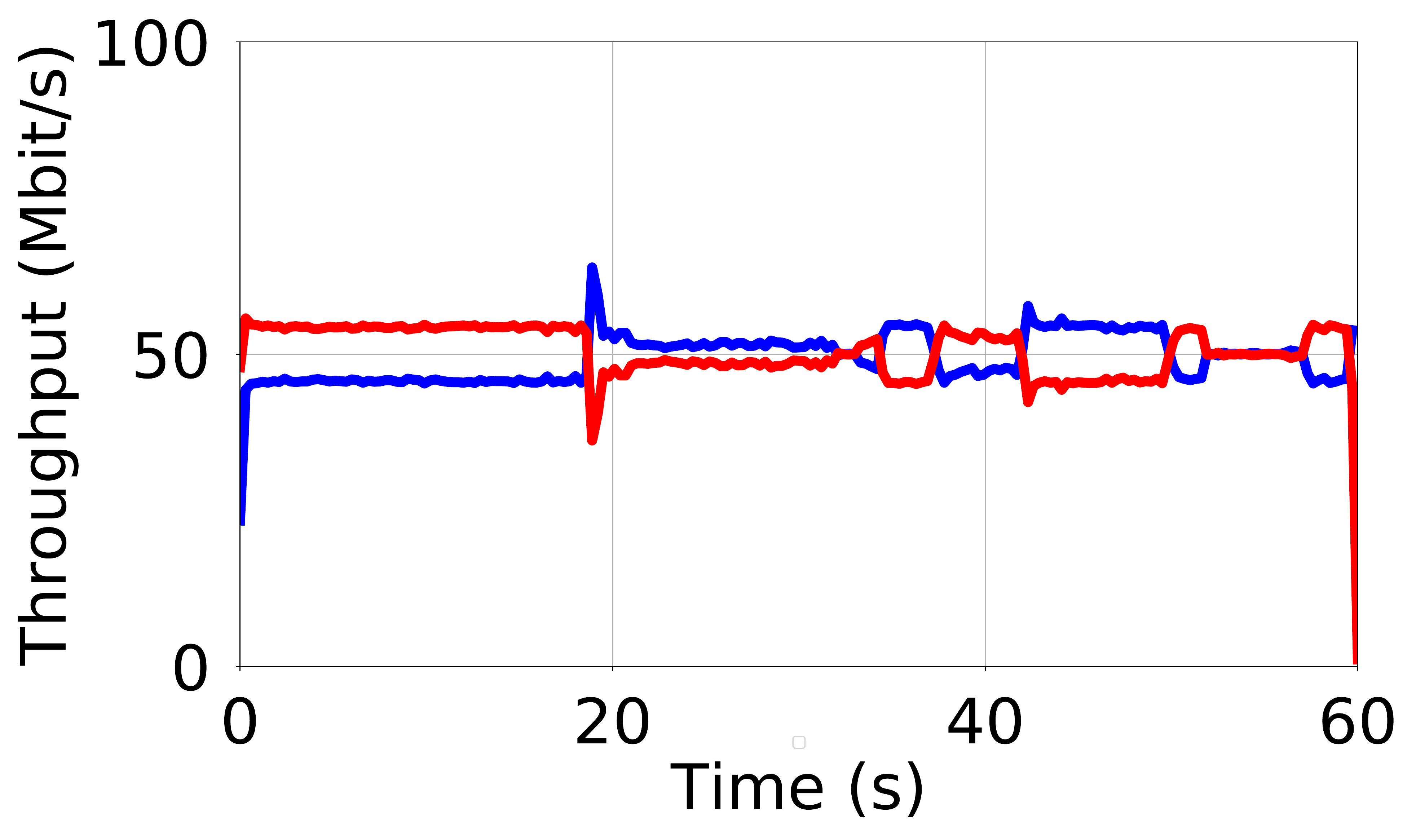} }}%
    \subfloat[]{{\includegraphics[width=\textwidth/3]{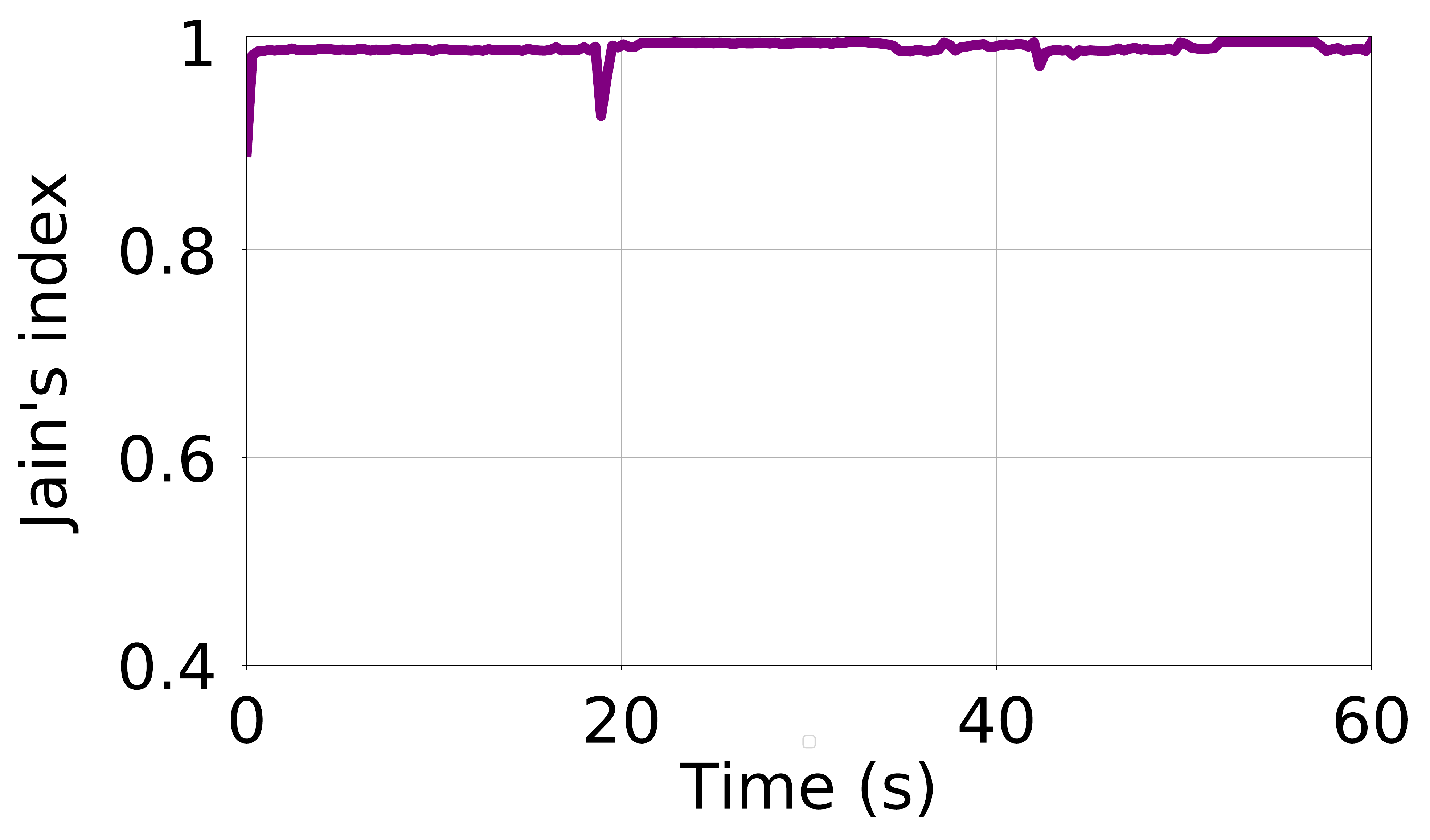} }}\\
    \vspace{-0.5cm}
    \caption{BW scenario: 2 Vegas flows. The aggregation interval is 300 ms.\\The top-row plots are by the testbed, the bottom-row -- by CoCo-Beholder.}%
    \label{fig:fig4212}
\end{figure}

\begin{figure}[h!]
\vspace*{-0.2cm}
\captionsetup[subfigure]{labelformat=empty}
    \centering
    \subfloat[]{{\includegraphics[width=\textwidth]{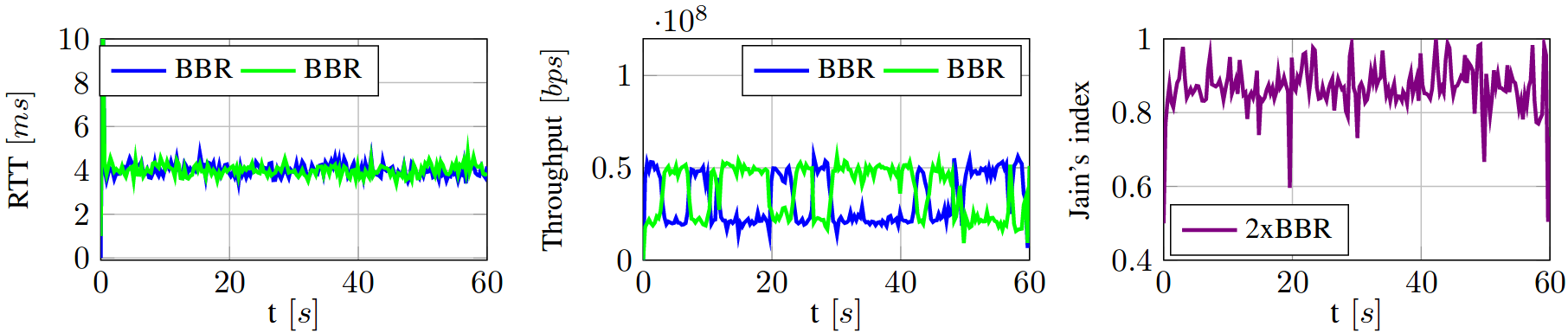} }}\\
    \vspace{-0.7cm}
    \subfloat[]{{\includegraphics[width=\textwidth/3]{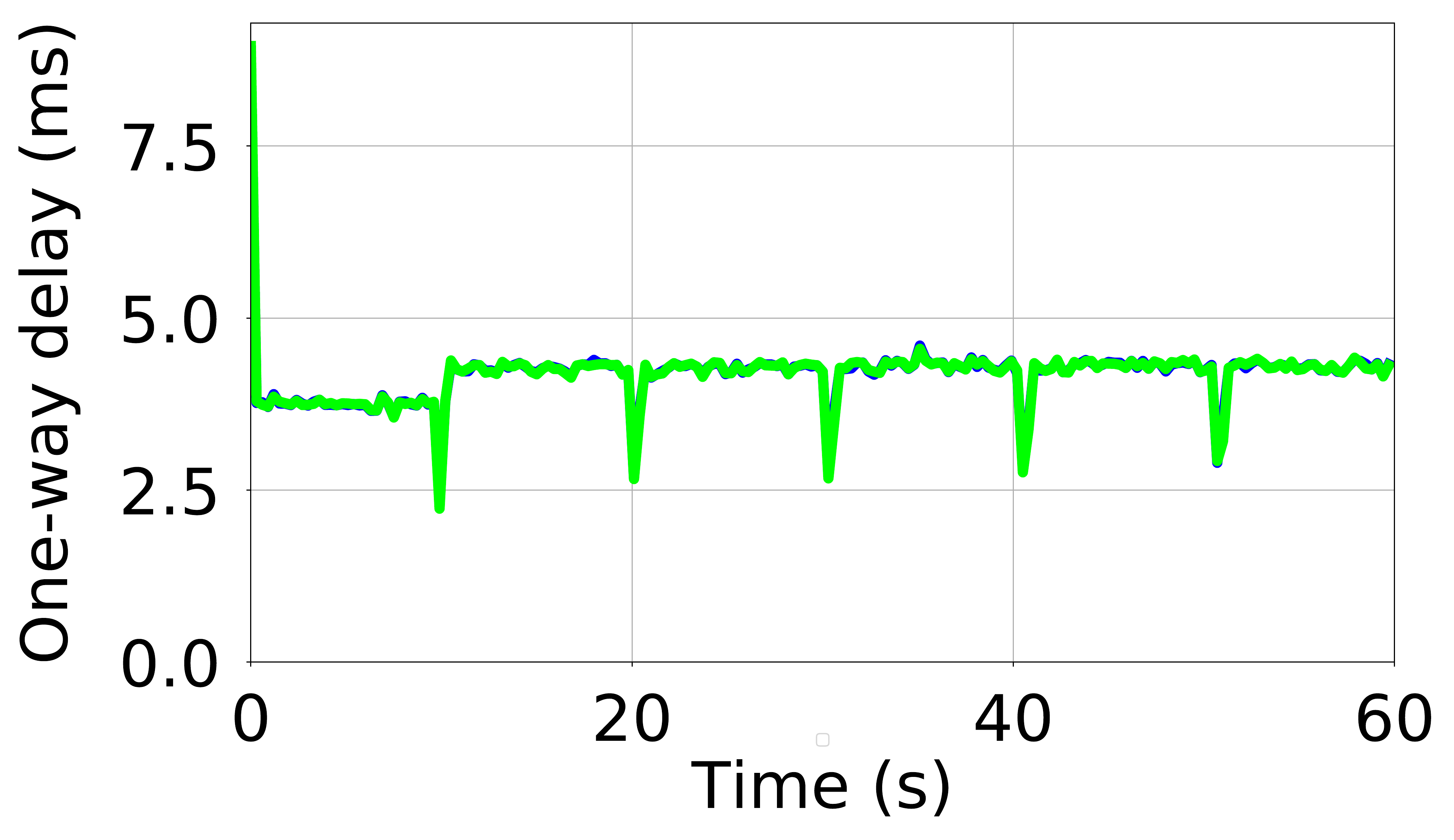} }}%
    \subfloat[]{{\includegraphics[width=\textwidth/3]{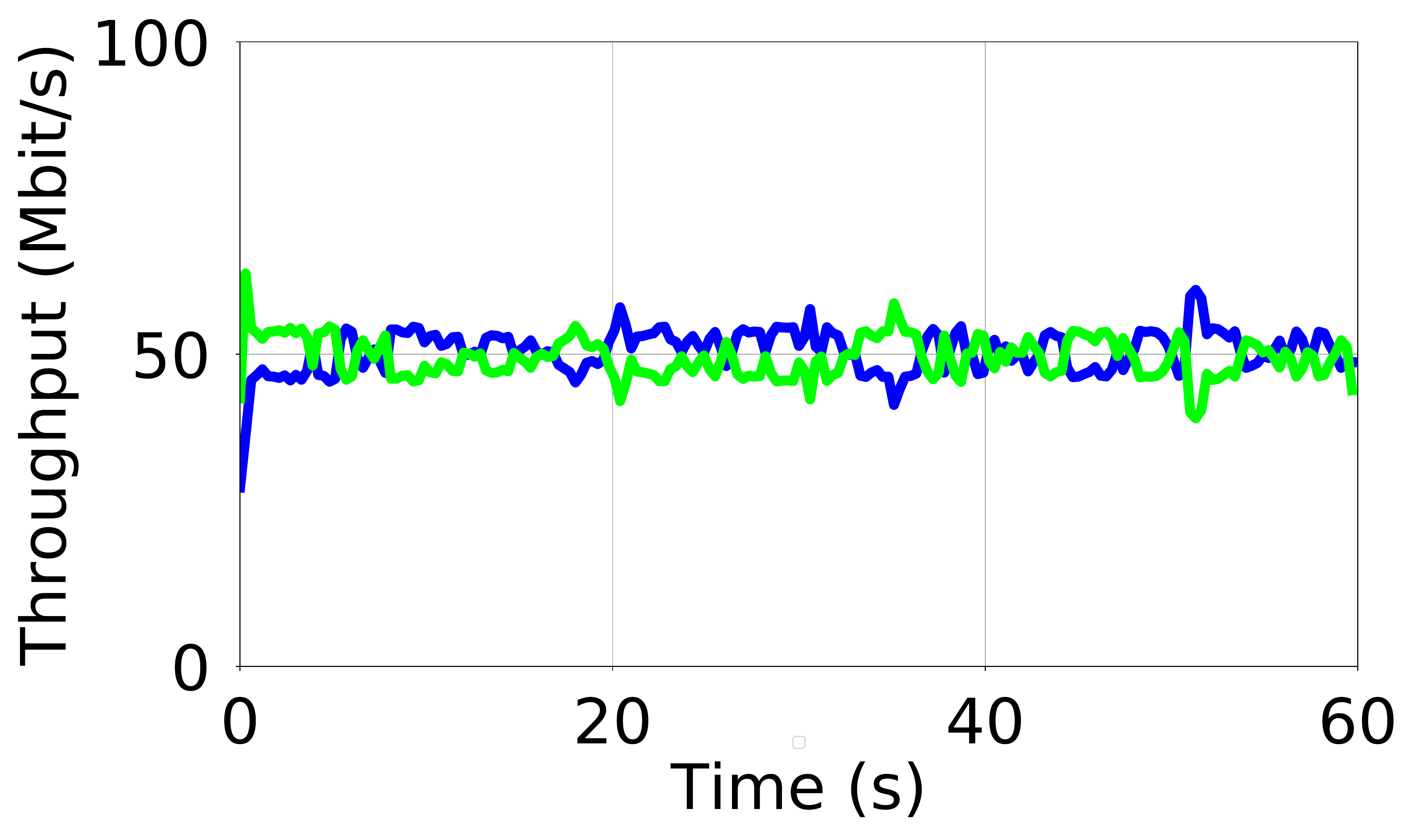} }}%
    \subfloat[]{{\includegraphics[width=\textwidth/3]{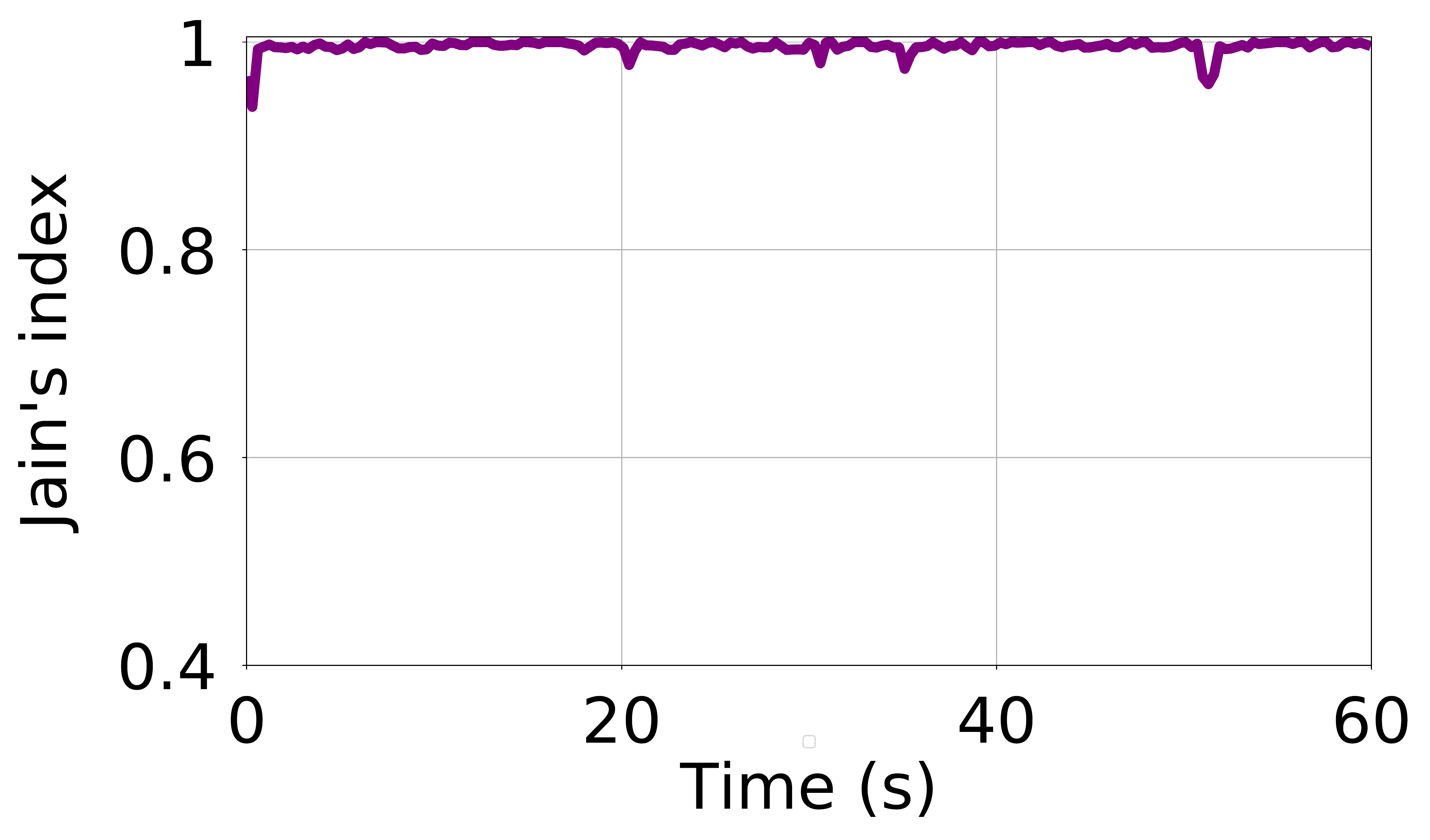} }}\\
    \vspace{-0.4cm}
    \caption{BW scenario: 2 BBR flows. The aggregation interval is 300 ms.\\The top-row plots are by the testbed, the bottom-row -- by CoCo-Beholder.}%
    \label{fig:fig4213}
\end{figure}

The results of CoCo-Beholder and the testbed~\cite{turkovic2019fifty} can be seen for Cubic in Table~\ref{tab:tab4211} and Figure~\ref{fig:fig4211}, for Vegas in Table~\ref{tab:tab4212} and Figure~\ref{fig:fig4212}, and for BBR in Table~\ref{tab:tab4213} and Figure~\ref{fig:fig4213}. All the tables are located on one page. All the figures are also on one page. Each CoCo-Beholder's one-way delay plot contains two curves: they just overlap.

The authors of the paper~\cite{turkovic2019fifty} write that in BW scenario with two flows, Vegas shows the best intra-fairness, BBR never converges to the same rate oscillating between 20 and 50 Mbit/s, and Cubic converges only after 15 seconds and its rate shows the greatest oscillation. This is indeed reflected in the corresponding plots from the paper. Meanwhile, CoCo-Beholder just shows the ideal intra-fairness for all the three schemes with their Jain's indices very close to 1, which can be easily spotted both in its tables and plots.

The paper~\cite{turkovic2019fifty} points out that Vegas shows a very small average RTT, which is the minimum of all the schemes, BBR shows also a very small RTT but a little bit greater than that of Vegas, and Cubic, on the contrary but as expected from a loss-based\nolinebreak[4] scheme, quickly fills up the queues and, therefore, shows a very big RTT. The results by CoCo-Beholder fully confirm these observations with one-way delays shown by the schemes.

\begin{table*}[h!]
\vspace{0.3cm}
\centering
\Large
\caption{BW scenario: 2 Cubic flows.}
\renewcommand{\arraystretch}{1.4} 
\resizebox*{\textwidth}{!}{\begin{tabu}{|c|cV{5}c|c|cV{5}c|cV{5}c|cV{5}c|c|cV{5}}
\cline{1-7}\cline{10-12}
\multirow{2}*{\parbox[c][2.5cm]{2.1cm}{\centering \bf \large Scheme}} &  \multicolumn{1}{c|}{\multirow{2}*{\parbox[c][2.5cm]{0.0cm}{}}} & \multicolumn{3}{c|}{\parbox[c][1cm]{2.7cm}{\bf \large \centering Rate (Mbps)}} & \parbox[c][1cm]{1.8cm}{\bf \large \centering Delay (ms)} & \multicolumn{1}{c|}{\parbox[c][1cm]{1.8cm}{\bf \large \centering RTT (ms)}}  & \multicolumn{1}{c}{} & \multicolumn{1}{c|}{} & \multicolumn{3}{c|}{\parbox[c][1cm]{2.5cm}{\bf \centering \large Jain's index}}\\ 
\cline{3-7}\cline{10-12}
 & \multicolumn{1}{c|}{} & \parbox[c][1.5cm]{1.8cm}{\centering \bf \normalsize CoCo-Beholder}  & \parbox[1cm]{1.8cm}{\centering \bf \normalsize Testbed} & \multicolumn{1}{c|}{\parbox{1.8cm}{\centering \pmb{$d_r$}}} & \parbox{1.8cm}{\centering \bf \normalsize CoCo-Beholder} & \multicolumn{1}{c|}{\parbox{1.8cm}{\centering \bf \normalsize Testbed}} &\multicolumn{1}{c}{} & \multicolumn{1}{c|}{}   & \parbox{1.8cm}{\centering \bf \normalsize CoCo-Beholder}& \parbox{1.8cm}{\centering \bf \normalsize Testbed} &\multicolumn{1}{c|}{ \parbox{1.8cm}{\centering \pmb{$d_r$}}}\\
\cline{1-2}\noalign{\vskip-1pt}\tabucline[2pt]{3-7}\noalign{\vskip-2pt}\cline{9-9}\tabucline[2pt]{10-12}
\multirow{2}*{\parbox[c][1cm]{2.1cm}{\centering cubic}} & \large \pmb{$\mu$} & \cellcolor{myg}49.59 & \cellcolor{myr}40.58 & 19.99\% & 103.83
 & 481.54 & & \large \pmb{$\mu$} & 0.99 & 0.93 & 6.48\%\\
\cline{2-7}\cline{9-12}
& \large \pmb{$\sigma$} & 2.67 &  &  & 1.35 &  &   &  \large \pmb{$\sigma$}& 0.01 & &\\
\cline{1-2}\tabucline[2pt]{3-7}\noalign{\vskip-2pt}\cline{9-9}\tabucline[2pt]{10-12}
\multirow{2}*{\parbox[c][1cm]{2.1cm}{\centering cubic}} & \large \pmb{$\mu$} & 50.25 & 45.52  & 9.88\% & 103.85 &  497.35  \\
\cline{2-7}
& \large \pmb{$\sigma$} & 2.67 &  &  & 1.34 &    \\
\cline{1-2}\tabucline[2pt]{3-7}
\end{tabu}}
\label{tab:tab4211}
\end{table*}

\begin{table*}[h!]
\vspace{0.3cm}
\centering
\Large
\caption{BW scenario: 2 Vegas flows.}
\renewcommand{\arraystretch}{1.4} 
\resizebox*{\textwidth}{!}{\begin{tabu}{|c|cV{5}c|c|cV{5}c|cV{5}c|cV{5}c|c|cV{5}}
\cline{1-7}\cline{10-12}
\multirow{2}*{\parbox[c][2.5cm]{2.1cm}{\centering \bf \large Scheme}} &  \multicolumn{1}{c|}{\multirow{2}*{\parbox[c][2.5cm]{0.0cm}{}}} & \multicolumn{3}{c|}{\parbox[c][1cm]{2.7cm}{\bf \large \centering Rate (Mbps)}} & \parbox[c][1cm]{1.8cm}{\bf \large \centering Delay (ms)} & \multicolumn{1}{c|}{\parbox[c][1cm]{1.8cm}{\bf \large \centering RTT (ms)}}  & \multicolumn{1}{c}{} & \multicolumn{1}{c|}{} & \multicolumn{3}{c|}{\parbox[c][1cm]{2.5cm}{\bf \centering \large Jain's index}}\\ 
\cline{3-7}\cline{10-12}
 & \multicolumn{1}{c|}{} & \parbox[c][1.5cm]{1.8cm}{\centering \bf \normalsize CoCo-Beholder}  & \parbox[1cm]{1.8cm}{\centering \bf \normalsize Testbed} & \multicolumn{1}{c|}{\parbox{1.8cm}{\centering \pmb{$d_r$}}} & \parbox{1.8cm}{\centering \bf \normalsize CoCo-Beholder} & \multicolumn{1}{c|}{\parbox{1.8cm}{\centering \bf \normalsize Testbed}} &\multicolumn{1}{c}{} & \multicolumn{1}{c|}{}   & \parbox{1.8cm}{\centering \bf \normalsize CoCo-Beholder}& \parbox{1.8cm}{\centering \bf \normalsize Testbed} &\multicolumn{1}{c|}{ \parbox{1.8cm}{\centering \pmb{$d_r$}}}\\
\cline{1-2}\noalign{\vskip-1pt}\tabucline[2pt]{3-7}\noalign{\vskip-2pt}\cline{9-9}\tabucline[2pt]{10-12}
\multirow{2}*{\parbox[c][1cm]{2.1cm}{\centering vegas}} & \large \pmb{$\mu$} & \cellcolor{myg}49.47 & \cellcolor{myr}36.48 & 30.23\% & 1.28 & 2.18  & & \large \pmb{$\mu$} &0.99 &0.98 & 0.80\%\\
\cline{2-7}\cline{9-12}
& \large \pmb{$\sigma$} & 1.63 &  &  & 0.21 &  &   &  \large \pmb{$\sigma$}& 0.01& &\\
\cline{1-2}\tabucline[2pt]{3-7}\noalign{\vskip-2pt}\cline{9-9}\tabucline[2pt]{10-12}
\multirow{2}*{\parbox[c][1cm]{2.1cm}{\centering vegas}} & \large \pmb{$\mu$} & \cellcolor{myg}50.37 & \cellcolor{myr}36.52 & 31.88\% &1.28  &  2.18  \\
\cline{2-7}
& \large \pmb{$\sigma$} & 1.65 &  &  & 0.20 &    \\
\cline{1-2}\tabucline[2pt]{3-7}
\end{tabu}}
\label{tab:tab4212}
\end{table*}

\begin{table*}[h!]
\vspace{0.3cm}
\centering
\Large
\caption{BW scenario: 2 BBR flows.}
\renewcommand{\arraystretch}{1.4} 
\resizebox*{\textwidth}{!}{\begin{tabu}{|c|cV{5}c|c|cV{5}c|cV{5}c|cV{5}c|c|cV{5}}
\cline{1-7}\cline{10-12}
\multirow{2}*{\parbox[c][2.5cm]{2.1cm}{\centering \bf \large Scheme}} &  \multicolumn{1}{c|}{\multirow{2}*{\parbox[c][2.5cm]{0.0cm}{}}} & \multicolumn{3}{c|}{\parbox[c][1cm]{2.7cm}{\bf \large \centering Rate (Mbps)}} & \parbox[c][1cm]{1.8cm}{\bf \large \centering Delay (ms)} & \multicolumn{1}{c|}{\parbox[c][1cm]{1.8cm}{\bf \large \centering RTT (ms)}}  & \multicolumn{1}{c}{} & \multicolumn{1}{c|}{} & \multicolumn{3}{c|}{\parbox[c][1cm]{2.5cm}{\bf \centering \large Jain's index}}\\ 
\cline{3-7}\cline{10-12}
 & \multicolumn{1}{c|}{} & \parbox[c][1.5cm]{1.8cm}{\centering \bf \normalsize CoCo-Beholder}  & \parbox[1cm]{1.8cm}{\centering \bf \normalsize Testbed} & \multicolumn{1}{c|}{\parbox{1.8cm}{\centering \pmb{$d_r$}}} & \parbox{1.8cm}{\centering \bf \normalsize CoCo-Beholder} & \multicolumn{1}{c|}{\parbox{1.8cm}{\centering \bf \normalsize Testbed}} &\multicolumn{1}{c}{} & \multicolumn{1}{c|}{}   & \parbox{1.8cm}{\centering \bf \normalsize CoCo-Beholder}& \parbox{1.8cm}{\centering \bf \normalsize Testbed} &\multicolumn{1}{c|}{ \parbox{1.8cm}{\centering \pmb{$d_r$}}}\\
\cline{1-2}\noalign{\vskip-1pt}\tabucline[2pt]{3-7}\noalign{\vskip-2pt}\cline{9-9}\tabucline[2pt]{10-12}
\multirow{2}*{\parbox[c][1cm]{2.1cm}{\centering bbr}} & \large \pmb{$\mu$} & \cellcolor{myg}50.35 & \cellcolor{myr}37.15 & 30.17\% & 4.17 & 4.08 & & \large \pmb{$\mu$} &\cellcolor{myg}0.99 & \cellcolor{myr}0.87 & 13.21\%\\
\cline{2-7}\cline{9-12}
& \large \pmb{$\sigma$} & 1.32 &  &  & 0.01 &  &   &  \large \pmb{$\sigma$}& 0.01& &\\
\cline{1-2}\tabucline[2pt]{3-7}\noalign{\vskip-2pt}\cline{9-9}\tabucline[2pt]{10-12}
\multirow{2}*{\parbox[c][1cm]{2.1cm}{\centering bbr}} & \large \pmb{$\mu$} &\cellcolor{myg}49.50 & \cellcolor{myr}32.64 & 41.05\% & 4.17 & 4.05   \\
\cline{2-7}
& \large \pmb{$\sigma$} & 1.32 &  &  & 0.01 &    \\
\cline{1-2}\tabucline[2pt]{3-7}
\end{tabu}}
\label{tab:tab4213}
\end{table*}

\FloatBarrier
\subsection{Inter-Fairness For Two Flows}
\label{subsec:ssec2}

\begin{figure}[p!]
\vspace*{-0.3cm}
\captionsetup[subfigure]{labelformat=empty}
    \centering
    \subfloat[]{{\includegraphics[width=\textwidth]{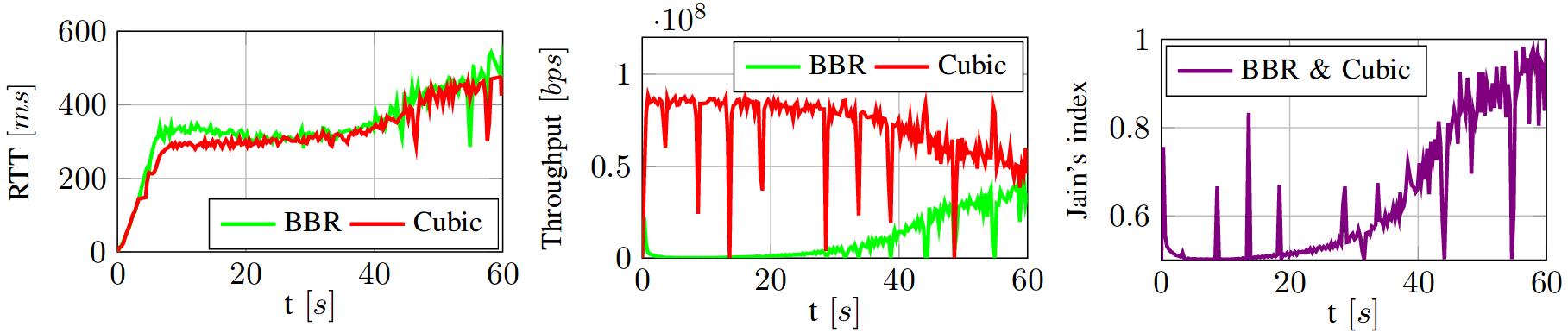} }}\\
    \vspace{-0.7cm}
    \subfloat[]{{\includegraphics[width=\textwidth/3]{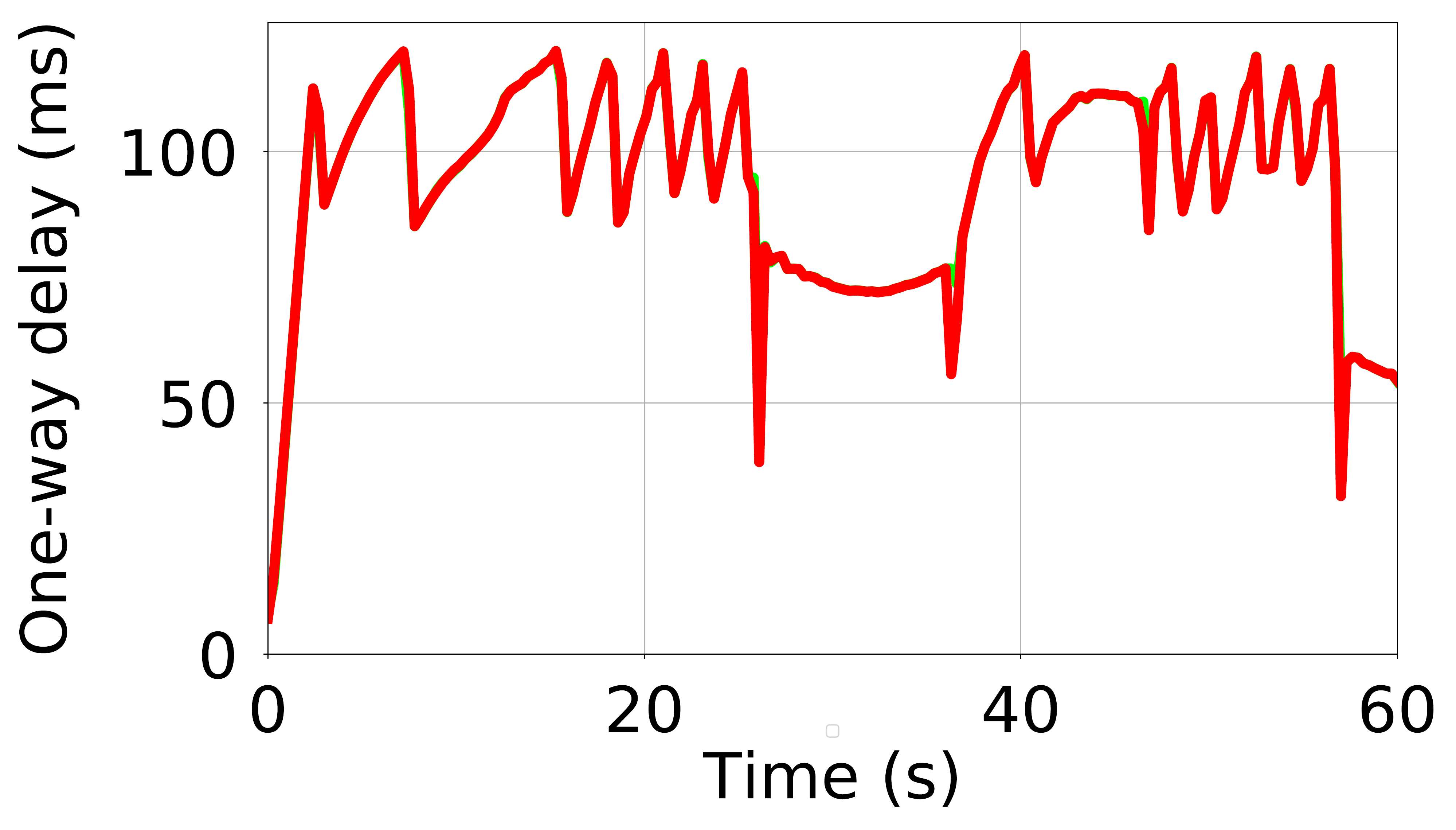} }}%
    \subfloat[]{{\includegraphics[width=\textwidth/3]{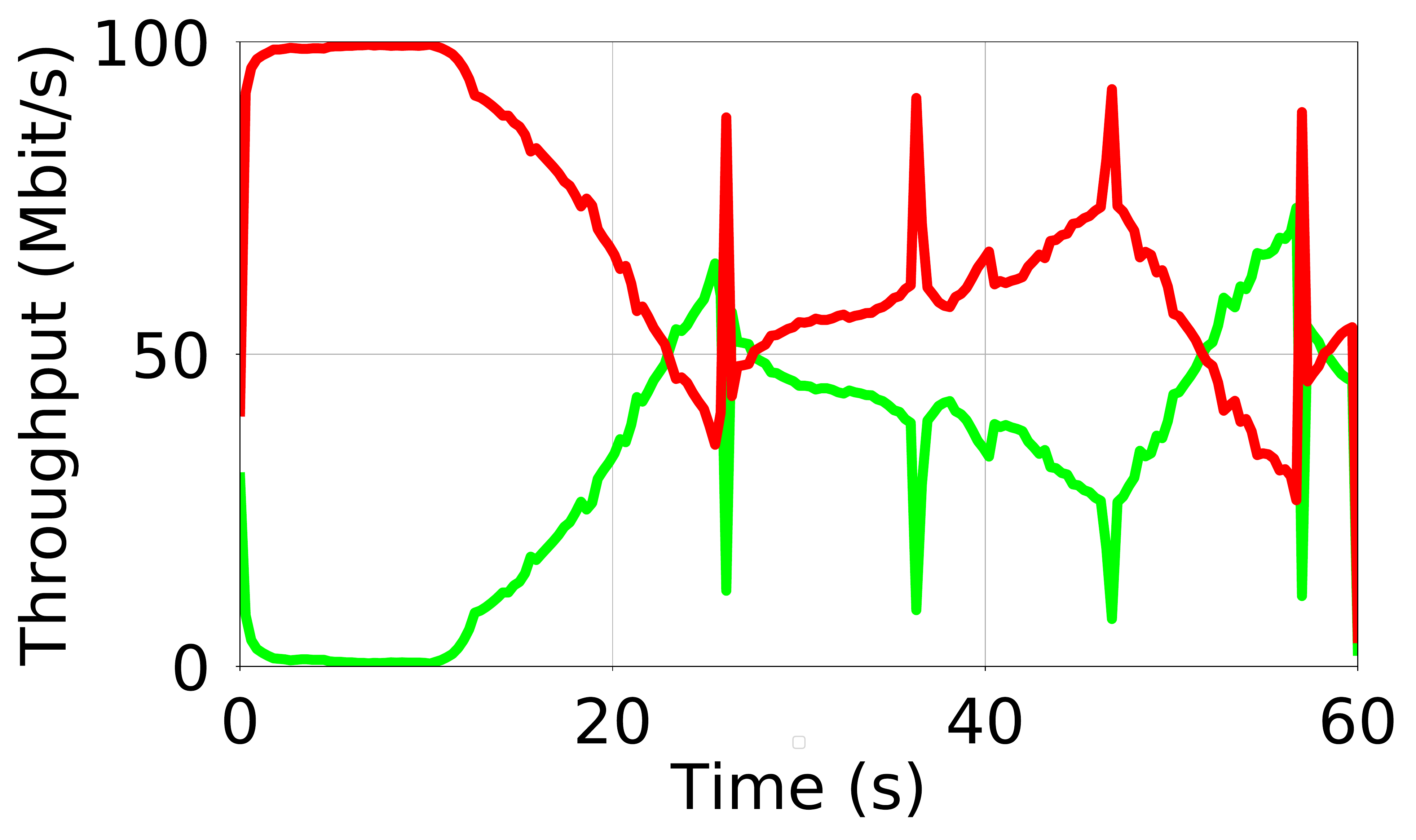} }}%
    \subfloat[]{{\includegraphics[width=\textwidth/3]{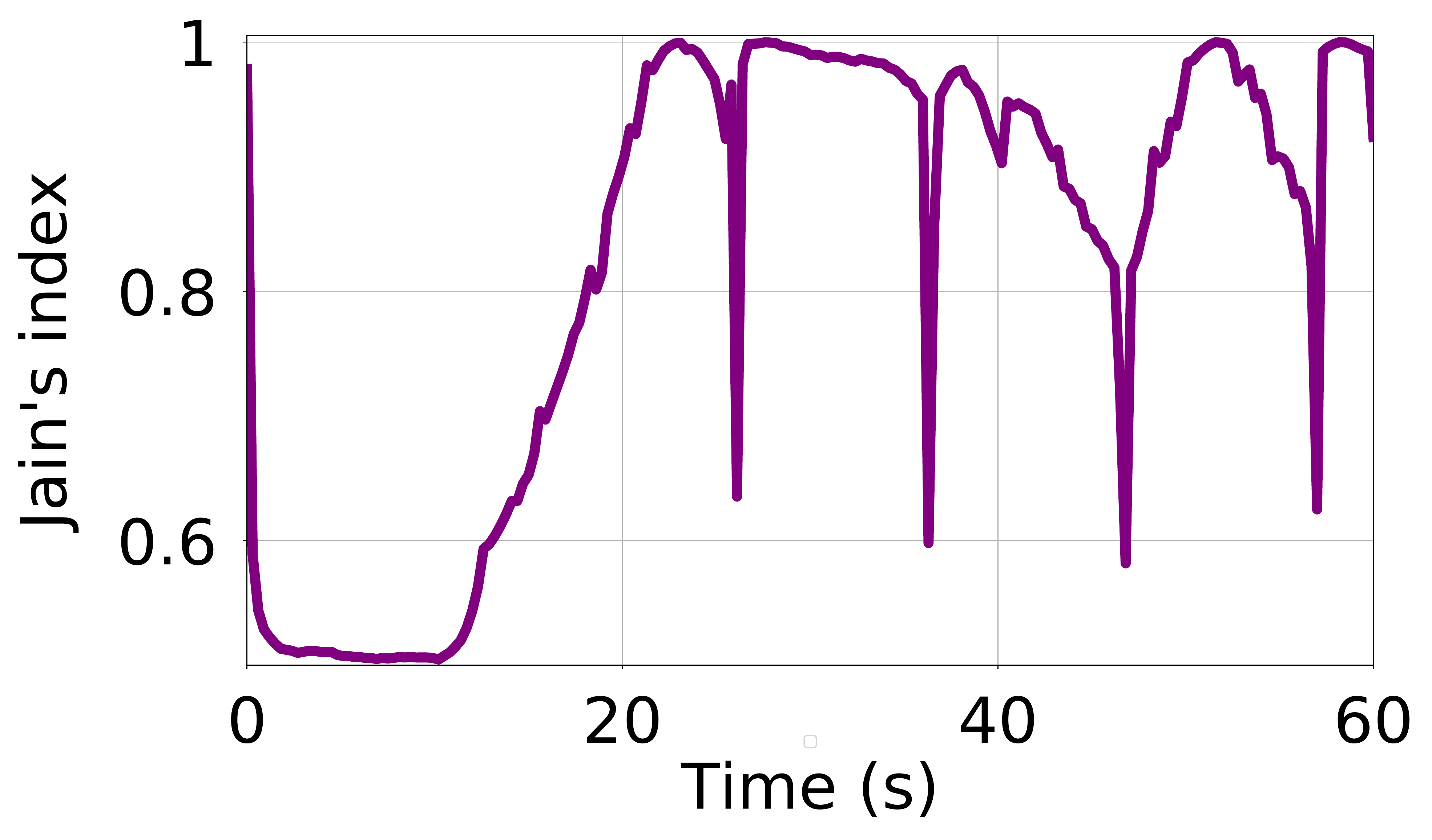} }}\\
    \vspace{-0.5cm}
    \caption{BW scenario: BBR \& Cubic. The aggregation interval is 300 ms.\\The top-row plots are by the testbed, the bottom-row -- by CoCo-Beholder.}%
    \label{fig:fig4221}
\end{figure}

\begin{figure}[h!]
\vspace*{-0.2cm}
\captionsetup[subfigure]{labelformat=empty}
    \centering
    \subfloat[]{{\includegraphics[width=\textwidth]{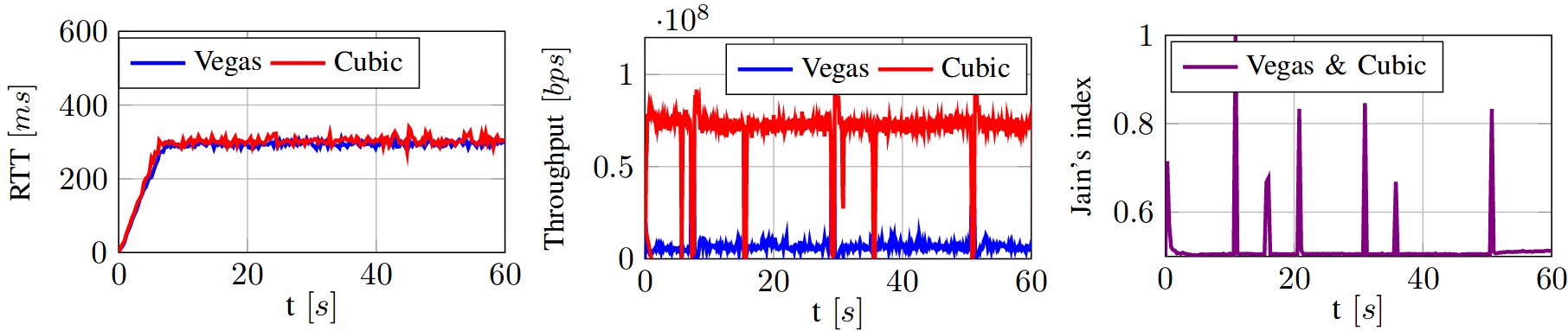} }}\\
    \vspace{-0.7cm}
    \subfloat[]{{\includegraphics[width=\textwidth/3]{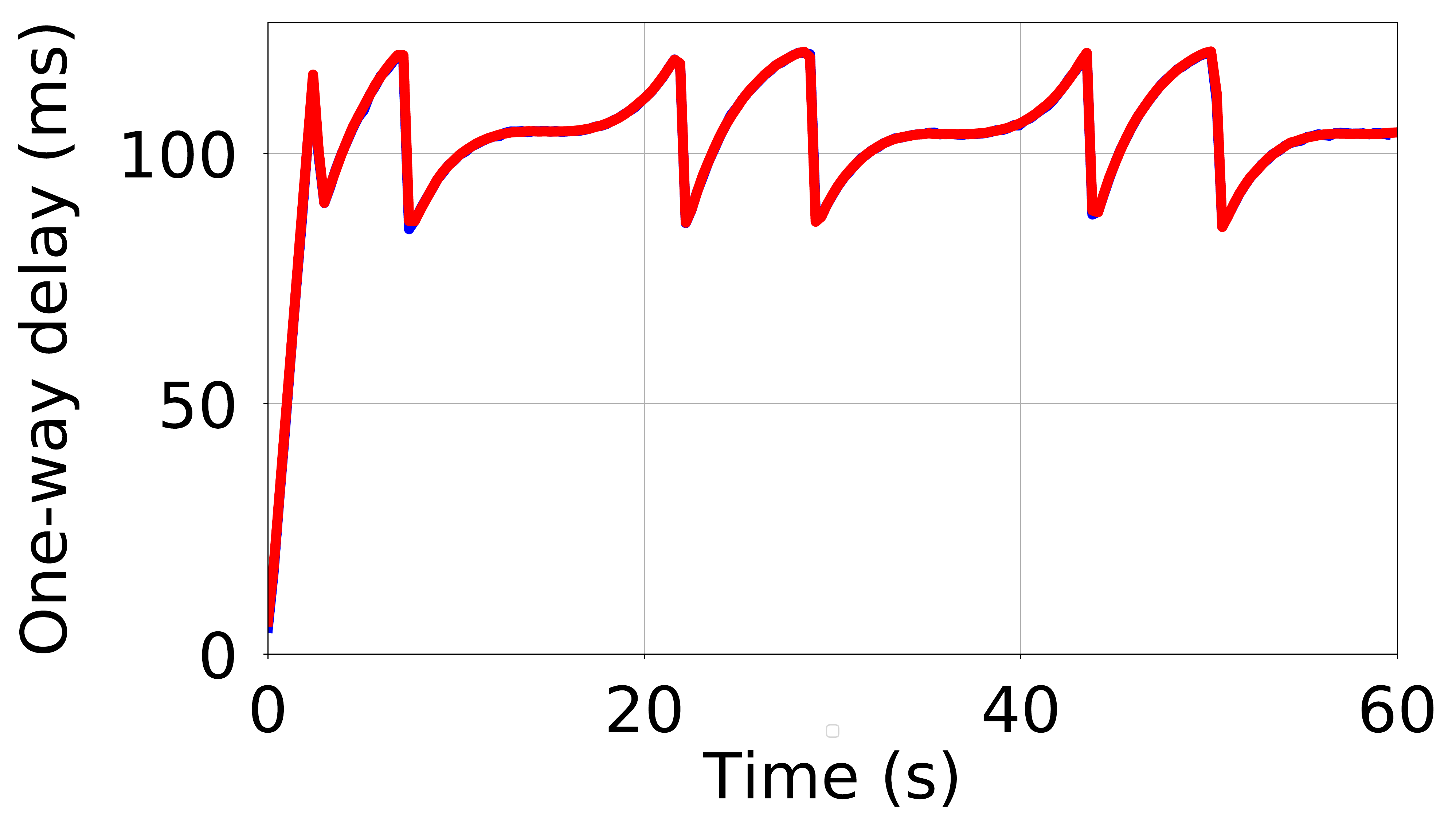} }}%
    \subfloat[]{{\includegraphics[width=\textwidth/3]{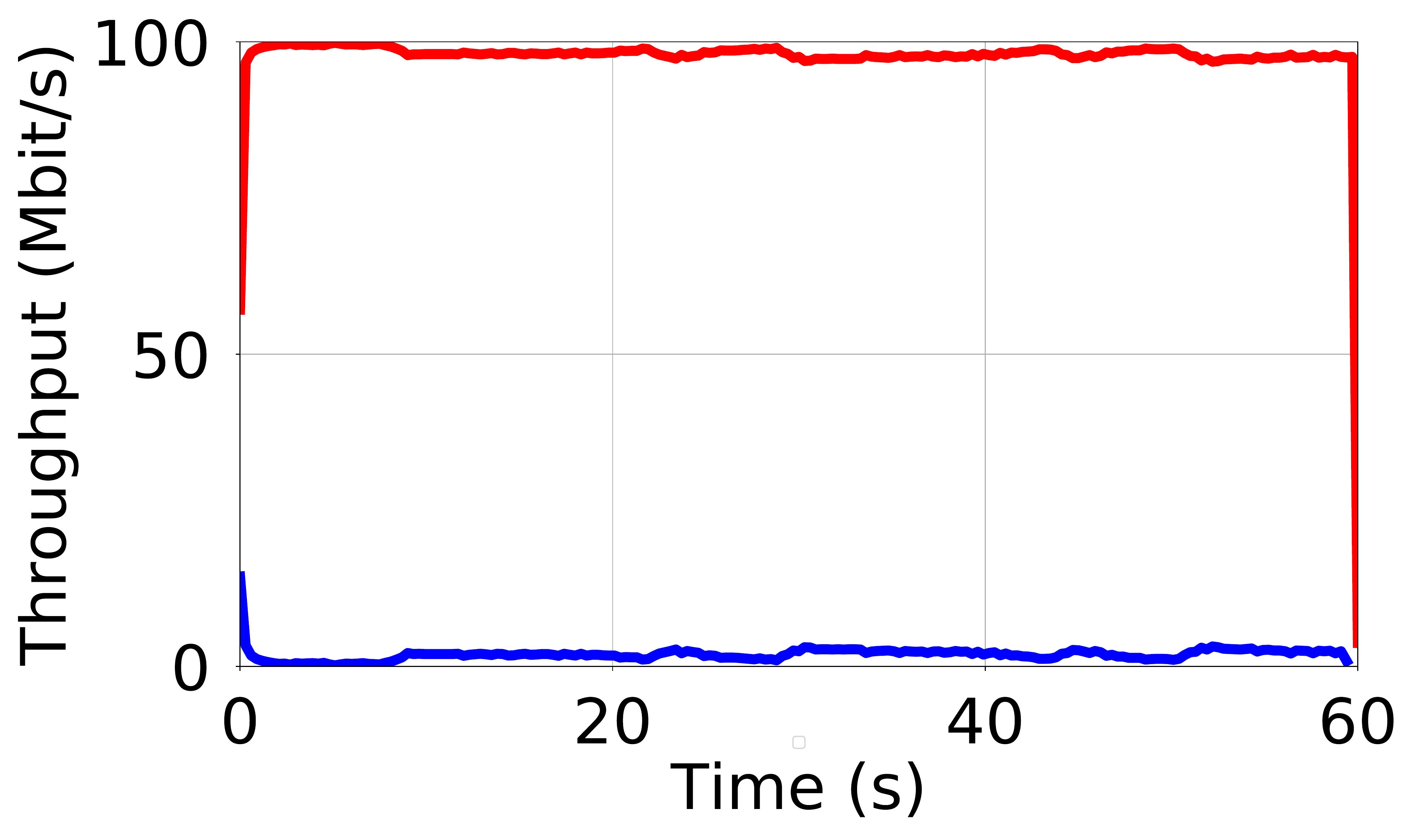} }}%
    \subfloat[]{{\includegraphics[width=\textwidth/3]{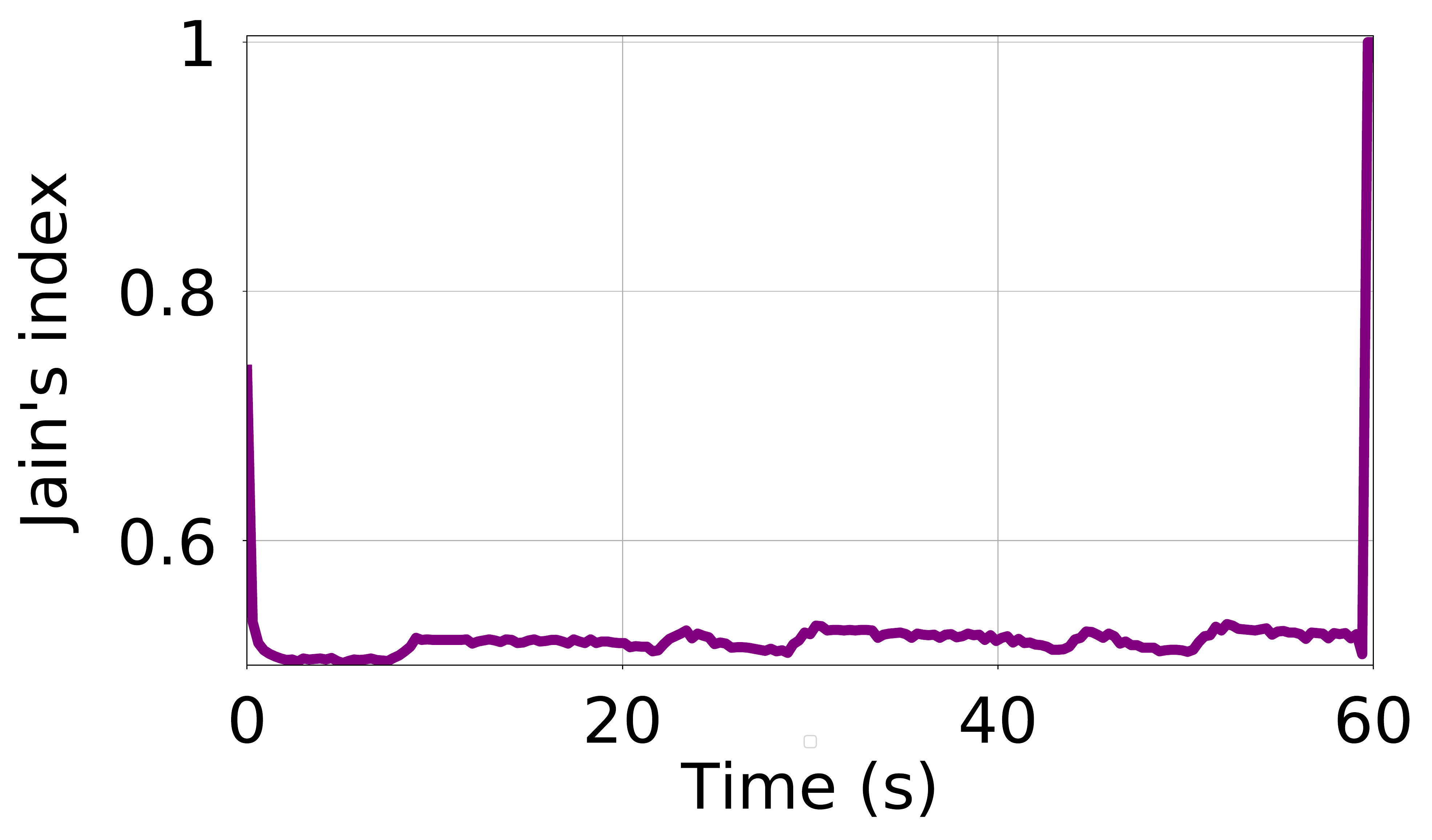} }}\\
    \vspace{-0.5cm}
    \caption{BW scenario: Vegas \& Cubic. The aggregation interval is 300 ms.\\The top-row plots are by the testbed, the bottom-row -- by CoCo-Beholder.}%
    \label{fig:fig4222}
\end{figure}

\begin{figure}[h!]
\vspace*{-0.2cm}
\captionsetup[subfigure]{labelformat=empty}
    \centering
    \subfloat[]{{\includegraphics[width=\textwidth]{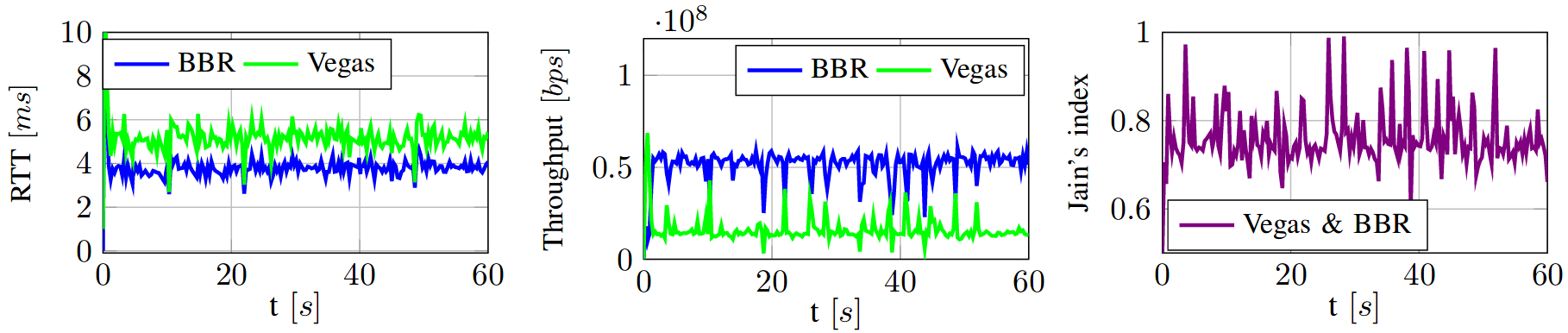} }}\\
    \vspace{-0.7cm}
    \subfloat[]{{\includegraphics[width=\textwidth/3]{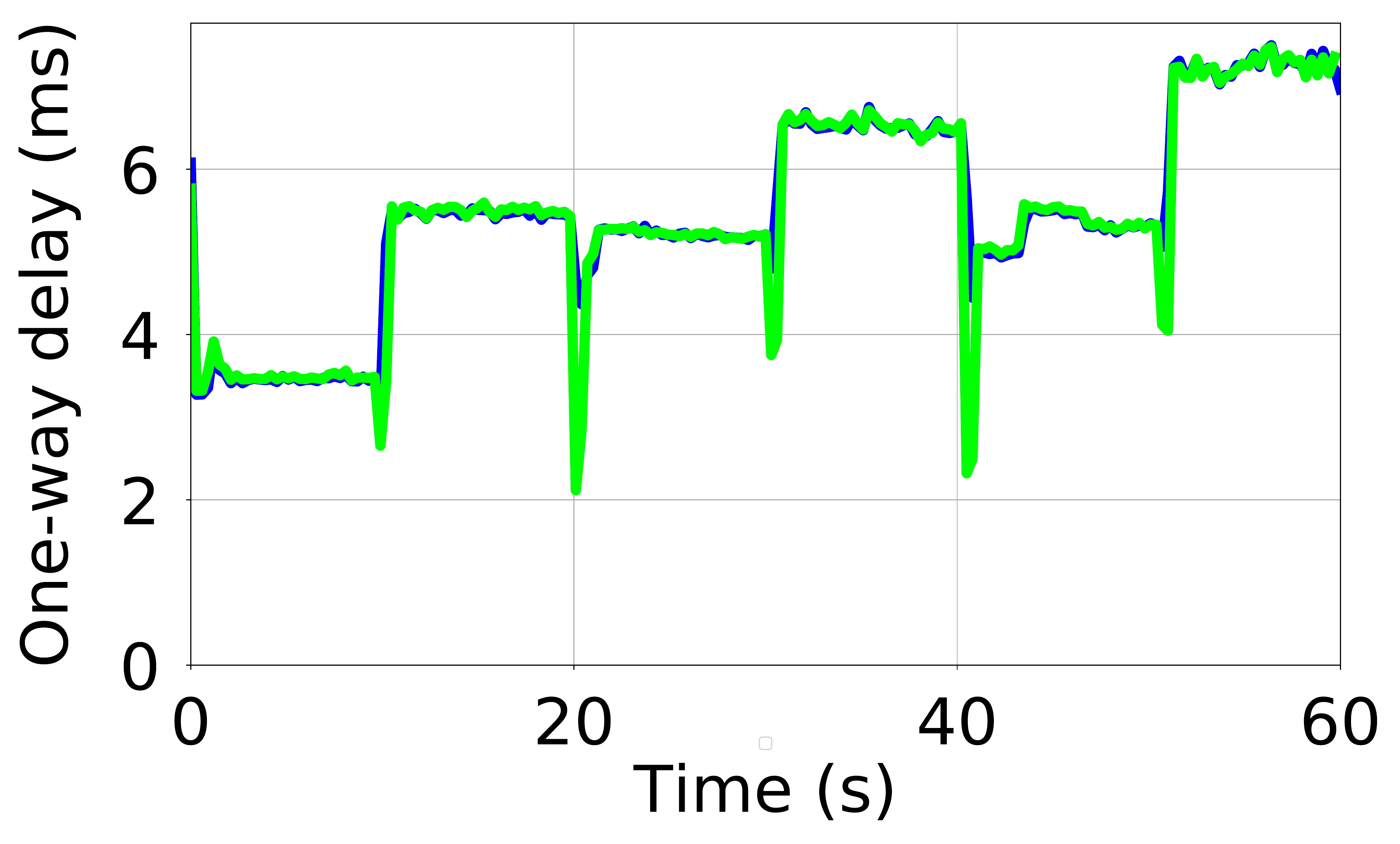} }}%
    \subfloat[]{{\includegraphics[width=\textwidth/3]{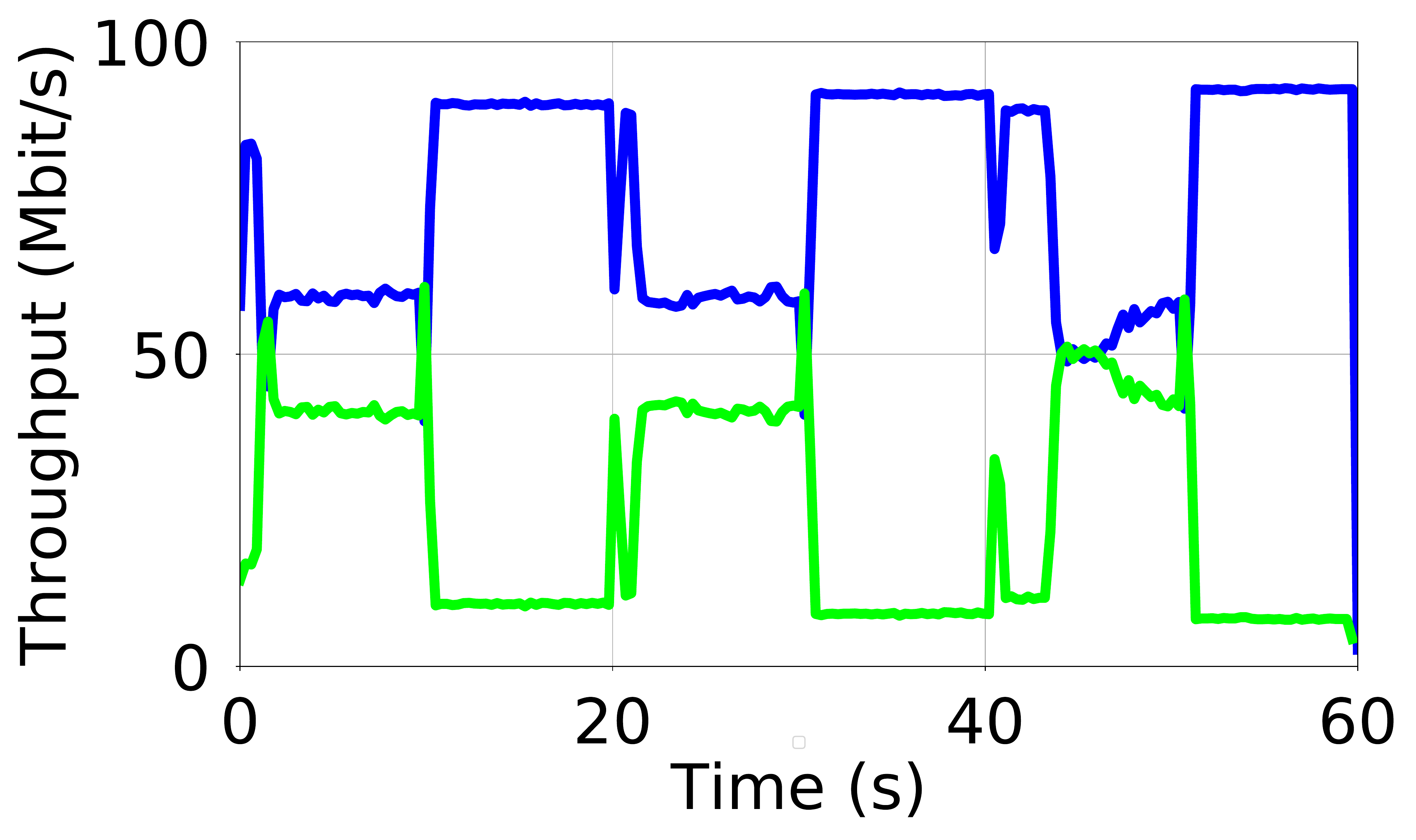} }}%
    \subfloat[]{{\includegraphics[width=\textwidth/3]{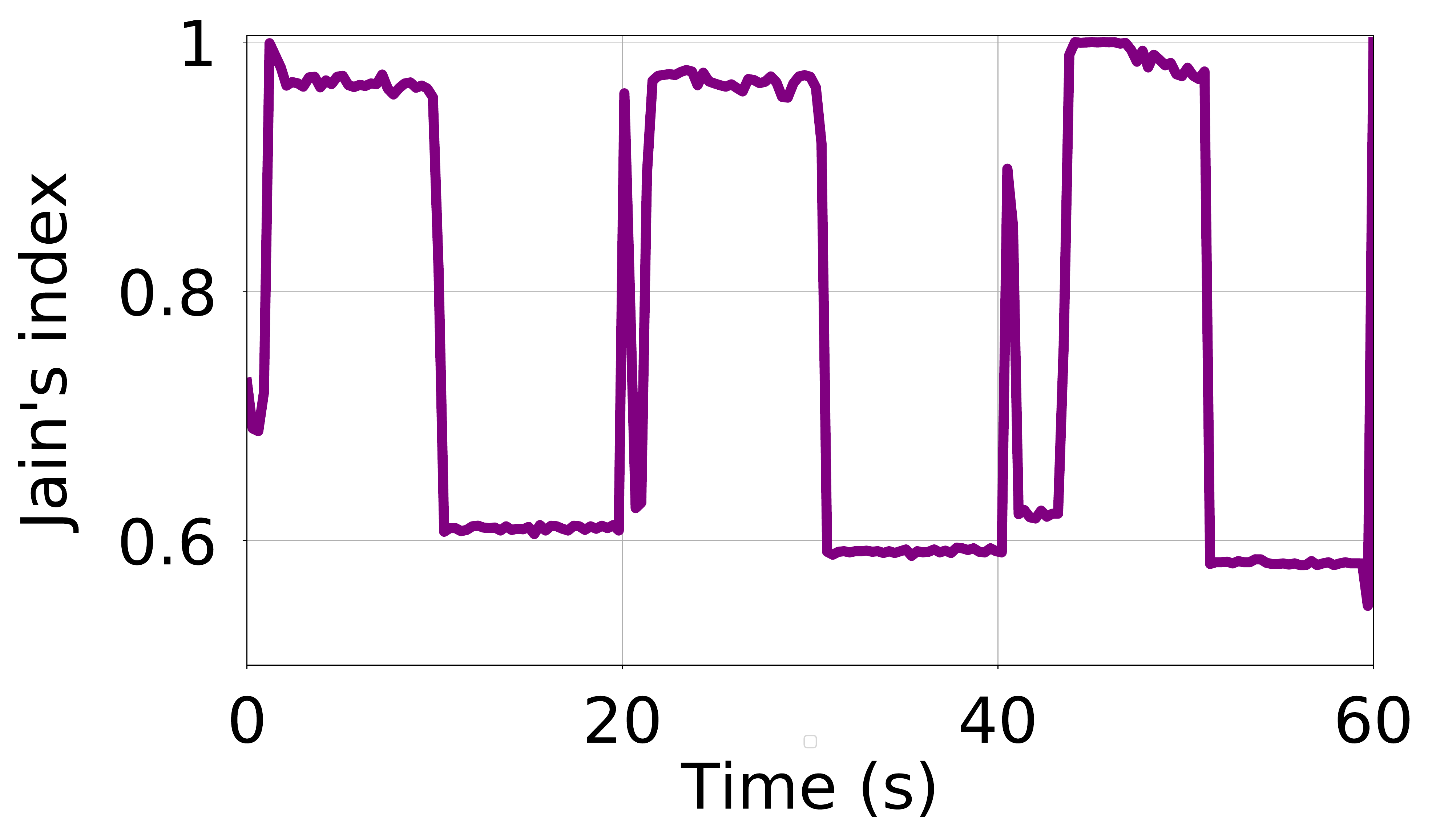} }}\\
    \vspace{-0.4cm}
    \caption{BW scenario: BBR \& Vegas. The aggregation interval is 300 ms.\\The top-row plots are by the testbed, the bottom-row -- by CoCo-Beholder.}%
    \label{fig:fig4223}
\end{figure}

The results of CoCo-Beholder and the testbed~\cite{turkovic2019fifty} can be seen for BBR\&Cubic in Table~\ref{tab:tab4221} and Figure~\ref{fig:fig4221}, for Vegas\&Cubic in Table~\ref{tab:tab4222} and Figure~\ref{fig:fig4222}, and for BBR\&Vegas in Table~\ref{tab:tab4223} and Figure~\ref{fig:fig4223}. Again, each CoCo-Beholder's one-way delay plot contains two curves: one of the curves overlaps the other.

Both CoCo-Beholder and the testbed~\cite{turkovic2019fifty} showed
that delay-based Vegas has a very low rate when sharing a bottleneck link with a loss-based or a hybrid scheme. While no loss is detected, loss-based Cubic continues to increase the congestion window, so the queues and, thus, the RTT observed by Vegas grow excessively preventing it from\nolinebreak[4] \mbox{raising}\nolinebreak[4] the\nolinebreak[4] rate.

\begin{table*}[h!]
\vspace{0.3cm}
\centering
\Large
\caption{BW scenario: BBR \& Cubic.}
\renewcommand{\arraystretch}{1.4} 
\resizebox*{\textwidth}{!}{\begin{tabu}{|c|cV{5}c|c|cV{5}c|cV{5}c|cV{5}c|c|cV{5}}
\cline{1-7}\cline{10-12}
\multirow{2}*{\parbox[c][2.5cm]{2.1cm}{\centering \bf \large Scheme}} &  \multicolumn{1}{c|}{\multirow{2}*{\parbox[c][2.5cm]{0.0cm}{}}} & \multicolumn{3}{c|}{\parbox[c][1cm]{2.7cm}{\bf \large \centering Rate (Mbps)}} & \parbox[c][1cm]{1.8cm}{\bf \large \centering Delay (ms)} & \multicolumn{1}{c|}{\parbox[c][1cm]{1.8cm}{\bf \large \centering RTT (ms)}}  & \multicolumn{1}{c}{} & \multicolumn{1}{c|}{} & \multicolumn{3}{c|}{\parbox[c][1cm]{2.5cm}{\bf \centering \large Jain's index}}\\ 
\cline{3-7}\cline{10-12}
 & \multicolumn{1}{c|}{} & \parbox[c][1.5cm]{1.8cm}{\centering \bf \normalsize CoCo-Beholder}  & \parbox[1cm]{1.8cm}{\centering \bf \normalsize Testbed} & \multicolumn{1}{c|}{\parbox{1.8cm}{\centering \pmb{$d_r$}}} & \parbox{1.8cm}{\centering \bf \normalsize CoCo-Beholder} & \multicolumn{1}{c|}{\parbox{1.8cm}{\centering \bf \normalsize Testbed}} &\multicolumn{1}{c}{} & \multicolumn{1}{c|}{}   & \parbox{1.8cm}{\centering \bf \normalsize CoCo-Beholder}& \parbox{1.8cm}{\centering \bf \normalsize Testbed} &\multicolumn{1}{c|}{ \parbox{1.8cm}{\centering \pmb{$d_r$}}}\\
\cline{1-2}\noalign{\vskip-1pt}\tabucline[2pt]{3-7}\noalign{\vskip-2pt}\cline{9-9}\tabucline[2pt]{10-12}
\multirow{2}*{\parbox[c][1cm]{2.1cm}{\centering bbr}} & \large \pmb{$\mu$} & \cellcolor{myg}32.25 & \cellcolor{myr}10.53 & 101.54\% & 91.55 & 308.34 &  & \large \pmb{$\mu$} &\cellcolor{myg}0.80 & \cellcolor{myr}0.62 & 25.91\%\\
\cline{2-7}\cline{9-12}
& \large \pmb{$\sigma$} & 3.53 &  &  & 6.83 &  &   &  \large \pmb{$\sigma$}& 0.03& &\\
\cline{1-2}\tabucline[2pt]{3-7}\noalign{\vskip-2pt}\cline{9-9}\tabucline[2pt]{10-12}
\multirow{2}*{\parbox[c][1cm]{2.1cm}{\centering cubic}} & \large \pmb{$\mu$} & 67.60 & 73.22 & 7.98\% & 91.12 &  379.90 \\
\cline{2-7}
& \large \pmb{$\sigma$} & 3.53 &  &  & 6.71 &    \\
\cline{1-2}\tabucline[2pt]{3-7}
\end{tabu}}
\label{tab:tab4221}
\end{table*}

\begin{table*}[h!]
\vspace{0.3cm}
\centering
\Large
\caption{BW scenario: Vegas \& Cubic.}
\renewcommand{\arraystretch}{1.4} 
\resizebox*{\textwidth}{!}{\begin{tabu}{|c|cV{5}c|c|cV{5}c|cV{5}c|cV{5}c|c|cV{5}}
\cline{1-7}\cline{10-12}
\multirow{2}*{\parbox[c][2.5cm]{2.1cm}{\centering \bf \large Scheme}} &  \multicolumn{1}{c|}{\multirow{2}*{\parbox[c][2.5cm]{0.0cm}{}}} & \multicolumn{3}{c|}{\parbox[c][1cm]{2.7cm}{\bf \large \centering Rate (Mbps)}} & \parbox[c][1cm]{1.8cm}{\bf \large \centering Delay (ms)} & \multicolumn{1}{c|}{\parbox[c][1cm]{1.8cm}{\bf \large \centering RTT (ms)}}  & \multicolumn{1}{c}{} & \multicolumn{1}{c|}{} & \multicolumn{3}{c|}{\parbox[c][1cm]{2.5cm}{\bf \centering \large Jain's index}}\\ 
\cline{3-7}\cline{10-12}
 & \multicolumn{1}{c|}{} & \parbox[c][1.5cm]{1.8cm}{\centering \bf \normalsize CoCo-Beholder}  & \parbox[1cm]{1.8cm}{\centering \bf \normalsize Testbed} & \multicolumn{1}{c|}{\parbox{1.8cm}{\centering \pmb{$d_r$}}} & \parbox{1.8cm}{\centering \bf \normalsize CoCo-Beholder} & \multicolumn{1}{c|}{\parbox{1.8cm}{\centering \bf \normalsize Testbed}} &\multicolumn{1}{c}{} & \multicolumn{1}{c|}{}   & \parbox{1.8cm}{\centering \bf \normalsize CoCo-Beholder}& \parbox{1.8cm}{\centering \bf \normalsize Testbed} &\multicolumn{1}{c|}{ \parbox{1.8cm}{\centering \pmb{$d_r$}}}\\
\cline{1-2}\noalign{\vskip-1pt}\tabucline[2pt]{3-7}\noalign{\vskip-2pt}\cline{9-9}\tabucline[2pt]{10-12}
\multirow{2}*{\parbox[c][1cm]{2.1cm}{\centering vegas}} & \large \pmb{$\mu$} & \cellcolor{myg}1.79 & \cellcolor{myr}0.71 & 86.37\% & 102.05 & 279.39 & & \large \pmb{$\mu$} & 0.52 & 0.51  & 2.41\%\\
\cline{2-7}\cline{9-12}
& \large \pmb{$\sigma$} &0.40  &  &  & 0.65 &  &   &  \large \pmb{$\sigma$}& 0.01& &\\
\cline{1-2}\tabucline[2pt]{3-7}\noalign{\vskip-2pt}\cline{9-9}\tabucline[2pt]{10-12}
\multirow{2}*{\parbox[c][1cm]{2.1cm}{\centering cubic}} & \large \pmb{$\mu$} & \cellcolor{myg}98.04 & \cellcolor{myr}82.63 & 17.06\% & 102.79 & 228.79\\
\cline{2-7}
& \large \pmb{$\sigma$} & 0.41 &  &  & 0.73 &    \\
\cline{1-2}\tabucline[2pt]{3-7}
\end{tabu}}
\label{tab:tab4222}
\end{table*}

\begin{table*}[h!]
\vspace{0.3cm}
\centering
\Large
\caption{BW scenario: BBR \& Vegas.}
\renewcommand{\arraystretch}{1.4} 
\resizebox*{\textwidth}{!}{\begin{tabu}{|c|cV{5}c|c|cV{5}c|cV{5}c|cV{5}c|c|cV{5}}
\cline{1-7}\cline{10-12}
\multirow{2}*{\parbox[c][2.5cm]{2.1cm}{\centering \bf \large Scheme}} &  \multicolumn{1}{c|}{\multirow{2}*{\parbox[c][2.5cm]{0.0cm}{}}} & \multicolumn{3}{c|}{\parbox[c][1cm]{2.7cm}{\bf \large \centering Rate (Mbps)}} & \parbox[c][1cm]{1.8cm}{\bf \large \centering Delay (ms)} & \multicolumn{1}{c|}{\parbox[c][1cm]{1.8cm}{\bf \large \centering RTT (ms)}}  & \multicolumn{1}{c}{} & \multicolumn{1}{c|}{} & \multicolumn{3}{c|}{\parbox[c][1cm]{2.5cm}{\bf \centering \large Jain's index}}\\ 
\cline{3-7}\cline{10-12}
 & \multicolumn{1}{c|}{} & \parbox[c][1.5cm]{1.8cm}{\centering \bf \normalsize CoCo-Beholder}  & \parbox[1cm]{1.8cm}{\centering \bf \normalsize Testbed} & \multicolumn{1}{c|}{\parbox{1.8cm}{\centering \pmb{$d_r$}}} & \parbox{1.8cm}{\centering \bf \normalsize CoCo-Beholder} & \multicolumn{1}{c|}{\parbox{1.8cm}{\centering \bf \normalsize Testbed}} &\multicolumn{1}{c}{} & \multicolumn{1}{c|}{}   & \parbox{1.8cm}{\centering \bf \normalsize CoCo-Beholder}& \parbox{1.8cm}{\centering \bf \normalsize Testbed} &\multicolumn{1}{c|}{ \parbox{1.8cm}{\centering \pmb{$d_r$}}}\\
\cline{1-2}\noalign{\vskip-1pt}\tabucline[2pt]{3-7}\noalign{\vskip-2pt}\cline{9-9}\tabucline[2pt]{10-12}
\multirow{2}*{\parbox[c][1cm]{2.1cm}{\centering bbr}} & \large \pmb{$\mu$} & \cellcolor{myg}75.29 & \cellcolor{myr}51.62 & 37.31\% & 5.81 & 3.82 & & \large \pmb{$\mu$} & 0.73 & 0.76 &4.69\% \\
\cline{2-7}\cline{9-12}
& \large \pmb{$\sigma$} &4.45  &  &  & 0.72 &  &   &  \large \pmb{$\sigma$}&0.03 & &\\
\cline{1-2}\tabucline[2pt]{3-7}\noalign{\vskip-2pt}\cline{9-9}\tabucline[2pt]{10-12}
\multirow{2}*{\parbox[c][1cm]{2.1cm}{\centering vegas}} & \large \pmb{$\mu$} &\cellcolor{myg}24.54 &\cellcolor{myr}16.26 & 40.58\% & 5.75 &  5.18 \\
\cline{2-7}
& \large \pmb{$\sigma$} & 4.45 &  &  & 0.71 &    \\
\cline{1-2}\tabucline[2pt]{3-7}
\end{tabu}}
\label{tab:tab4223}
\end{table*}

\FloatBarrier

Hybrid BBR tries not to fill up the queues but the too conservative nature of Vegas does not allow the rates of the two flows to converge to a common value. 

The results of CoCo-Beholder and the testbed~\cite{turkovic2019fifty} show that BBR tries to compete against Cubic and Vegas tries to compete against BBR for the bandwidth. However, as seen in the rate and Jain's index plots, comparing to the testbed, CoCo-Beholder witnessed the more distinct competition -- both for BBR\&Cubic and BBR\&Vegas. 

Another important difference is that the testbed~\cite{turkovic2019fifty} demonstrated the best fairness for BBR\&Vegas, while CoCo-Beholder -- for BBR\&Cubic, though the fairness shown by CoCo-Beholder for BBR\&Vegas was still high: 0.73 Jain's index on average.

The results of CoCo-Beholder confirm that, as highlighted in the paper~\cite{turkovic2019fifty}, if a flow of a loss-based scheme is present in the bottleneck, the observed RTT will be high, regardless of the efforts of the other delay-based or hybrid scheme to keep it low. At the same time, \mbox{a hybrid} and a delay-based scheme can share the bottleneck maintaining the \mbox{RTT low}. As mentioned in the paper~\cite{turkovic2019fifty} and appears in Table~\ref{tab:tab4223}, the testbed showed that BBR flow even had the RTT smaller than that of Vegas flow. However, the same table indicates that, for CoCo-Beholder, this was not true and the one-way delays of the two flows were roughly the same. 

\vspace{-0.1cm}
\subsection{Intra-fairness For Four Flows}
\vspace{-0.2cm}
\label{subsec:ssec3}

The three figures and three tables with the results can be found on pages~\pageref{fig:fig4231} and~\pageref{tab:tab4231}. Each CoCo-Beholder's one-way delay plot contains four curves: they just overlap.

The testbed~\cite{turkovic2019fifty} showed the very high and the emulator showed just the ideal (0.97 and\nolinebreak[4] 1) intra-fairness for Cubic and BBR. Also, both the testbed and the emulator showed the low delays for Vegas and BBR, while the expectedly high delays for Cubic.

However, the testbed and CoCo-Beholder showed very different intra-fairness of Vegas. For the testbed, Vegas turned out to be the champion of the intra-fairness with its Jain's index equal to 0.97. On the contrary, for CoCo-Beholder, Vegas' fairness is only 0.73. The difference is clearly reflected by the rate curves in the rate plots in Figure~\ref{fig:fig4232}. Moreover, the rates by CoCo-Beholder differed much for the ten runs of the experiment, as it can be seen in Table~\ref{tab:tab4232} from the remarkably big sample standard deviations computed over the flow rates output in the ten runs. This is in contrast to all other experiments of BW scenario having very small sample standard deviations of the rates. The thesis author ran this experiment much more than ten times on the Debian machine, and the results were always unstable and showed poor fairness. The thesis author also ran this experiment on the Ubuntu machine (with kernel 4.13 as in the testbed nodes). There, the expected ideal fairness was being achieved but only occasionally. 

\begin{figure}[p!]
\vspace*{-0.3cm}
\captionsetup[subfigure]{labelformat=empty}
    \centering
    \subfloat[]{{\includegraphics[width=\textwidth]{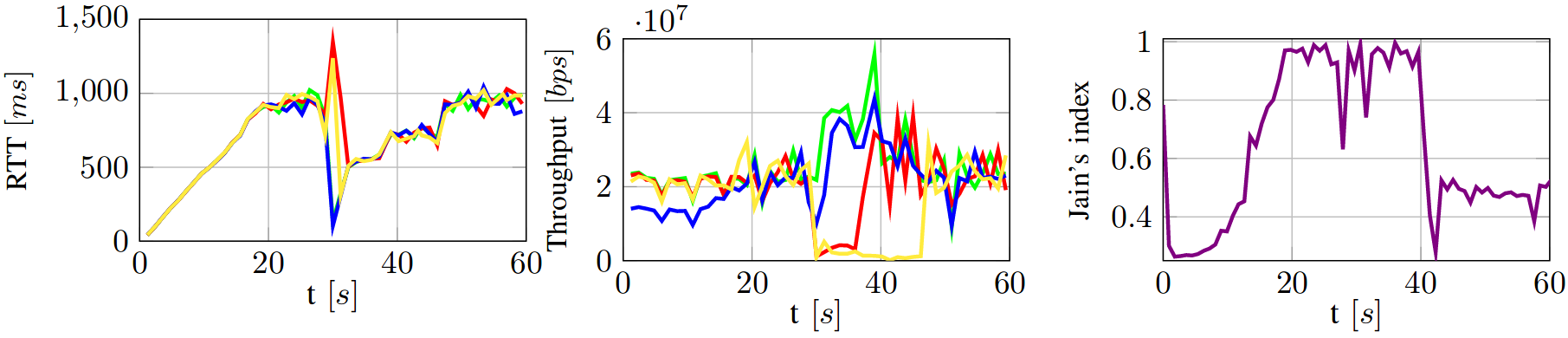} }}\\
    \vspace{-0.7cm}
    \subfloat[]{{\includegraphics[width=\textwidth/3]{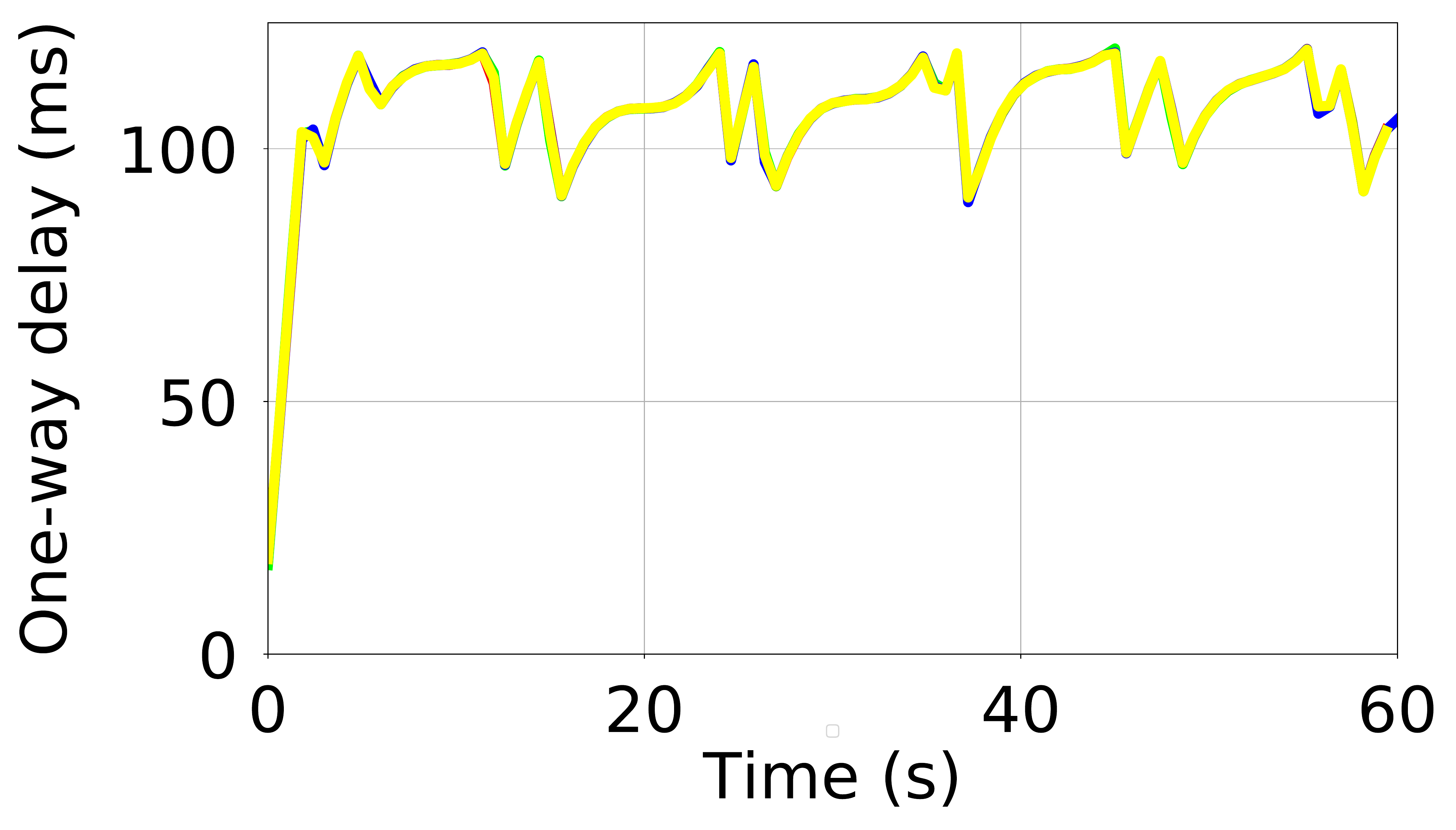} }}%
    \subfloat[]{{\includegraphics[width=\textwidth/3]{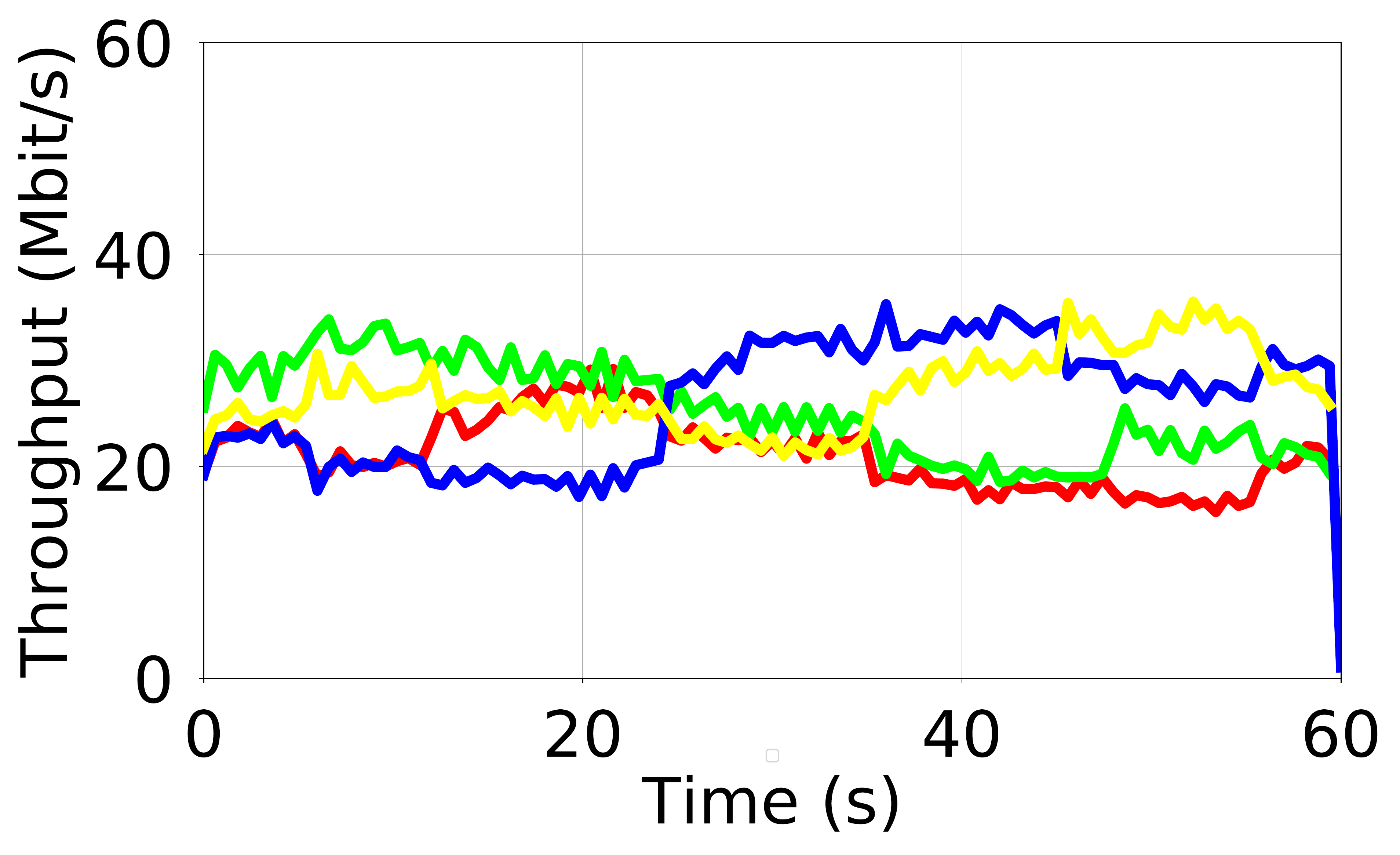} }}%
    \subfloat[]{{\includegraphics[width=\textwidth/3]{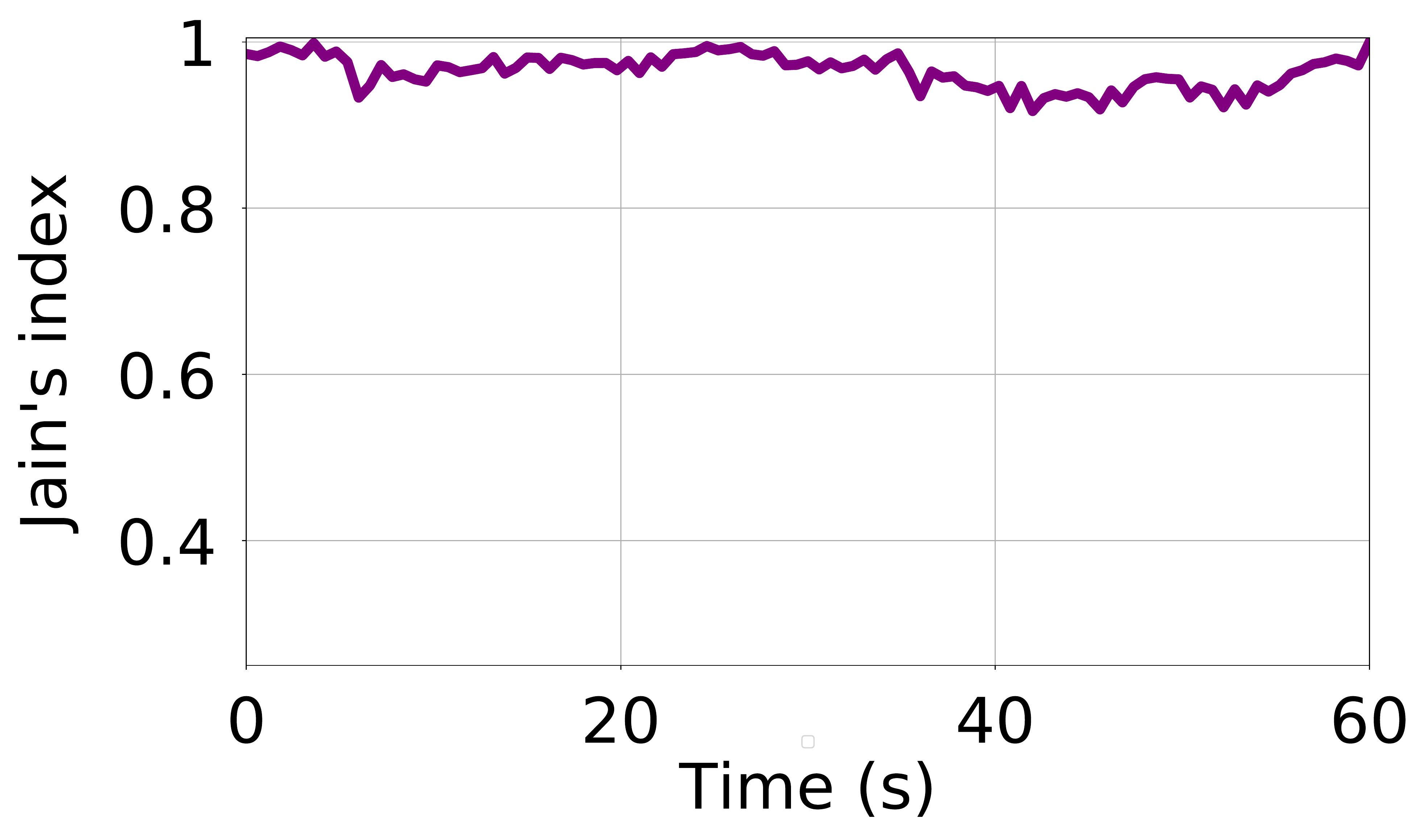} }}\\
    \vspace{-0.5cm}
    \caption{BW scenario: 4 Cubic flows. The aggregation interval is 600 ms.\\The top-row plots are by the testbed, the bottom-row -- by CoCo-Beholder.}%
    \label{fig:fig4231}
\end{figure}

\begin{figure}[h!]
\vspace*{-0.2cm}
\captionsetup[subfigure]{labelformat=empty}
    \centering
    \subfloat[]{{\includegraphics[width=\textwidth]{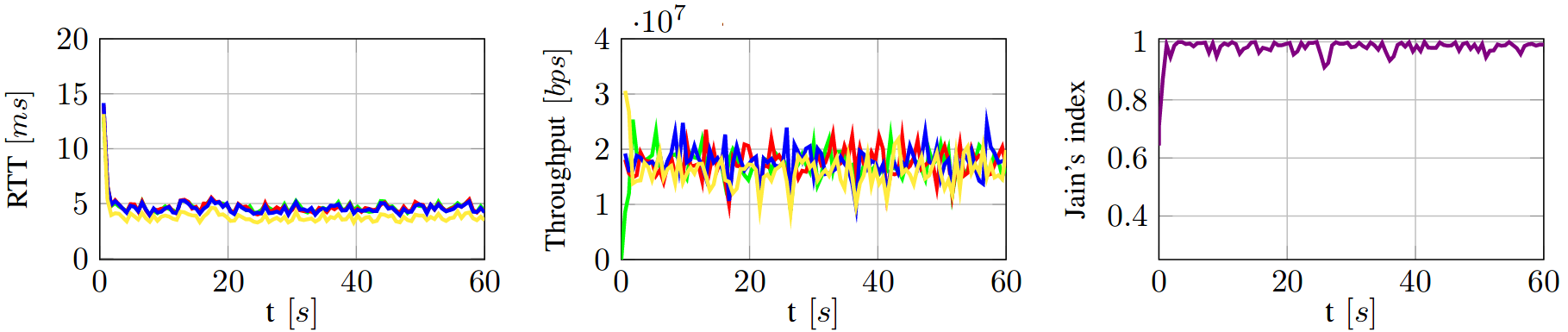} }}\\
    \vspace{-0.7cm}
    \subfloat[]{{\includegraphics[width=\textwidth/3]{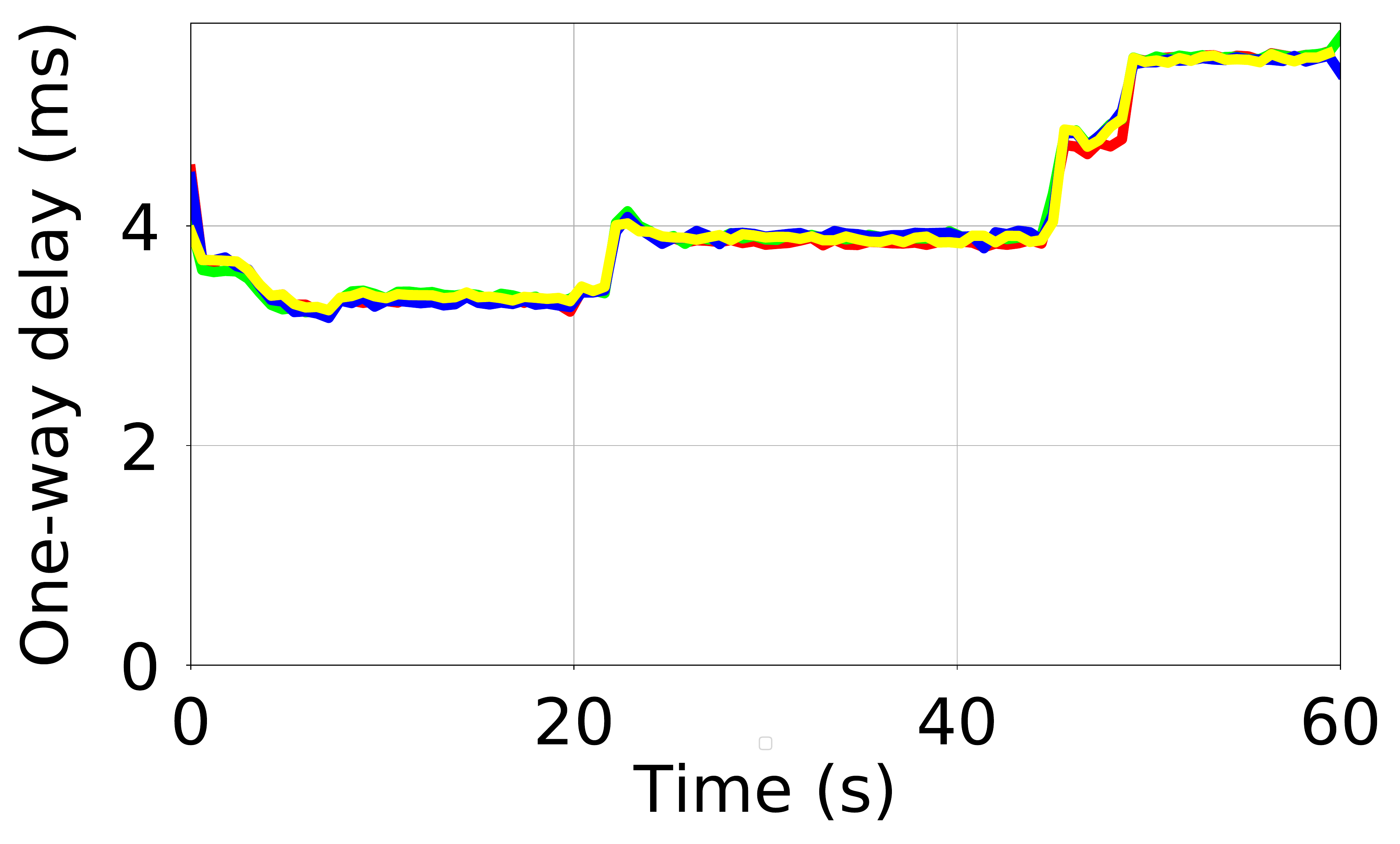} }}%
    \subfloat[]{{\includegraphics[width=\textwidth/3]{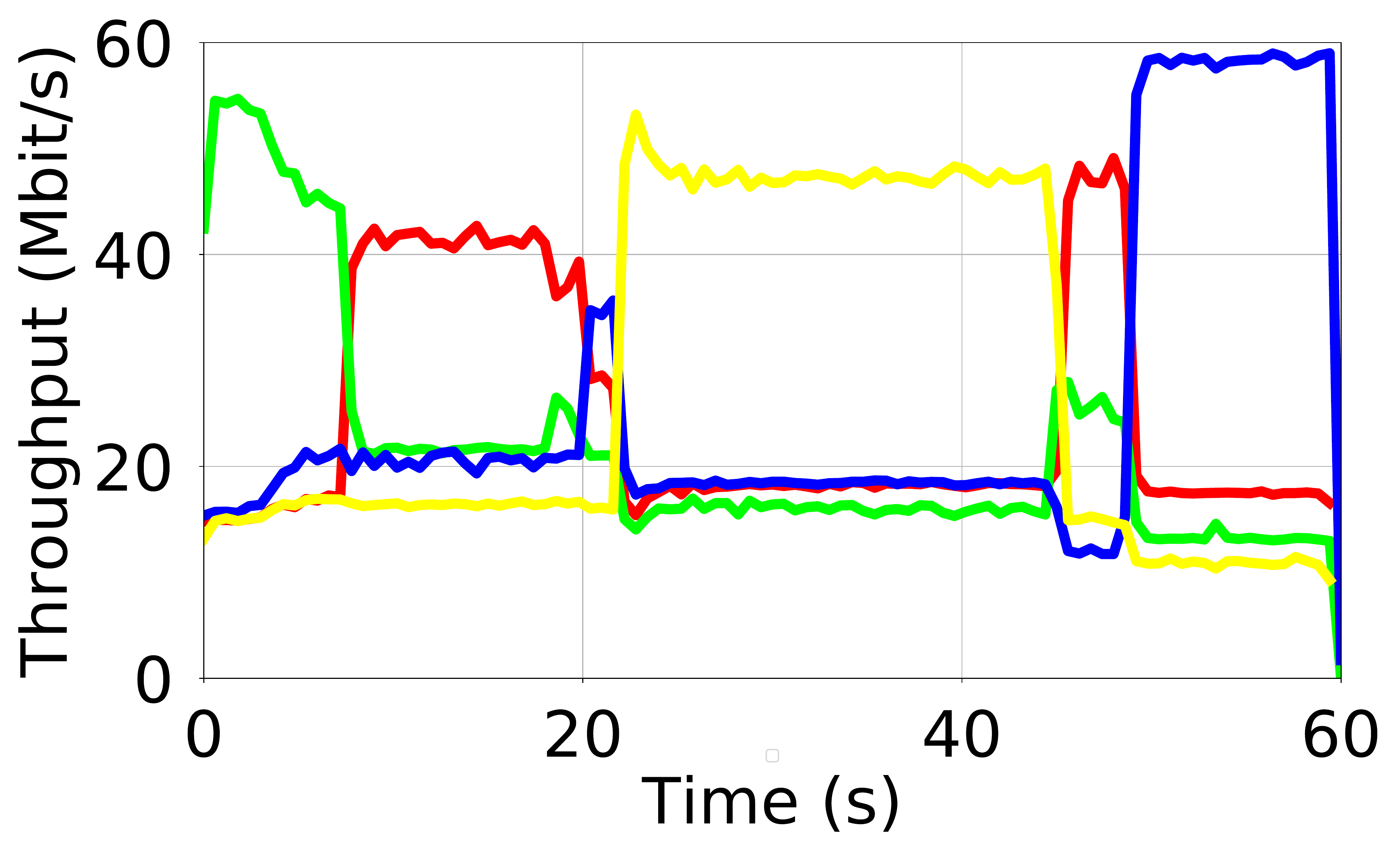} }}%
    \subfloat[]{{\includegraphics[width=\textwidth/3]{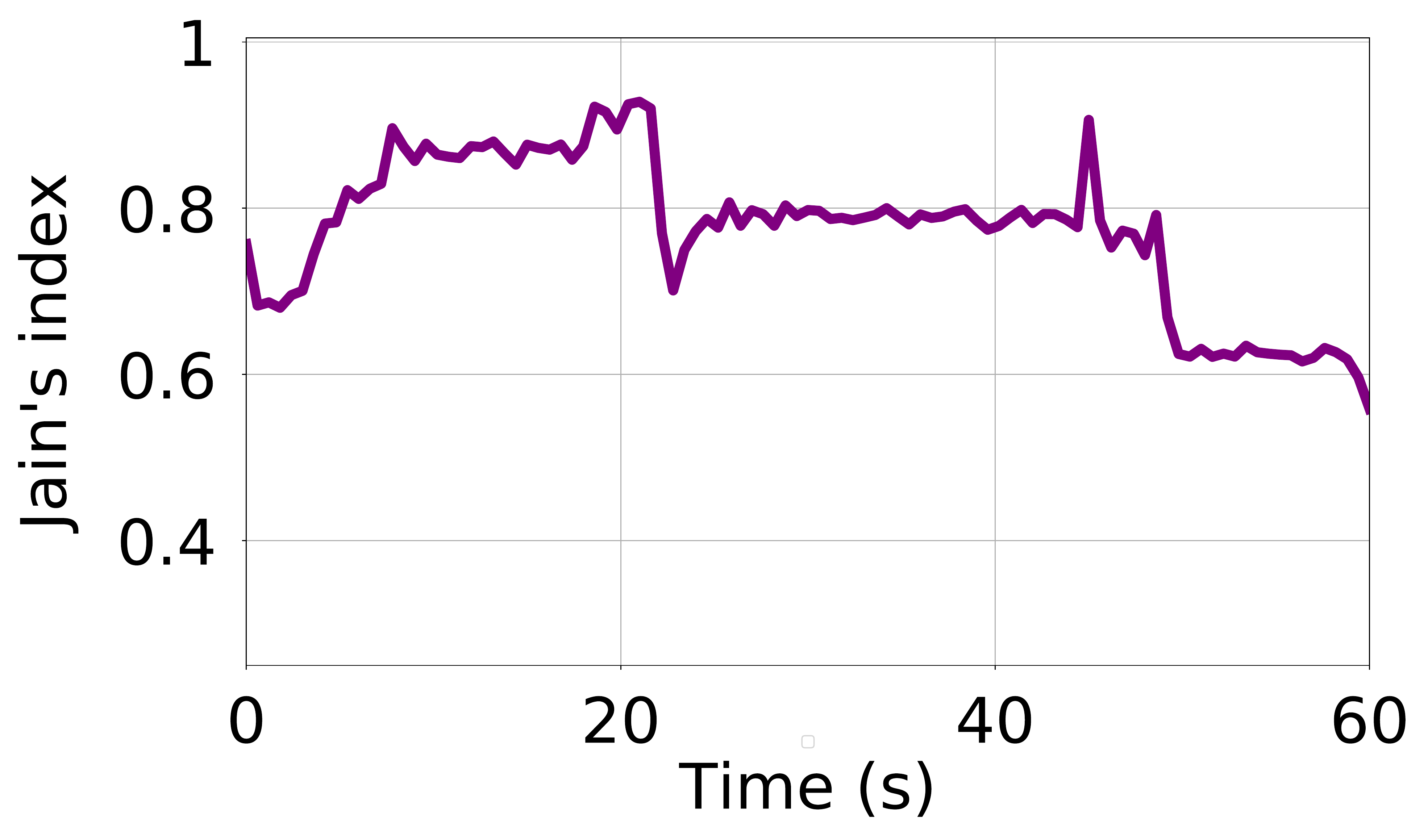} }}\\
    \vspace{-0.5cm}
    \caption{BW scenario: 4 Vegas flows. The aggregation interval is 600 ms.\\The top-row plots are by the testbed, the bottom-row -- by CoCo-Beholder.}%
    \label{fig:fig4232}
\end{figure}

\begin{figure}[h!]
\vspace*{-0.2cm}
\captionsetup[subfigure]{labelformat=empty}
    \centering
    \subfloat[]{{\includegraphics[width=\textwidth]{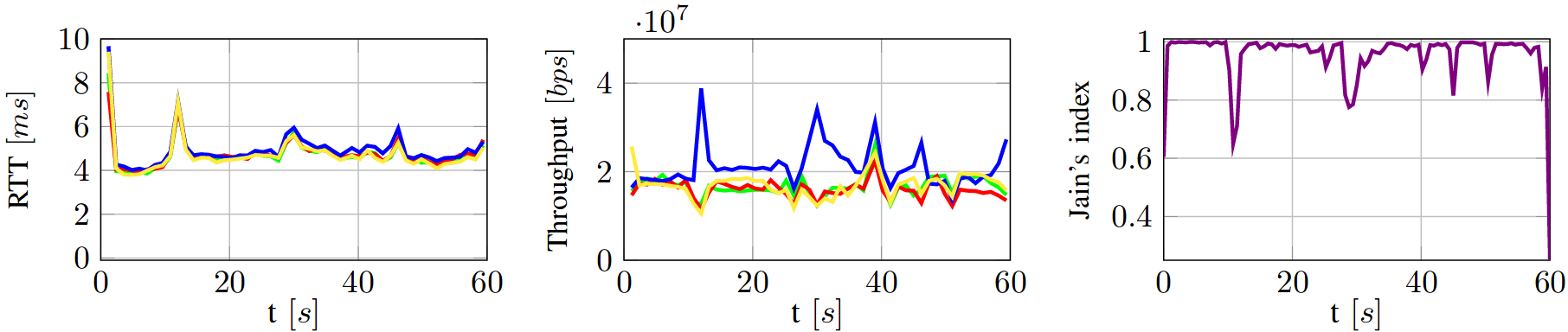} }}\\
    \vspace{-0.7cm}
    \subfloat[]{{\includegraphics[width=\textwidth/3]{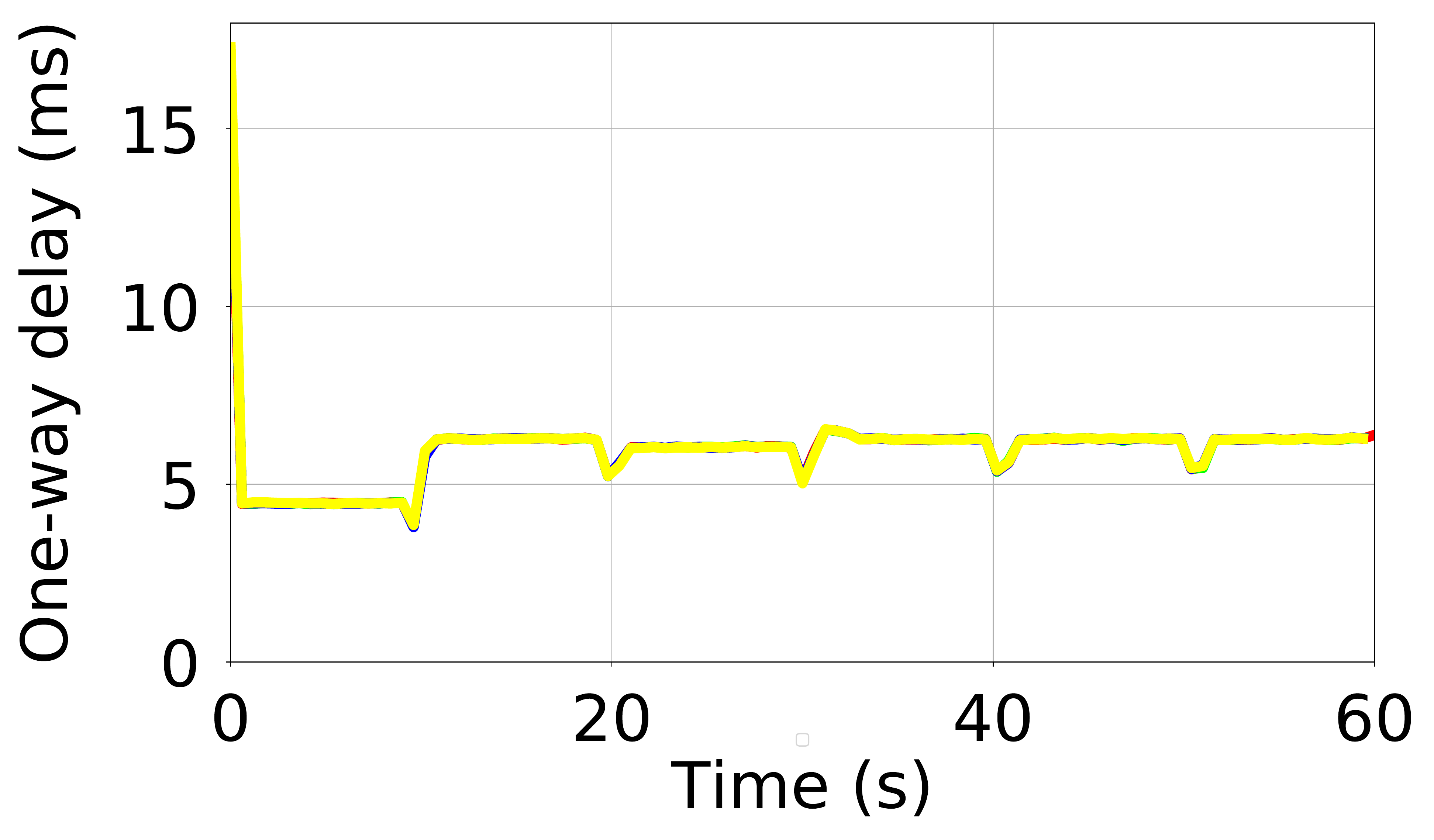} }}%
    \subfloat[]{{\includegraphics[width=\textwidth/3]{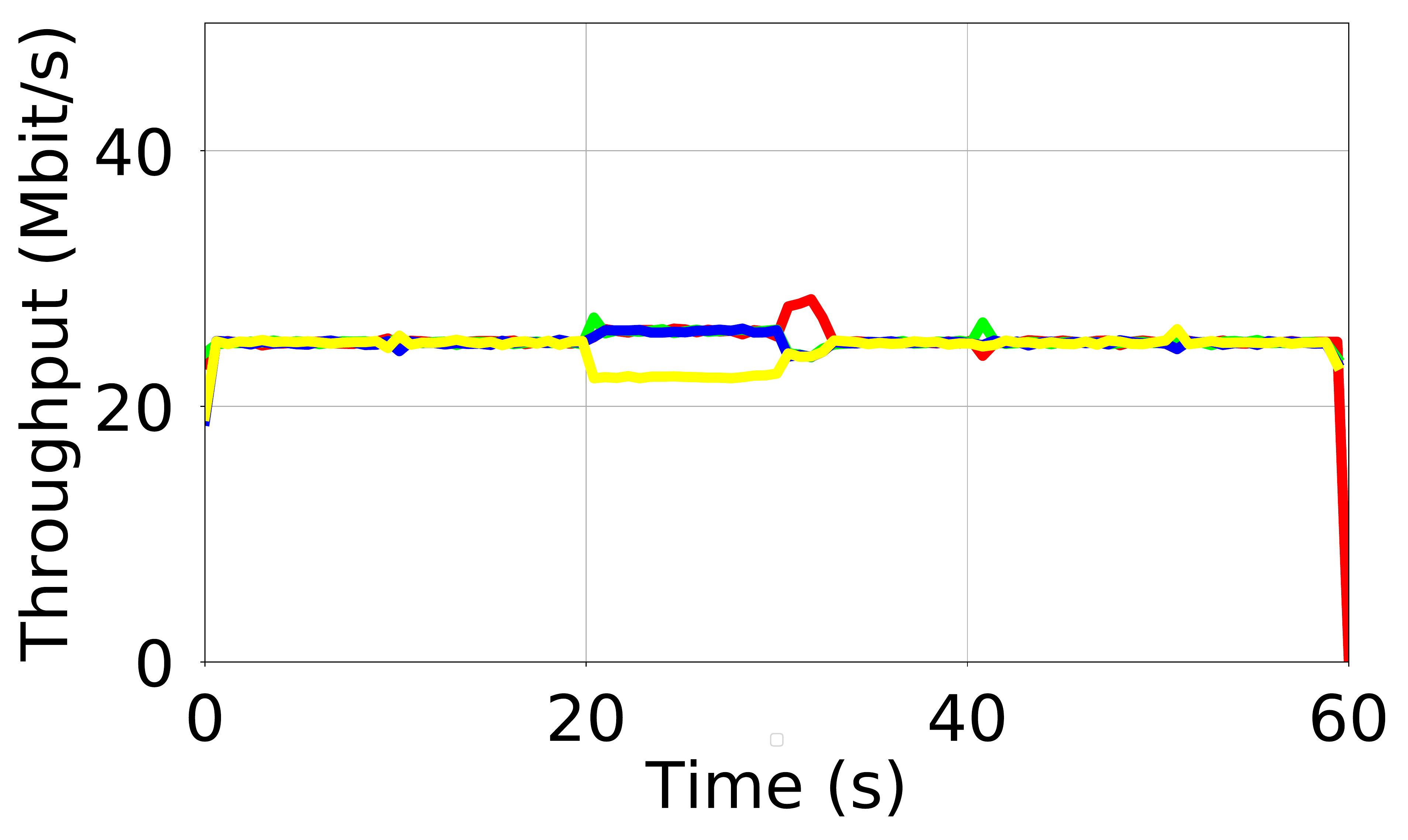} }}%
    \subfloat[]{{\includegraphics[width=\textwidth/3]{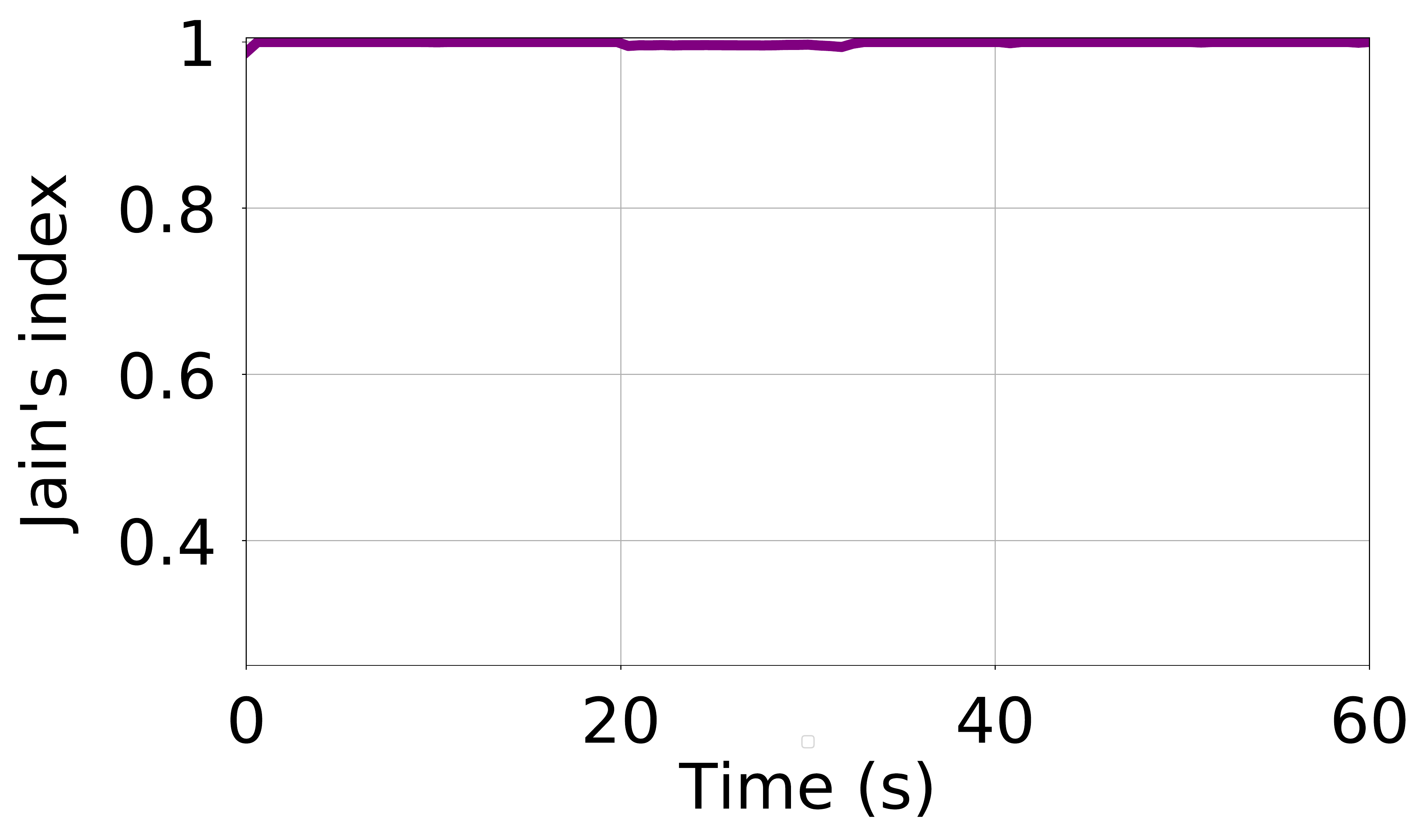} }}\\
    \vspace{-0.4cm}
    \caption{BW scenario: 4 BBR flows. The aggregation interval is 600 ms.\\The top-row plots are by the testbed, the bottom-row -- by CoCo-Beholder.}%
    \label{fig:fig4233}
\end{figure}

\newpage

\begin{table*}[p!]
\vspace{0.3cm}
\centering
\Large
\caption{BW scenario: 4 Cubic flows.}
\renewcommand{\arraystretch}{1.4} 
\resizebox*{\textwidth}{!}{\begin{tabu}{|c|cV{5}c|c|cV{5}c|cV{5}c|cV{5}c|c|cV{5}}
\cline{1-7}\cline{10-12}
\multirow{2}*{\parbox[c][2.5cm]{2.1cm}{\centering \bf \large Scheme}} &  \multicolumn{1}{c|}{\multirow{2}*{\parbox[c][2.5cm]{0.0cm}{}}} & \multicolumn{3}{c|}{\parbox[c][1cm]{2.7cm}{\bf \large \centering Rate (Mbps)}} & \parbox[c][1cm]{1.8cm}{\bf \large \centering Delay (ms)} & \multicolumn{1}{c|}{\parbox[c][1cm]{1.8cm}{\bf \large \centering RTT (ms)}}  & \multicolumn{1}{c}{} & \multicolumn{1}{c|}{} & \multicolumn{3}{c|}{\parbox[c][1cm]{2.5cm}{\bf \centering \large Jain's index}}\\ 
\cline{3-7}\cline{10-12}
 & \multicolumn{1}{c|}{} & \parbox[c][1.5cm]{1.8cm}{\centering \bf \normalsize CoCo-Beholder}  & \parbox[1cm]{1.8cm}{\centering \bf \normalsize Testbed} & \multicolumn{1}{c|}{\parbox{1.8cm}{\centering \pmb{$d_r$}}} & \parbox{1.8cm}{\centering \bf \normalsize CoCo-Beholder} & \multicolumn{1}{c|}{\parbox{1.8cm}{\centering \bf \normalsize Testbed}} &\multicolumn{1}{c}{} & \multicolumn{1}{c|}{}   & \parbox{1.8cm}{\centering \bf \normalsize CoCo-Beholder}& \parbox{1.8cm}{\centering \bf \normalsize Testbed} &\multicolumn{1}{c|}{ \parbox{1.8cm}{\centering \pmb{$d_r$}}}\\
\cline{1-2}\noalign{\vskip-1pt}\tabucline[2pt]{3-7}\noalign{\vskip-2pt}\cline{9-9}\tabucline[2pt]{10-12}
\multirow{2}*{\parbox[c][1cm]{2.1cm}{\centering cubic}} & \large \pmb{$\mu$} & \cellcolor{myg}24.68 &\cellcolor{myr}20.35 & 19.25\% & 107.04 & 720.88 & & \large \pmb{$\mu$} & \cellcolor{myg}0.97 & \cellcolor{myr}0.82 & 16.73\%\\
\cline{2-7}\cline{9-12}
& \large \pmb{$\sigma$} & 2.03 &  &  & 1.26 &  &   &  \large \pmb{$\sigma$}& 0.02 & &\\
\cline{1-2}\tabucline[2pt]{3-7}\noalign{\vskip-2pt}\cline{9-9}\tabucline[2pt]{10-12}
\multirow{2}*{\parbox[c][1cm]{2.1cm}{\centering cubic}} & \large \pmb{$\mu$} &\cellcolor{myg}25.40 & \cellcolor{myr}21.84 & 15.09\% & 107.06 & 689.73 \\
\cline{2-7}
& \large \pmb{$\sigma$} &1.96 &  &  &1.25 &    \\
\cline{1-2}\tabucline[2pt]{3-7}
\multirow{2}*{\parbox[c][1cm]{2.1cm}{\centering cubic}} & \large \pmb{$\mu$} & \cellcolor{myg}23.34 & \cellcolor{myr}16.37 & 35.11\% &107.06 & 705.50 \\
\cline{2-7}
& \large \pmb{$\sigma$} & 3.26&  &  & 1.26&    \\
\cline{1-2}\tabucline[2pt]{3-7}
\multirow{2}*{\parbox[c][1cm]{2.1cm}{\centering cubic}} & \large \pmb{$\mu$} & 26.38 & 25.16 & 4.73\% &107.05 & 697.81 \\
\cline{2-7}
& \large \pmb{$\sigma$} & 4.09&  &  & 1.27&    \\
\cline{1-2}\tabucline[2pt]{3-7}
\end{tabu}}
\label{tab:tab4231}
\end{table*}

\begin{table*}[h!]
\vspace{0.3cm}
\centering
\Large
\caption{BW scenario: 4 Vegas flows.}
\renewcommand{\arraystretch}{1.4} 
\resizebox*{\textwidth}{!}{\begin{tabu}{|c|cV{5}c|c|cV{5}c|cV{5}c|cV{5}c|c|cV{5}}
\cline{1-7}\cline{10-12}
\multirow{2}*{\parbox[c][2.5cm]{2.1cm}{\centering \bf \large Scheme}} &  \multicolumn{1}{c|}{\multirow{2}*{\parbox[c][2.5cm]{0.0cm}{}}} & \multicolumn{3}{c|}{\parbox[c][1cm]{2.7cm}{\bf \large \centering Rate (Mbps)}} & \parbox[c][1cm]{1.8cm}{\bf \large \centering Delay (ms)} & \multicolumn{1}{c|}{\parbox[c][1cm]{1.8cm}{\bf \large \centering RTT (ms)}}  & \multicolumn{1}{c}{} & \multicolumn{1}{c|}{} & \multicolumn{3}{c|}{\parbox[c][1cm]{2.5cm}{\bf \centering \large Jain's index}}\\ 
\cline{3-7}\cline{10-12}
 & \multicolumn{1}{c|}{} & \parbox[c][1.5cm]{1.8cm}{\centering \bf \normalsize CoCo-Beholder}  & \parbox[1cm]{1.8cm}{\centering \bf \normalsize Testbed} & \multicolumn{1}{c|}{\parbox{1.8cm}{\centering \pmb{$d_r$}}} & \parbox{1.8cm}{\centering \bf \normalsize CoCo-Beholder} & \multicolumn{1}{c|}{\parbox{1.8cm}{\centering \bf \normalsize Testbed}} &\multicolumn{1}{c}{} & \multicolumn{1}{c|}{}   & \parbox{1.8cm}{\centering \bf \normalsize CoCo-Beholder}& \parbox{1.8cm}{\centering \bf \normalsize Testbed} &\multicolumn{1}{c|}{ \parbox{1.8cm}{\centering \pmb{$d_r$}}}\\
\cline{1-2}\noalign{\vskip-1pt}\tabucline[2pt]{3-7}\noalign{\vskip-2pt}\cline{9-9}\tabucline[2pt]{10-12}
\multirow{2}*{\parbox[c][1cm]{2.1cm}{\centering vegas}} & \large \pmb{$\mu$} & \cellcolor{myg}27.70 &\cellcolor{myr}17.66 & 44.25\% & 5.69 & 4.66 & & \large \pmb{$\mu$} & \cellcolor{myr}0.73 & \cellcolor{myg}0.97 &27.89\%\\
\cline{2-7}\cline{9-12}
& \large \pmb{$\sigma$} & 9.61 &  &  & 1.43 &  &   &  \large \pmb{$\sigma$}& 0.07 & &\\
\cline{1-2}\tabucline[2pt]{3-7}\noalign{\vskip-2pt}\cline{9-9}\tabucline[2pt]{10-12}
\multirow{2}*{\parbox[c][1cm]{2.1cm}{\centering vegas}} & \large \pmb{$\mu$} &\cellcolor{myg}23.57  &\cellcolor{myr}16.04  & 38.02\% & 5.71& 3.85 \\
\cline{2-7}
& \large \pmb{$\sigma$} & 8.82 &  &  & 1.44&    \\
\cline{1-2}\tabucline[2pt]{3-7}
\multirow{2}*{\parbox[c][1cm]{2.1cm}{\centering vegas}} & \large \pmb{$\mu$} & \cellcolor{myg}23.98 & \cellcolor{myr}17.08 & 33.59\% & 5.71& 4.73 \\
\cline{2-7}
& \large \pmb{$\sigma$} &6.68 &  &  &1.45 &    \\
\cline{1-2}\tabucline[2pt]{3-7}
\multirow{2}*{\parbox[c][1cm]{2.1cm}{\centering vegas}} & \large \pmb{$\mu$} & \cellcolor{myg}24.57 & \cellcolor{myr}17.42 & 34.04\% & 5.69& 4.72 \\
\cline{2-7}
& \large \pmb{$\sigma$} & 8.21&  &  &1.42 &    \\
\cline{1-2}\tabucline[2pt]{3-7}
\end{tabu}}
\label{tab:tab4232}
\end{table*}

\begin{table*}[h!]
\vspace{0.3cm}
\centering
\Large
\caption{BW scenario: 4 BBR flows.}
\renewcommand{\arraystretch}{1.4} 
\resizebox*{\textwidth}{!}{\begin{tabu}{|c|cV{5}c|c|cV{5}c|cV{5}c|cV{5}c|c|cV{5}}
\cline{1-7}\cline{10-12}
\multirow{2}*{\parbox[c][2.5cm]{2.1cm}{\centering \bf \large Scheme}} &  \multicolumn{1}{c|}{\multirow{2}*{\parbox[c][2.5cm]{0.0cm}{}}} & \multicolumn{3}{c|}{\parbox[c][1cm]{2.7cm}{\bf \large \centering Rate (Mbps)}} & \parbox[c][1cm]{1.8cm}{\bf \large \centering Delay (ms)} & \multicolumn{1}{c|}{\parbox[c][1cm]{1.8cm}{\bf \large \centering RTT (ms)}}  & \multicolumn{1}{c}{} & \multicolumn{1}{c|}{} & \multicolumn{3}{c|}{\parbox[c][1cm]{2.5cm}{\bf \centering \large Jain's index}}\\ 
\cline{3-7}\cline{10-12}
 & \multicolumn{1}{c|}{} & \parbox[c][1.5cm]{1.8cm}{\centering \bf \normalsize CoCo-Beholder}  & \parbox[1cm]{1.8cm}{\centering \bf \normalsize Testbed} & \multicolumn{1}{c|}{\parbox{1.8cm}{\centering \pmb{$d_r$}}} & \parbox{1.8cm}{\centering \bf \normalsize CoCo-Beholder} & \multicolumn{1}{c|}{\parbox{1.8cm}{\centering \bf \normalsize Testbed}} &\multicolumn{1}{c}{} & \multicolumn{1}{c|}{}   & \parbox{1.8cm}{\centering \bf \normalsize CoCo-Beholder}& \parbox{1.8cm}{\centering \bf \normalsize Testbed} &\multicolumn{1}{c|}{ \parbox{1.8cm}{\centering \pmb{$d_r$}}}\\
\cline{1-2}\noalign{\vskip-1pt}\tabucline[2pt]{3-7}\noalign{\vskip-2pt}\cline{9-9}\tabucline[2pt]{10-12}
\multirow{2}*{\parbox[c][1cm]{2.1cm}{\centering bbr}} & \large \pmb{$\mu$} & \cellcolor{myg}25.06 &\cellcolor{myr}20.95 & 17.87\% & 5.98 & 4.94 &  & \large \pmb{$\mu$} & 1.00
 & 0.95 & 4.86\%\\
\cline{2-7}\cline{9-12}
& \large \pmb{$\sigma$} & 0.36 &  &  & 0.06 &  &   &  \large \pmb{$\sigma$}&0.00  & &\\
\cline{1-2}\tabucline[2pt]{3-7}\noalign{\vskip-2pt}\cline{9-9}\tabucline[2pt]{10-12}
\multirow{2}*{\parbox[c][1cm]{2.1cm}{\centering bbr}} & \large \pmb{$\mu$} & \cellcolor{myg}24.98 & \cellcolor{myr}16.69 & 39.78\% & 5.98&4.70  \\
\cline{2-7}
& \large \pmb{$\sigma$} & 0.41&  &  &0.04 &    \\
\cline{1-2}\tabucline[2pt]{3-7}
\multirow{2}*{\parbox[c][1cm]{2.1cm}{\centering bbr}} & \large \pmb{$\mu$} & \cellcolor{myg}25.00 & \cellcolor{myr}16.31 & 42.08\% &5.99 & 4.72 \\
\cline{2-7}
& \large \pmb{$\sigma$} &0.51 &  &  & 0.05&    \\
\cline{1-2}\tabucline[2pt]{3-7}
\multirow{2}*{\parbox[c][1cm]{2.1cm}{\centering bbr}} & \large \pmb{$\mu$} & \cellcolor{myg}24.76 & \cellcolor{myr}15.76 &44.43\%  & 5.98& 4.75 \\
\cline{2-7}
& \large \pmb{$\sigma$} &0.41 &  &  &0.06 &    \\
\cline{1-2}\tabucline[2pt]{3-7}
\end{tabu}}
\label{tab:tab4233}
\end{table*}

\FloatBarrier

\subsection{Inter-fairness For Four Flows}
\label{subsec:ssec4}

The tables and figures with the results for Cubic\&3Vegas, Cubic\&3BBR, Vegas\&3Cubic can be found on pages~\pageref{tab:tab4241} and~\pageref{fig:fig4241}. The tables and figures with the results for BBR\&3Cubic, BBR\&3Vegas, Vegas\&3BBR can be found on pages~\pageref{tab:tab4244} and~\pageref{fig:fig4244}. Again, each CoCo-Beholder's one-way delay plot contains four curves: they just overlap.

Both the testbed~\cite{turkovic2019fifty} and CoCo-Beholder showed the similar results for Cubic when it shares the bottleneck link with one flow of a delay-based/hybrid scheme and when it shares the link with three flows of this delay-based/hybrid scheme (for comparison, the plots for Cubic\&Vegas and Cubic\&BBR are on page~\pageref{fig:fig4221}). That is, for Cubic\&3Vegas and Cubic\&3BBR, both the testbed and the emulator showed that one Cubic flow suppresses all the flows of a delay-based/hybrid scheme. Though, the same as in the case with two flows, comparing to the testbed, CoCo-Beholder witnessed the more distinct competition of BBR flows against Cubic flow, as seen in the rate and fairness plots in\nolinebreak[4] \mbox{Figure}~\ref{fig:fig4242}.

For Vegas\&3Cubic, the results of the testbed and CoCo-Beholder, are nearly identical: the three Cubic flows share all the bandwidth between themselves, in equal proportion. 

For BBR\&3Cubic, the testbed shows the result analogous to Vegas\&3Cubic (the top-row rate and fairness plots look alike in Figures~\ref{fig:fig4243} and~\ref{fig:fig4244}), with the whole bandwidth of the link fully shared between the three Cubic flows in equal proportion. At the same time, the result by CoCo-Beholder is different: the one BBR flow persistently fights against the three Cubic flows raising the Jain's index\nolinebreak[4] to\nolinebreak[4] 0.82.

When BBR and Vegas share the bottleneck (BBR\&3Vegas and Vegas\&3BBR), the plots and the statistics by the testbed claim the great inter-fairness: 0.9 and 0.94 Jain's\nolinebreak[4] indices. On the contrary, CoCo-Beholder demonstrates that BBR always suppresses Vegas. This can be easily spotted in Figures~\ref{fig:fig4245} and ~\ref{fig:fig4246}. 

Nevertheless, both the testbed and the emulator proved again that BBR and Vegas is the only combination that is able to maintain the low RTTs/one-way delays when working together. The only discrepancy is the same as it was in the case with two flows (see Table~\ref{tab:tab4223} for BBR\&Vegas): the testbed showed that the RTT of BBR was even a little bit smaller than Vegas', while for CoCo-Beholder, this was not true and the one-way delays of all the four flows were roughly the same for both the cases BBR\&3Vegas and Vegas\&3BBR, as appears in Tables~\ref{tab:tab4245} and~\ref{tab:tab4246}.

If at least one Cubic flow is present in the bottleneck, all the other flows in the link experience very high RTTs, regardless of the efforts of the delay-based or hybrid schemes to keep it low. This can be observed for the testbed~\cite{turkovic2019fifty} and CoCo-Beholder in the first four tables of this subsection (see Tables~\ref{tab:tab4241}--~\ref{tab:tab4244}).

\begin{table*}[p!]
\vspace{0.3cm}
\centering
\Large
\caption{BW scenario: Cubic \& 3 Vegas.}
\renewcommand{\arraystretch}{1.4} 
\resizebox*{\textwidth}{!}{\begin{tabu}{|c|cV{5}c|c|cV{5}c|cV{5}c|cV{5}c|c|cV{5}}
\cline{1-7}\cline{10-12}
\multirow{2}*{\parbox[c][2.5cm]{2.1cm}{\centering \bf \large Scheme}} &  \multicolumn{1}{c|}{\multirow{2}*{\parbox[c][2.5cm]{0.0cm}{}}} & \multicolumn{3}{c|}{\parbox[c][1cm]{2.7cm}{\bf \large \centering Rate (Mbps)}} & \parbox[c][1cm]{1.8cm}{\bf \large \centering Delay (ms)} & \multicolumn{1}{c|}{\parbox[c][1cm]{1.8cm}{\bf \large \centering RTT (ms)}}  & \multicolumn{1}{c}{} & \multicolumn{1}{c|}{} & \multicolumn{3}{c|}{\parbox[c][1cm]{2.5cm}{\bf \centering \large Jain's index}}\\ 
\cline{3-7}\cline{10-12}
 & \multicolumn{1}{c|}{} & \parbox[c][1.5cm]{1.8cm}{\centering \bf \normalsize CoCo-Beholder}  & \parbox[1cm]{1.8cm}{\centering \bf \normalsize Testbed} & \multicolumn{1}{c|}{\parbox{1.8cm}{\centering \pmb{$d_r$}}} & \parbox{1.8cm}{\centering \bf \normalsize CoCo-Beholder} & \multicolumn{1}{c|}{\parbox{1.8cm}{\centering \bf \normalsize Testbed}} &\multicolumn{1}{c}{} & \multicolumn{1}{c|}{}   & \parbox{1.8cm}{\centering \bf \normalsize CoCo-Beholder}& \parbox{1.8cm}{\centering \bf \normalsize Testbed} &\multicolumn{1}{c|}{ \parbox{1.8cm}{\centering \pmb{$d_r$}}}\\
\cline{1-2}\noalign{\vskip-1pt}\tabucline[2pt]{3-7}\noalign{\vskip-2pt}\cline{9-9}\tabucline[2pt]{10-12}
\multirow{2}*{\parbox[c][1cm]{2.1cm}{\centering cubic}} & \large \pmb{$\mu$} & \cellcolor{myg}94.38 & \cellcolor{myr}75.30 & 22.49\% & 104.33 & 287.69 &  & \large \pmb{$\mu$} & 0.28 & 0.28 & 1.48\%\\
\cline{2-7}\cline{9-12}
& \large \pmb{$\sigma$} & 0.96 &  &  & 1.55 &  & &  \large \pmb{$\sigma$}& 0.01 & &\\
\cline{1-2}\tabucline[2pt]{3-7}\noalign{\vskip-2pt}\cline{9-9}\tabucline[2pt]{10-12}
\multirow{2}*{\parbox[c][1cm]{2.1cm}{\centering vegas}} & \large \pmb{$\mu$} & 1.72 & 1.70 &1.04\% & 103.55 & 282.98   \\
\cline{2-7}
& \large \pmb{$\sigma$} & 0.45 &  &  & 1.73 &    \\
\cline{1-2}\tabucline[2pt]{3-7}
\multirow{2}*{\parbox[c][1cm]{2.1cm}{\centering vegas}} & \large \pmb{$\mu$} & \cellcolor{myg}2.16 & \cellcolor{myr}1.87 & 14.20\% & 103.72 & 283.50 \\
\cline{2-7}
& \large \pmb{$\sigma$} & 0.53 &  &  & 1.73 &    \\
\cline{1-2}\tabucline[2pt]{3-7}
\multirow{2}*{\parbox[c][1cm]{2.1cm}{\centering vegas}} & \large \pmb{$\mu$} & \cellcolor{myg}1.56 & \cellcolor{myr}1.19 & 26.90\% & 103.59 & 281.92 \\
\cline{2-7}
& \large \pmb{$\sigma$} & 0.48 &  &  & 1.63 &    \\
\cline{1-2}\tabucline[2pt]{3-7}
\end{tabu}}
\label{tab:tab4241}
\end{table*}

\begin{table*}[p!]
\vspace{0.3cm}
\centering
\Large
\caption{BW scenario: Cubic \& 3 BBR.}
\renewcommand{\arraystretch}{1.4} 
\resizebox*{\textwidth}{!}{\begin{tabu}{|c|cV{5}c|c|cV{5}c|cV{5}c|cV{5}c|c|cV{5}}
\cline{1-7}\cline{10-12}
\multirow{2}*{\parbox[c][2.5cm]{2.1cm}{\centering \bf \large Scheme}} &  \multicolumn{1}{c|}{\multirow{2}*{\parbox[c][2.5cm]{0.0cm}{}}} & \multicolumn{3}{c|}{\parbox[c][1cm]{2.7cm}{\bf \large \centering Rate (Mbps)}} & \parbox[c][1cm]{1.8cm}{\bf \large \centering Delay (ms)} & \multicolumn{1}{c|}{\parbox[c][1cm]{1.8cm}{\bf \large \centering RTT (ms)}}  & \multicolumn{1}{c}{} & \multicolumn{1}{c|}{} & \multicolumn{3}{c|}{\parbox[c][1cm]{2.5cm}{\bf \centering \large Jain's index}}\\ 
\cline{3-7}\cline{10-12}
 & \multicolumn{1}{c|}{} & \parbox[c][1.5cm]{1.8cm}{\centering \bf \normalsize CoCo-Beholder}  & \parbox[1cm]{1.8cm}{\centering \bf \normalsize Testbed} & \multicolumn{1}{c|}{\parbox{1.8cm}{\centering \pmb{$d_r$}}} & \parbox{1.8cm}{\centering \bf \normalsize CoCo-Beholder} & \multicolumn{1}{c|}{\parbox{1.8cm}{\centering \bf \normalsize Testbed}} &\multicolumn{1}{c}{} & \multicolumn{1}{c|}{}   & \parbox{1.8cm}{\centering \bf \normalsize CoCo-Beholder}& \parbox{1.8cm}{\centering \bf \normalsize Testbed} &\multicolumn{1}{c|}{ \parbox{1.8cm}{\centering \pmb{$d_r$}}}\\
\cline{1-2}\noalign{\vskip-1pt}\tabucline[2pt]{3-7}\noalign{\vskip-2pt}\cline{9-9}\tabucline[2pt]{10-12}
\multirow{2}*{\parbox[c][1cm]{2.1cm}{\centering cubic}} & \large \pmb{$\mu$} & 66.26 & 63.14 & 4.82\% & 82.87 & 358.66 &  & \large \pmb{$\mu$} & \cellcolor{myg}0.52 & \cellcolor{myr}0.42 & 21.59\% \\
\cline{2-7}\cline{9-12}
& \large \pmb{$\sigma$} & 1.48 &  &  & 1.66 &  & &  \large \pmb{$\sigma$}& 0.01 & &\\
\cline{1-2}\tabucline[2pt]{3-7}\noalign{\vskip-2pt}\cline{9-9}\tabucline[2pt]{10-12}
\multirow{2}*{\parbox[c][1cm]{2.1cm}{\centering bbr}} & \large \pmb{$\mu$} & \cellcolor{myg}10.76 & \cellcolor{myr}4.21 & 87.51\% & 83.30 & 368.90 \\
\cline{2-7}
& \large \pmb{$\sigma$} & 0.96 &  &  & 1.58 &    \\
\cline{1-2}\tabucline[2pt]{3-7}
\multirow{2}*{\parbox[c][1cm]{2.1cm}{\centering bbr}} & \large \pmb{$\mu$} & \cellcolor{myg}10.78 & \cellcolor{myr}5.41 & 66.31\% & 83.25 & 379.34 \\
\cline{2-7}
& \large \pmb{$\sigma$} & 0.70 &  &  & 1.58 &    \\
\cline{1-2}\tabucline[2pt]{3-7}
\multirow{2}*{\parbox[c][1cm]{2.1cm}{\centering bbr}} & \large \pmb{$\mu$} & \cellcolor{myg}12.00 & \cellcolor{myr}7.69 & 43.86\% & 83.38 & 375.20
 \\
\cline{2-7}
& \large \pmb{$\sigma$} & 1.28 &  &  & 1.58 &    \\
\cline{1-2}\tabucline[2pt]{3-7}
\end{tabu}}
\label{tab:tab4242}
\end{table*}

\begin{table*}[p!]
\vspace{0.3cm}
\centering
\Large
\caption{BW scenario: Vegas \& 3 Cubic.}
\renewcommand{\arraystretch}{1.4} 
\resizebox*{\textwidth}{!}{\begin{tabu}{|c|cV{5}c|c|cV{5}c|cV{5}c|cV{5}c|c|cV{5}}
\cline{1-7}\cline{10-12}
\multirow{2}*{\parbox[c][2.5cm]{2.1cm}{\centering \bf \large Scheme}} &  \multicolumn{1}{c|}{\multirow{2}*{\parbox[c][2.5cm]{0.0cm}{}}} & \multicolumn{3}{c|}{\parbox[c][1cm]{2.7cm}{\bf \large \centering Rate (Mbps)}} & \parbox[c][1cm]{1.8cm}{\bf \large \centering Delay (ms)} & \multicolumn{1}{c|}{\parbox[c][1cm]{1.8cm}{\bf \large \centering RTT (ms)}}  & \multicolumn{1}{c}{} & \multicolumn{1}{c|}{} & \multicolumn{3}{c|}{\parbox[c][1cm]{2.5cm}{\bf \centering \large Jain's index}}\\ 
\cline{3-7}\cline{10-12}
 & \multicolumn{1}{c|}{} & \parbox[c][1.5cm]{1.8cm}{\centering \bf \normalsize CoCo-Beholder}  & \parbox[1cm]{1.8cm}{\centering \bf \normalsize Testbed} & \multicolumn{1}{c|}{\parbox{1.8cm}{\centering \pmb{$d_r$}}} & \parbox{1.8cm}{\centering \bf \normalsize CoCo-Beholder} & \multicolumn{1}{c|}{\parbox{1.8cm}{\centering \bf \normalsize Testbed}} &\multicolumn{1}{c}{} & \multicolumn{1}{c|}{}   & \parbox{1.8cm}{\centering \bf \normalsize CoCo-Beholder}& \parbox{1.8cm}{\centering \bf \normalsize Testbed} &\multicolumn{1}{c|}{ \parbox{1.8cm}{\centering \pmb{$d_r$}}}\\
\cline{1-2}\noalign{\vskip-1pt}\tabucline[2pt]{3-7}\noalign{\vskip-2pt}\cline{9-9}\tabucline[2pt]{10-12}
\multirow{2}*{\parbox[c][1cm]{2.1cm}{\centering vegas}} & \large \pmb{$\mu$} & \cellcolor{myg}1.46 & \cellcolor{myr}0.29 & 133.78\% & 107.34 & 576.67 &  & \large \pmb{$\mu$} & \cellcolor{myg}0.75 & \cellcolor{myr}0.65 & 14.55\% \\
\cline{2-7}\cline{9-12}
& \large \pmb{$\sigma$} & 0.16 &  &  & 0.84 &  & &  \large \pmb{$\sigma$}& 0.02 & &\\
\cline{1-2}\tabucline[2pt]{3-7}\noalign{\vskip-2pt}\cline{9-9}\tabucline[2pt]{10-12}
\multirow{2}*{\parbox[c][1cm]{2.1cm}{\centering cubic}} & \large \pmb{$\mu$} & \cellcolor{myg}32.12 & \cellcolor{myr}28.02 & 13.65\% & 108.05 & 669.35 \\
\cline{2-7}
& \large \pmb{$\sigma$} & 1.58 &  &  & 0.84 &    \\
\cline{1-2}\tabucline[2pt]{3-7}
\multirow{2}*{\parbox[c][1cm]{2.1cm}{\centering cubic}} & \large \pmb{$\mu$} & \cellcolor{myg}35.92 & \cellcolor{myr}30.82 & 15.28\% & 108.06 & 665.12 \\
\cline{2-7}
& \large \pmb{$\sigma$} & 3.76 &  &  &  0.85 &    \\
\cline{1-2}\tabucline[2pt]{3-7}
\multirow{2}*{\parbox[c][1cm]{2.1cm}{\centering cubic}} & \large \pmb{$\mu$} & \cellcolor{myg}30.31 & \cellcolor{myr}26.28 & 14.26\% & 108.02 & 668.63 \\
\cline{2-7}
& \large \pmb{$\sigma$} & 3.57 &  &  & 0.83 &    \\
\cline{1-2}\tabucline[2pt]{3-7}
\end{tabu}}
\label{tab:tab4243}
\end{table*}

\FloatBarrier

\begin{figure}[p!]
\vspace*{-0.3cm}
\captionsetup[subfigure]{labelformat=empty}
    \centering
    \subfloat[]{{\includegraphics[width=\textwidth]{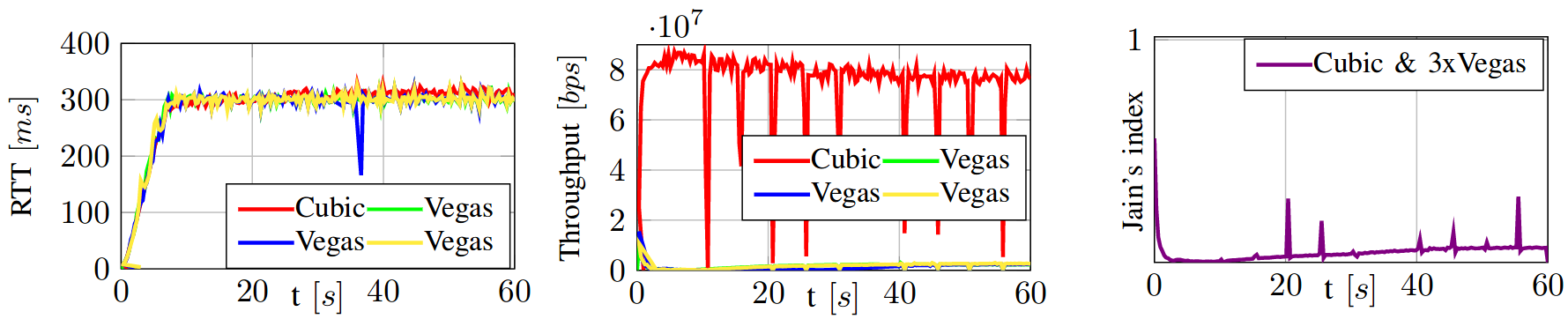} }}\\
    \vspace{-0.7cm}
    \subfloat[]{{\includegraphics[width=\textwidth/3]{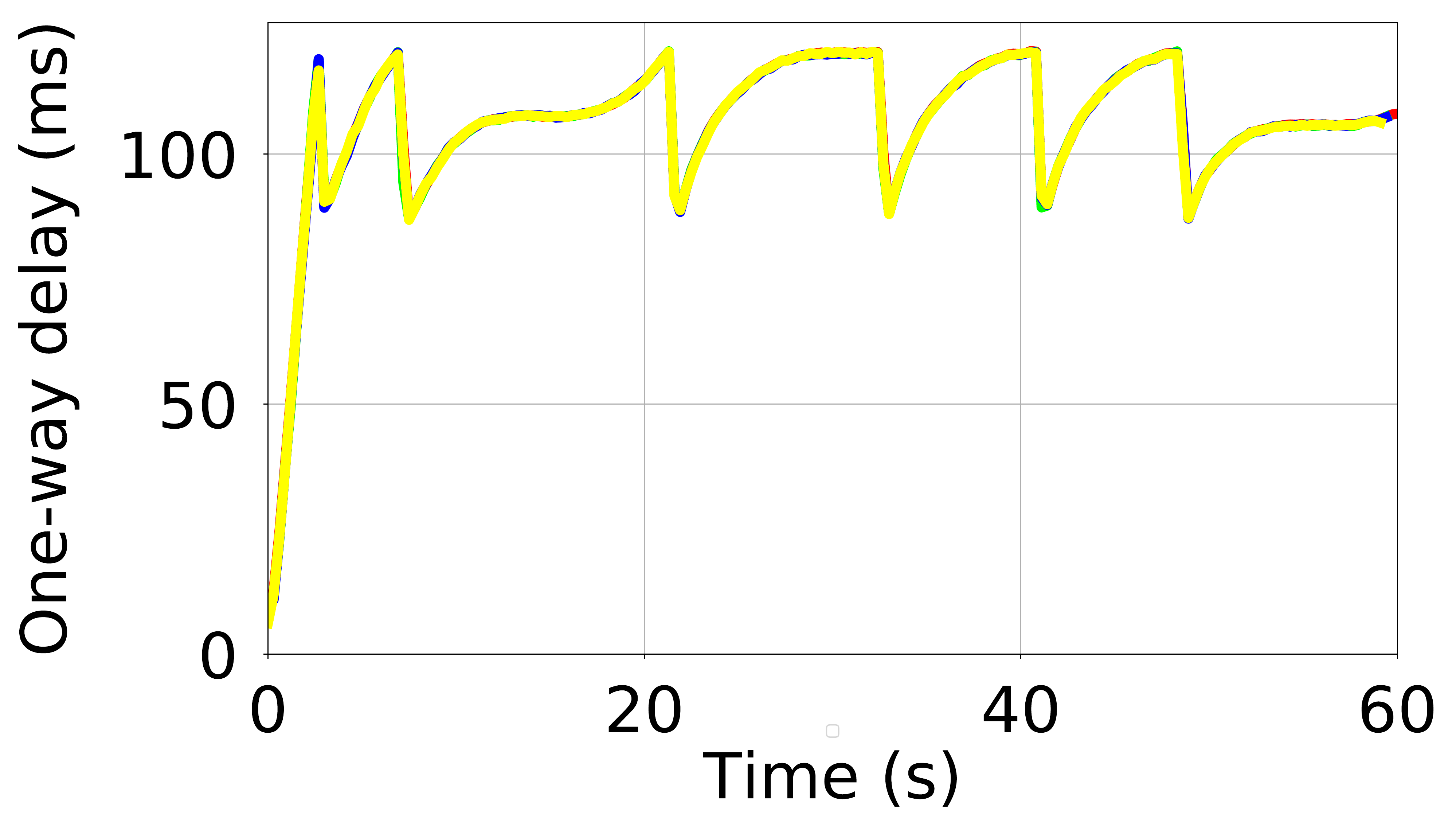} }}%
    \subfloat[]{{\includegraphics[width=\textwidth/3]{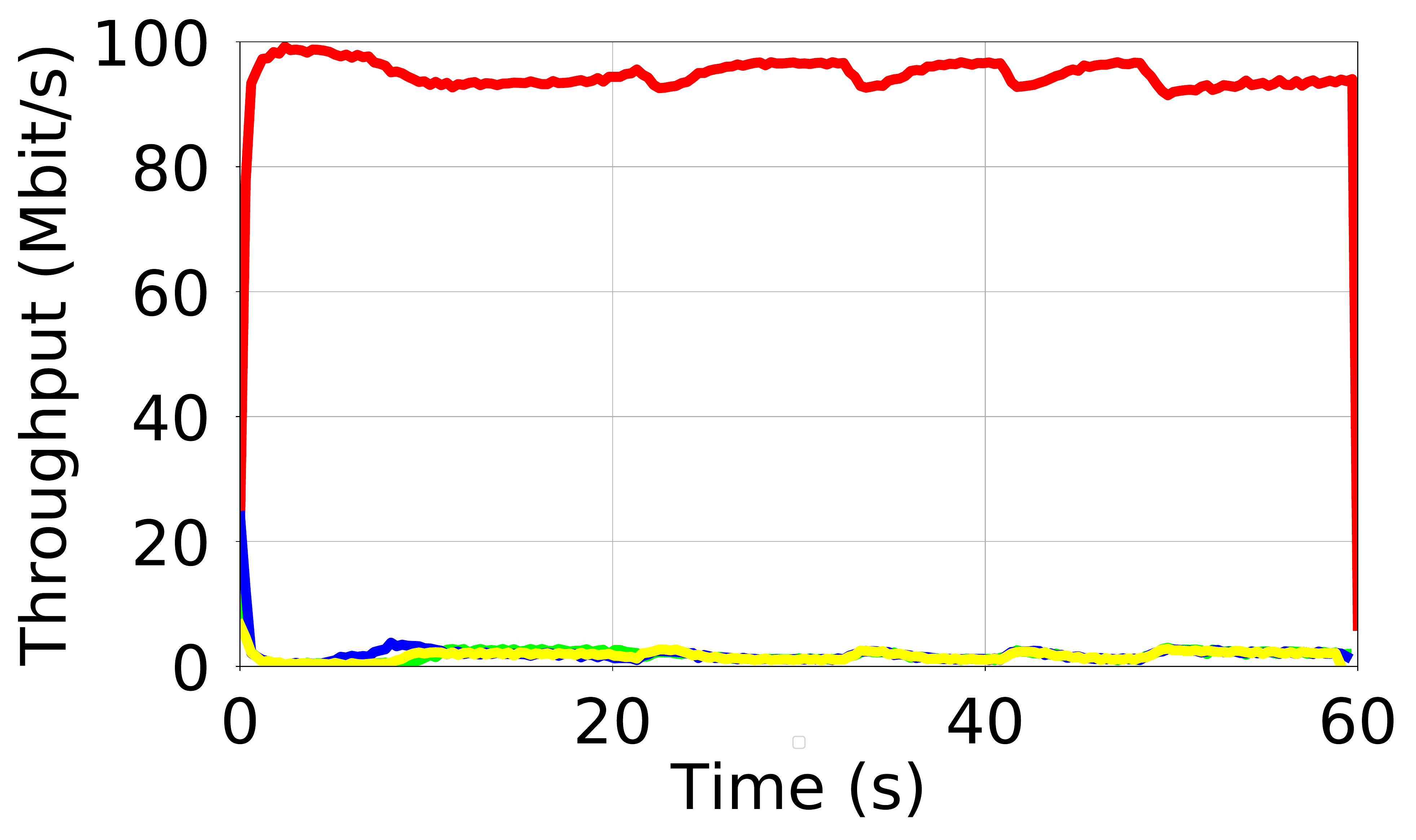} }}%
    \subfloat[]{{\includegraphics[width=\textwidth/3]{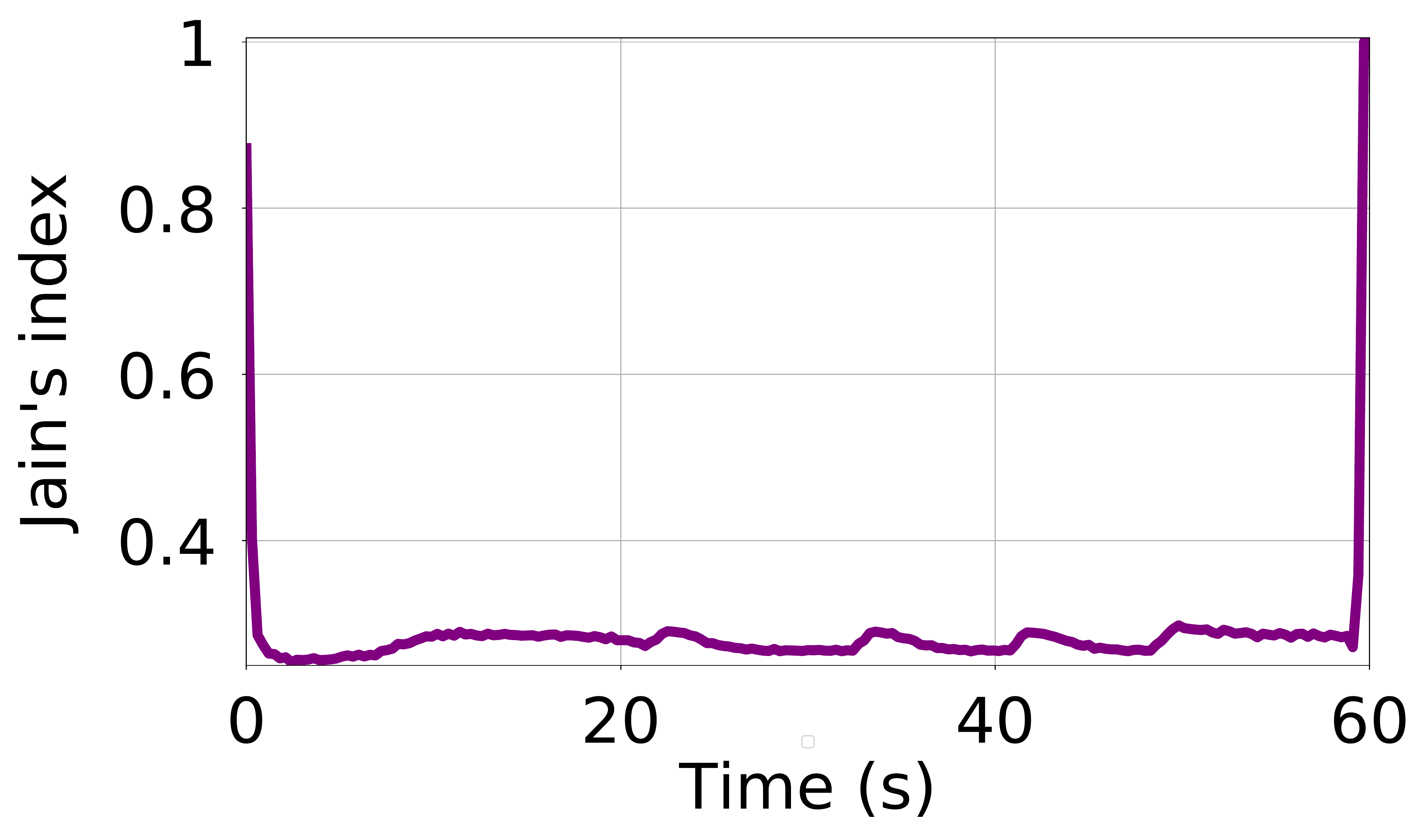} }}\\
    \vspace{-0.5cm}
    \caption{BW scenario: Cubic \& 3 Vegas. The aggregation interval is 300 ms.\\The top-row plots are by the testbed, the bottom-row -- by CoCo-Beholder.}%
    \label{fig:fig4241}
\end{figure}

\begin{figure}[h!]
\vspace*{-0.2cm}
\captionsetup[subfigure]{labelformat=empty}
    \centering
    \subfloat[]{{\includegraphics[width=\textwidth]{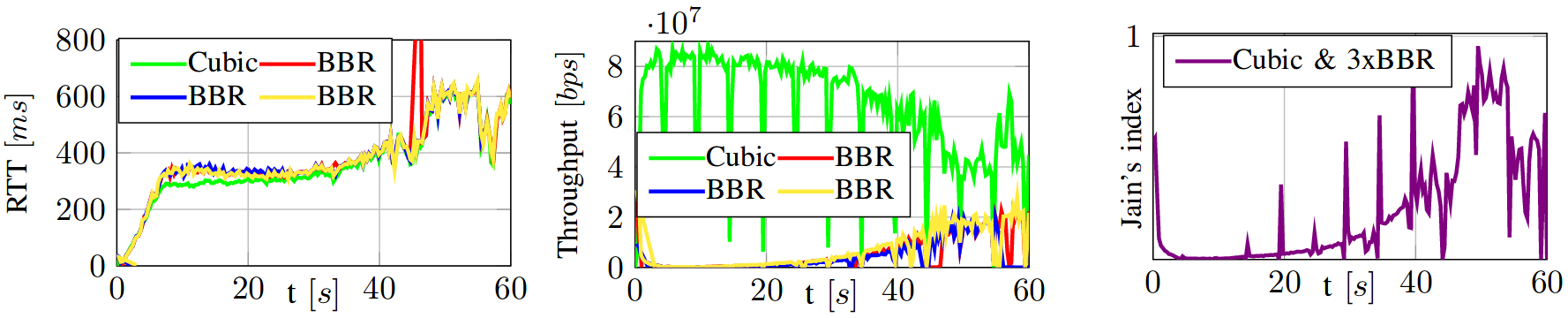} }}\\
    \vspace{-0.7cm}
    \subfloat[]{{\includegraphics[width=\textwidth/3]{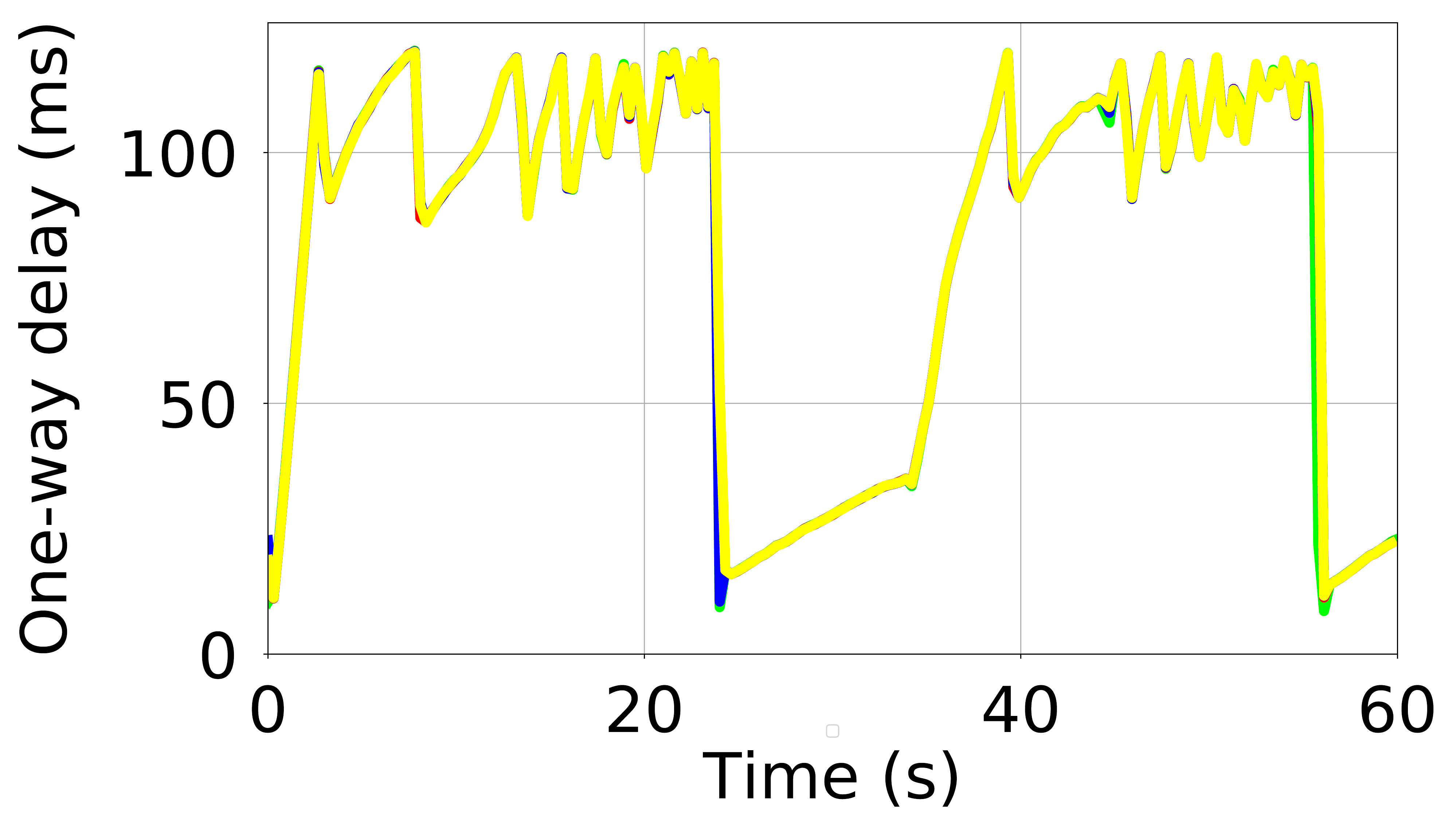} }}%
    \subfloat[]{{\includegraphics[width=\textwidth/3]{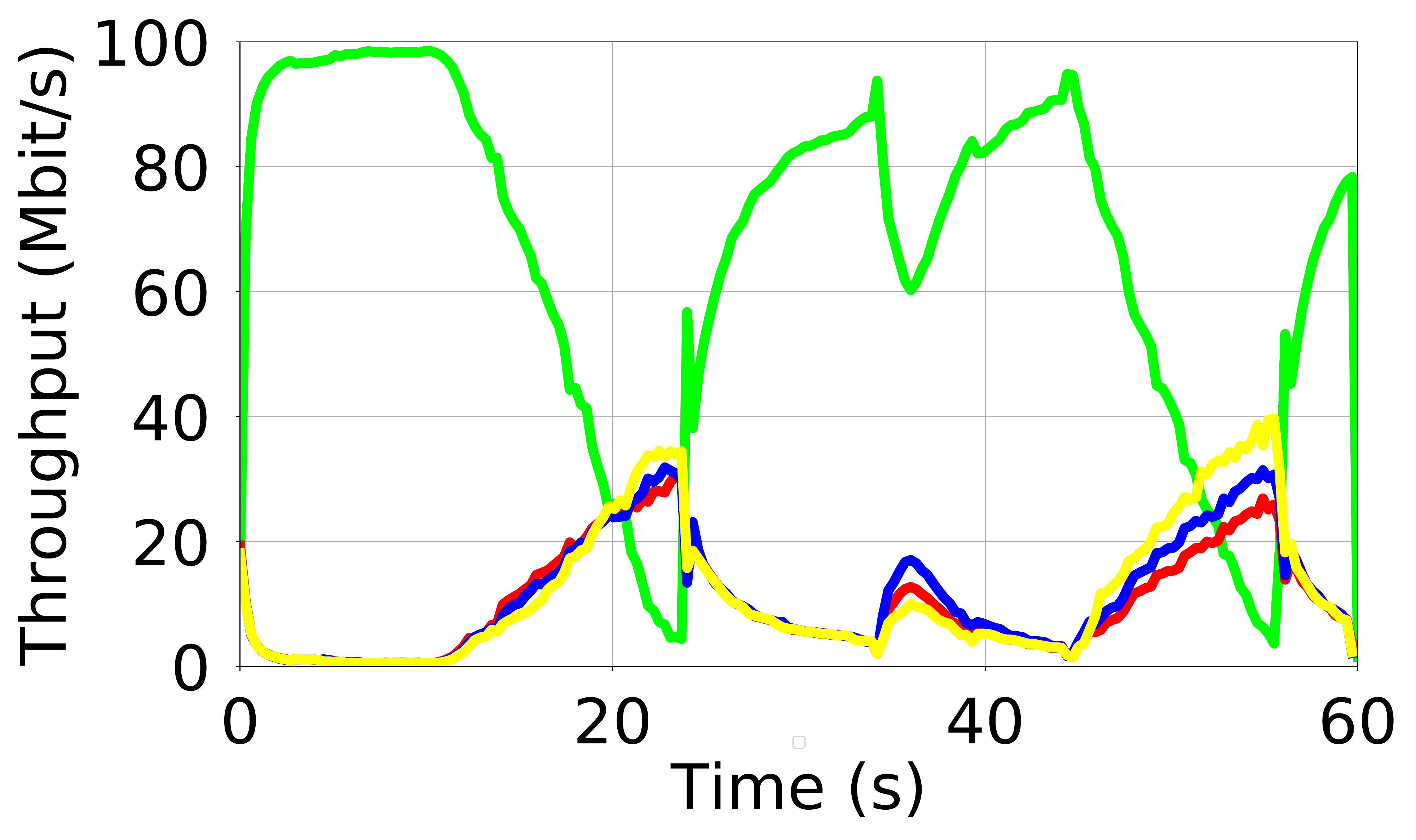} }}%
    \subfloat[]{{\includegraphics[width=\textwidth/3]{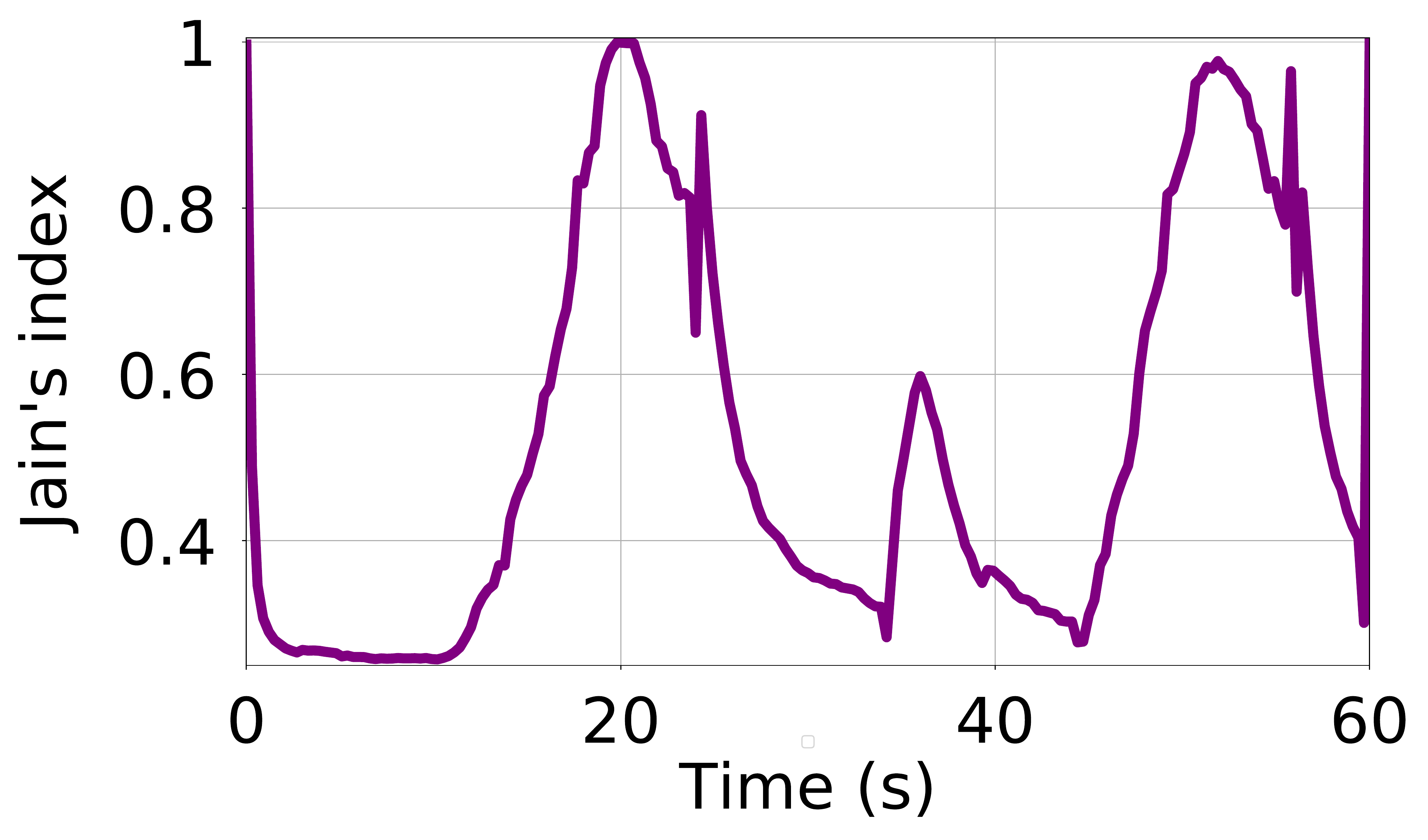} }}\\
    \vspace{-0.5cm}
    \caption{BW scenario: Cubic \& 3 BBR. The aggregation interval is 300 ms.\\The top-row plots are by the testbed, the bottom-row -- by CoCo-Beholder.}%
    \label{fig:fig4242}
\end{figure}

\begin{figure}[h!]
\vspace*{-0.2cm}
\captionsetup[subfigure]{labelformat=empty}
    \centering
    \subfloat[]{{\includegraphics[width=\textwidth]{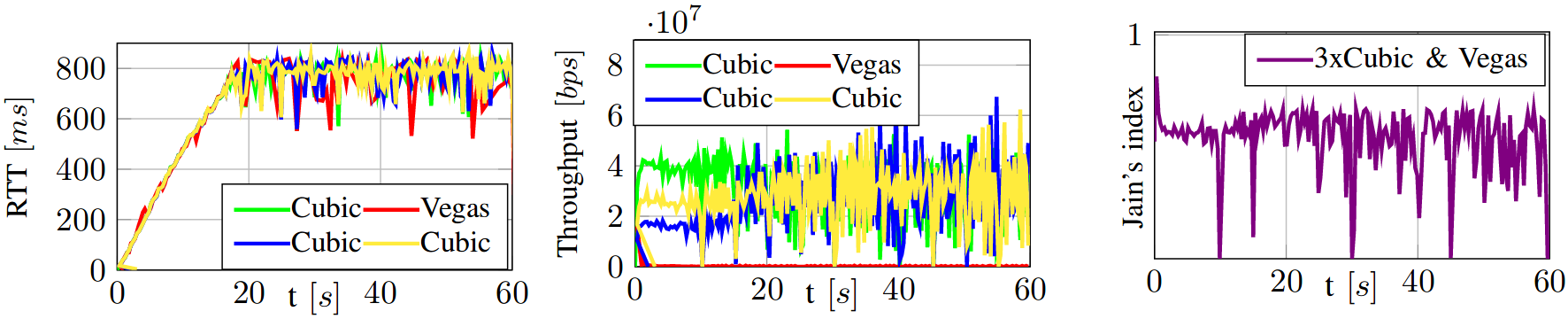} }}\\
    \vspace{-0.7cm}
    \subfloat[]{{\includegraphics[width=\textwidth/3]{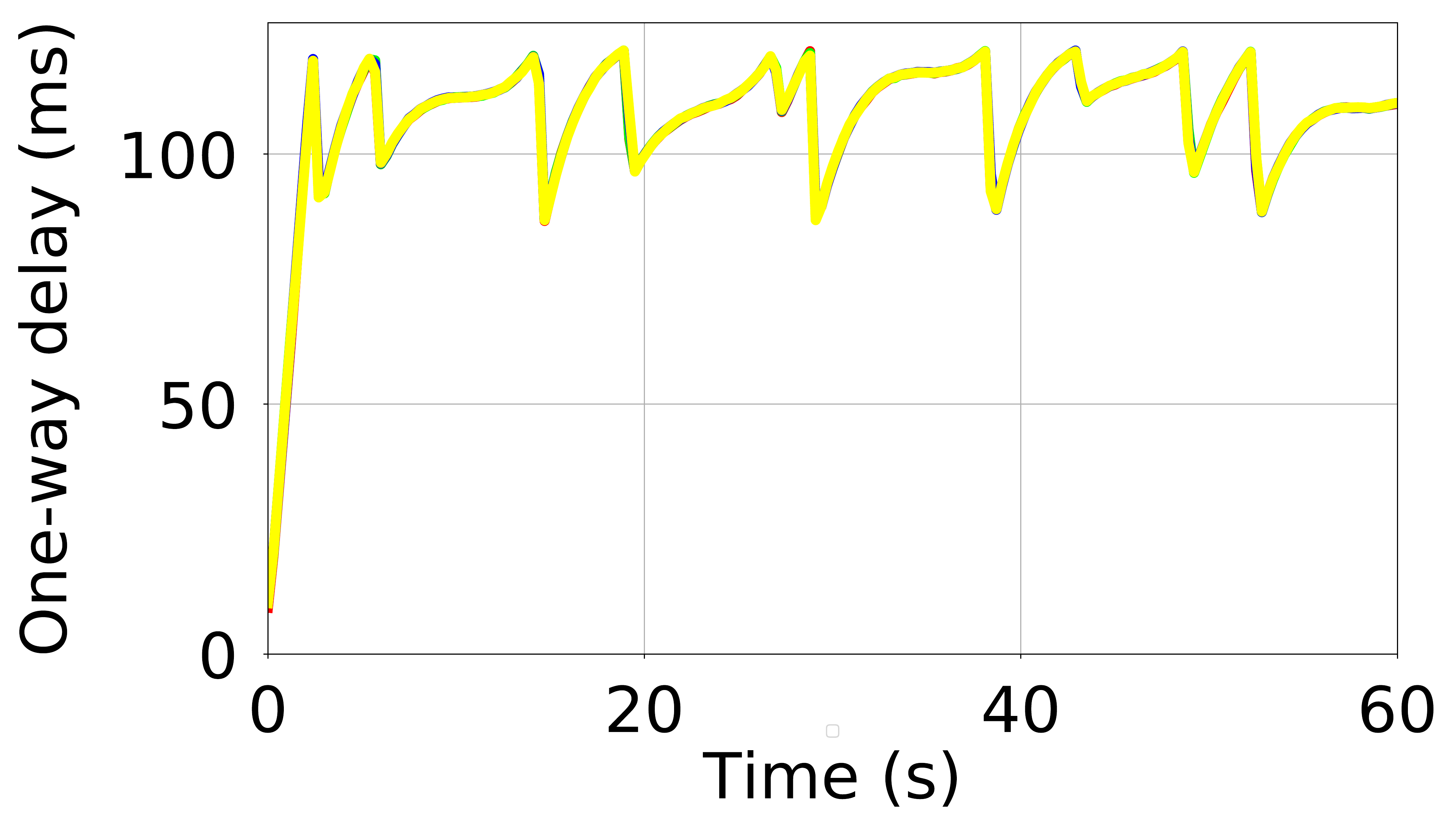} }}%
    \subfloat[]{{\includegraphics[width=\textwidth/3]{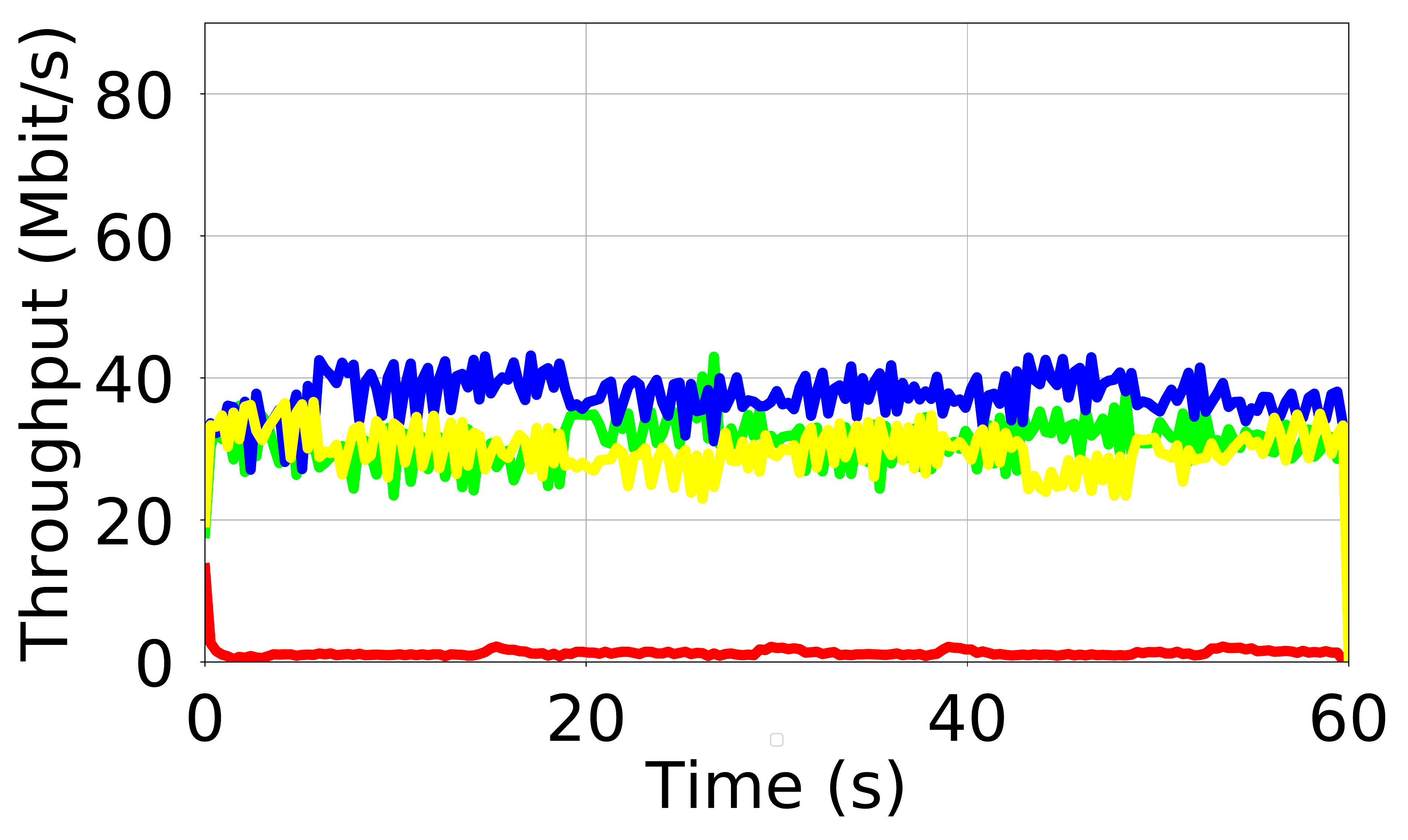} }}%
    \subfloat[]{{\includegraphics[width=\textwidth/3]{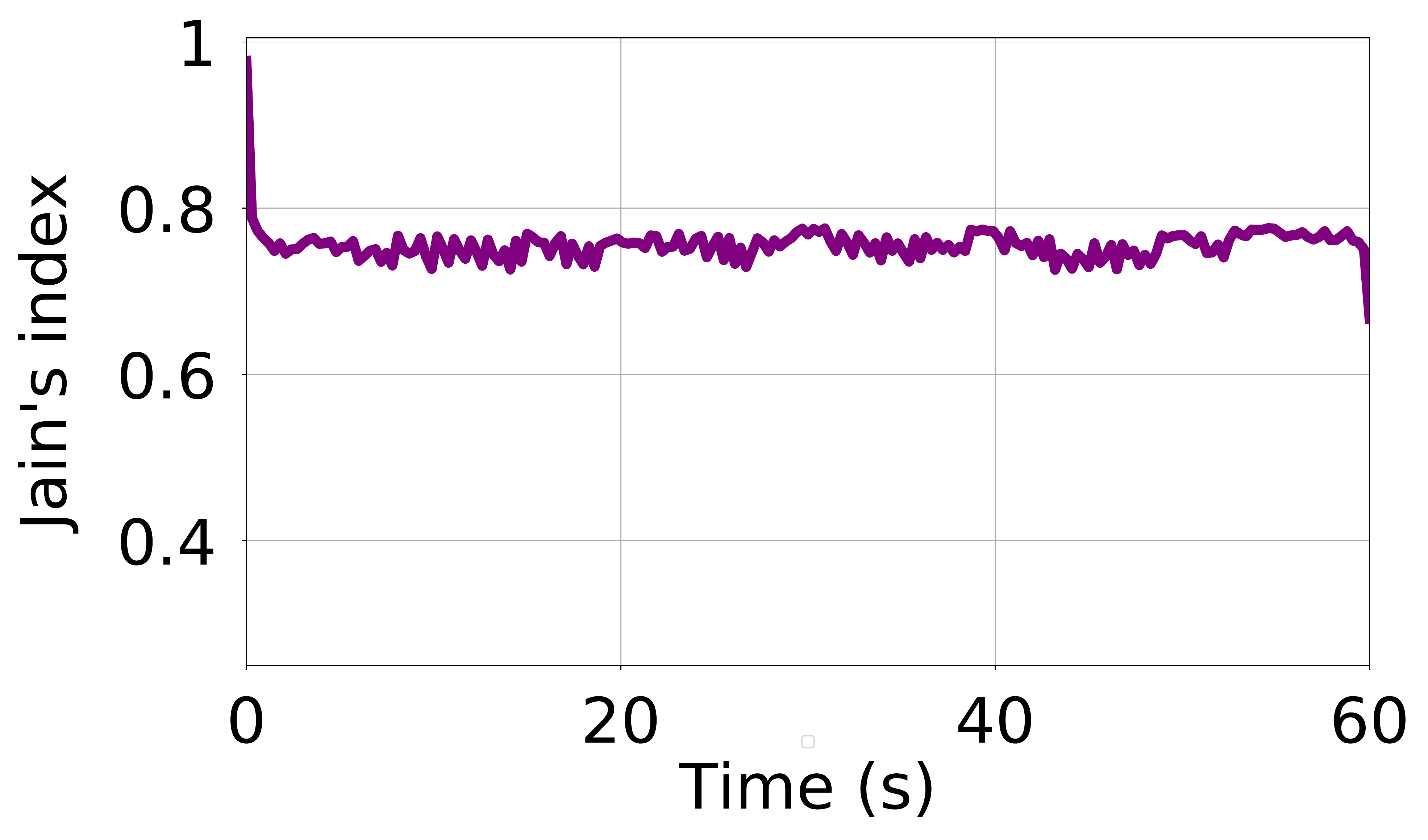} }}\\
    \vspace{-0.4cm}
    \caption{BW scenario: Vegas \& 3 Cubic. The aggregation interval is 300 ms.\\The top-row plots are by the testbed, the bottom-row -- by CoCo-Beholder.}%
    \label{fig:fig4243}
\end{figure}

\FloatBarrier

\begin{table*}[p!]
\vspace{0.3cm}
\centering
\Large
\caption{BW scenario: BBR \& 3 Cubic.}
\renewcommand{\arraystretch}{1.4} 
\resizebox*{\textwidth}{!}{\begin{tabu}{|c|cV{5}c|c|cV{5}c|cV{5}c|cV{5}c|c|cV{5}}
\cline{1-7}\cline{10-12}
\multirow{2}*{\parbox[c][2.5cm]{2.1cm}{\centering \bf \large Scheme}} &  \multicolumn{1}{c|}{\multirow{2}*{\parbox[c][2.5cm]{0.0cm}{}}} & \multicolumn{3}{c|}{\parbox[c][1cm]{2.7cm}{\bf \large \centering Rate (Mbps)}} & \parbox[c][1cm]{1.8cm}{\bf \large \centering Delay (ms)} & \multicolumn{1}{c|}{\parbox[c][1cm]{1.8cm}{\bf \large \centering RTT (ms)}}  & \multicolumn{1}{c}{} & \multicolumn{1}{c|}{} & \multicolumn{3}{c|}{\parbox[c][1cm]{2.5cm}{\bf \centering \large Jain's index}}\\ 
\cline{3-7}\cline{10-12}
 & \multicolumn{1}{c|}{} & \parbox[c][1.5cm]{1.8cm}{\centering \bf \normalsize CoCo-Beholder}  & \parbox[1cm]{1.8cm}{\centering \bf \normalsize Testbed} & \multicolumn{1}{c|}{\parbox{1.8cm}{\centering \pmb{$d_r$}}} & \parbox{1.8cm}{\centering \bf \normalsize CoCo-Beholder} & \multicolumn{1}{c|}{\parbox{1.8cm}{\centering \bf \normalsize Testbed}} &\multicolumn{1}{c}{} & \multicolumn{1}{c|}{}   & \parbox{1.8cm}{\centering \bf \normalsize CoCo-Beholder}& \parbox{1.8cm}{\centering \bf \normalsize Testbed} &\multicolumn{1}{c|}{ \parbox{1.8cm}{\centering \pmb{$d_r$}}}\\
\cline{1-2}\noalign{\vskip-1pt}\tabucline[2pt]{3-7}\noalign{\vskip-2pt}\cline{9-9}\tabucline[2pt]{10-12}
\multirow{2}*{\parbox[c][1cm]{2.1cm}{\centering bbr}} & \large \pmb{$\mu$} & \cellcolor{myg}29.50 & \cellcolor{myr}0.34 & 195.44\% & 94.68 & 775.87 &  & \large \pmb{$\mu$} & \cellcolor{myg}0.82 & \cellcolor{myr}0.67 & 20.11\%\\
\cline{2-7}\cline{9-12}
& \large \pmb{$\sigma$} & 1.41 &  &  & 2.29 &  & &  \large \pmb{$\sigma$}& 0.03 & &\\
\cline{1-2}\tabucline[2pt]{3-7}\noalign{\vskip-2pt}\cline{9-9}\tabucline[2pt]{10-12}
\multirow{2}*{\parbox[c][1cm]{2.1cm}{\centering cubic}} & \large \pmb{$\mu$} & \cellcolor{myr}23.14 & \cellcolor{myg}25.74 & 10.63\% & 94.12 & 790.97 \\
\cline{2-7}
& \large \pmb{$\sigma$} & 1.54 &  &  & 2.23 &    \\
\cline{1-2}\tabucline[2pt]{3-7}
\multirow{2}*{\parbox[c][1cm]{2.1cm}{\centering cubic}} & \large \pmb{$\mu$} & \cellcolor{myr}23.35 & \cellcolor{myg}30.83 & 27.62\% & 94.12 & 749.91 \\
\cline{2-7}
& \large \pmb{$\sigma$} & 1.50 &  &  & 2.22 &    \\
\cline{1-2}\tabucline[2pt]{3-7}
\multirow{2}*{\parbox[c][1cm]{2.1cm}{\centering cubic}} & \large \pmb{$\mu$} & \cellcolor{myr}23.83 & \cellcolor{myg}31.04 & 26.29\%  & 94.12 & 747.01 \\
\cline{2-7}
& \large \pmb{$\sigma$} & 1.11 &  &  & 2.21 &    \\
\cline{1-2}\tabucline[2pt]{3-7}
\end{tabu}}
\label{tab:tab4244}
\end{table*}

\begin{table*}[p!]
\vspace{0.3cm}
\centering
\Large
\caption{BW scenario: BBR \& 3 Vegas.}
\renewcommand{\arraystretch}{1.4} 
\resizebox*{\textwidth}{!}{\begin{tabu}{|c|cV{5}c|c|cV{5}c|cV{5}c|cV{5}c|c|cV{5}}
\cline{1-7}\cline{10-12}
\multirow{2}*{\parbox[c][2.5cm]{2.1cm}{\centering \bf \large Scheme}} &  \multicolumn{1}{c|}{\multirow{2}*{\parbox[c][2.5cm]{0.0cm}{}}} & \multicolumn{3}{c|}{\parbox[c][1cm]{2.7cm}{\bf \large \centering Rate (Mbps)}} & \parbox[c][1cm]{1.8cm}{\bf \large \centering Delay (ms)} & \multicolumn{1}{c|}{\parbox[c][1cm]{1.8cm}{\bf \large \centering RTT (ms)}}  & \multicolumn{1}{c}{} & \multicolumn{1}{c|}{} & \multicolumn{3}{c|}{\parbox[c][1cm]{2.5cm}{\bf \centering \large Jain's index}}\\ 
\cline{3-7}\cline{10-12}
 & \multicolumn{1}{c|}{} & \parbox[c][1.5cm]{1.8cm}{\centering \bf \normalsize CoCo-Beholder}  & \parbox[1cm]{1.8cm}{\centering \bf \normalsize Testbed} & \multicolumn{1}{c|}{\parbox{1.8cm}{\centering \pmb{$d_r$}}} & \parbox{1.8cm}{\centering \bf \normalsize CoCo-Beholder} & \multicolumn{1}{c|}{\parbox{1.8cm}{\centering \bf \normalsize Testbed}} &\multicolumn{1}{c}{} & \multicolumn{1}{c|}{}   & \parbox{1.8cm}{\centering \bf \normalsize CoCo-Beholder}& \parbox{1.8cm}{\centering \bf \normalsize Testbed} &\multicolumn{1}{c|}{ \parbox{1.8cm}{\centering \pmb{$d_r$}}}\\
\cline{1-2}\noalign{\vskip-1pt}\tabucline[2pt]{3-7}\noalign{\vskip-2pt}\cline{9-9}\tabucline[2pt]{10-12}
\multirow{2}*{\parbox[c][1cm]{2.1cm}{\centering bbr}} & \large \pmb{$\mu$} & \cellcolor{myg}63.37 & \cellcolor{myr}21.35 & 99.20\% & 9.71 & 3.85 & 0.50 & \large \pmb{$\mu$} & \cellcolor{myr}0.50 & \cellcolor{myg}0.90 & 57.44\%  \\
\cline{2-7}\cline{9-12}
& \large \pmb{$\sigma$} & 2.72 &  &  & 0.76 &  & &  \large \pmb{$\sigma$}& 0.03 & &\\
\cline{1-2}\tabucline[2pt]{3-7}\noalign{\vskip-2pt}\cline{9-9}\tabucline[2pt]{10-12}
\multirow{2}*{\parbox[c][1cm]{2.1cm}{\centering vegas}} & \large \pmb{$\mu$} & \cellcolor{myr}10.71 & \cellcolor{myg}13.99 & 26.52\% & 9.59 & 5.39 \\
\cline{2-7}
& \large \pmb{$\sigma$} & 2.45 &  &  & 0.75 &    \\
\cline{1-2}\tabucline[2pt]{3-7}
\multirow{2}*{\parbox[c][1cm]{2.1cm}{\centering vegas}} & \large \pmb{$\mu$} & \cellcolor{myr}13.30 & \cellcolor{myg}15.29 & 13.93\% & 9.57 & 5.26 \\
\cline{2-7}
& \large \pmb{$\sigma$} & 2.65 &  &  & 0.74 &    \\
\cline{1-2}\tabucline[2pt]{3-7}
\multirow{2}*{\parbox[c][1cm]{2.1cm}{\centering vegas}} & \large \pmb{$\mu$} & \cellcolor{myr}12.41 & \cellcolor{myg}15.22 & 20.33\% & 9.59 & 5.32 \\
\cline{2-7}
& \large \pmb{$\sigma$} & 2.49 &  &  & 0.76 &    \\
\cline{1-2}\tabucline[2pt]{3-7}
\end{tabu}}
\label{tab:tab4245}
\end{table*}

\begin{table*}[p!]
\vspace{0.3cm}
\centering
\Large
\caption{BW scenario: Vegas \& 3 BBR.}
\renewcommand{\arraystretch}{1.4} 
\resizebox*{\textwidth}{!}{\begin{tabu}{|c|cV{5}c|c|cV{5}c|cV{5}c|cV{5}c|c|cV{5}}
\cline{1-7}\cline{10-12}
\multirow{2}*{\parbox[c][2.5cm]{2.1cm}{\centering \bf \large Scheme}} &  \multicolumn{1}{c|}{\multirow{2}*{\parbox[c][2.5cm]{0.0cm}{}}} & \multicolumn{3}{c|}{\parbox[c][1cm]{2.7cm}{\bf \large \centering Rate (Mbps)}} & \parbox[c][1cm]{1.8cm}{\bf \large \centering Delay (ms)} & \multicolumn{1}{c|}{\parbox[c][1cm]{1.8cm}{\bf \large \centering RTT (ms)}}  & \multicolumn{1}{c}{} & \multicolumn{1}{c|}{} & \multicolumn{3}{c|}{\parbox[c][1cm]{2.5cm}{\bf \centering \large Jain's index}}\\ 
\cline{3-7}\cline{10-12}
 & \multicolumn{1}{c|}{} & \parbox[c][1.5cm]{1.8cm}{\centering \bf \normalsize CoCo-Beholder}  & \parbox[1cm]{1.8cm}{\centering \bf \normalsize Testbed} & \multicolumn{1}{c|}{\parbox{1.8cm}{\centering \pmb{$d_r$}}} & \parbox{1.8cm}{\centering \bf \normalsize CoCo-Beholder} & \multicolumn{1}{c|}{\parbox{1.8cm}{\centering \bf \normalsize Testbed}} &\multicolumn{1}{c}{} & \multicolumn{1}{c|}{}   & \parbox{1.8cm}{\centering \bf \normalsize CoCo-Beholder}& \parbox{1.8cm}{\centering \bf \normalsize Testbed} &\multicolumn{1}{c|}{ \parbox{1.8cm}{\centering \pmb{$d_r$}}}\\
\cline{1-2}\noalign{\vskip-1pt}\tabucline[2pt]{3-7}\noalign{\vskip-2pt}\cline{9-9}\tabucline[2pt]{10-12}
\multirow{2}*{\parbox[c][1cm]{2.1cm}{\centering vegas}} & \large \pmb{$\mu$} & \cellcolor{myg}17.28 & 1\cellcolor{myr}3.30 & 26.04\% & 9.83 & 5.75 &  & \large \pmb{$\mu$} & \cellcolor{myr}0.84 & \cellcolor{myg}0.94 & 10.72\%\\
\cline{2-7}\cline{9-12}
& \large \pmb{$\sigma$} & 3.96 &  &  & 1.52 &  & &  \large \pmb{$\sigma$}& 0.05 & &\\
\cline{1-2}\tabucline[2pt]{3-7}\noalign{\vskip-2pt}\cline{9-9}\tabucline[2pt]{10-12}
\multirow{2}*{\parbox[c][1cm]{2.1cm}{\centering bbr}} & \large \pmb{$\mu$} & \cellcolor{myg}26.94 & \cellcolor{myr}17.21 & 44.07\% & 9.93 & 4.12 \\
\cline{2-7}
& \large \pmb{$\sigma$} & 2.19 &  &  & 1.51 &    \\
\cline{1-2}\tabucline[2pt]{3-7}
\multirow{2}*{\parbox[c][1cm]{2.1cm}{\centering bbr}} & \large \pmb{$\mu$} & \cellcolor{myg}28.37 & \cellcolor{myr}19.07 & 39.20\% & 9.92 & 4.04 \\
\cline{2-7}
& \large \pmb{$\sigma$} & 4.53 &  &  & 1.51 &    \\
\cline{1-2}\tabucline[2pt]{3-7}
\multirow{2}*{\parbox[c][1cm]{2.1cm}{\centering bbr}} & \large \pmb{$\mu$} & \cellcolor{myg}27.24 & \cellcolor{myr}19.01 & 35.58\% & 9.93 & 4.04 \\
\cline{2-7}
& \large \pmb{$\sigma$} & 1.91 &  &  & 1.52 &    \\
\cline{1-2}\tabucline[2pt]{3-7}
\end{tabu}}
\label{tab:tab4246}
\end{table*}

\FloatBarrier

\begin{figure}[p!]
\vspace*{-0.3cm}
\captionsetup[subfigure]{labelformat=empty}
    \centering
    \subfloat[]{{\includegraphics[width=\textwidth]{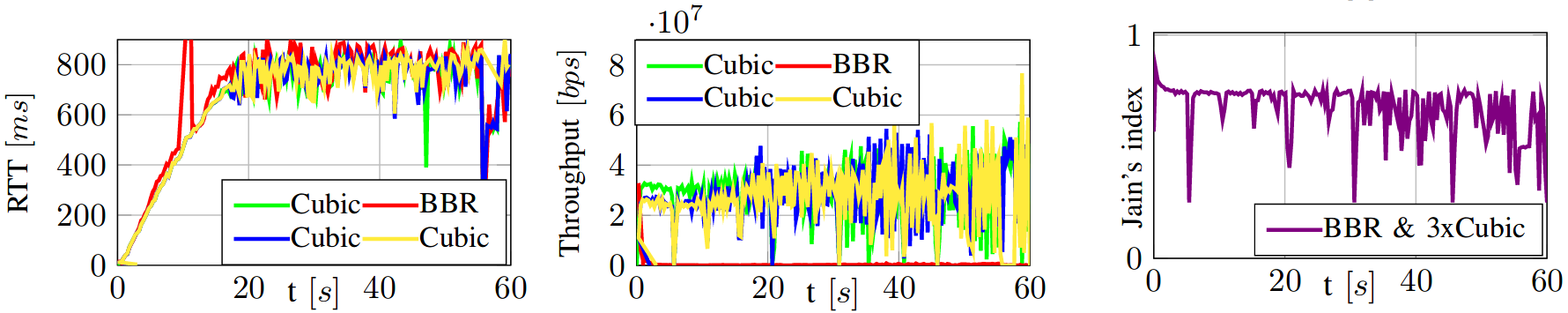} }}\\
    \vspace{-0.7cm}
    \subfloat[]{{\includegraphics[width=\textwidth/3]{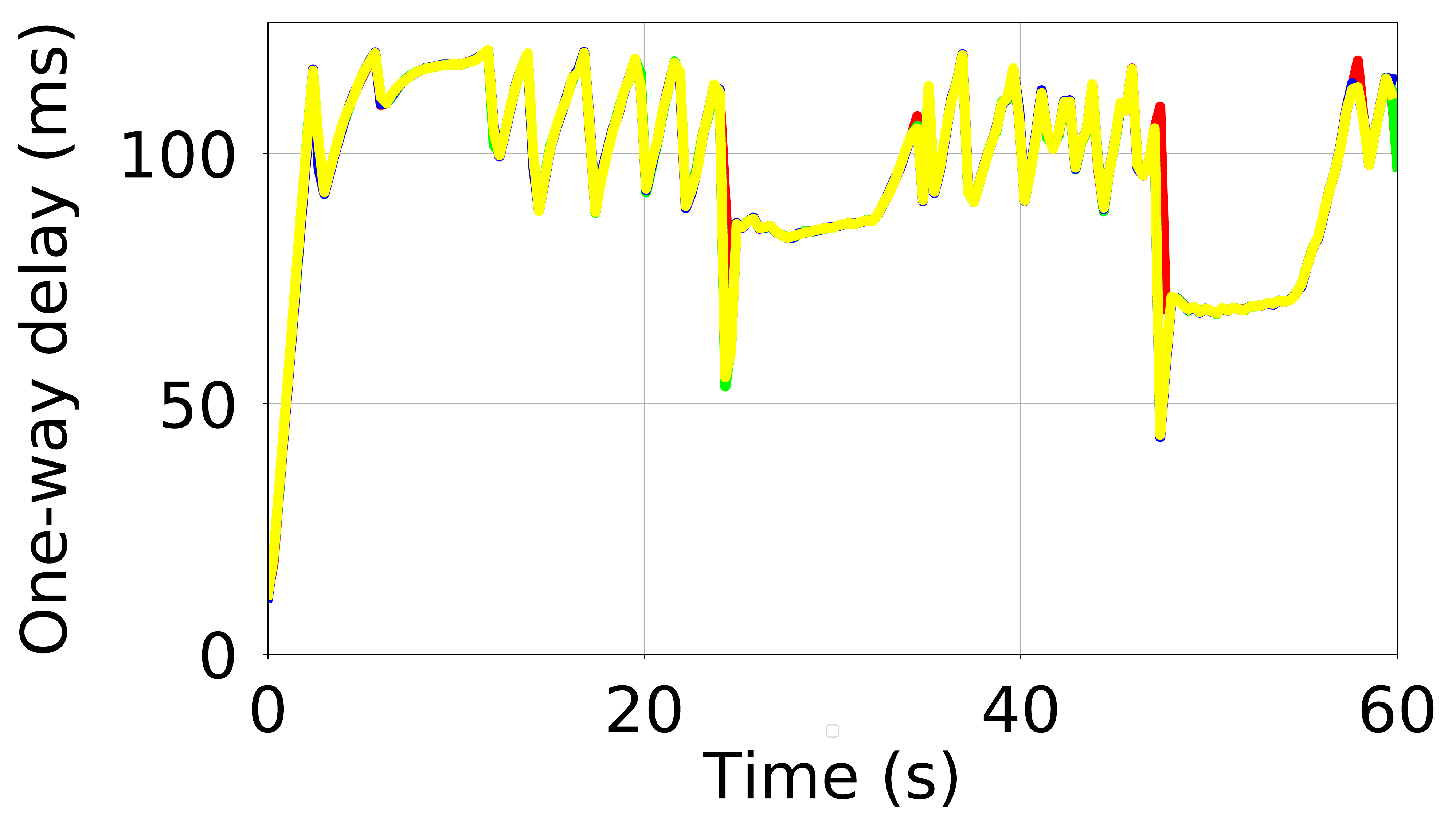} }}%
    \subfloat[]{{\includegraphics[width=\textwidth/3]{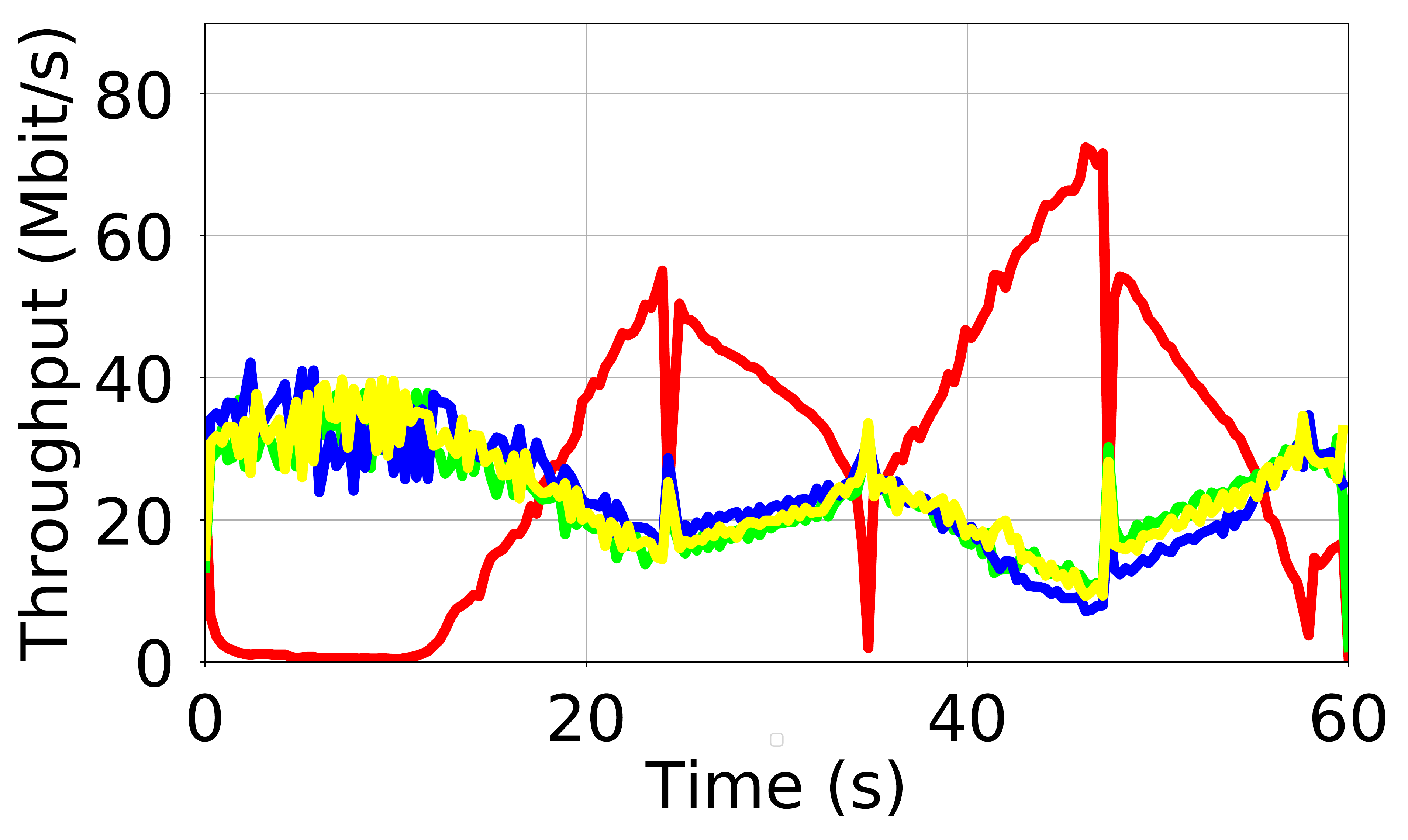} }}%
    \subfloat[]{{\includegraphics[width=\textwidth/3]{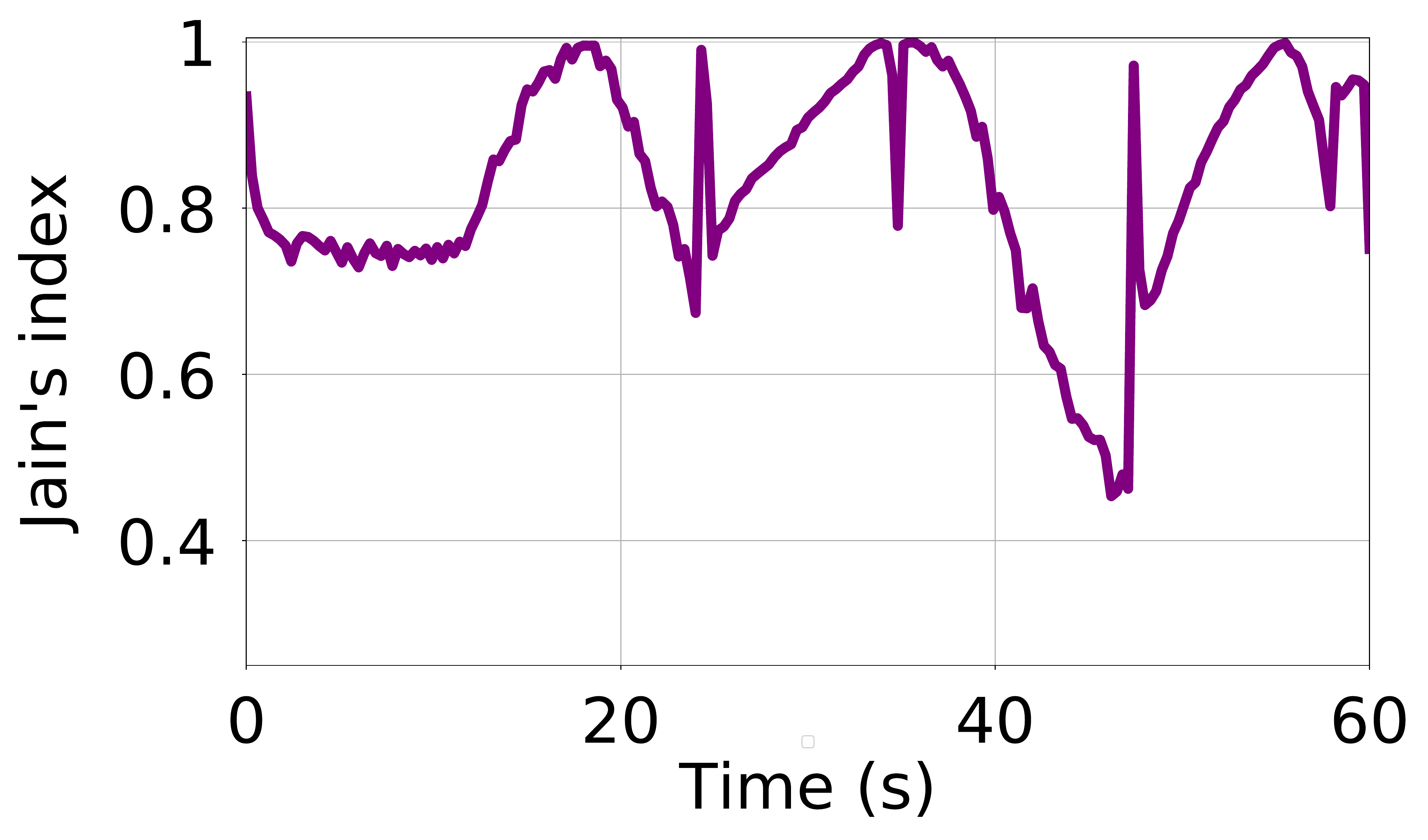} }}\\
    \vspace{-0.5cm}
    \caption{BW scenario: BBR \& 3 Cubic. The aggregation interval is 300 ms.\\The top-row plots are by the testbed, the bottom-row -- by CoCo-Beholder.}%
    \label{fig:fig4244}
\end{figure}

\begin{figure}[h!]
\vspace*{-0.2cm}
\captionsetup[subfigure]{labelformat=empty}
    \centering
    \subfloat[]{{\includegraphics[width=\textwidth]{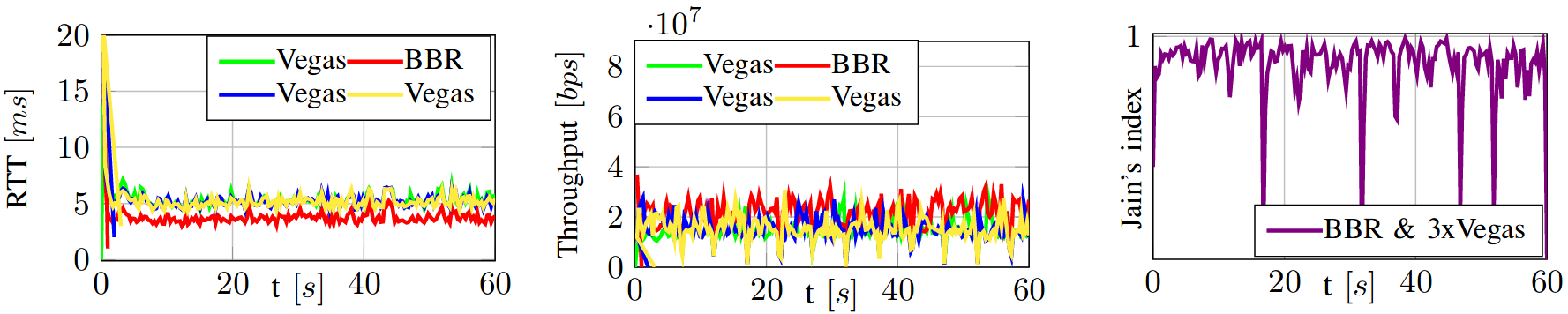} }}\\
    \vspace{-0.7cm}
    \subfloat[]{{\includegraphics[width=\textwidth/3]{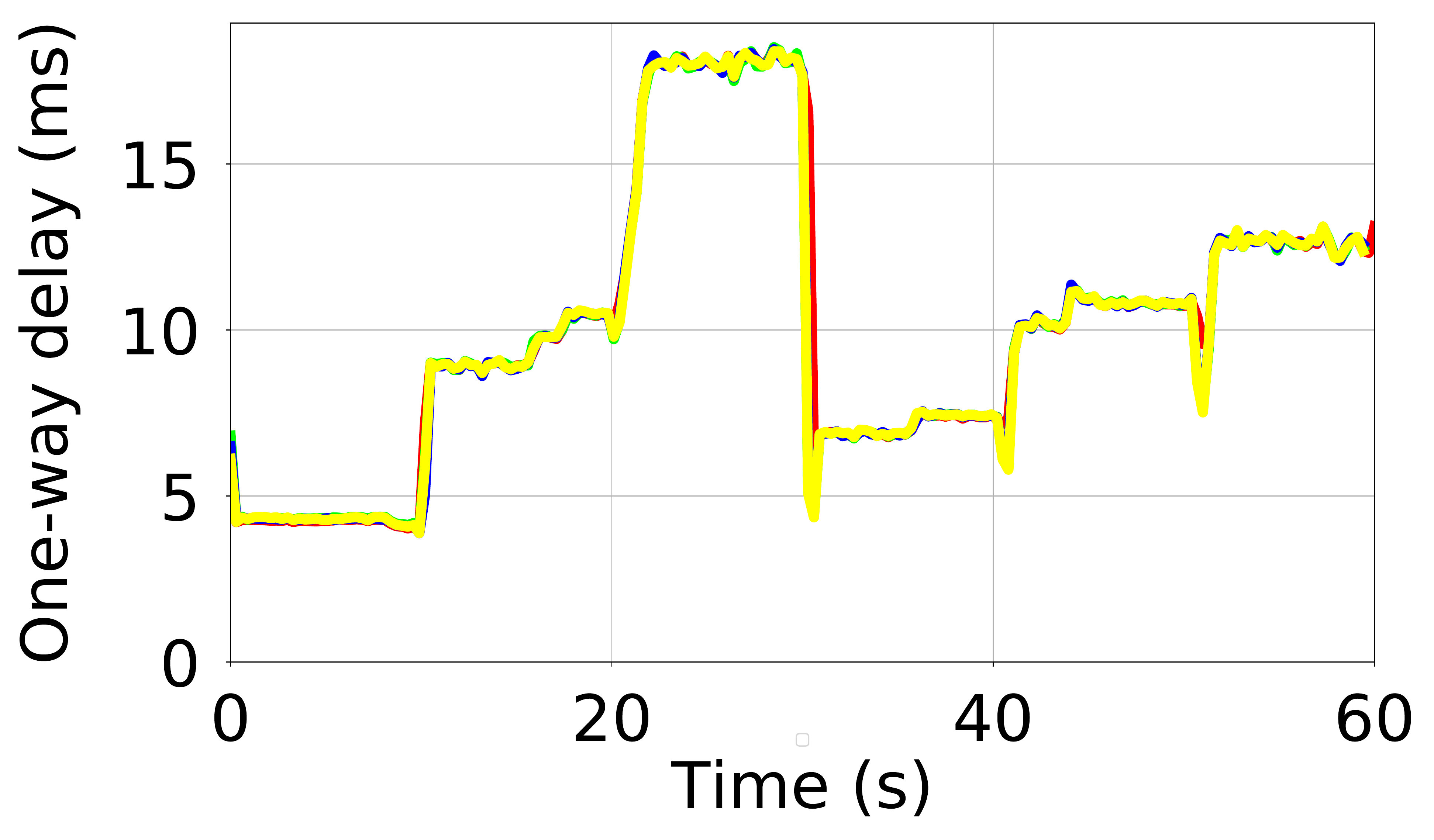} }}%
    \subfloat[]{{\includegraphics[width=\textwidth/3]{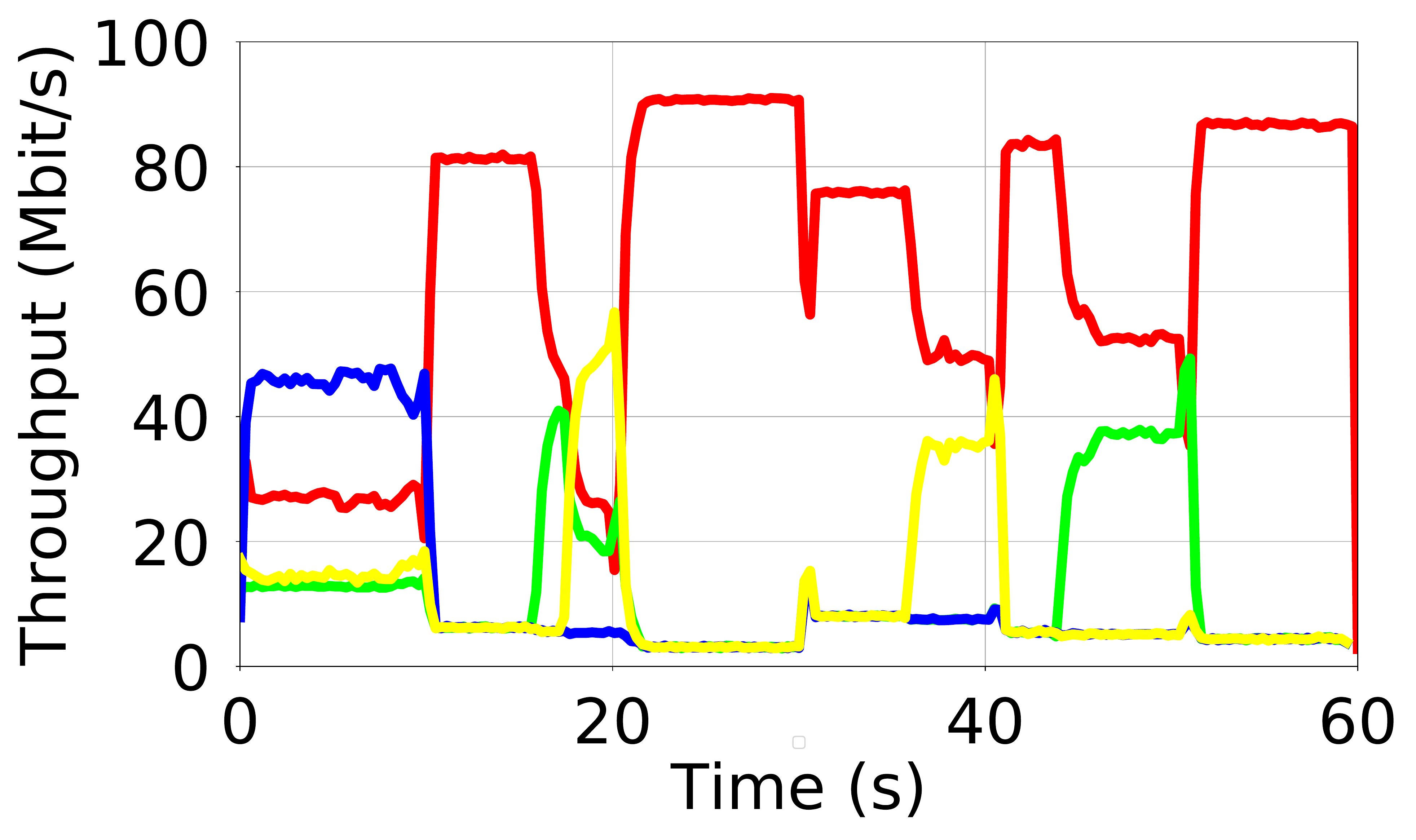} }}%
    \subfloat[]{{\includegraphics[width=\textwidth/3]{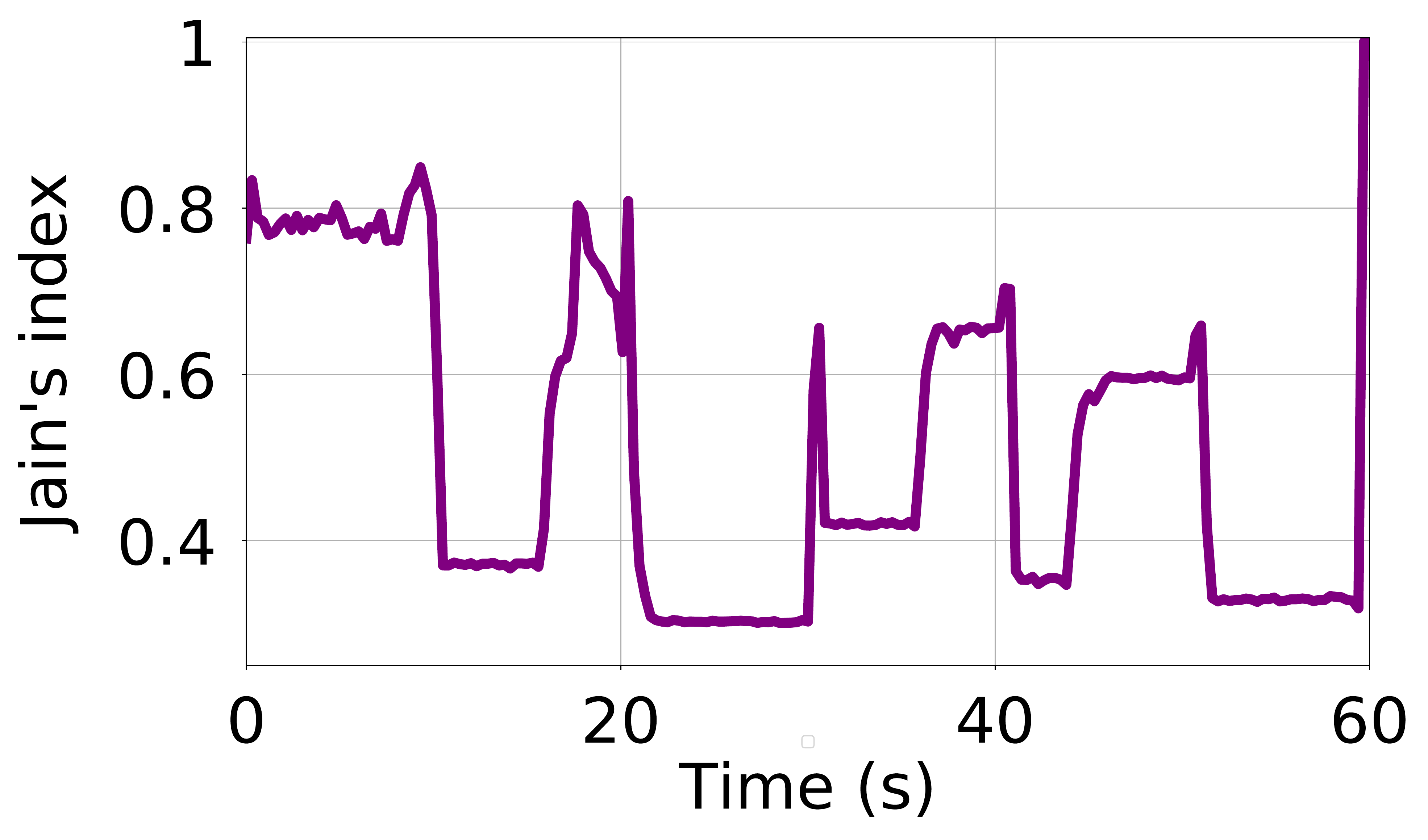} }}\\
    \vspace{-0.5cm}
    \caption{BW scenario: BBR \& 3 Vegas. The aggregation interval is 300 ms.\\The top-row plots are by the testbed, the bottom-row -- by CoCo-Beholder.}%
    \label{fig:fig4245}
\end{figure}

\begin{figure}[h!]
\vspace*{-0.2cm}
\captionsetup[subfigure]{labelformat=empty}
    \centering
    \subfloat[]{{\includegraphics[width=\textwidth]{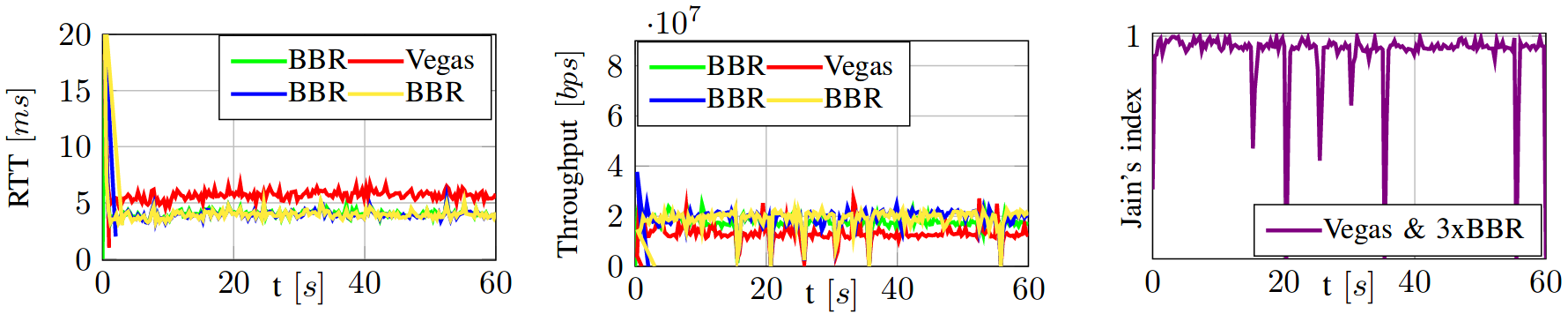} }}\\
    \vspace{-0.7cm}
    \subfloat[]{{\includegraphics[width=\textwidth/3]{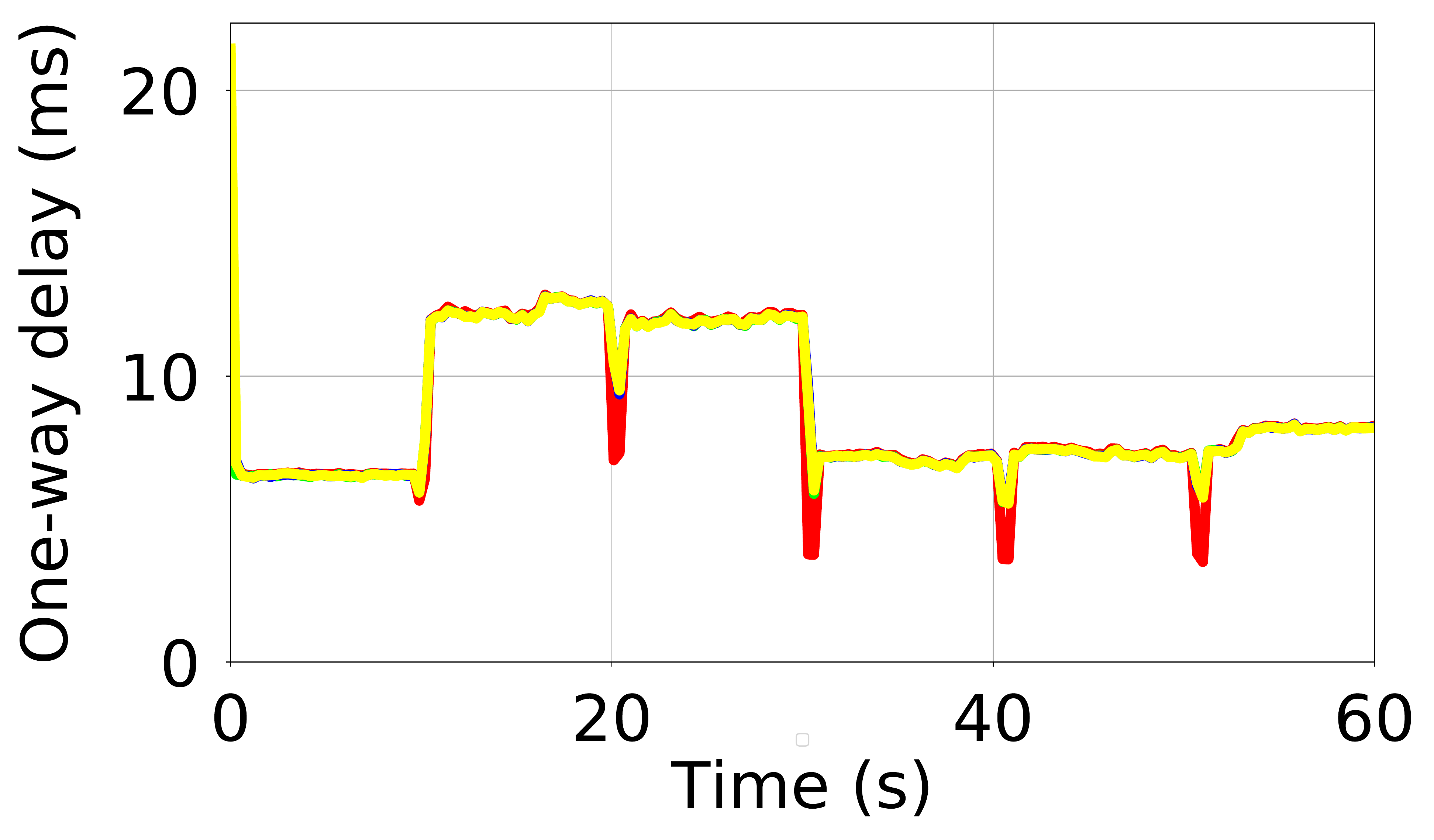} }}%
    \subfloat[]{{\includegraphics[width=\textwidth/3]{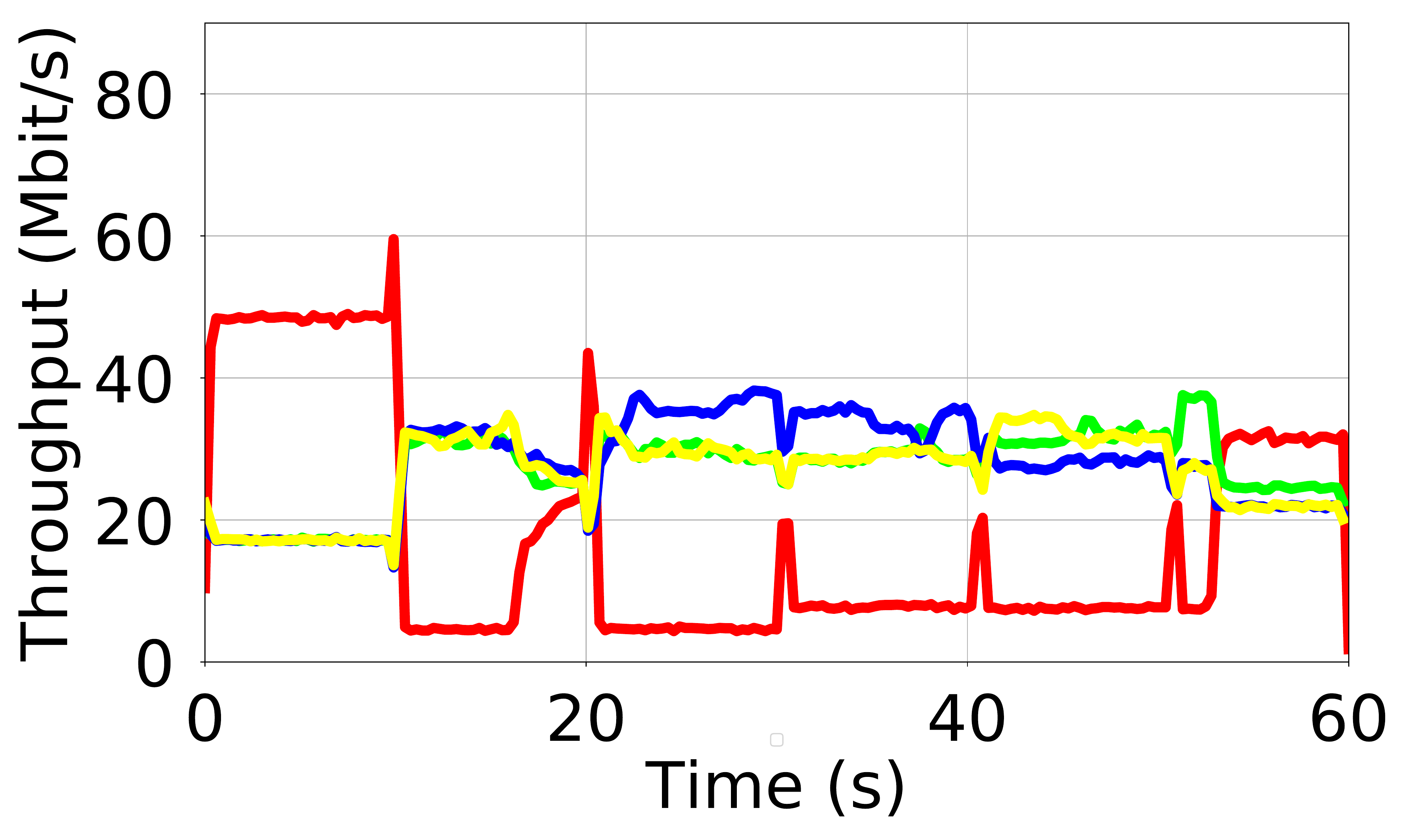} }}%
    \subfloat[]{{\includegraphics[width=\textwidth/3]{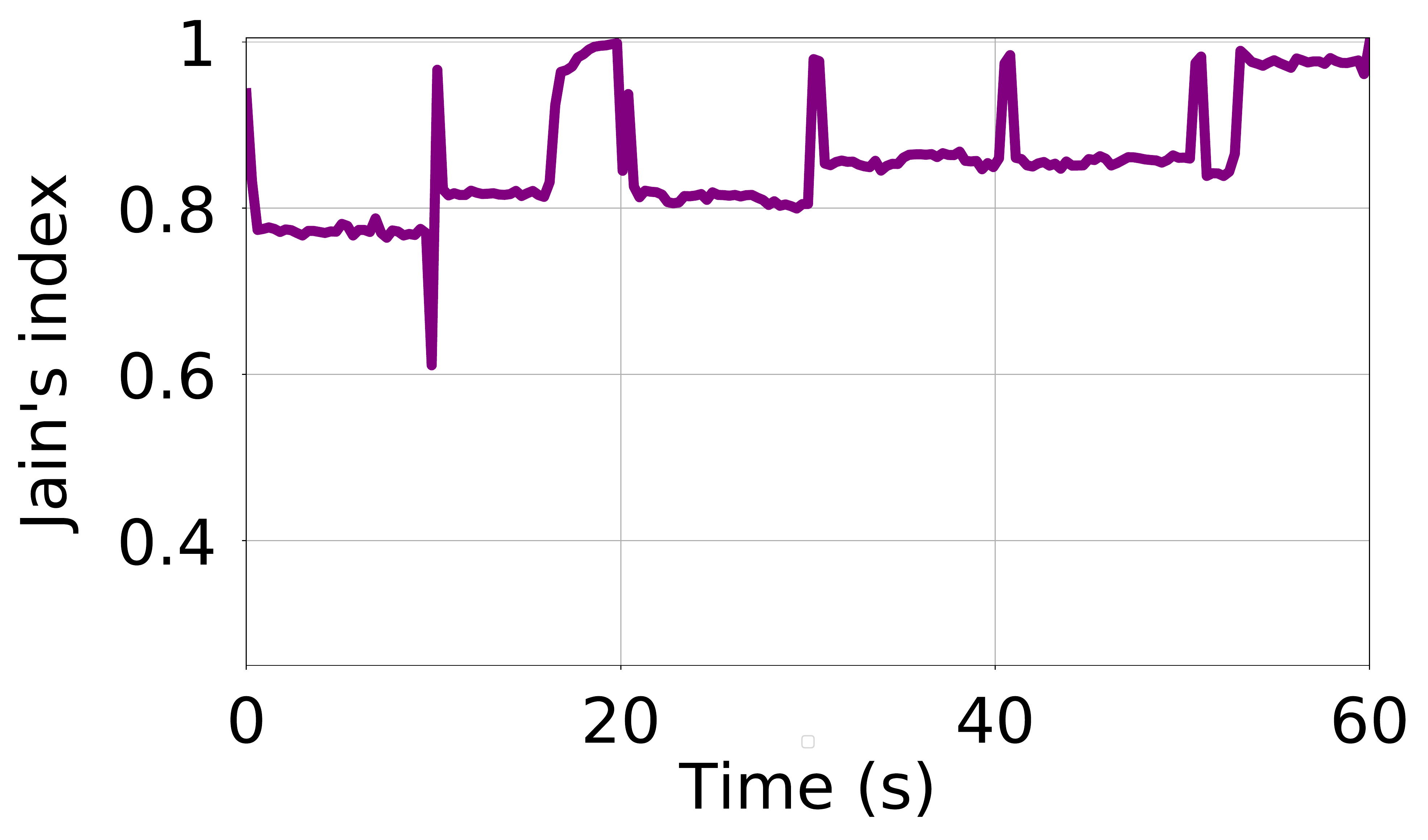} }}\\
    \vspace{-0.4cm}
    \caption{BW scenario: Vegas \& 3 BBR. The aggregation interval is 300 ms.\\The top-row plots are by the testbed, the bottom-row -- by CoCo-Beholder.}%
    \label{fig:fig4246}
\end{figure}

\FloatBarrier

\subsection{Intra-RTT-fairness For Two Flows}
\label{subsec:ssec5}

In RTT scenario, two flows of the same scheme have individual RTTs, 0 ms and 200\nolinebreak[4] ms, (i.e., one-way delays, 0 ms and 100 ms) set up at the links in the right half of the dumbbell topology. The resulting tables and figures can be found on pages~\pageref{tab:tab4251} and~\pageref{fig:fig4251}.

The RTT plots by the testbed~\cite{turkovic2019fifty} and the one-way delay plots by CoCo-Beholder look similar and according to expectations. In particular, for Vegas, the curves in the RTT plot in Figure~\ref{fig:fig4252} have close to perfect 0 ms and 250 ms values, while the curves in the one-way delay plot have the perfect 0 ms and 100 ms values.

For Cubic, the rate and fairness plots by the testbed and by CoCo-Beholder look very much alike: as highlighted in paper~\cite{turkovic2019fifty}, the flow with the lower RTT is favored, even though Cubic claims to be RTT-fair. The difference is, however, that in the rate plot of the testbed, the rate curves lie a little bit closer to each other, which indicates better fairness, and, also, the rate curves oscillate much more.

For two Vegas flows, the testbed witnessed the great fairness, which was only a little bit worse than the fairness shown by the testbed in BW scenario for two Vegas flows: the flows converged not instantly but only after five seconds of the runtime. On the contrary, when launched in CoCo-Beholder, the flow with the \emph{higher} RTT was favored, and the rates of the two flows differed by a factor more than two during the whole runtime.

For two BBR flows, the rate plots by the testbed and CoCo-Beholder look similar: the flow with the \emph{higher} RTT is favored, and every ten seconds the flow with the lower RTT makes an attempt to get its share of the bandwidth reducing the rate of the higher-RTT flow significantly but only for a moment. However, for CoCo-Beholder, the domination of the flow with the higher RTT was more evident: the two curves lie very close to \mbox{0 Mbit/s} and 100 Mbit/s in the rate plot, which is why the Jain's index by CoCo-Beholder approaches the minimum and the fairness plots by the testbed and CoCo-Beholder are noticeably different.

\subsection{Intra-RTT-fairness For Four Flows}
\label{subsec:ssec6}

In RTT scenario, four flows of a scheme have individual 100, 200, 300, and 400 ms RTTs (i.e., 50, 100, 150, 200 ms  one-way delays), set up at the links in the right half of the dumbbell topology. The resulting tables and figures can be found on pages~\pageref{tab:tab4254} and~\pageref{fig:fig4254}.

The curves in the RTT plot by the testbed~\cite{turkovic2019fifty} and in the one-way delay plot by CoCo-Beholder for Vegas look nearly identical, and Table~\ref{tab:tab4255} indicates that in both cases the RTTs of four flows were rather close to the perfect 100, 200, 300, and 400\nolinebreak[4] ms. \mbox{For Cubic} and BBR, the curves in the one-way delay plots by CoCo-Beholder are noticeably periodic, as opposed to the curves in the RTT plots by the testbed. Both the testbed and the emulator showed that, as highlighted in the paper~\cite{turkovic2019fifty}, for BBR, the increase in the number of flows led to the increase of the average RTTs of the flows to the level of loss-based Cubic (see Tables~\ref{tab:tab4254} and~\ref{tab:tab4256} to compare these statistics for Cubic and\nolinebreak[4] BBR).  

Concerning the rate, for Cubic, the testbed~\cite{turkovic2019fifty} and CoCo-Beholder demonstrated that, the same as for the case with two flows discussed in the previous section, the lower-RTT flow is favored. However, in the testbed, the other three flows competed for the bandwidth much more aggressively, which is reflected in the higher Jain's index statistic comparing to that of CoCo-Beholder, as seen in Table~\ref{tab:tab4254}.   

For Vegas, the rate and fairness plots by the testbed and the emulator look very much alike,  with the lower-RTT flow being favored. This is in contrast to the case with two flows discussed in the previous section, where CoCo-Beholder showed that the higher-RTT flow was favored instead and the testbed showed that the rates of the two flows converged to a common value after several seconds. It should be also noted that the sample standard deviations of the rates by CoCo-Beholder are big (see Table~\ref{tab:tab4255}). 

For BBR, analogously to the case with two flows, the results by the testbed~\cite{turkovic2019fifty} showed that the two higher-RTT flows outperformed the two lower-RTT flows. On the contrary, CoCo-Beholder showed the good RTT-fairness: the four BBR flows converged after around ten seconds of the runtime. This is why the Jain's index statistic by CoCo-Beholder is 0.85 against 0.7 by the testbed, as appears in Table~\ref{tab:tab4256}.

\subsection{Summary of BW and RTT Scenarios}
\label{subsec:ssec7}

\textbf{BW scenario:}
\begin{itemize}
\item The testbed~\cite{turkovic2019fifty} showed that Vegas is most intra-fair, while CoCo-Beholder -- that BBR is most intra-fair and Vegas' results were unfair and unstable for four flows.

\item Both the testbed and CoCo-Beholder proved that if at least one flow of loss-based Cubic is present in the bottleneck link, all other flows experience big delays.

\item CoCo-Beholder showed that one BBR flow tried to fight with three Cubic flows for bandwidth, while in the testbed all the bandwidth was captured by Cubic flows.

\item While the testbed witnessed that BBR and Vegas work well together with high fairness, CoCo-Beholder showed that Vegas always looses to both BBR and Cubic.
\end{itemize}

\FloatBarrier

\begin{table*}[h!]
\vspace{0.3cm}
\centering
\Large
\caption{RTT scenario: 2 Cubic flows.}
\renewcommand{\arraystretch}{1.4} 
\resizebox*{\textwidth}{!}{\begin{tabu}{|c|cV{5}c|c|cV{5}c|cV{5}c|cV{5}c|c|cV{5}}
\cline{1-7}\cline{10-12}
\multirow{2}*{\parbox[c][2.5cm]{2.1cm}{\centering \bf \large Scheme}} &  \multicolumn{1}{c|}{\multirow{2}*{\parbox[c][2.5cm]{0.0cm}{}}} & \multicolumn{3}{c|}{\parbox[c][1cm]{2.7cm}{\bf \large \centering Rate (Mbps)}} & \parbox[c][1cm]{1.8cm}{\bf \large \centering Delay (ms)} & \multicolumn{1}{c|}{\parbox[c][1cm]{1.8cm}{\bf \large \centering RTT (ms)}}  & \multicolumn{1}{c}{} & \multicolumn{1}{c|}{} & \multicolumn{3}{c|}{\parbox[c][1cm]{2.5cm}{\bf \centering \large Jain's index}}\\ 
\cline{3-7}\cline{10-12}
 & \multicolumn{1}{c|}{} & \parbox[c][1.5cm]{1.8cm}{\centering \bf \normalsize CoCo-Beholder}  & \parbox[1cm]{1.8cm}{\centering \bf \normalsize Testbed} & \multicolumn{1}{c|}{\parbox{1.8cm}{\centering \pmb{$d_r$}}} & \parbox{1.8cm}{\centering \bf \normalsize CoCo-Beholder} & \multicolumn{1}{c|}{\parbox{1.8cm}{\centering \bf \normalsize Testbed}} &\multicolumn{1}{c}{} & \multicolumn{1}{c|}{}   & \parbox{1.8cm}{\centering \bf \normalsize CoCo-Beholder}& \parbox{1.8cm}{\centering \bf \normalsize Testbed} &\multicolumn{1}{c|}{ \parbox{1.8cm}{\centering \pmb{$d_r$}}}\\
\cline{1-2}\noalign{\vskip-1pt}\tabucline[2pt]{3-7}\noalign{\vskip-2pt}\cline{9-9}\tabucline[2pt]{10-12}
\multirow{2}*{\parbox[c][1cm]{2.1cm}{\centering cubic}} & \large \pmb{$\mu$} & \cellcolor{myg}82.30 & \cellcolor{myr}64.20 & 24.71\% & 101.39 & 368.25 & & \large \pmb{$\mu$} & 0.71 & 0.72 & 2.06\%\\
\cline{2-7}\cline{9-12}
& \large \pmb{$\sigma$} & 3.37 &  &  & 0.94  &  &   &  \large \pmb{$\sigma$}& 0.04 & &\\
\cline{1-2}\tabucline[2pt]{3-7}\noalign{\vskip-2pt}\cline{9-9}\tabucline[2pt]{10-12}
\multirow{2}*{\parbox[c][1cm]{2.1cm}{\centering cubic}} & \large \pmb{$\mu$} & \cellcolor{myr}17.33 & \cellcolor{myg}25.05 & 36.42\% & 201.20 & 570.07 \\
\cline{2-7}
& \large \pmb{$\sigma$} & 2.98 &  &  & 0.89 &    \\
\cline{1-2}\tabucline[2pt]{3-7}
\end{tabu}}
\label{tab:tab4251}
\end{table*}

\begin{table*}[h!]
\vspace{3cm}
\centering
\Large
\caption{RTT scenario: 2 Vegas flows.}
\renewcommand{\arraystretch}{1.4} 
\resizebox*{\textwidth}{!}{\begin{tabu}{|c|cV{5}c|c|cV{5}c|cV{5}c|cV{5}c|c|cV{5}}
\cline{1-7}\cline{10-12}
\multirow{2}*{\parbox[c][2.5cm]{2.1cm}{\centering \bf \large Scheme}} &  \multicolumn{1}{c|}{\multirow{2}*{\parbox[c][2.5cm]{0.0cm}{}}} & \multicolumn{3}{c|}{\parbox[c][1cm]{2.7cm}{\bf \large \centering Rate (Mbps)}} & \parbox[c][1cm]{1.8cm}{\bf \large \centering Delay (ms)} & \multicolumn{1}{c|}{\parbox[c][1cm]{1.8cm}{\bf \large \centering RTT (ms)}}  & \multicolumn{1}{c}{} & \multicolumn{1}{c|}{} & \multicolumn{3}{c|}{\parbox[c][1cm]{2.5cm}{\bf \centering \large Jain's index}}\\ 
\cline{3-7}\cline{10-12}
 & \multicolumn{1}{c|}{} & \parbox[c][1.5cm]{1.8cm}{\centering \bf \normalsize CoCo-Beholder}  & \parbox[1cm]{1.8cm}{\centering \bf \normalsize Testbed} & \multicolumn{1}{c|}{\parbox{1.8cm}{\centering \pmb{$d_r$}}} & \parbox{1.8cm}{\centering \bf \normalsize CoCo-Beholder} & \multicolumn{1}{c|}{\parbox{1.8cm}{\centering \bf \normalsize Testbed}} &\multicolumn{1}{c}{} & \multicolumn{1}{c|}{}   & \parbox{1.8cm}{\centering \bf \normalsize CoCo-Beholder}& \parbox{1.8cm}{\centering \bf \normalsize Testbed} &\multicolumn{1}{c|}{ \parbox{1.8cm}{\centering \pmb{$d_r$}}}\\
\cline{1-2}\noalign{\vskip-1pt}\tabucline[2pt]{3-7}\noalign{\vskip-2pt}\cline{9-9}\tabucline[2pt]{10-12}
\multirow{2}*{\parbox[c][1cm]{2.1cm}{\centering vegas}} & \large \pmb{$\mu$} & \cellcolor{myr}29.06 & \cellcolor{myg}34.62 & 17.45\% & 2.51 & 2.37 & & \large \pmb{$\mu$} & 0.80 & 0.87 & 8.40\%\\
\cline{2-7}\cline{9-12}
& \large \pmb{$\sigma$} & 3.96 &  &  & 0.30 &  &   &  \large \pmb{$\sigma$}& 0.05 & &\\
\cline{1-2}\tabucline[2pt]{3-7}\noalign{\vskip-2pt}\cline{9-9}\tabucline[2pt]{10-12}
\multirow{2}*{\parbox[c][1cm]{2.1cm}{\centering vegas}} & \large \pmb{$\mu$} & \cellcolor{myg}70.65 & \cellcolor{myr}36.72 & 63.20\% & 103.35 & 249.33   \\
\cline{2-7}
& \large \pmb{$\sigma$} & 3.94 &  &  & 0.43 &    \\
\cline{1-2}\tabucline[2pt]{3-7}
\end{tabu}}
\label{tab:tab4252}
\end{table*}

\begin{table*}[h!]
\vspace{3cm}
\centering
\Large
\caption{RTT scenario: 2 BBR flows.}
\renewcommand{\arraystretch}{1.4} 
\resizebox*{\textwidth}{!}{\begin{tabu}{|c|cV{5}c|c|cV{5}c|cV{5}c|cV{5}c|c|cV{5}}
\cline{1-7}\cline{10-12}
\multirow{2}*{\parbox[c][2.5cm]{2.1cm}{\centering \bf \large Scheme}} &  \multicolumn{1}{c|}{\multirow{2}*{\parbox[c][2.5cm]{0.0cm}{}}} & \multicolumn{3}{c|}{\parbox[c][1cm]{2.7cm}{\bf \large \centering Rate (Mbps)}} & \parbox[c][1cm]{1.8cm}{\bf \large \centering Delay (ms)} & \multicolumn{1}{c|}{\parbox[c][1cm]{1.8cm}{\bf \large \centering RTT (ms)}}  & \multicolumn{1}{c}{} & \multicolumn{1}{c|}{} & \multicolumn{3}{c|}{\parbox[c][1cm]{2.5cm}{\bf \centering \large Jain's index}}\\ 
\cline{3-7}\cline{10-12}
 & \multicolumn{1}{c|}{} & \parbox[c][1.5cm]{1.8cm}{\centering \bf \normalsize CoCo-Beholder}  & \parbox[1cm]{1.8cm}{\centering \bf \normalsize Testbed} & \multicolumn{1}{c|}{\parbox{1.8cm}{\centering \pmb{$d_r$}}} & \parbox{1.8cm}{\centering \bf \normalsize CoCo-Beholder} & \multicolumn{1}{c|}{\parbox{1.8cm}{\centering \bf \normalsize Testbed}} &\multicolumn{1}{c}{} & \multicolumn{1}{c|}{}   & \parbox{1.8cm}{\centering \bf \normalsize CoCo-Beholder}& \parbox{1.8cm}{\centering \bf \normalsize Testbed} &\multicolumn{1}{c|}{ \parbox{1.8cm}{\centering \pmb{$d_r$}}}\\
\cline{1-2}\noalign{\vskip-1pt}\tabucline[2pt]{3-7}\noalign{\vskip-2pt}\cline{9-9}\tabucline[2pt]{10-12}
\multirow{2}*{\parbox[c][1cm]{2.1cm}{\centering bbr}} & \large \pmb{$\mu$} & \cellcolor{myr}10.95 & \cellcolor{myg}30.21 & 93.59\% & 43.81 & 15.91 & & \large \pmb{$\mu$} & \cellcolor{myr}0.55 & \cellcolor{myg}0.76 & 31.58\%\\
\cline{2-7}\cline{9-12}
& \large \pmb{$\sigma$} & 0.11&  &  & 0.48 &  &   &  \large \pmb{$\sigma$}&0.00 & &\\
\cline{1-2}\tabucline[2pt]{3-7}\noalign{\vskip-2pt}\cline{9-9}\tabucline[2pt]{10-12}
\multirow{2}*{\parbox[c][1cm]{2.1cm}{\centering bbr}} & \large \pmb{$\mu$} & \cellcolor{myg}88.50& \cellcolor{myr}50.98 & 53.80\% & 147.11 & 268.30  \\
\cline{2-7}
& \large \pmb{$\sigma$} & 0.06 &  &  & 0.65 &    \\
\cline{1-2}\tabucline[2pt]{3-7}
\end{tabu}}
\label{tab:tab4253}
\end{table*}

\FloatBarrier

\begin{figure}[p!]
\vspace*{-0.3cm}
\captionsetup[subfigure]{labelformat=empty}
    \centering
    \subfloat[]{{\includegraphics[width=\textwidth]{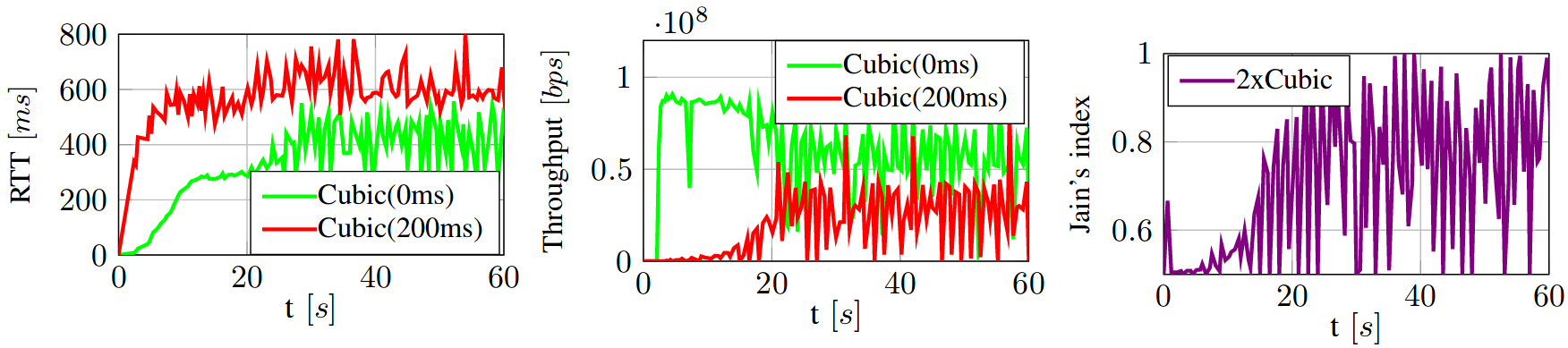} }}\\
    \vspace{-0.7cm}
    \subfloat[]{{\includegraphics[width=\textwidth/3]{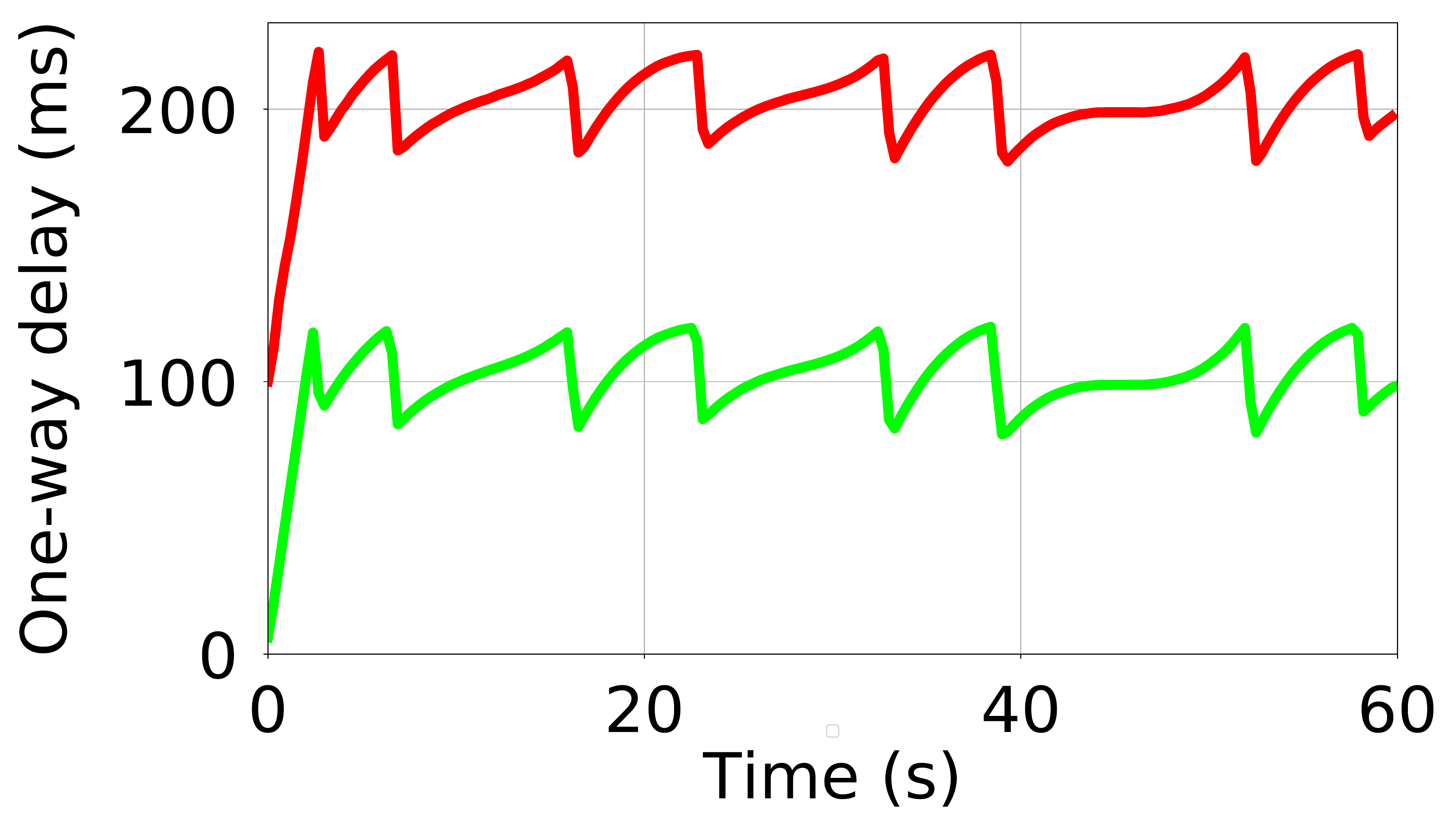} }}%
    \subfloat[]{{\includegraphics[width=\textwidth/3]{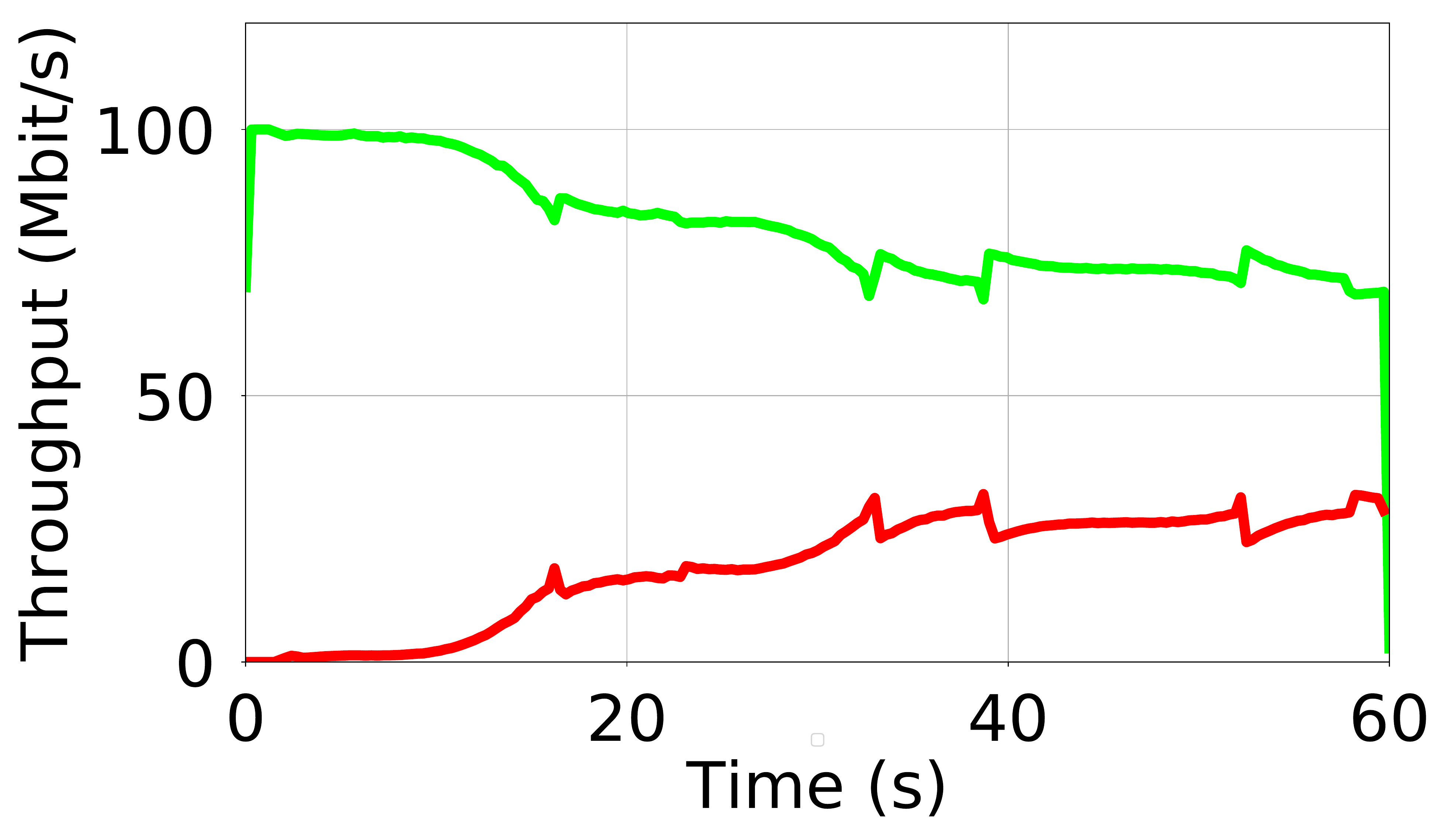} }}%
    \subfloat[]{{\includegraphics[width=\textwidth/3]{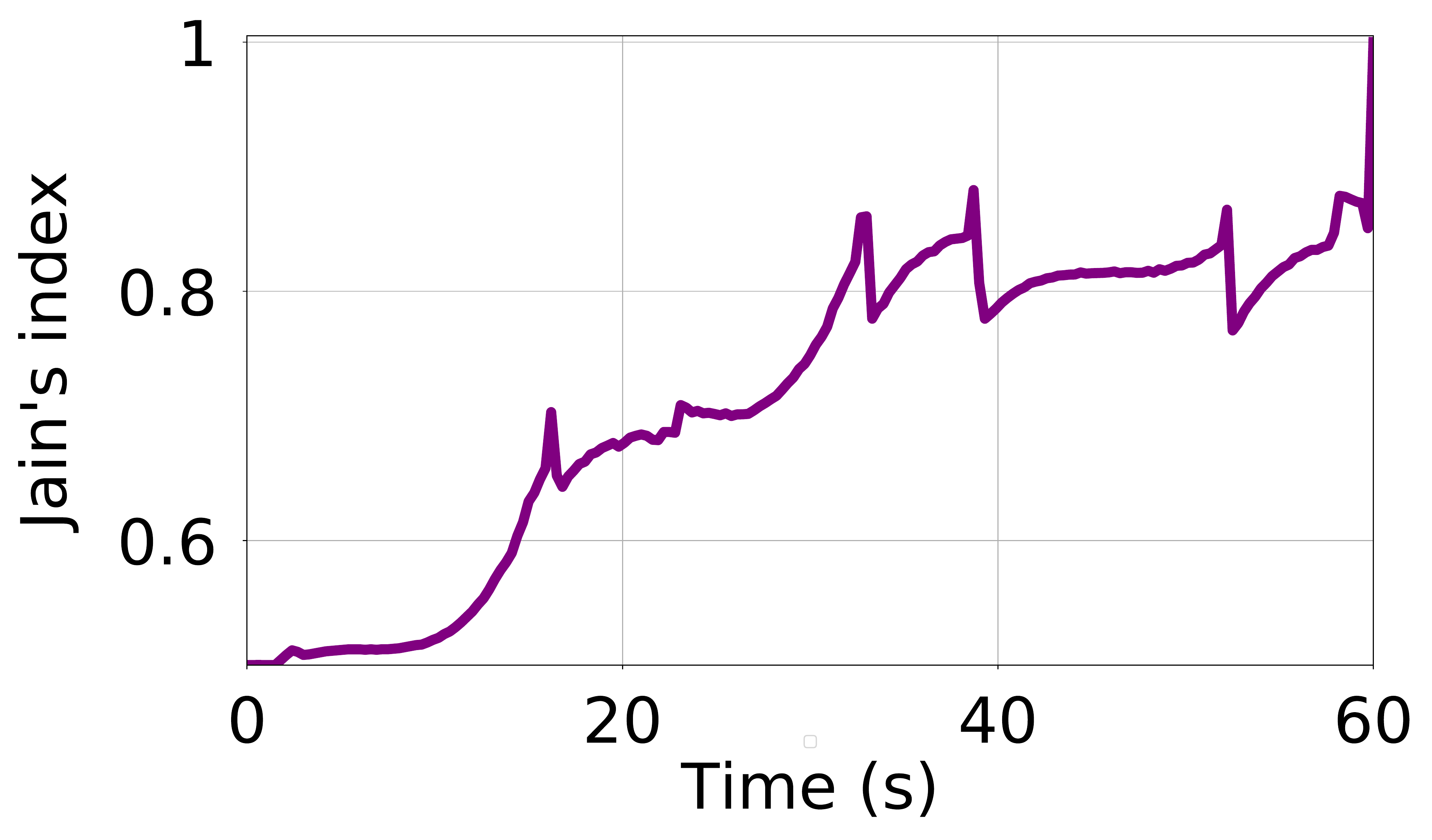} }}\\
    \vspace{-0.5cm}
    \caption{RTT scenario: 2 Cubic flows. The aggregation interval is 300 ms.\\The top-row plots are by the testbed, the bottom-row -- by CoCo-Beholder.}%
    \label{fig:fig4251}
\end{figure}

\begin{figure}[h!]
\vspace*{-0.2cm}
\captionsetup[subfigure]{labelformat=empty}
    \centering
    \subfloat[]{{\includegraphics[width=\textwidth]{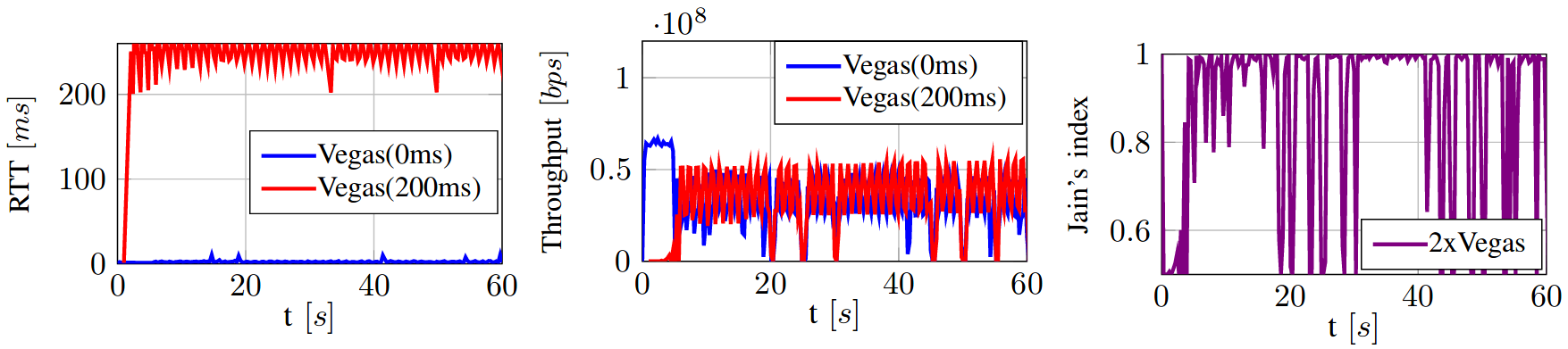} }}\\
    \vspace{-0.7cm}
    \subfloat[]{{\includegraphics[width=\textwidth/3]{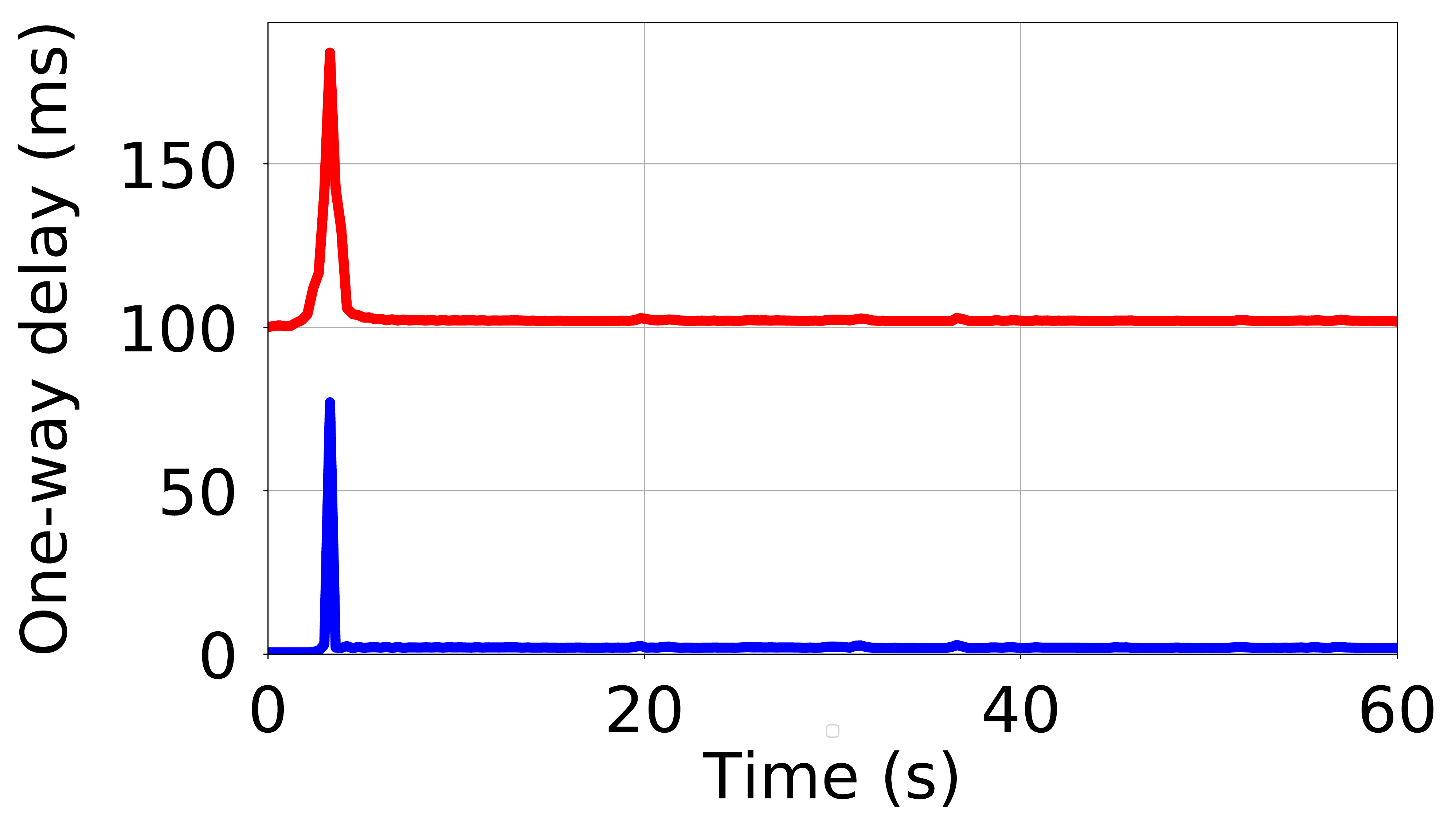} }}%
    \subfloat[]{{\includegraphics[width=\textwidth/3]{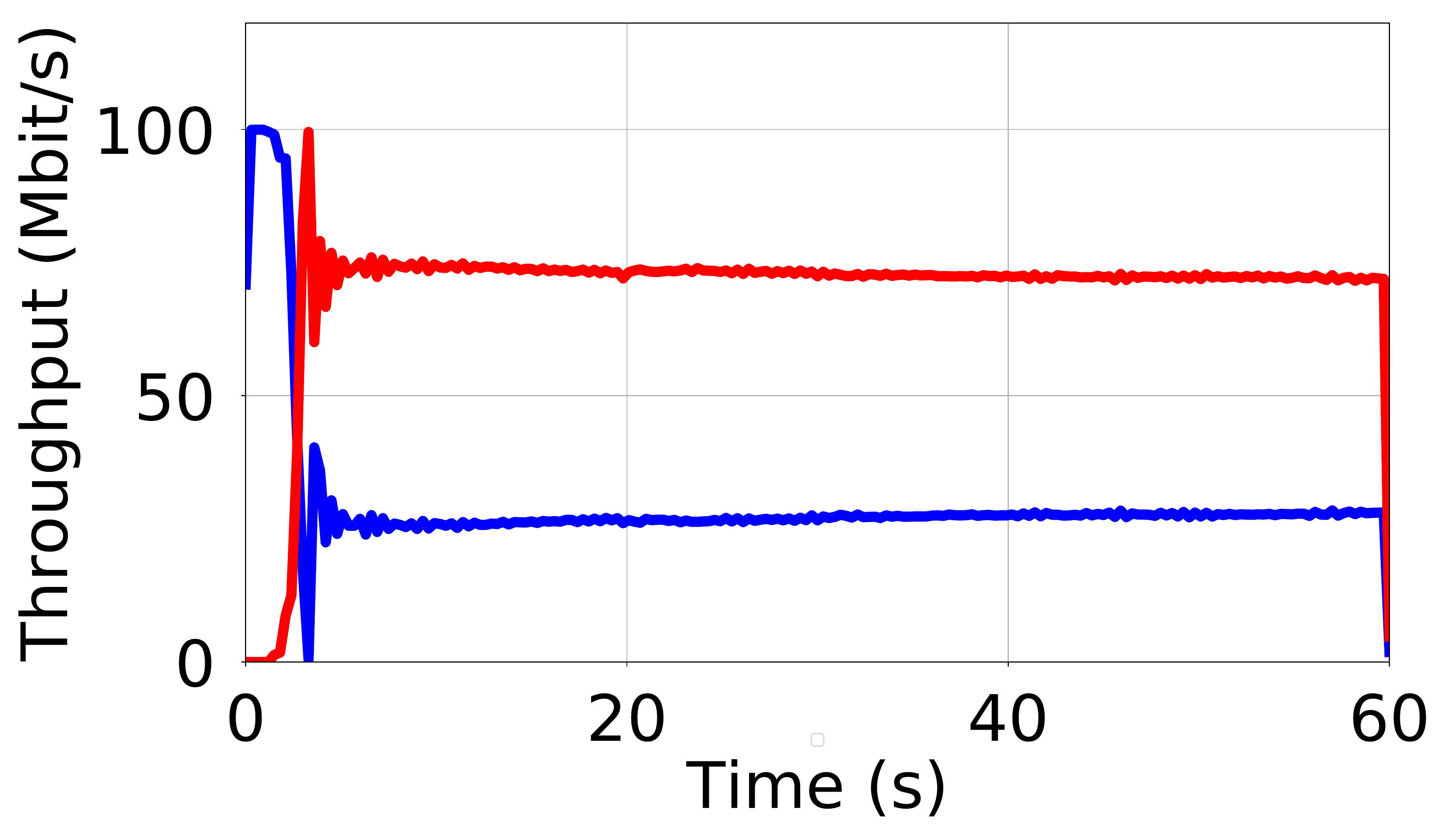} }}%
    \subfloat[]{{\includegraphics[width=\textwidth/3]{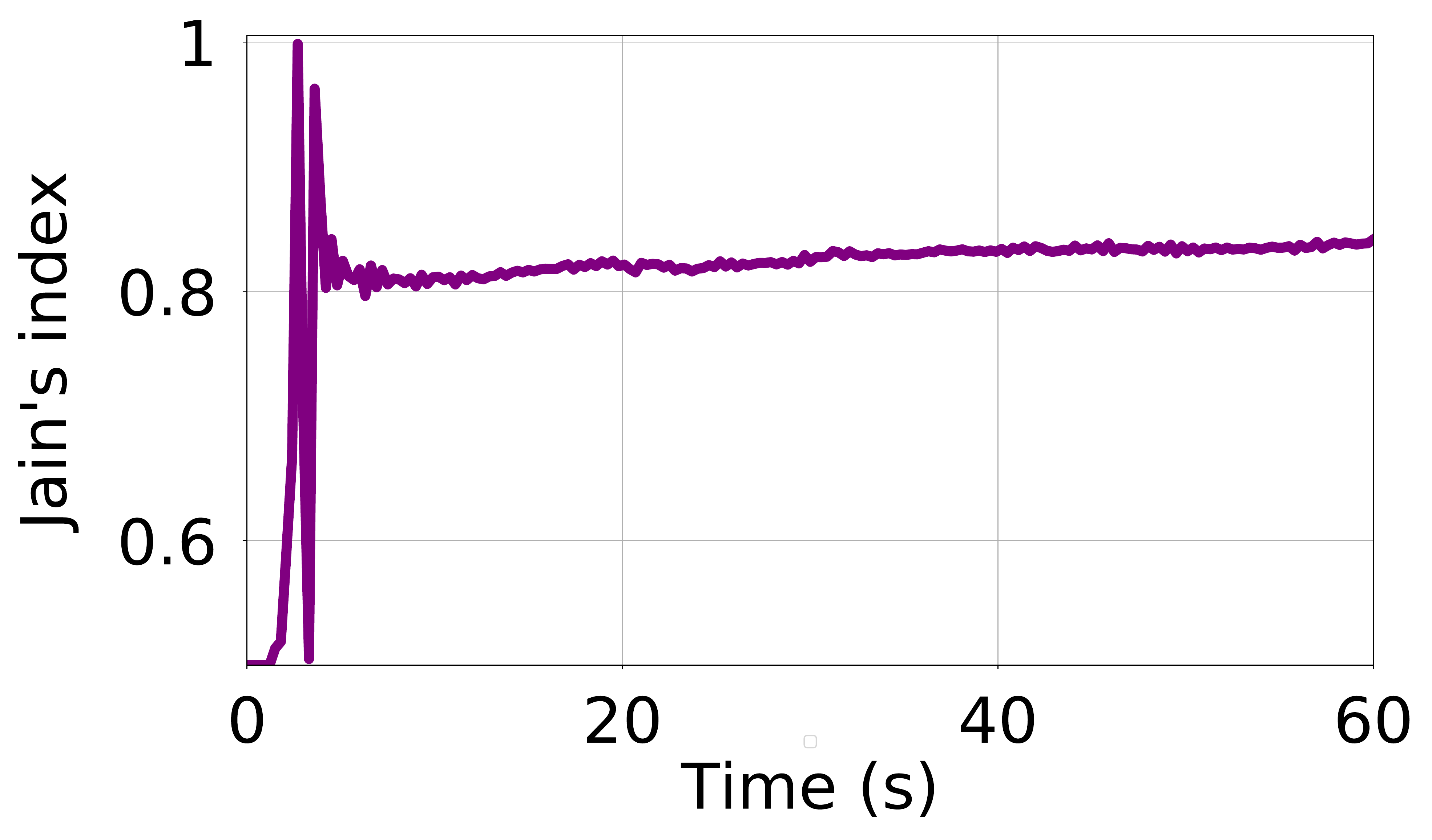} }}\\
    \vspace{-0.5cm}
    \caption{RTT scenario: 2 Vegas flows. The aggregation interval is 300 ms.\\The top-row plots are by the testbed, the bottom-row -- by CoCo-Beholder.}%
    \label{fig:fig4252}
\end{figure}

\begin{figure}[h!]
\vspace*{-0.2cm}
\captionsetup[subfigure]{labelformat=empty}
    \centering
    \subfloat[]{{\includegraphics[width=\textwidth]{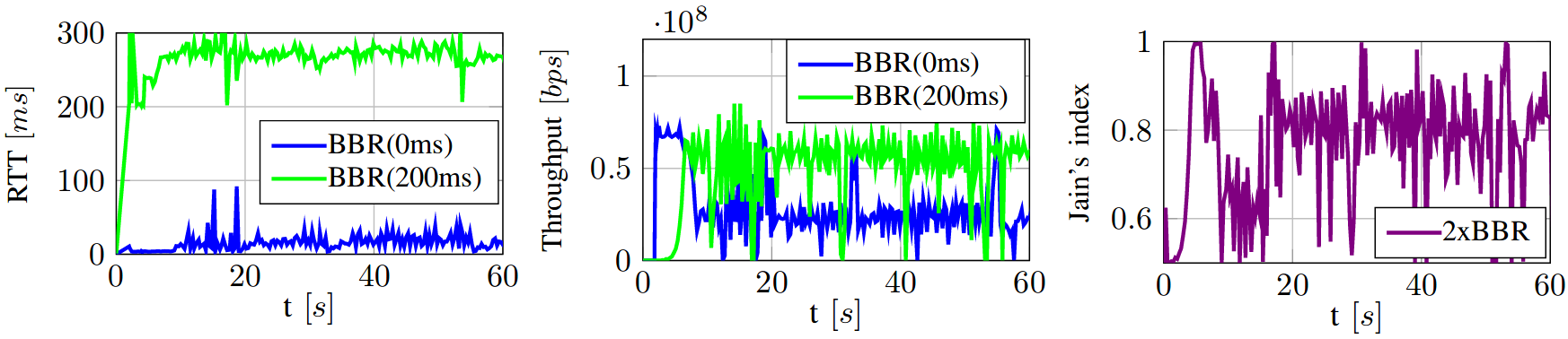} }}\\
    \vspace{-0.7cm}
    \subfloat[]{{\includegraphics[width=\textwidth/3]{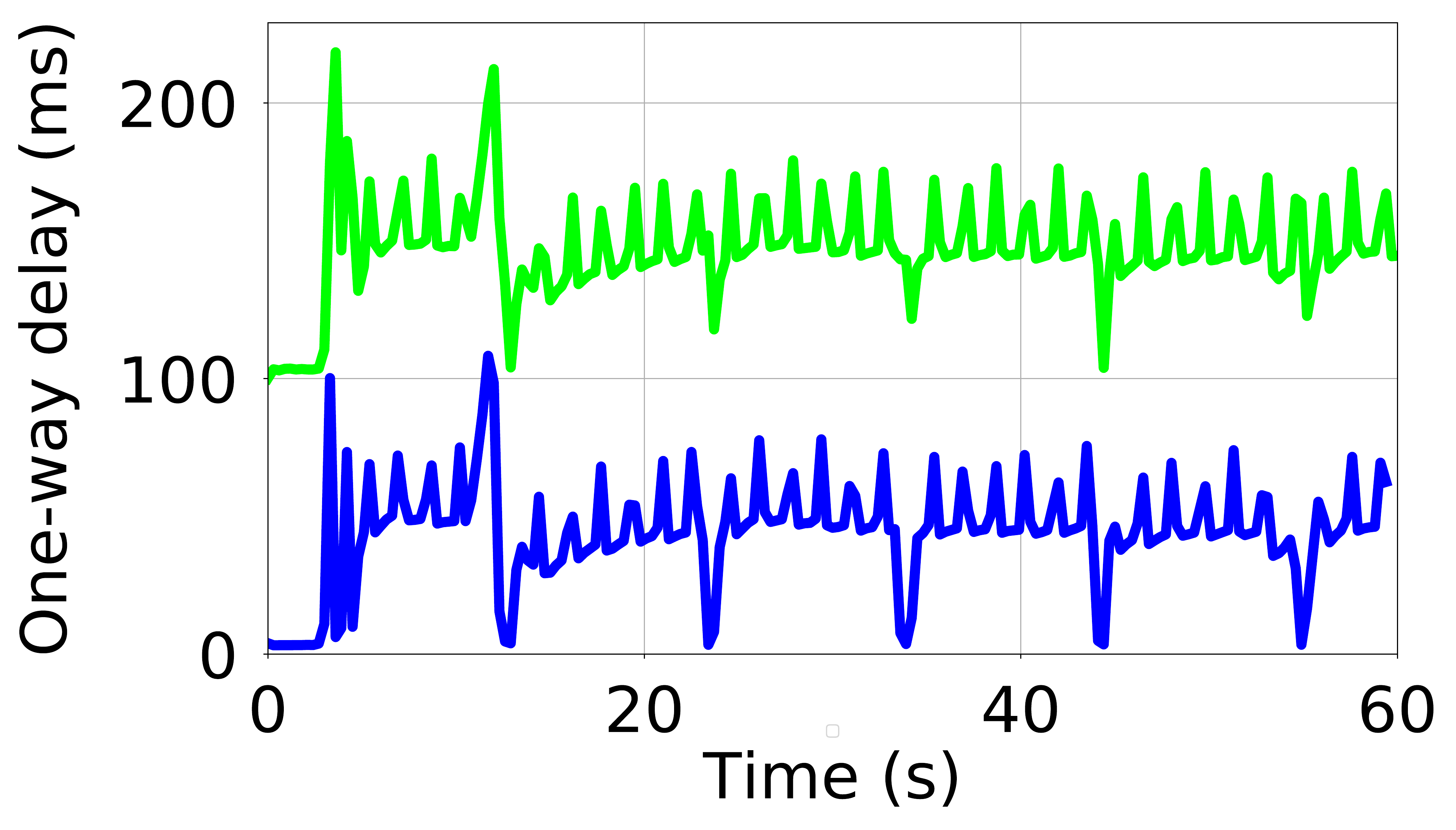} }}%
    \subfloat[]{{\includegraphics[width=\textwidth/3]{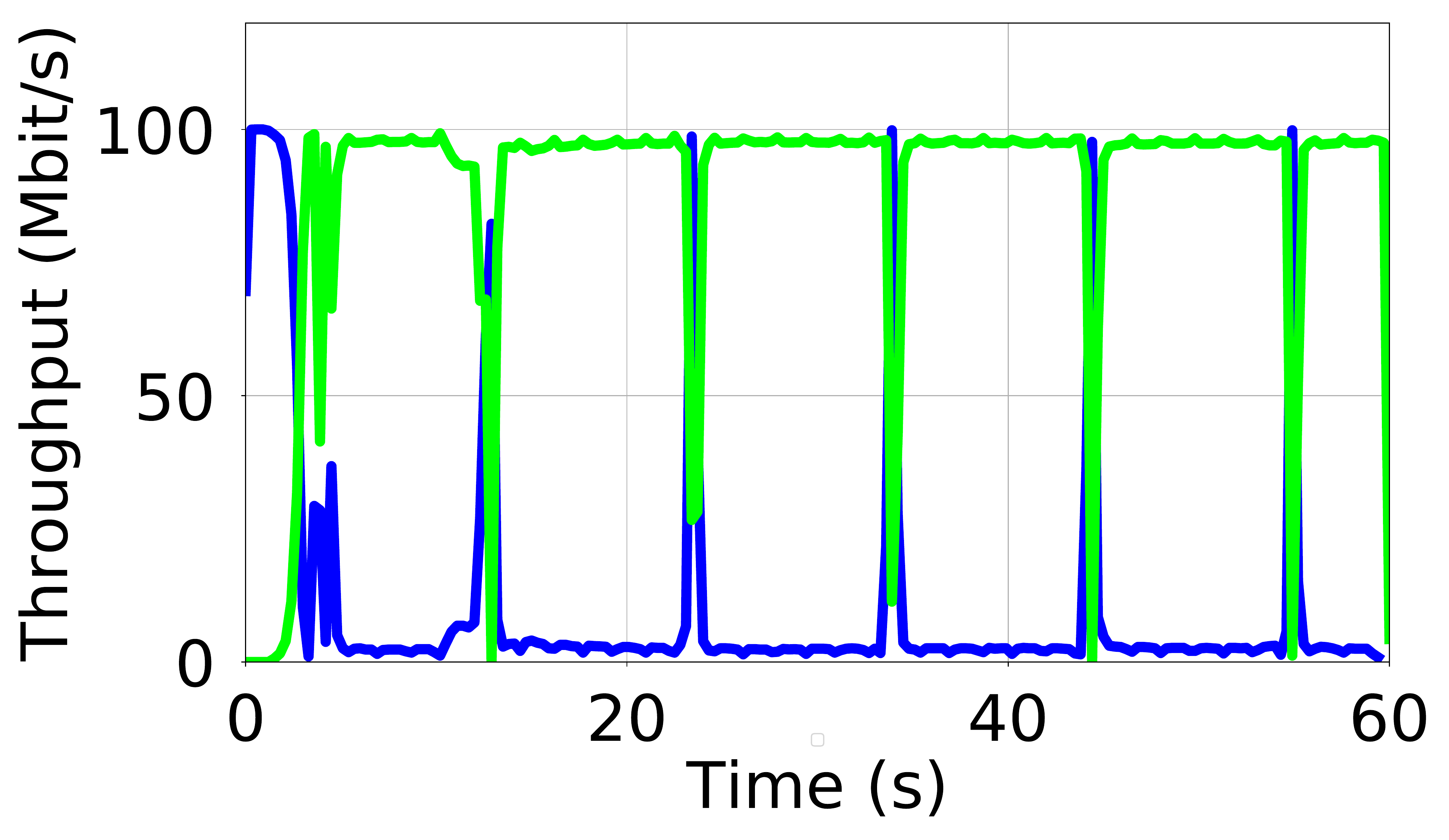} }}%
    \subfloat[]{{\includegraphics[width=\textwidth/3]{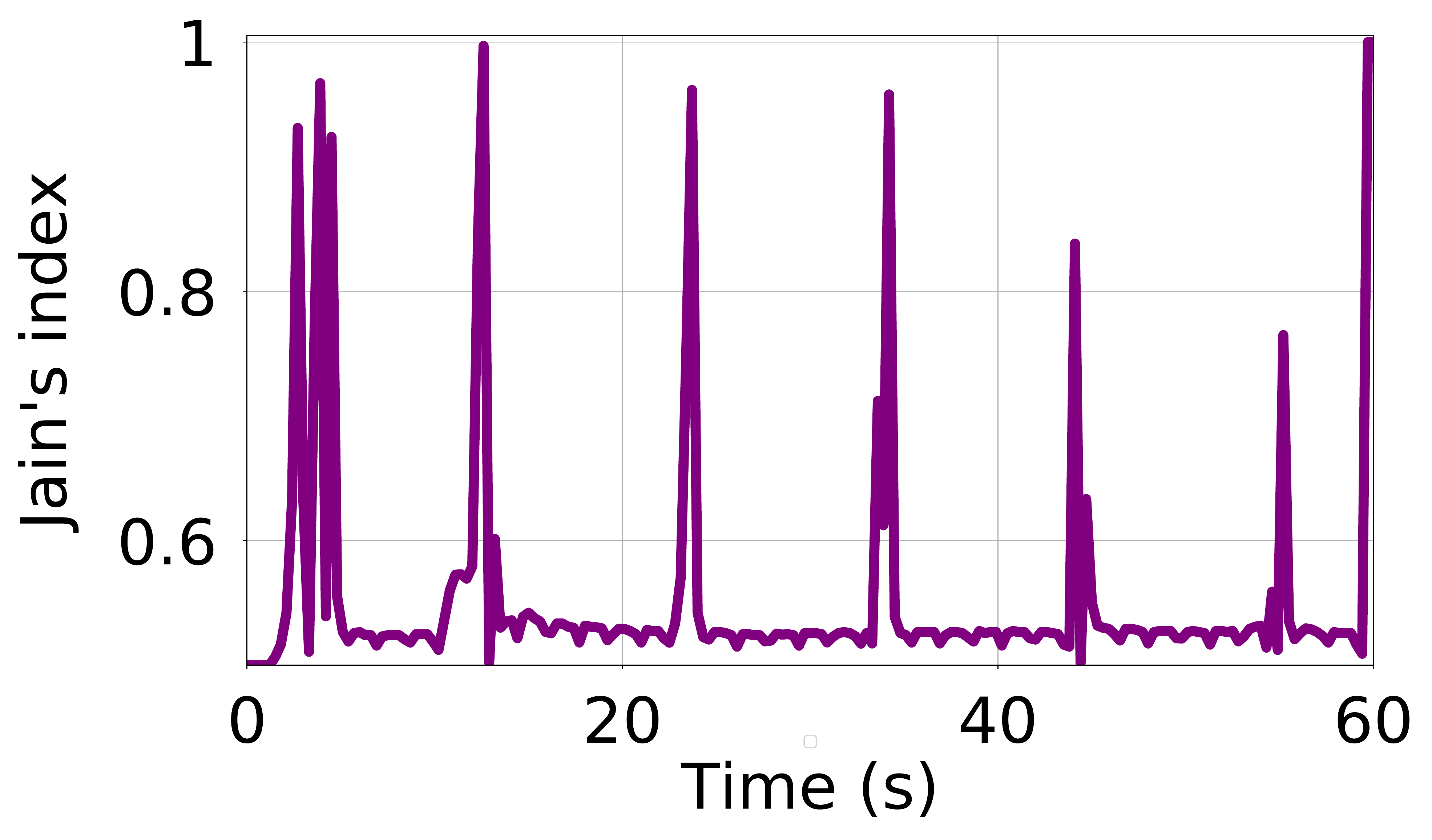} }}\\
    \vspace{-0.4cm}
    \caption{RTT scenario: 2 BBR flows. The aggregation interval is 300 ms.\\The top-row plots are by the testbed, the bottom-row -- by CoCo-Beholder.}%
    \label{fig:fig4253}
\end{figure}

\FloatBarrier

\begin{table*}[p!]
\vspace{0.3cm}
\centering
\Large
\caption{RTT scenario: 4 Cubic flows.}
\renewcommand{\arraystretch}{1.4} 
\resizebox*{\textwidth}{!}{\begin{tabu}{|c|cV{5}c|c|cV{5}c|cV{5}c|cV{5}c|c|cV{5}}
\cline{1-7}\cline{10-12}
\multirow{2}*{\parbox[c][2.5cm]{2.1cm}{\centering \bf \large Scheme}} &  \multicolumn{1}{c|}{\multirow{2}*{\parbox[c][2.5cm]{0.0cm}{}}} & \multicolumn{3}{c|}{\parbox[c][1cm]{2.7cm}{\bf \large \centering Rate (Mbps)}} & \parbox[c][1cm]{1.8cm}{\bf \large \centering Delay (ms)} & \multicolumn{1}{c|}{\parbox[c][1cm]{1.8cm}{\bf \large \centering RTT (ms)}}  & \multicolumn{1}{c}{} & \multicolumn{1}{c|}{} & \multicolumn{3}{c|}{\parbox[c][1cm]{2.5cm}{\bf \centering \large Jain's index}}\\ 
\cline{3-7}\cline{10-12}
 & \multicolumn{1}{c|}{} & \parbox[c][1.5cm]{1.8cm}{\centering \bf \normalsize CoCo-Beholder}  & \parbox[1cm]{1.8cm}{\centering \bf \normalsize Testbed} & \multicolumn{1}{c|}{\parbox{1.8cm}{\centering \pmb{$d_r$}}} & \parbox{1.8cm}{\centering \bf \normalsize CoCo-Beholder} & \multicolumn{1}{c|}{\parbox{1.8cm}{\centering \bf \normalsize Testbed}} &\multicolumn{1}{c}{} & \multicolumn{1}{c|}{}   & \parbox{1.8cm}{\centering \bf \normalsize CoCo-Beholder}& \parbox{1.8cm}{\centering \bf \normalsize Testbed} &\multicolumn{1}{c|}{ \parbox{1.8cm}{\centering \pmb{$d_r$}}}\\
\cline{1-2}\noalign{\vskip-1pt}\tabucline[2pt]{3-7}\noalign{\vskip-2pt}\cline{9-9}\tabucline[2pt]{10-12}
\multirow{2}*{\parbox[c][1cm]{2.1cm}{\centering cubic}} & \large \pmb{$\mu$} & \cellcolor{myg}63.29 & \cellcolor{myr}41.74 & 41.04\% & 133.13 & 550.68  & & \large \pmb{$\mu$} & \cellcolor{myr}0.56 & \cellcolor{myg}0.64 & 12.88\%\\
\cline{2-7}\cline{9-12}
& \large \pmb{$\sigma$} & 5.10 &  &  & 4.65 &  &   &  \large \pmb{$\sigma$}& 0.06& &\\
\cline{1-2}\tabucline[2pt]{3-7}\noalign{\vskip-2pt}\cline{9-9}\tabucline[2pt]{10-12}
\multirow{2}*{\parbox[c][1cm]{2.1cm}{\centering cubic}} & \large \pmb{$\mu$} & \cellcolor{myg}15.61 & \cellcolor{myr}11.86 & 27.29\% & 183.08 & 755.17 \\
\cline{2-7}
& \large \pmb{$\sigma$} & 4.70 &  &  & 4.65 &    \\
\cline{1-2}\tabucline[2pt]{3-7}
\multirow{2}*{\parbox[c][1cm]{2.1cm}{\centering cubic}} & \large \pmb{$\mu$} & \cellcolor{myr}9.99 & \cellcolor{myg}18.48 &59.63\%  &232.98 & 749.21 \\
\cline{2-7}
& \large \pmb{$\sigma$} & 2.97&  &  & 4.62&    \\
\cline{1-2}\tabucline[2pt]{3-7}
\multirow{2}*{\parbox[c][1cm]{2.1cm}{\centering cubic}} & \large \pmb{$\mu$} & 8.71 & 8.73 & 0.26\% & 282.62& 828.25 \\
\cline{2-7}
& \large \pmb{$\sigma$} & 2.43&  &  & 4.55&    \\
\cline{1-2}\tabucline[2pt]{3-7}
\end{tabu}}
\label{tab:tab4254}
\end{table*}

\begin{table*}[h!]
\vspace{0.3cm}
\centering
\Large
\caption{RTT scenario: 4 Vegas flows.}
\renewcommand{\arraystretch}{1.4} 
\resizebox*{\textwidth}{!}{\begin{tabu}{|c|cV{5}c|c|cV{5}c|cV{5}c|cV{5}c|c|cV{5}}
\cline{1-7}\cline{10-12}
\multirow{2}*{\parbox[c][2.5cm]{2.1cm}{\centering \bf \large Scheme}} &  \multicolumn{1}{c|}{\multirow{2}*{\parbox[c][2.5cm]{0.0cm}{}}} & \multicolumn{3}{c|}{\parbox[c][1cm]{2.7cm}{\bf \large \centering Rate (Mbps)}} & \parbox[c][1cm]{1.8cm}{\bf \large \centering Delay (ms)} & \multicolumn{1}{c|}{\parbox[c][1cm]{1.8cm}{\bf \large \centering RTT (ms)}}  & \multicolumn{1}{c}{} & \multicolumn{1}{c|}{} & \multicolumn{3}{c|}{\parbox[c][1cm]{2.5cm}{\bf \centering \large Jain's index}}\\ 
\cline{3-7}\cline{10-12}
 & \multicolumn{1}{c|}{} & \parbox[c][1.5cm]{1.8cm}{\centering \bf \normalsize CoCo-Beholder}  & \parbox[1cm]{1.8cm}{\centering \bf \normalsize Testbed} & \multicolumn{1}{c|}{\parbox{1.8cm}{\centering \pmb{$d_r$}}} & \parbox{1.8cm}{\centering \bf \normalsize CoCo-Beholder} & \multicolumn{1}{c|}{\parbox{1.8cm}{\centering \bf \normalsize Testbed}} &\multicolumn{1}{c}{} & \multicolumn{1}{c|}{}   & \parbox{1.8cm}{\centering \bf \normalsize CoCo-Beholder}& \parbox{1.8cm}{\centering \bf \normalsize Testbed} &\multicolumn{1}{c|}{ \parbox{1.8cm}{\centering \pmb{$d_r$}}}\\
\cline{1-2}\noalign{\vskip-1pt}\tabucline[2pt]{3-7}\noalign{\vskip-2pt}\cline{9-9}\tabucline[2pt]{10-12}
\multirow{2}*{\parbox[c][1cm]{2.1cm}{\centering vegas}} & \large \pmb{$\mu$} & \cellcolor{myg}52.50 & \cellcolor{myr}43.19 & 19.46\% & 58.56 & 125.87 & & \large \pmb{$\mu$} & 0.60 & 0.57 & 5.37\%\\
\cline{2-7}\cline{9-12}
& \large \pmb{$\sigma$} & 12.42 &  &  & 2.07 &  &   &  \large \pmb{$\sigma$}& 0.10& &\\
\cline{1-2}\tabucline[2pt]{3-7}\noalign{\vskip-2pt}\cline{9-9}\tabucline[2pt]{10-12}
\multirow{2}*{\parbox[c][1cm]{2.1cm}{\centering vegas}} & \large \pmb{$\mu$} & 7.26 & 7.10 & 2.29\% &108.11 & 226.26\\
\cline{2-7}
& \large \pmb{$\sigma$} & 1.26&  &  & 2.03&    \\
\cline{1-2}\tabucline[2pt]{3-7}
\multirow{2}*{\parbox[c][1cm]{2.1cm}{\centering vegas}} & \large \pmb{$\mu$} & \cellcolor{myg}13.54 & \cellcolor{myr}3.73 &113.61\% & 158.49& 326.22 \\
\cline{2-7}
& \large \pmb{$\sigma$} & 11.08&  &  &2.28 &    \\
\cline{1-2}\tabucline[2pt]{3-7}
\multirow{2}*{\parbox[c][1cm]{2.1cm}{\centering vegas}} & \large \pmb{$\mu$} & \cellcolor{myg}22.24 & \cellcolor{myr}16.32 & 30.70\% & 208.85 & 426.19\\
\cline{2-7}
& \large \pmb{$\sigma$} & 12.85&  &  &2.22 &    \\
\cline{1-2}\tabucline[2pt]{3-7}
\end{tabu}}
\label{tab:tab4255}
\end{table*}

\begin{table*}[h!]
\vspace{0.3cm}
\centering
\Large
\caption{RTT scenario: 4 BBR flows.}
\renewcommand{\arraystretch}{1.4} 
\resizebox*{\textwidth}{!}{\begin{tabu}{|c|cV{5}c|c|cV{5}c|cV{5}c|cV{5}c|c|cV{5}}
\cline{1-7}\cline{10-12}
\multirow{2}*{\parbox[c][2.5cm]{2.1cm}{\centering \bf \large Scheme}} &  \multicolumn{1}{c|}{\multirow{2}*{\parbox[c][2.5cm]{0.0cm}{}}} & \multicolumn{3}{c|}{\parbox[c][1cm]{2.7cm}{\bf \large \centering Rate (Mbps)}} & \parbox[c][1cm]{1.8cm}{\bf \large \centering Delay (ms)} & \multicolumn{1}{c|}{\parbox[c][1cm]{1.8cm}{\bf \large \centering RTT (ms)}}  & \multicolumn{1}{c}{} & \multicolumn{1}{c|}{} & \multicolumn{3}{c|}{\parbox[c][1cm]{2.5cm}{\bf \centering \large Jain's index}}\\ 
\cline{3-7}\cline{10-12}
 & \multicolumn{1}{c|}{} & \parbox[c][1.5cm]{1.8cm}{\centering \bf \normalsize CoCo-Beholder}  & \parbox[1cm]{1.8cm}{\centering \bf \normalsize Testbed} & \multicolumn{1}{c|}{\parbox{1.8cm}{\centering \pmb{$d_r$}}} & \parbox{1.8cm}{\centering \bf \normalsize CoCo-Beholder} & \multicolumn{1}{c|}{\parbox{1.8cm}{\centering \bf \normalsize Testbed}} &\multicolumn{1}{c}{} & \multicolumn{1}{c|}{}   & \parbox{1.8cm}{\centering \bf \normalsize CoCo-Beholder}& \parbox{1.8cm}{\centering \bf \normalsize Testbed} &\multicolumn{1}{c|}{ \parbox{1.8cm}{\centering \pmb{$d_r$}}}\\
\cline{1-2}\noalign{\vskip-1pt}\tabucline[2pt]{3-7}\noalign{\vskip-2pt}\cline{9-9}\tabucline[2pt]{10-12}
\multirow{2}*{\parbox[c][1cm]{2.1cm}{\centering bbr}} & \large \pmb{$\mu$} & 32.24 & 33.76 & 4.62\%&151.75 & 509.66&  & \large \pmb{$\mu$} &\cellcolor{myg}0.85 & \cellcolor{myr}0.7& 18.91\%\\
\cline{2-7}\cline{9-12}
& \large \pmb{$\sigma$} & 2.03&  &  &1.77 &  &   &  \large \pmb{$\sigma$}&0.02 & &\\
\cline{1-2}\tabucline[2pt]{3-7}\noalign{\vskip-2pt}\cline{9-9}\tabucline[2pt]{10-12}
\multirow{2}*{\parbox[c][1cm]{2.1cm}{\centering bbr}} & \large \pmb{$\mu$} &\cellcolor{myg}26.56 &\cellcolor{myr}6.23 & 124.01\%& 202.37&613.11 \\
\cline{2-7}
& \large \pmb{$\sigma$} & 2.60&  &  & 2.20 &    \\
\cline{1-2}\tabucline[2pt]{3-7}
\multirow{2}*{\parbox[c][1cm]{2.1cm}{\centering bbr}} & \large \pmb{$\mu$} & \cellcolor{myr}20.71 & \cellcolor{myg}23.46& 12.47\% & 252.66&722.18  \\
\cline{2-7}
& \large \pmb{$\sigma$} & 2.04&  &  & 1.98&    \\
\cline{1-2}\tabucline[2pt]{3-7}
\multirow{2}*{\parbox[c][1cm]{2.1cm}{\centering bbr}} & \large \pmb{$\mu$} & \cellcolor{myr}15.69 & \cellcolor{myg}20.23 & 25.26\%  & 302.72& 816.02\\
\cline{2-7}
& \large \pmb{$\sigma$} & 2.98&  &  & 1.55& \\
\cline{1-2}\tabucline[2pt]{3-7}
\end{tabu}}
\label{tab:tab4256}
\end{table*}

\FloatBarrier

\begin{figure}[p!]
\vspace*{-0.3cm}
\captionsetup[subfigure]{labelformat=empty}
    \centering
    \subfloat[]{{\includegraphics[width=\textwidth]{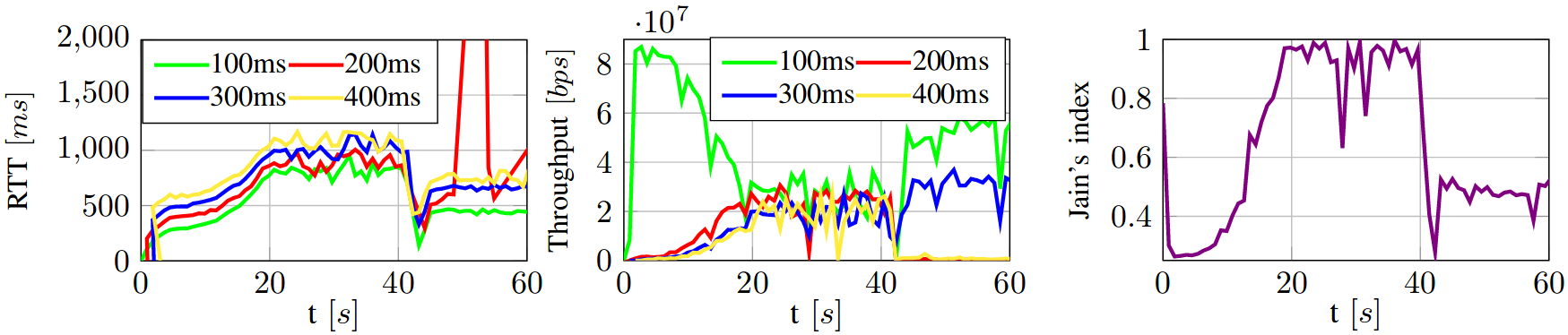} }}\\
    \vspace{-0.7cm}
    \subfloat[]{{\includegraphics[width=\textwidth/3]{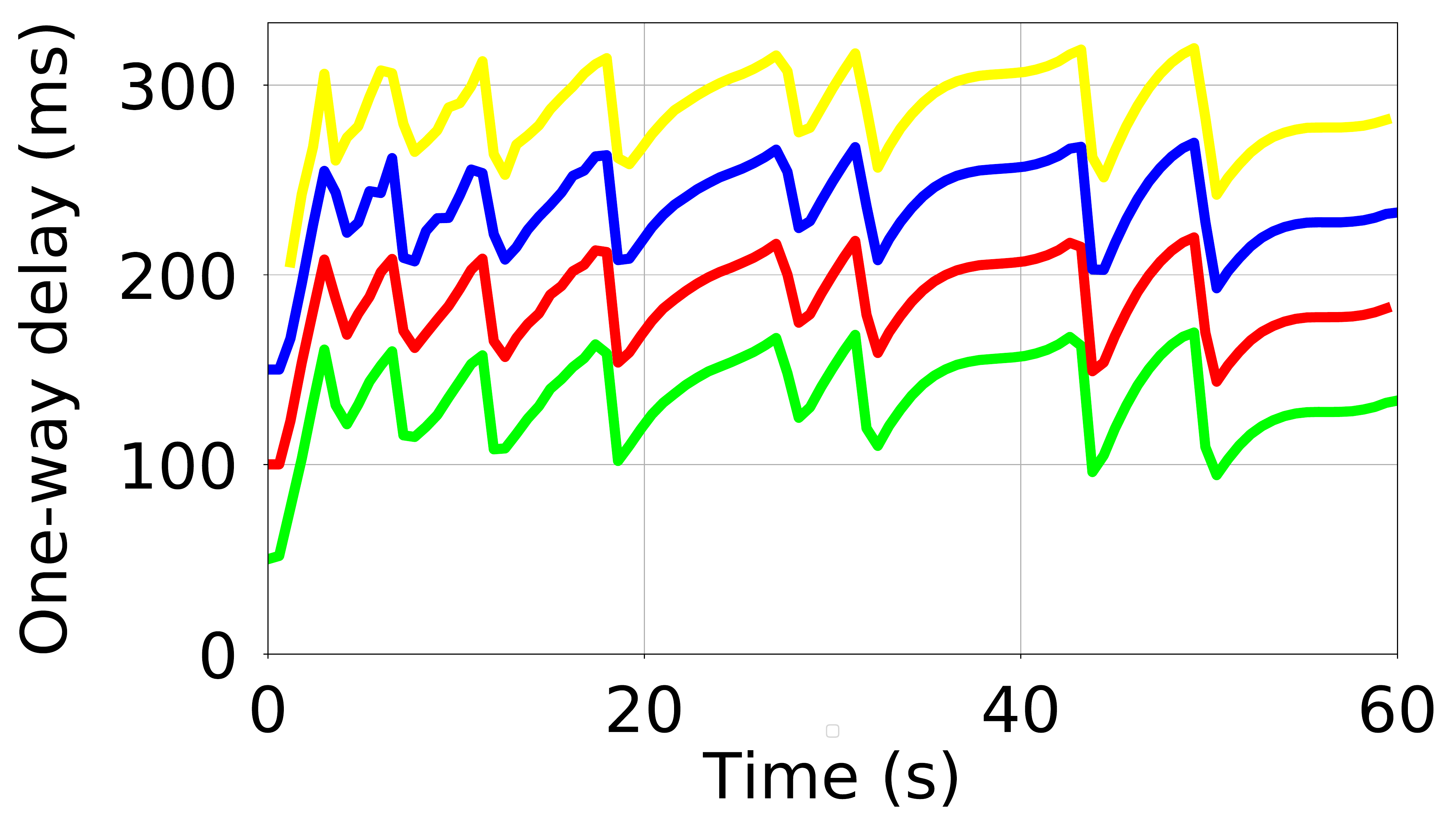} }}%
    \subfloat[]{{\includegraphics[width=\textwidth/3]{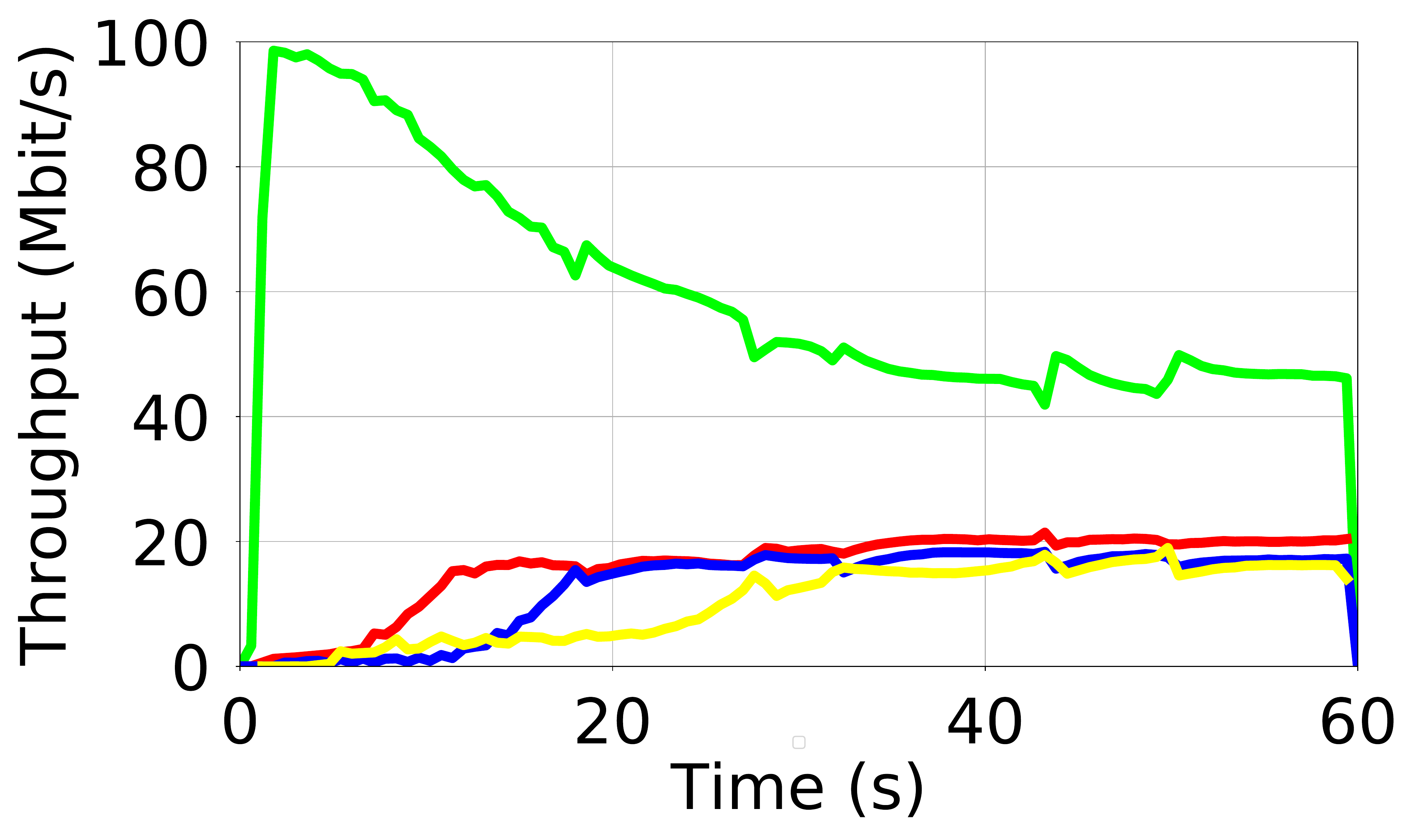} }}%
    \subfloat[]{{\includegraphics[width=\textwidth/3]{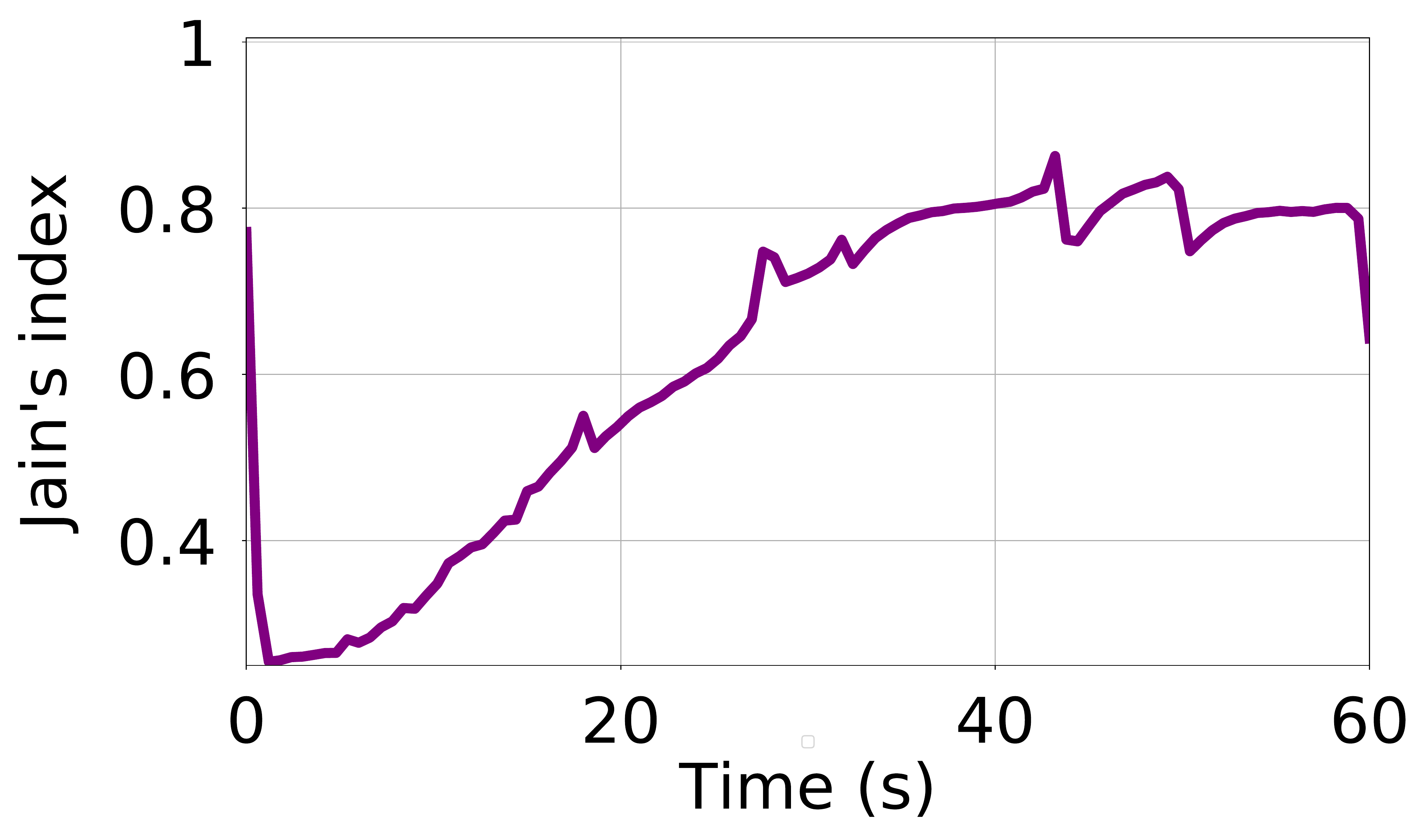} }}\\
    \vspace{-0.5cm}
    \caption{RTT scenario: 4 Cubic flows. The aggregation interval is 600 ms.\\The top-row plots are by the testbed, the bottom-row -- by CoCo-Beholder.}%
    \label{fig:fig4254}
\end{figure}

\begin{figure}[h!]
\vspace*{-0.2cm}
\captionsetup[subfigure]{labelformat=empty}
    \centering
    \subfloat[]{{\includegraphics[width=\textwidth]{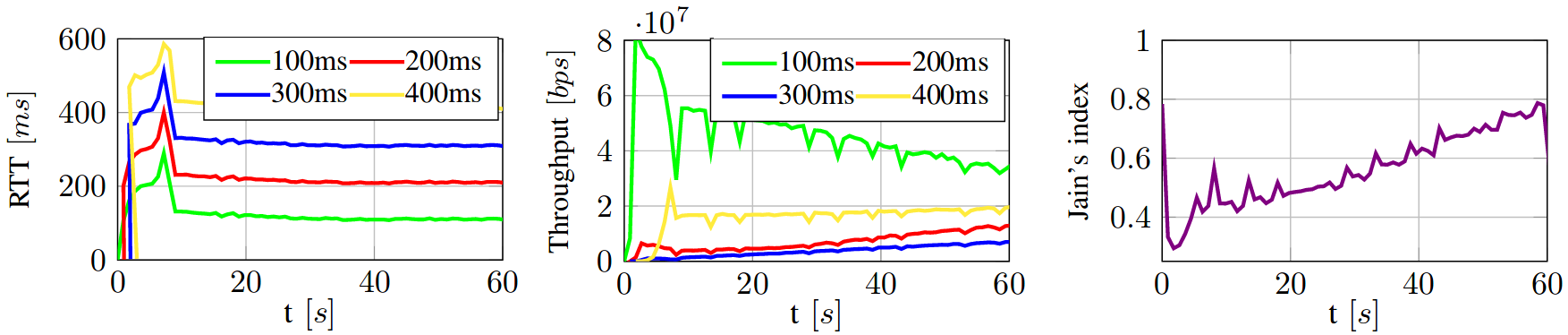} }}\\
    \vspace{-0.7cm}
    \subfloat[]{{\includegraphics[width=\textwidth/3]{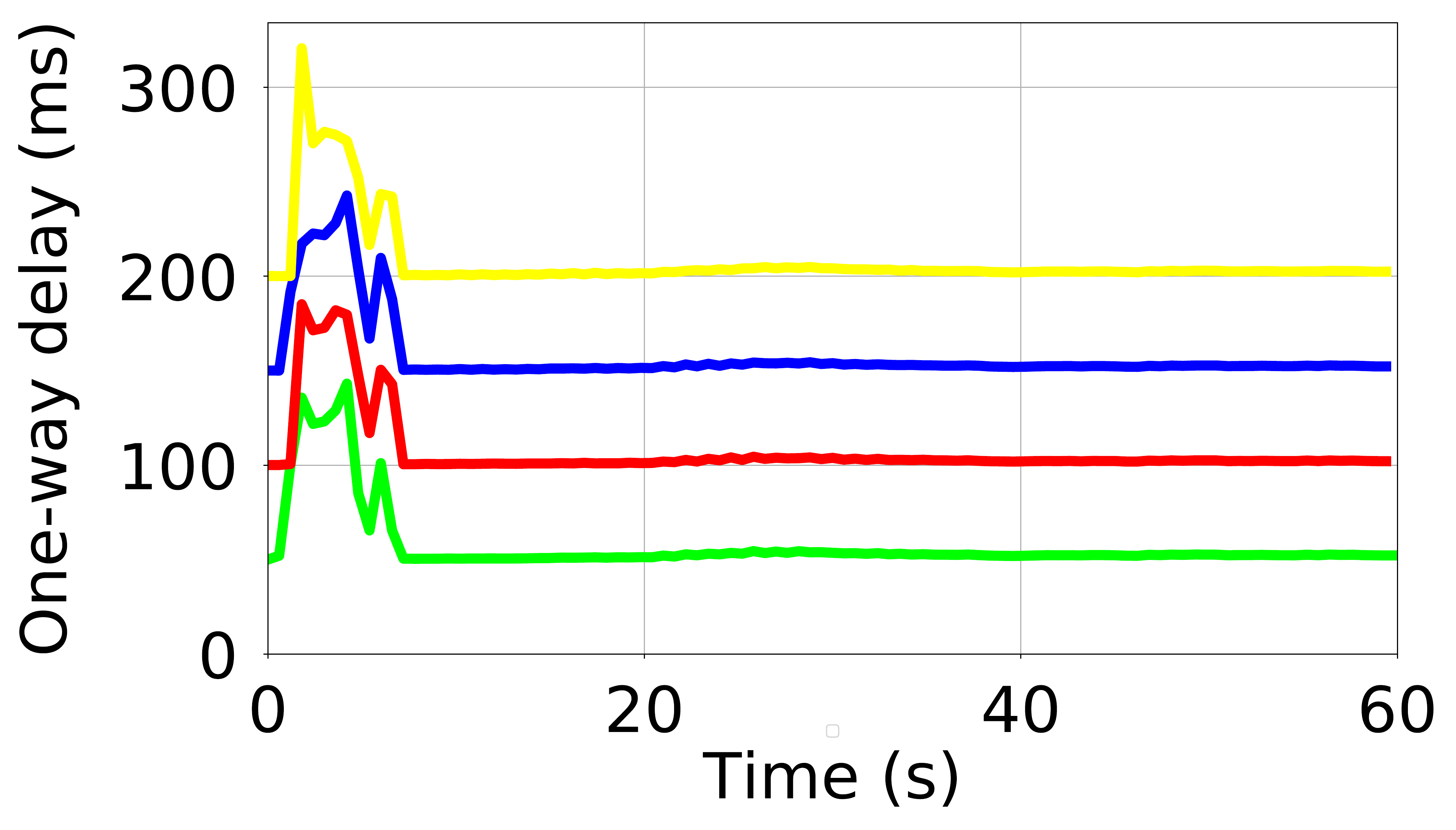} }}%
    \subfloat[]{{\includegraphics[width=\textwidth/3]{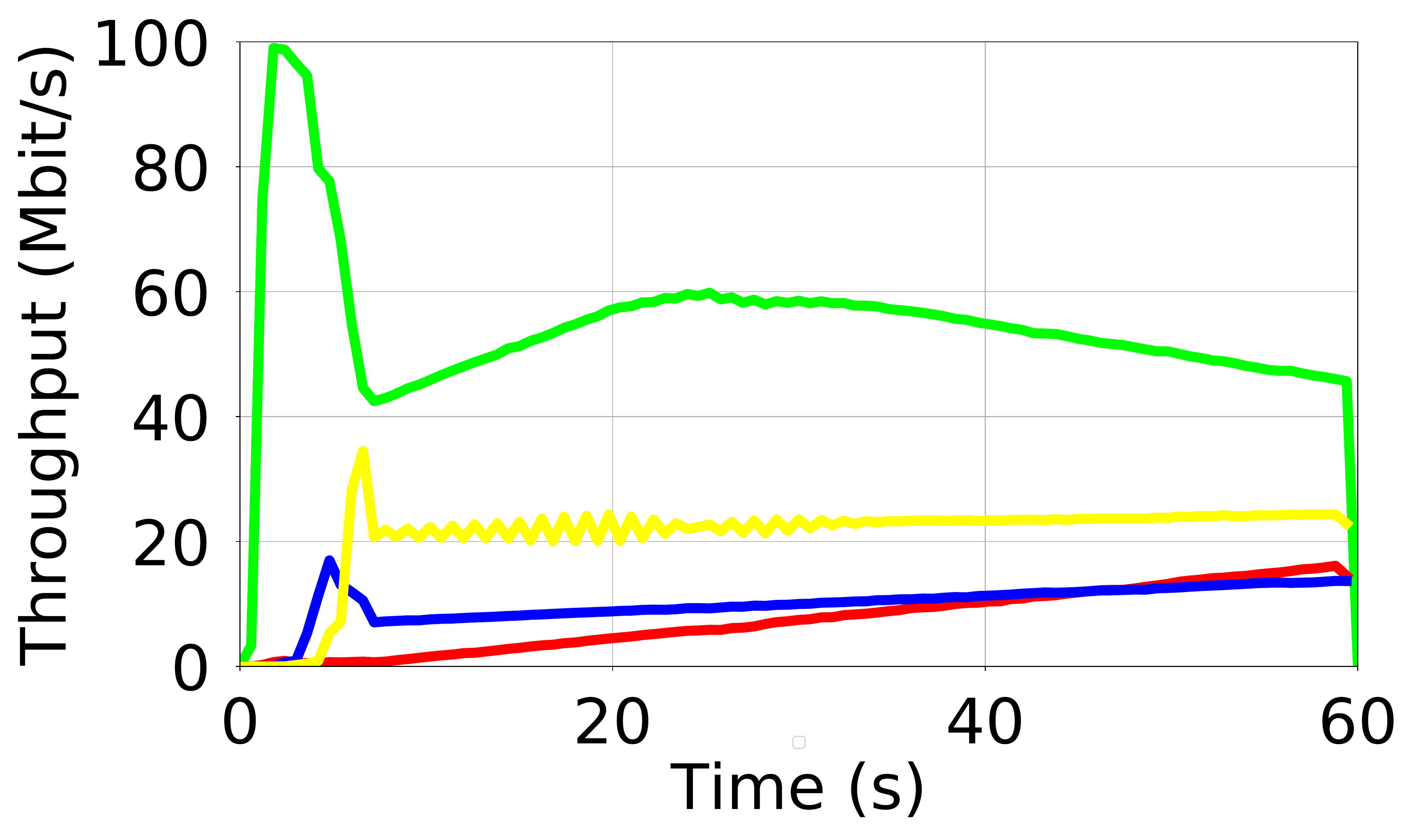} }}%
    \subfloat[]{{\includegraphics[width=\textwidth/3]{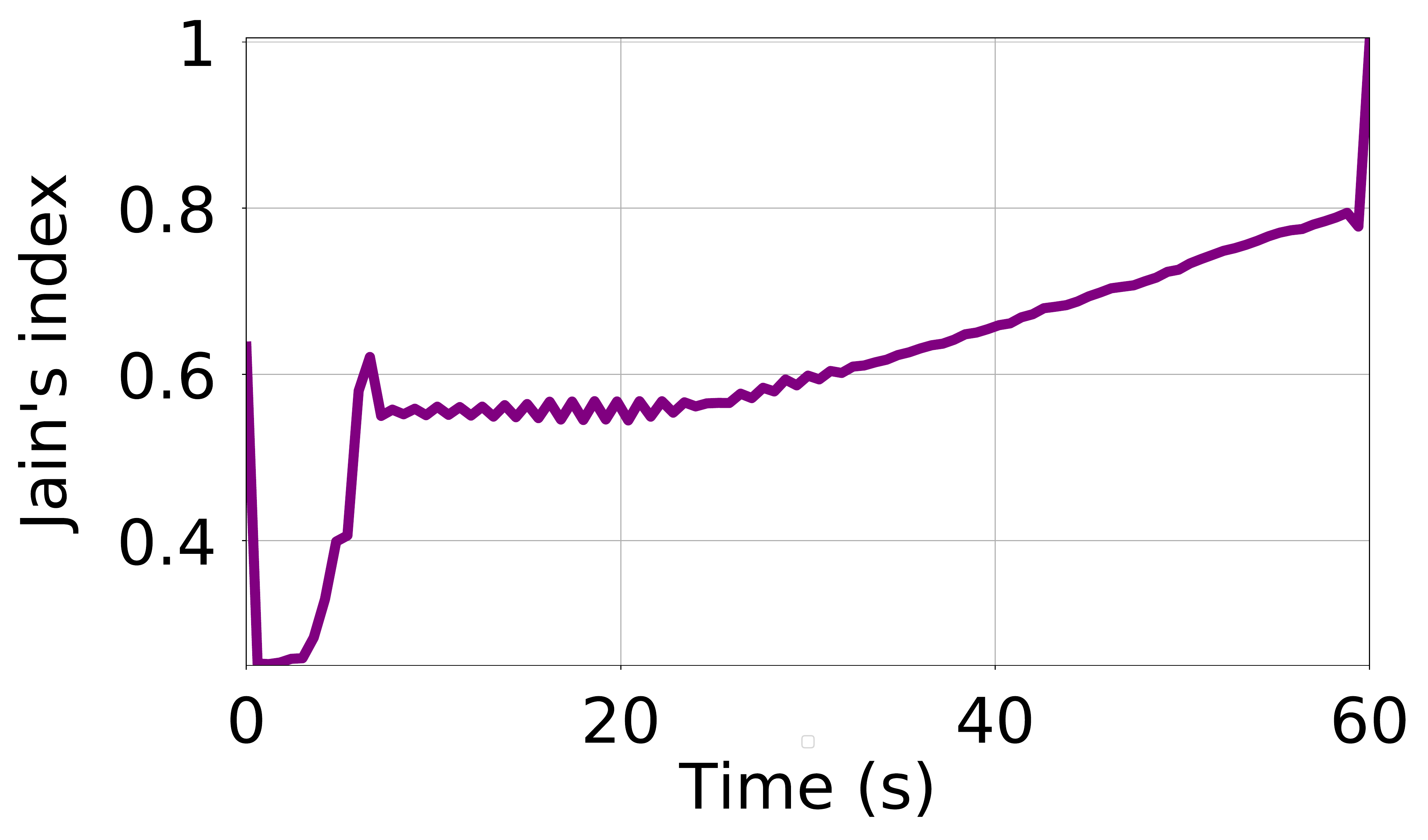} }}\\
    \vspace{-0.5cm}
    \caption{RTT scenario: 4 Vegas flows. The aggregation interval is 600 ms.\\The top-row plots are by the testbed, the bottom-row -- by CoCo-Beholder.}%
    \label{fig:fig4255}
\end{figure}

\begin{figure}[h!]
\vspace*{-0.2cm}
\captionsetup[subfigure]{labelformat=empty}
    \centering
    \subfloat[]{{\includegraphics[width=\textwidth]{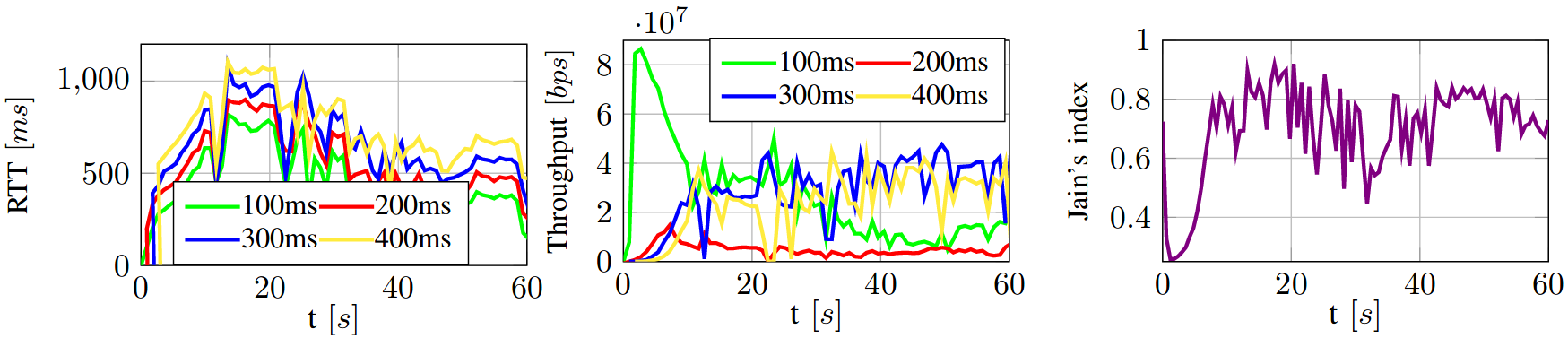} }}\\
    \vspace{-0.7cm}
    \subfloat[]{{\includegraphics[width=\textwidth/3]{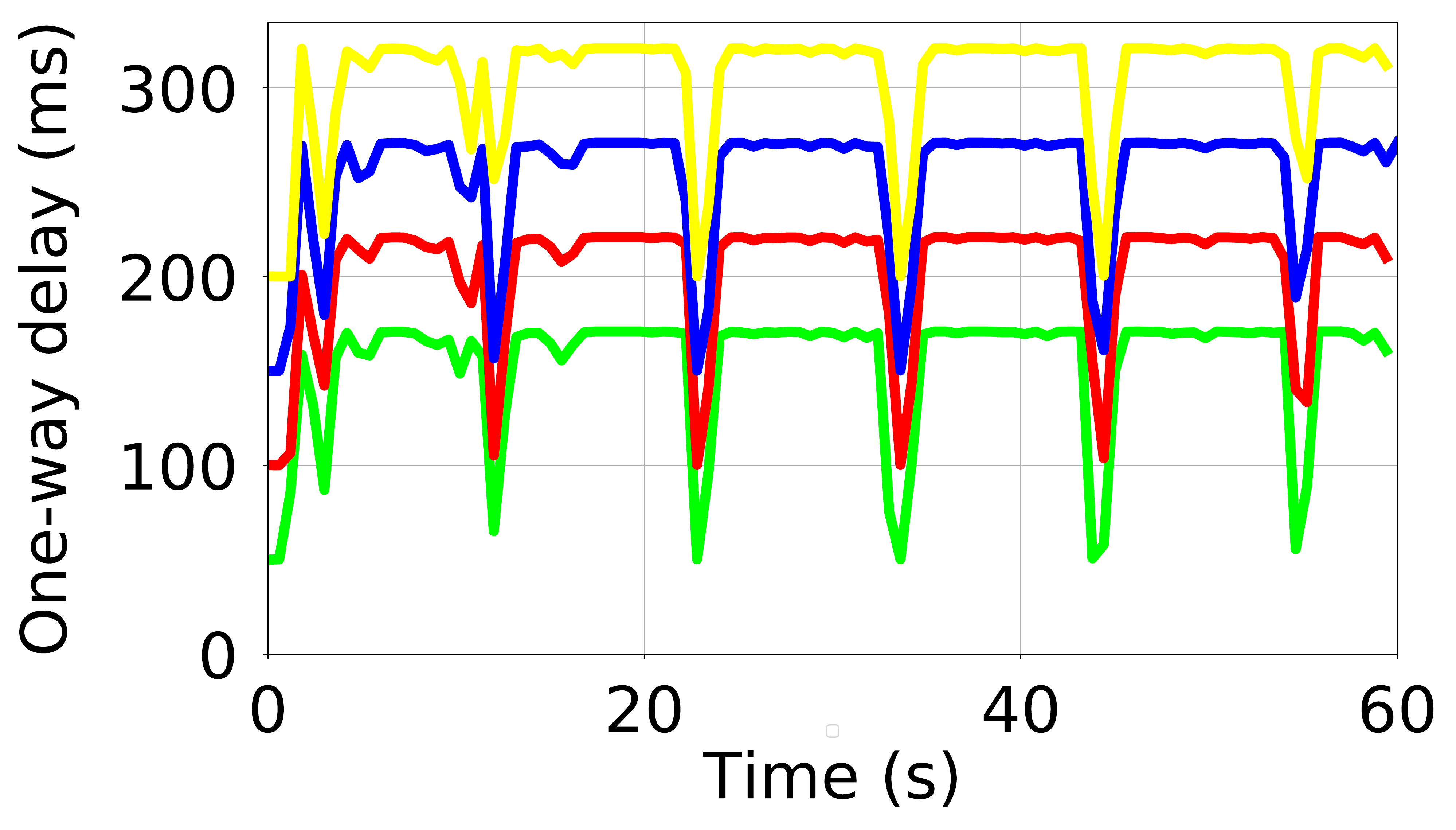} }}%
    \subfloat[]{{\includegraphics[width=\textwidth/3]{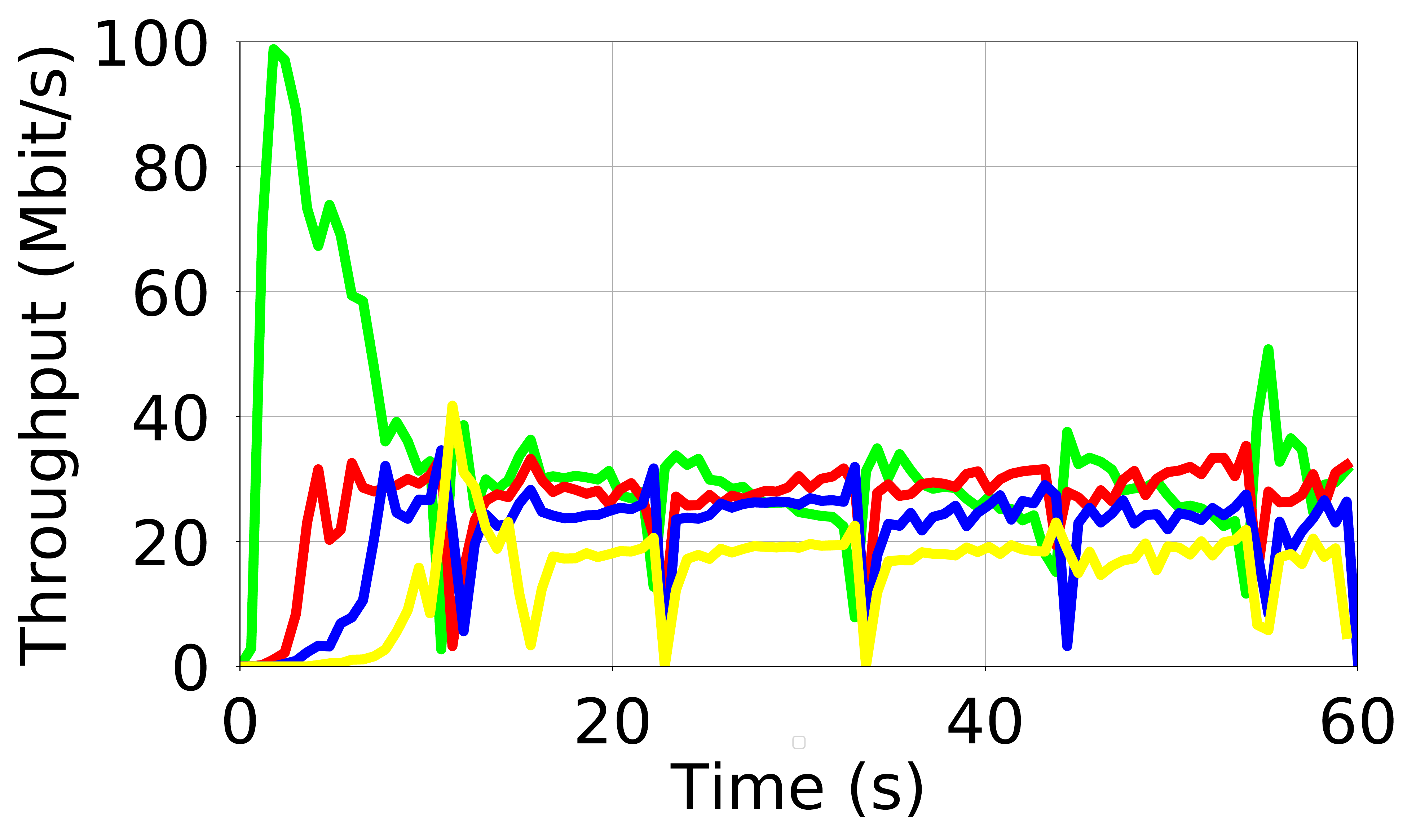} }}%
    \subfloat[]{{\includegraphics[width=\textwidth/3]{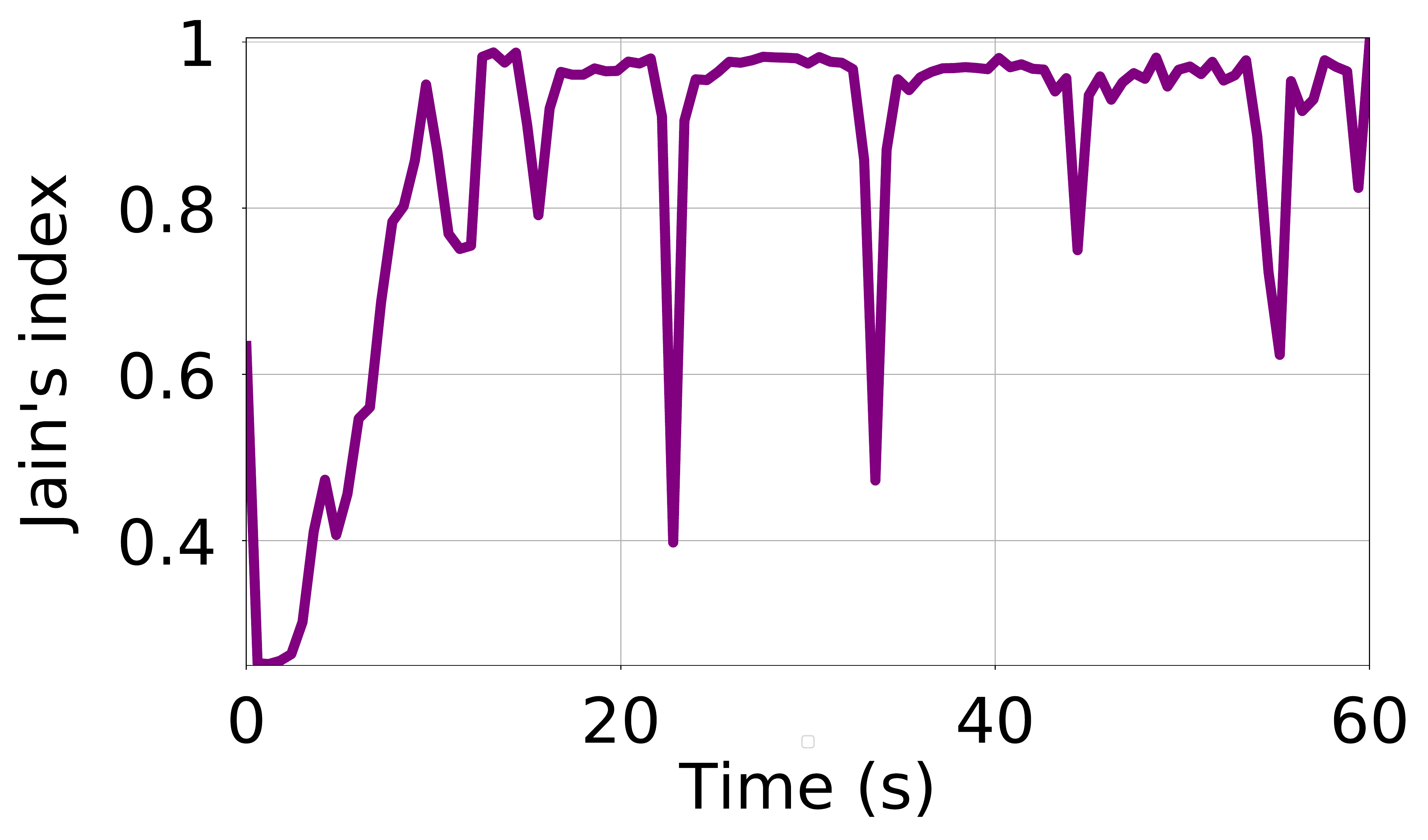} }}\\
    \vspace{-0.4cm}
    \caption{RTT scenario: 4 BBR flows. The aggregation interval is 600 ms.\\The top-row plots are by the testbed, the bottom-row -- by CoCo-Beholder.}%
    \label{fig:fig4256}
\end{figure}

\FloatBarrier

\textbf{RTT scenario:}
\begin{itemize}
\item Both the testbed~\cite{turkovic2019fifty} and CoCo-Beholder showed that none of the three schemes provide good intra-RTT-fairness, even though Cubic claims to do it.

\item As highlighted in the paper~\cite{turkovic2019fifty}, the testbed demonstrated that Vegas is most intra-RTT-fair. CoCo-Beholder did not confirm this conclusion. 

\item Four BBR flows converged to a common rate in CoCo-Beholder, while the two higher-RTT BBR flows outperformed the two lower-RTT flows in the testbed.
\end{itemize}

As quite many differences between the results by CoCo-Beholder and the paper~\cite{turkovic2019fifty} concerned BBR, there was an idea that it might happen because the thesis author and the authors of the paper~\cite{turkovic2019fifty} could use different versions of BBR (there exist at least BBR v1.0 and BBR v2.0~\cite{zhang2019evaluation}). Belma Turkovic, the author of the paper~\cite{turkovic2019fifty}, confirmed that in the testbed, BBR available as a module of Linux kernel 4.13 was used, that is, BBR v1.0. The thesis author also used BBR v1.0 in CoCo-Beholder.\\

\section{Congestion Control Schemes Under the Square-Wave Variable Delay}
\label{sec:tthird}

CoCo-Beholder enables the variable delay at the central link of the dumbbell topology, which is defined by specifying the base delay, delta, step, optional jitter, and seed. The randomization seed is used to decide if the delay at the central link should be decreased or increased after the next delta time interval, i.e., the seed defines the shape of the delay curve over time. Each delta time, the new delay is installed at both the interfaces at the ends of the central link using tc qdisc NetEm. This is being performed by one thread of CoCo-Beholder testing tool during the whole runtime of the experiment, while the other thread is launching new flows of congestion control schemes each second of the runtime (if any flows are scheduled to start at the second).

To improve the precision of the implementation, in general, and the implementation of the variable delay, in particular, everything that can be precomputed is precomputed before the two threads start their above-described cyclic activities. The array of all the future delays, installed at the central link of the dumbbell topology each delta time, is pregenerated with a function from the base delay, delta, step, seed, and maximum delay. The maximum delay is the maximum allowed delay (the jitter is not counted) both for the central link and for all the links in the left and right halves of the dumbbell topology. The delays in the pregenerated array always lie between zero and the maximum delay. The user can specify the maximum delay with \verb+-m+/\verb+--max-delay+ command-line argument. An example curve of the variable delay over time can be seen in the plan in Figure~\ref{fig:plan}.

The success of all the commands, used to build the dumbbell topology and configure the network conditions, is checked to ensure the consistency of the resulting virtual network. To improve the precision, the checks are not performed when the delay at the central link is set to a new value with tc qdisc NetEm. Still, such a change at one interface takes around 4 ms. The two interfaces at the ends of the central link are modified -- first, at the left router and then at the right router -- so it takes around 8 ms. This is why the minimum delta, which the user can specify, is 10 ms.

After a thread (one changing the delay or one starting new flows) completes another iteration, it sleeps for a (delta or second) time interval, reduced by an amount of time depending on the duration of this iteration in the way avoiding any drift.

The thesis author also created the variant of CoCo-Beholder testing tool utilizing the two processes, rather than two threads. However, the variant did not improve the precision noticeably, so the author of the thesis kept the implementation with the threads to keep it simple.

\begin{figure}[b!]
\centering
\includegraphics[width=\textwidth]{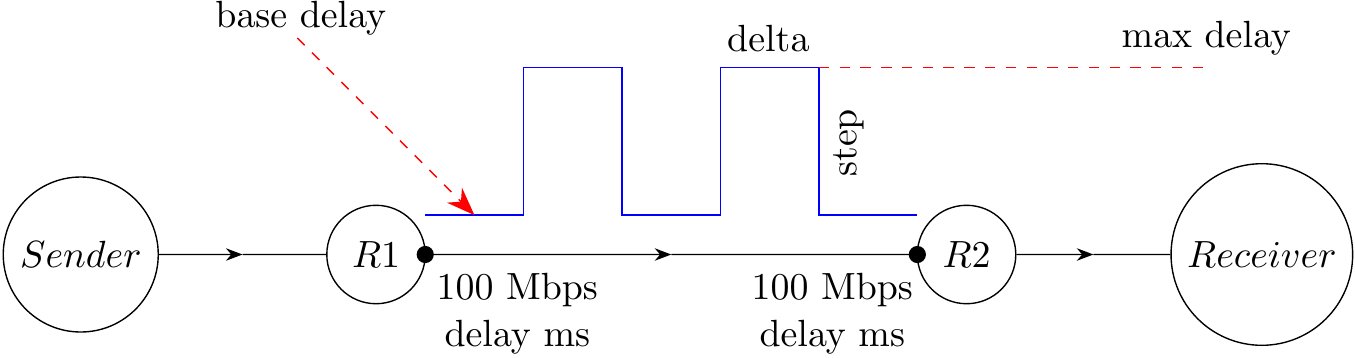}\vspace*{0.2cm}
\caption{The tested topology with the square-wave variable delay.}
\label{fig:vd-dumbbell}
\end{figure}

This section is devoted to the testing of congestion control schemes using CoCo-Beholder under the variable delay at the central link. The thesis author had to decide, which randomization seed to use and, thus, which variable delay curve to have for the testing. However, the thesis author solved the problem by having the maximum delay equal to the sum of the base delay and the step. Such a setting produces the square-wave looking variable delay that does not depend on the seed at all. The tested topology can be seen in Figure~\ref{fig:vd-dumbbell}. All the interfaces in the topology have the queue size of 1000 packets, so, in order not to overload the drawing, this is not indicated in it. 

The flow is rightward and runs for 10 seconds. The rate of the central link is 100 Mbit/s.  If the opposite is not stated, the base delay is 0 ms. 

The delta and step values lie in $[10, 20, 30, ..., 200, 210]$ ms, that is, for a chosen scheme, the thesis author performed $21 \cdot 21=441$ experiments. Each experiment was run only once because the two previous sections~\ref{sec:tfirst} and~\ref{sec:tsecond} of this chapter showed the good reproducibility of CoCo-Beholder's results: when an experiment is run ten times, the sample standard deviations of the resulting statistics are typically small. Same as for all other testing in this chapter, running the experiments was automated with a bash script wrapping over CoCo-Beholder's commands.

The thesis author tested the following schemes: TCP Cubic, TCP Vegas, TCP BBR, QUIC Cubic, TCP Veno, YeAH-TCP, TCP CDG, Indigo, PCC Allegro, Scalable TCP, TCP Westwood+. The idea of the results for all these schemes was the same, which is why the thesis includes the results only for eight of the schemes: loss-based TCP Cubic and QUIC Cubic, delay-based TCP Vegas and Copa, hybrid TCP BBR and TCP CDG, learned Indigo and PCC Allegro -- four of the eight schemes are UDP-based. 

The results are presented as the 3D throughput plots, where the overall average rate (i.e., overall average throughput) statistic value of the flow depends on the delta and\nolinebreak[4] step. 

\begin{figure}[h!]
\captionsetup[subfigure]{labelformat=empty}
    \begin{adjustbox}{minipage=\linewidth,totalheight=\textheight}
\vspace*{-0.9cm}
    \centering
    \subfloat[\bf TCP Cubic]{{\includegraphics[width=0.5\textwidth]{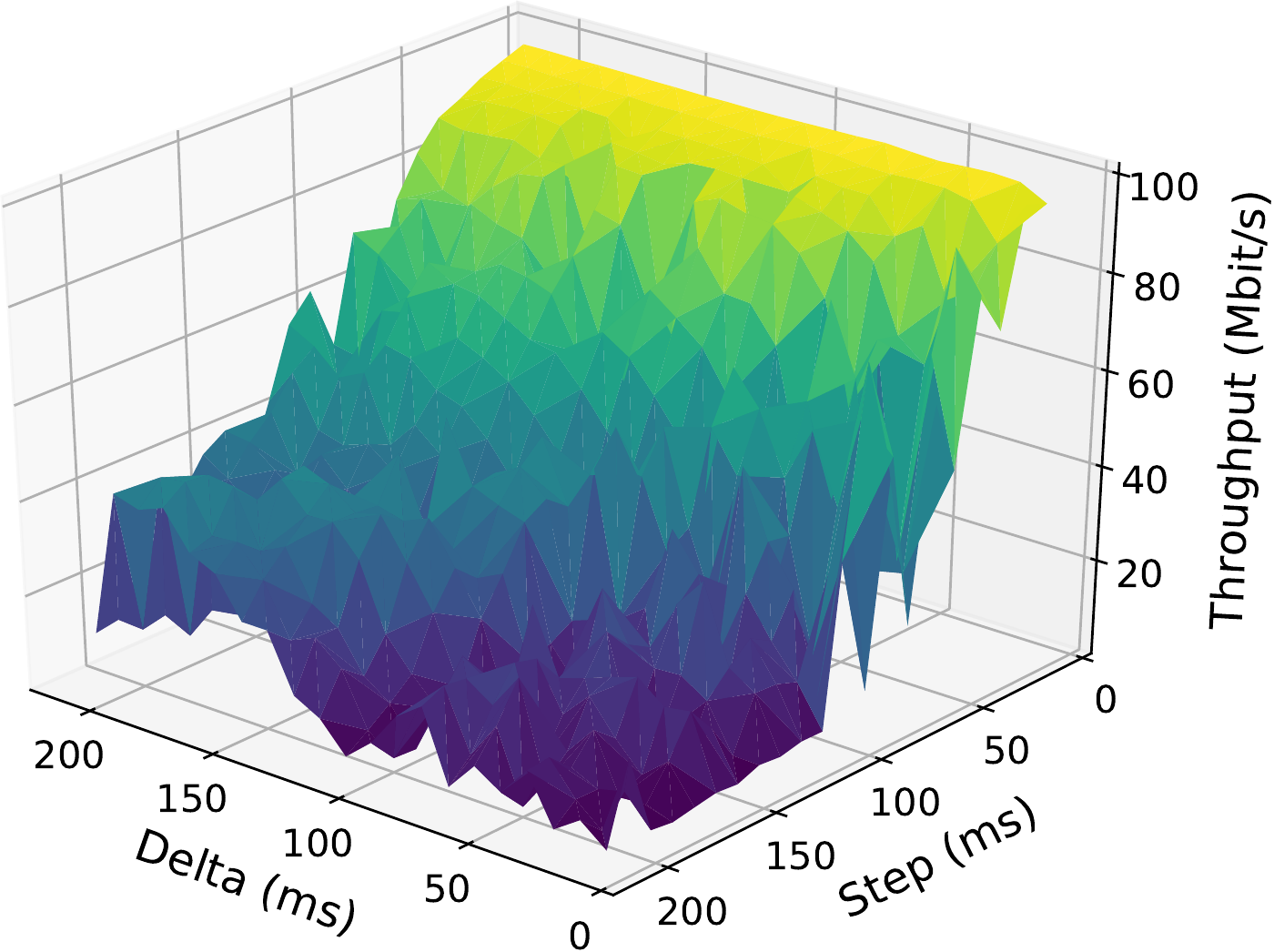} }}%
    \subfloat[\bf QUIC Cubic]{{\includegraphics[width=0.5\textwidth]{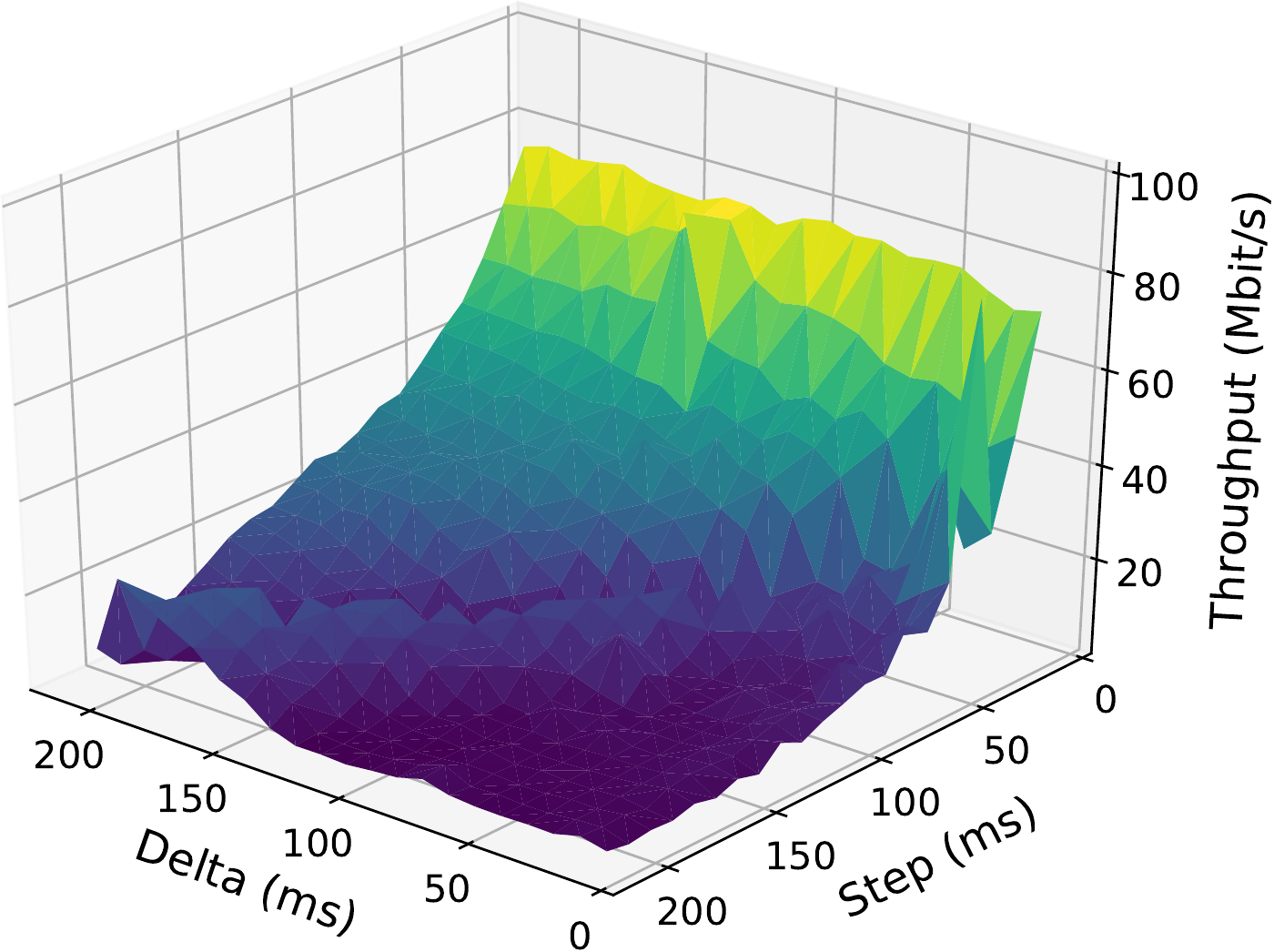} }}\\
    \vspace{-0.45cm}
    \subfloat[\bf TCP Vegas]{{\includegraphics[width=0.5\textwidth]{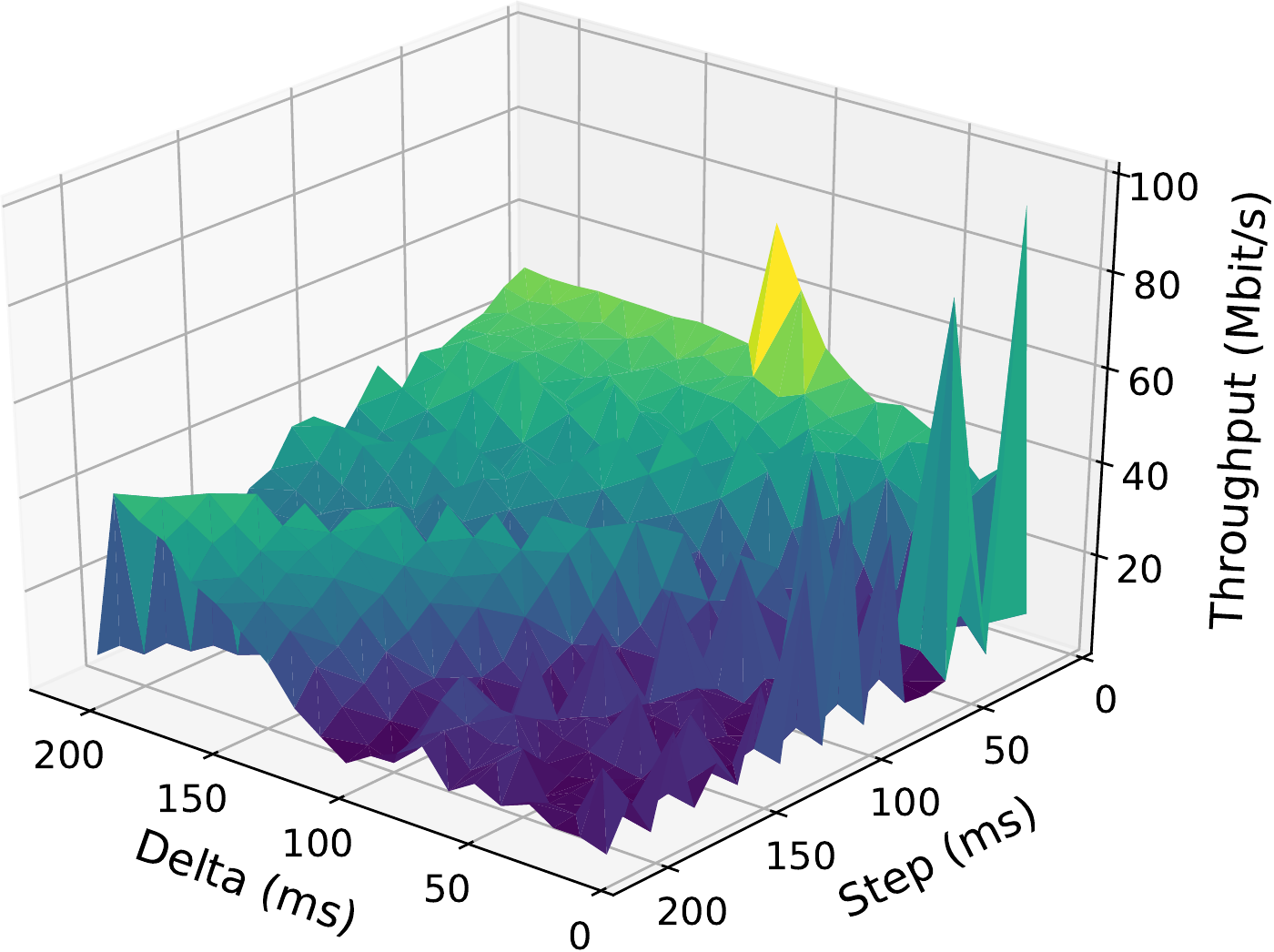} }}%
    \subfloat[\bf Copa]{{\includegraphics[width=0.5\textwidth]{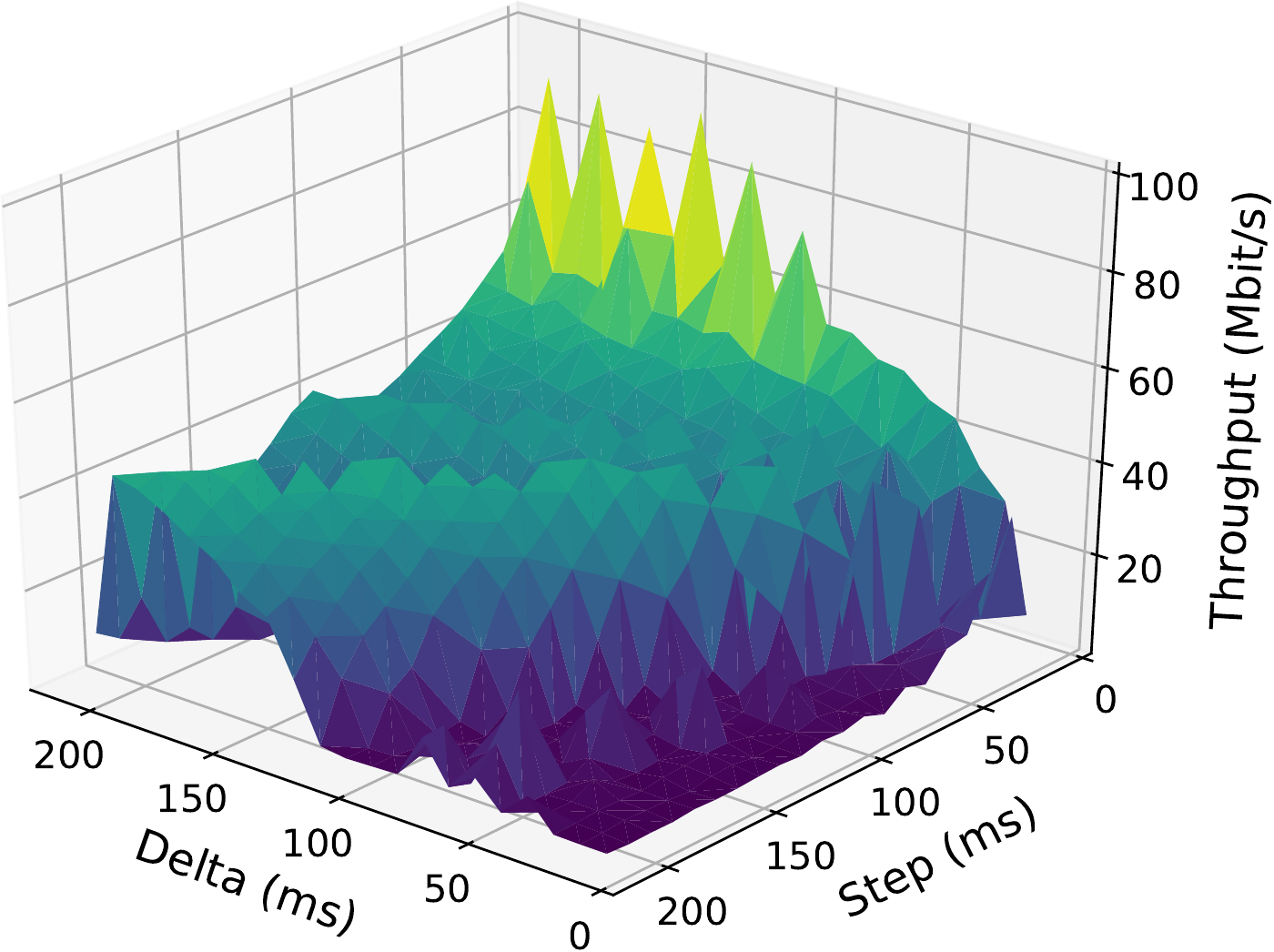} }}\\
    \vspace{-0.45cm}
    \subfloat[\bf TCP BBR]{{\includegraphics[width=0.5\textwidth]{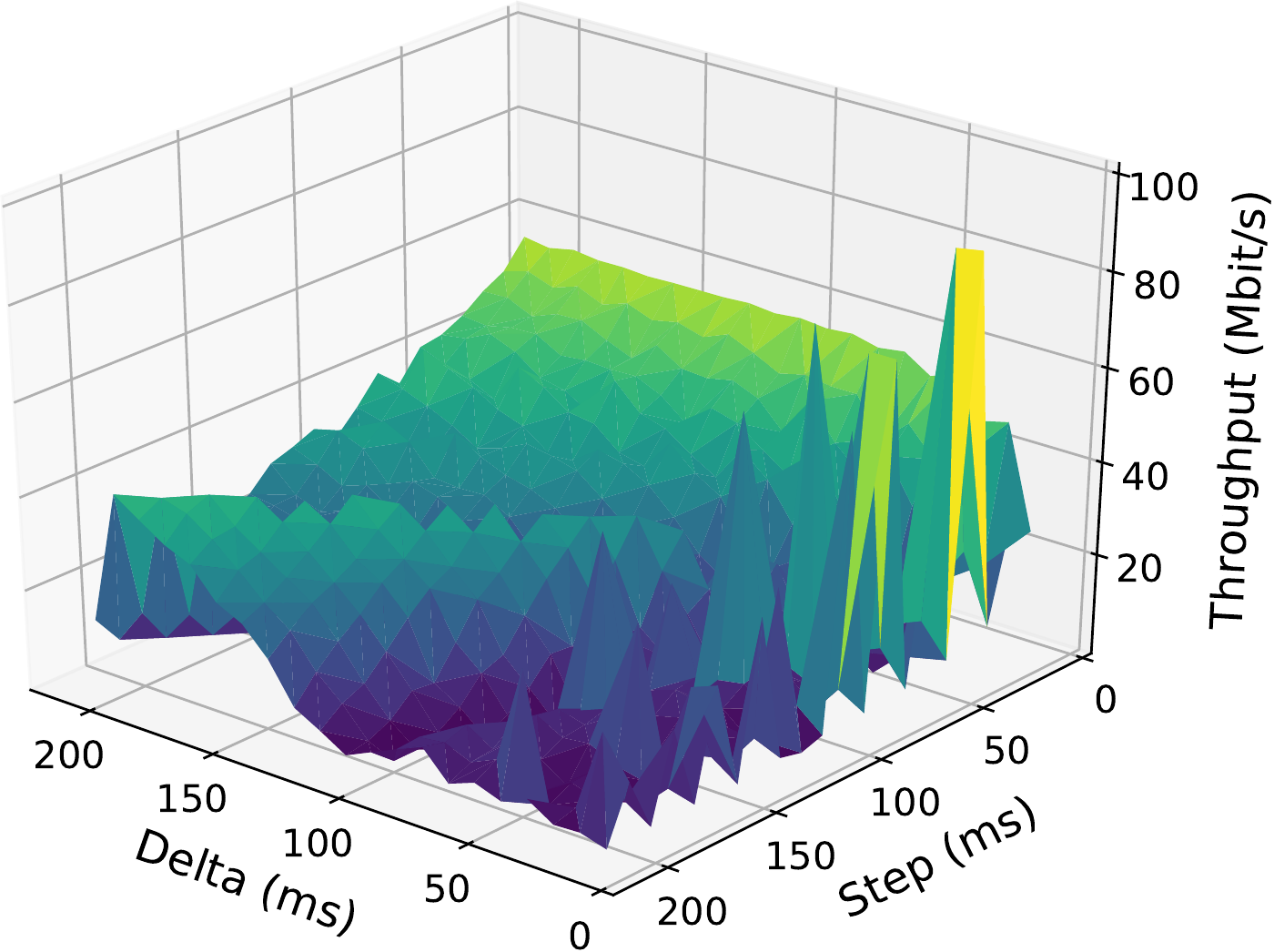} }}%
    \subfloat[\bf TCP CDG]{{\includegraphics[width=0.5\textwidth]{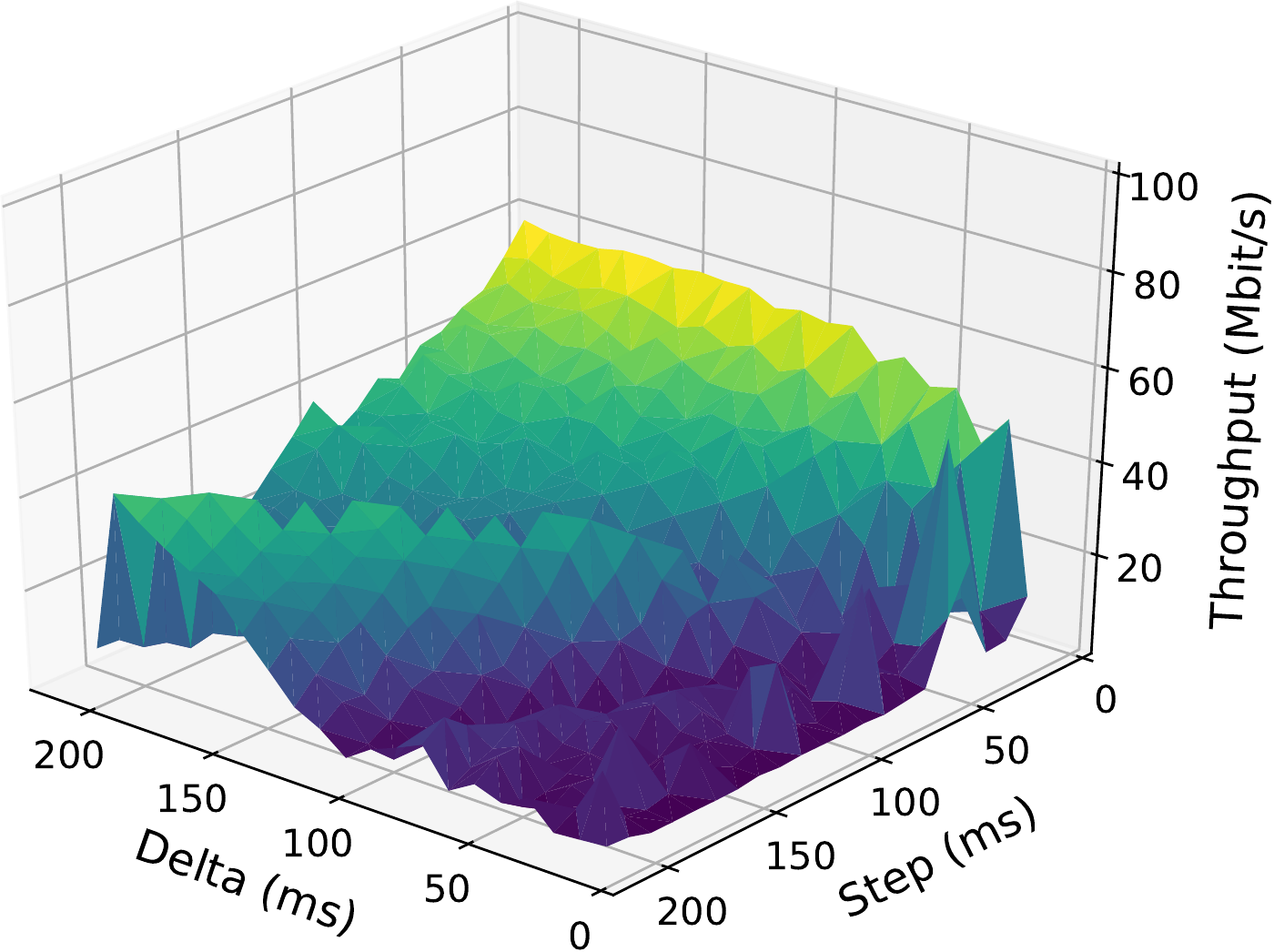} }}\\%
    \vspace{-0.45cm}
\subfloat[\bf Indigo]{{\includegraphics[width=0.5\textwidth]{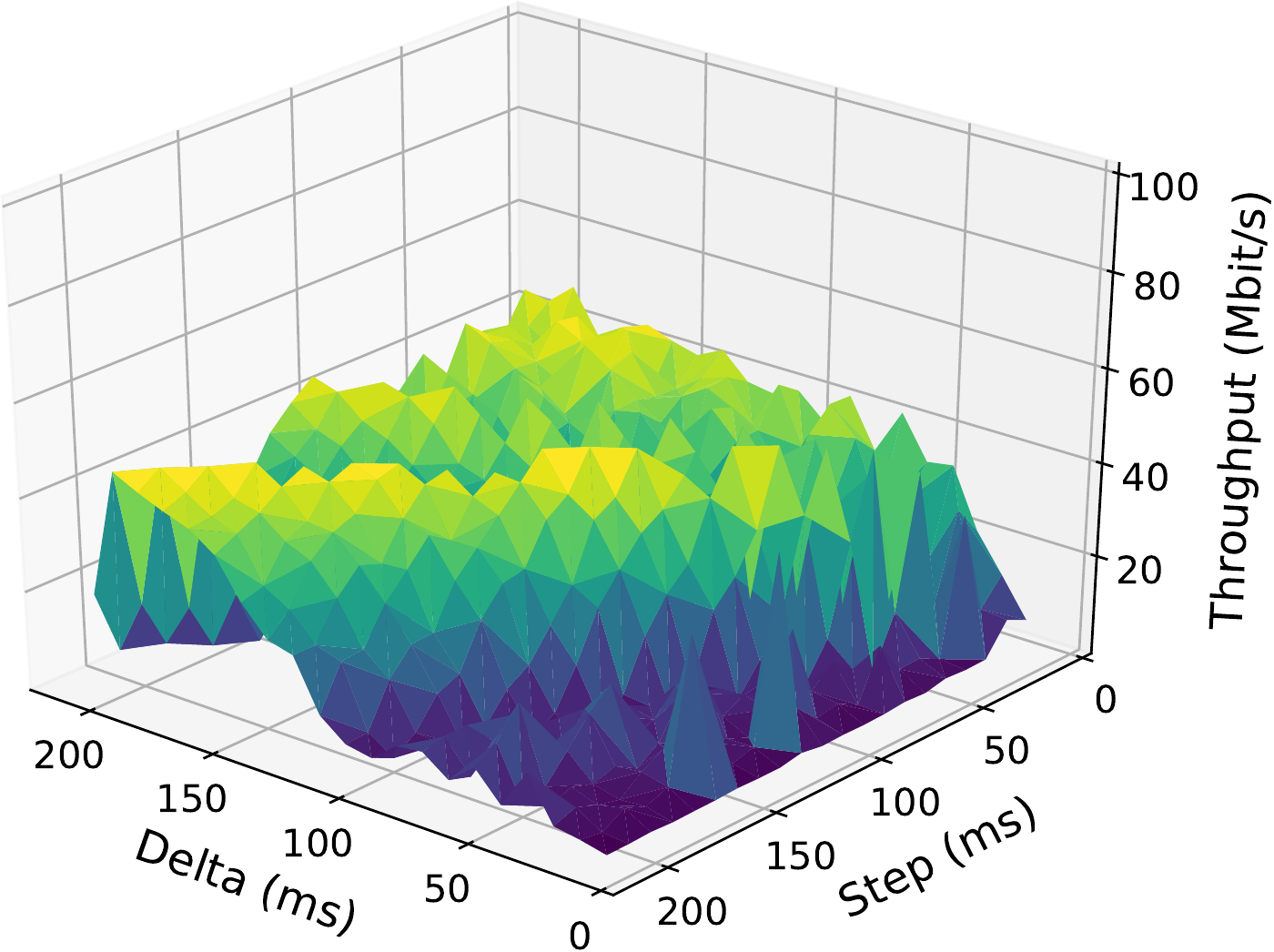} }}%
    \subfloat[\bf PCC]{{\includegraphics[width=0.5\textwidth]{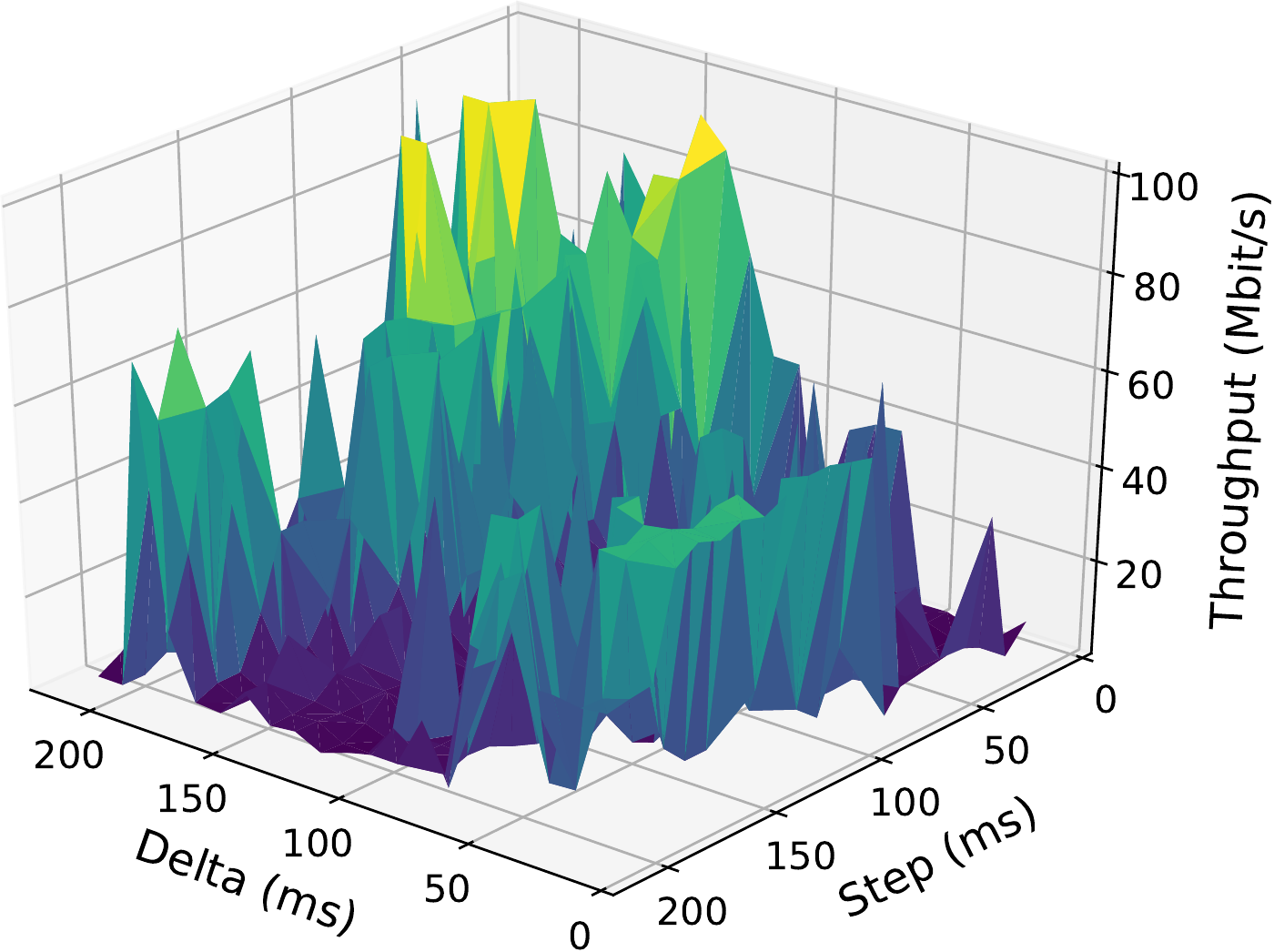} }}
    \caption{\mbox{Right side view rate plots for the schemes under the square-wave delay.}}%
    \label{fig:rightsideview}
    \end{adjustbox}
\end{figure}

\begin{figure}[h!]
\captionsetup[subfigure]{labelformat=empty}
    \begin{adjustbox}{minipage=\linewidth,totalheight=\textheight}
\vspace*{-0.9cm}
\hspace*{-0.4cm}
    \centering
    \subfloat[\bf TCP Cubic]{{\includegraphics[width=0.5\textwidth]{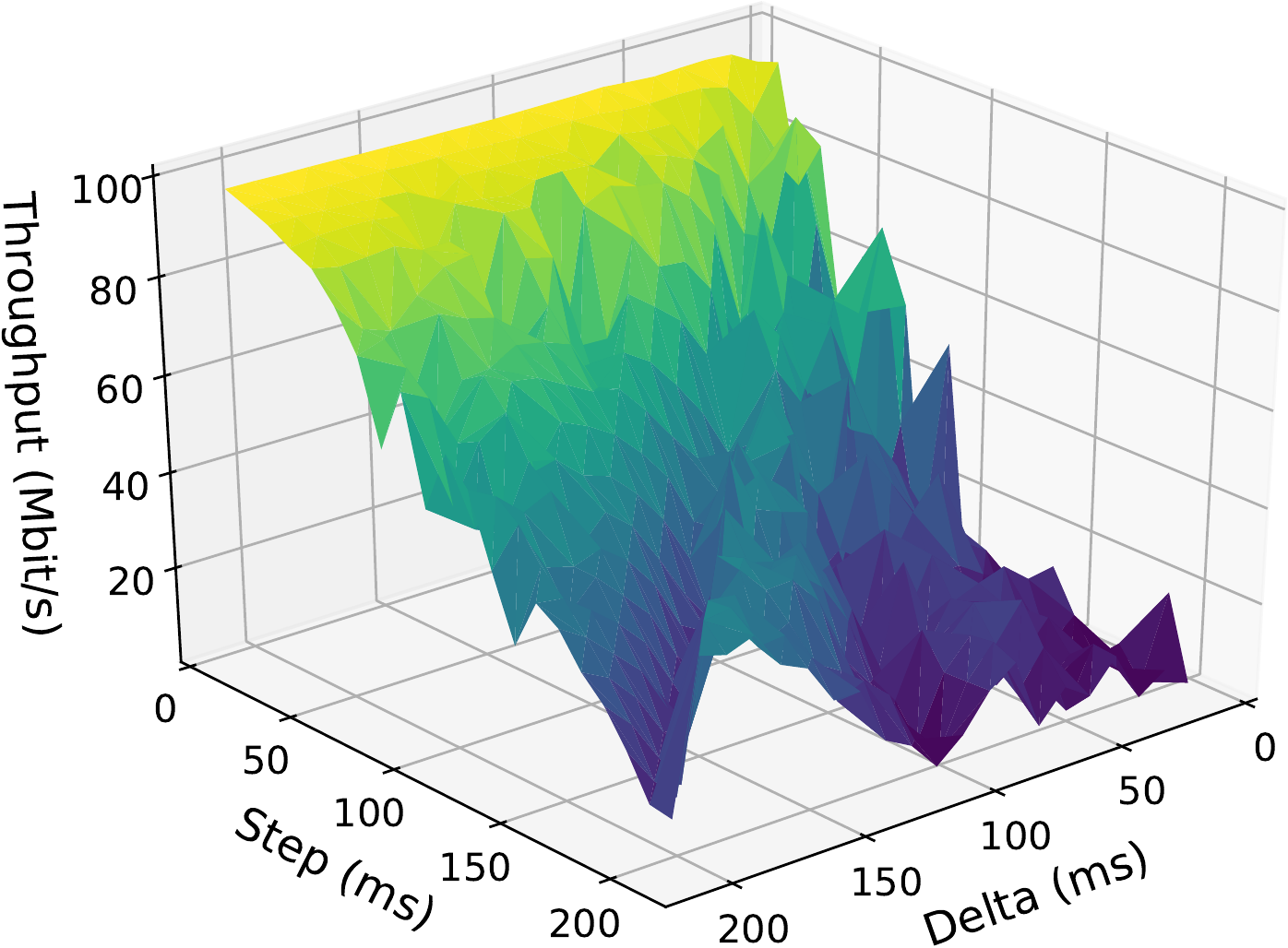} }}%
    \subfloat[\bf QUIC Cubic]{{\includegraphics[width=0.5\textwidth]{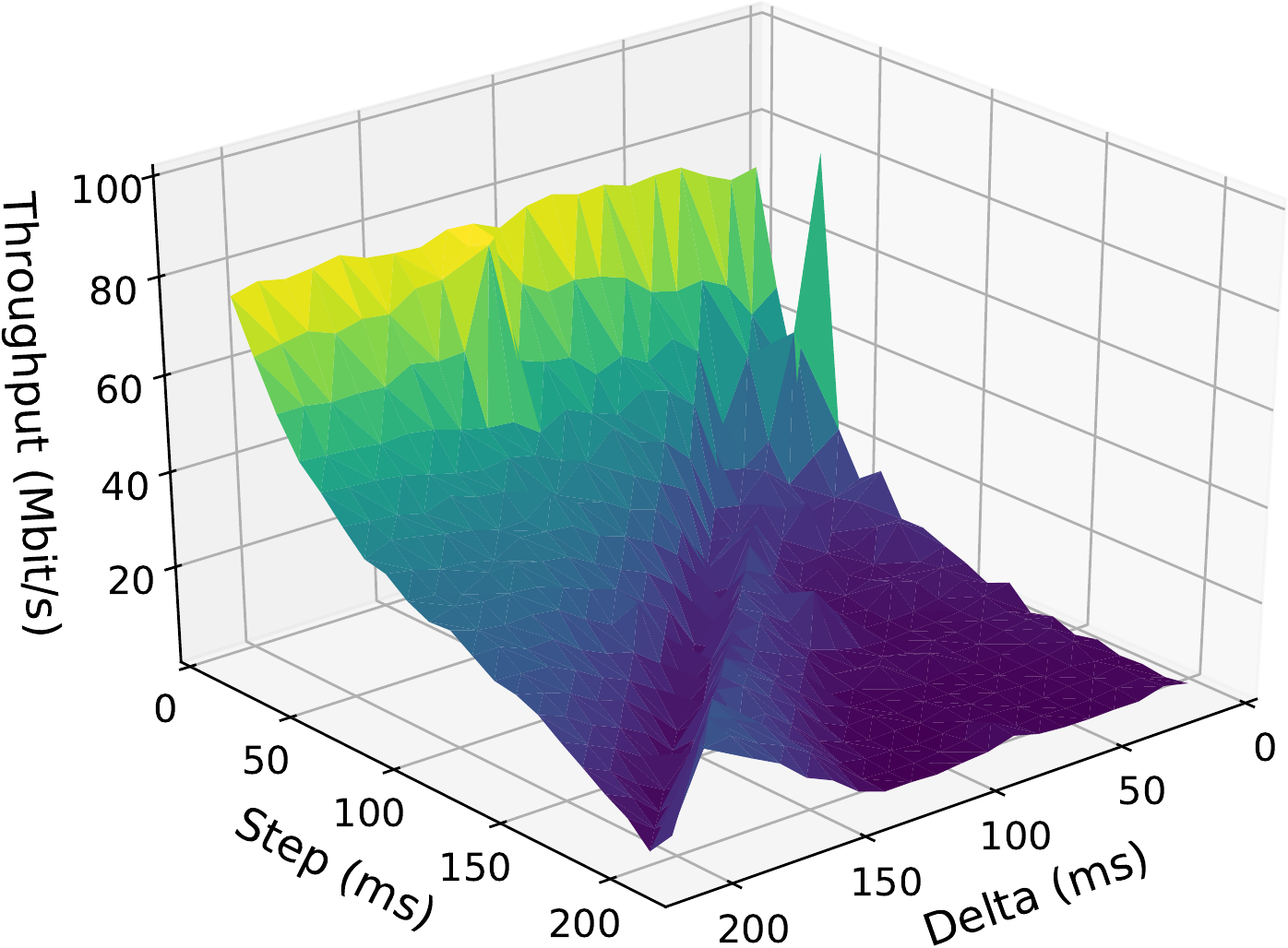} }}\\
    \vspace{-0.45cm}
    \hspace*{-0.4cm}
    \subfloat[\bf TCP Vegas]{{\includegraphics[width=0.5\textwidth]{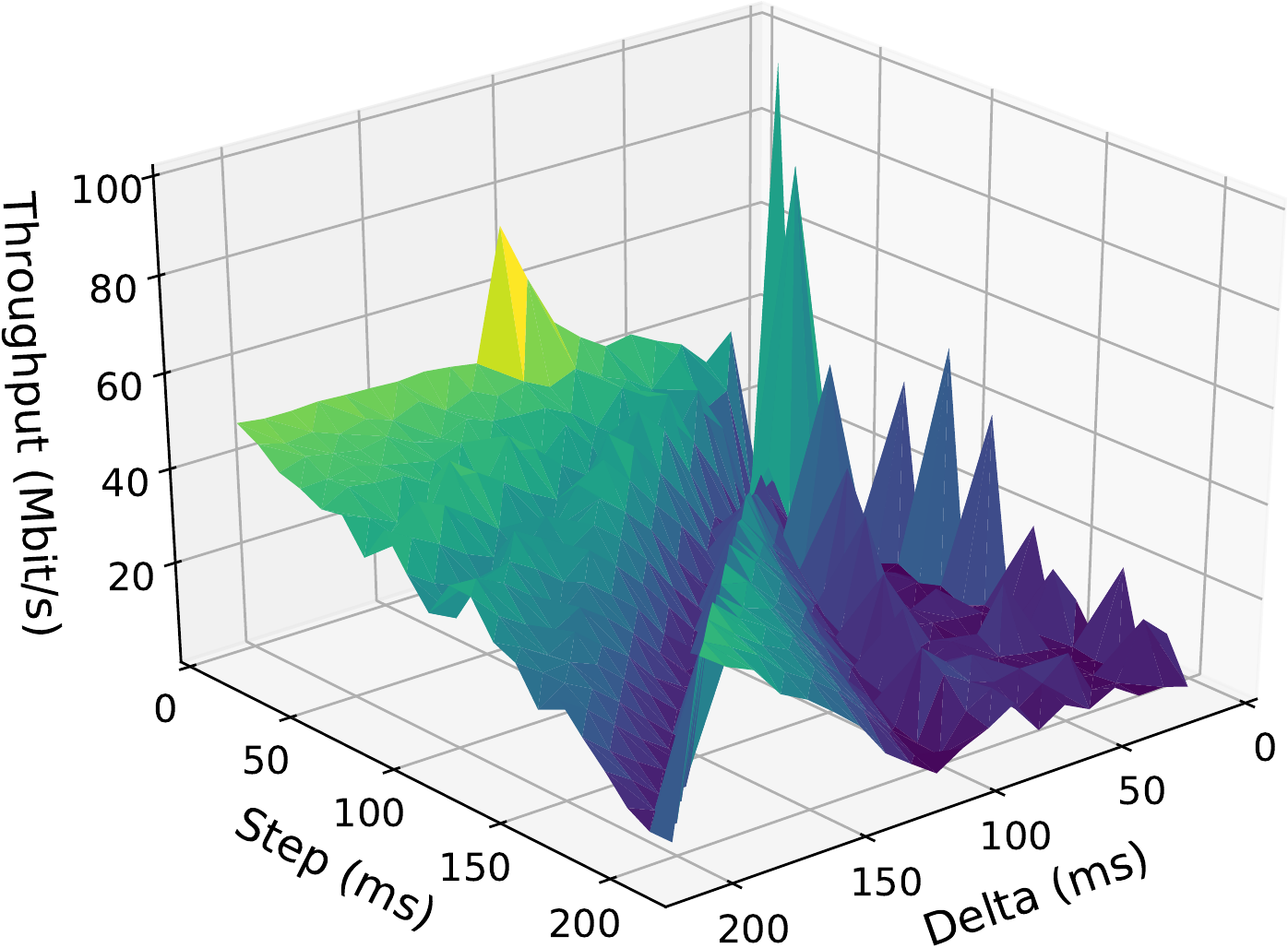} }}%
    \subfloat[\bf Copa]{{\includegraphics[width=0.5\textwidth]{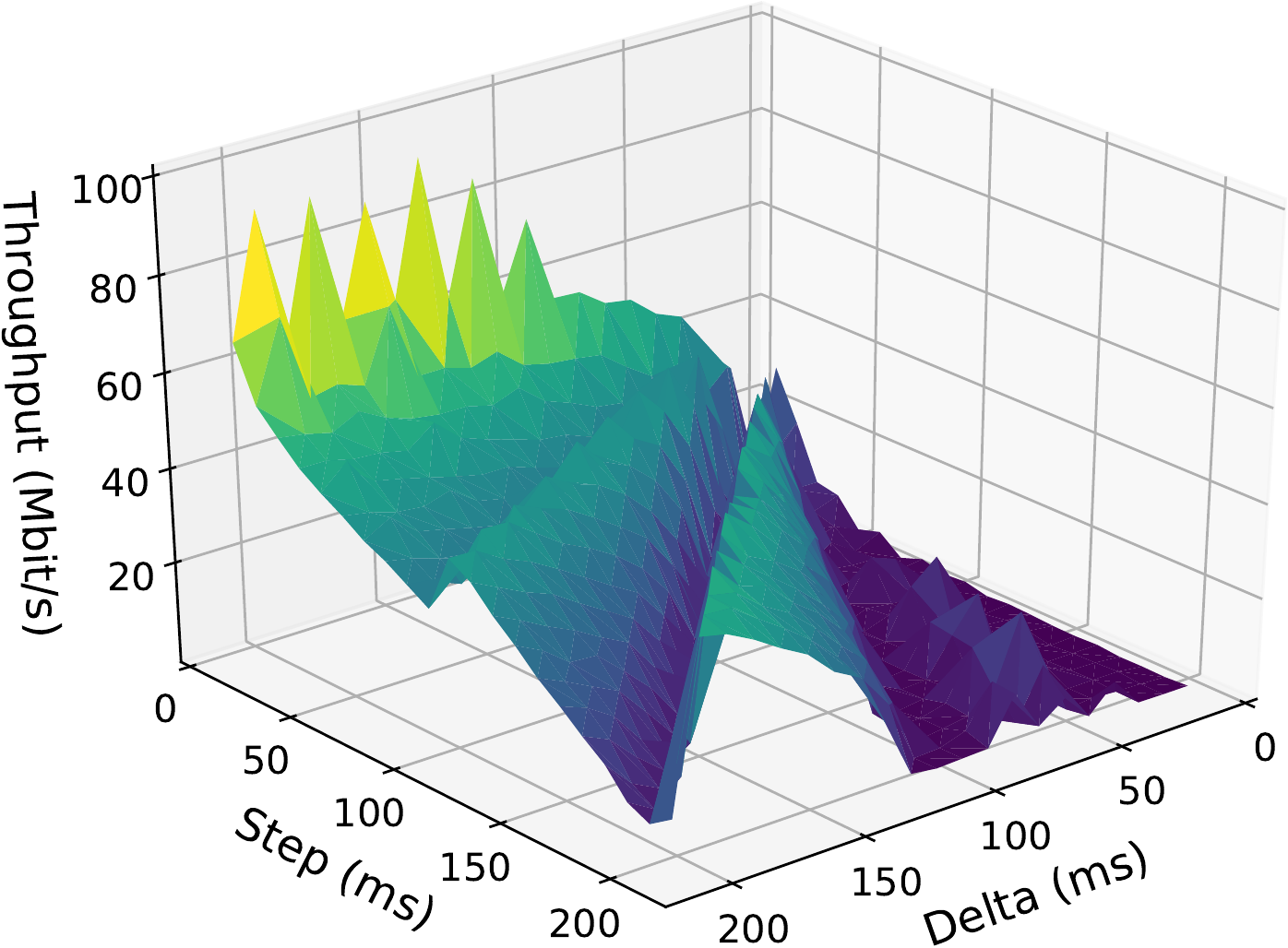} }}\\
    \vspace{-0.45cm}
    \hspace*{-0.4cm}
    \subfloat[\bf TCP BBR]{{\includegraphics[width=0.5\textwidth]{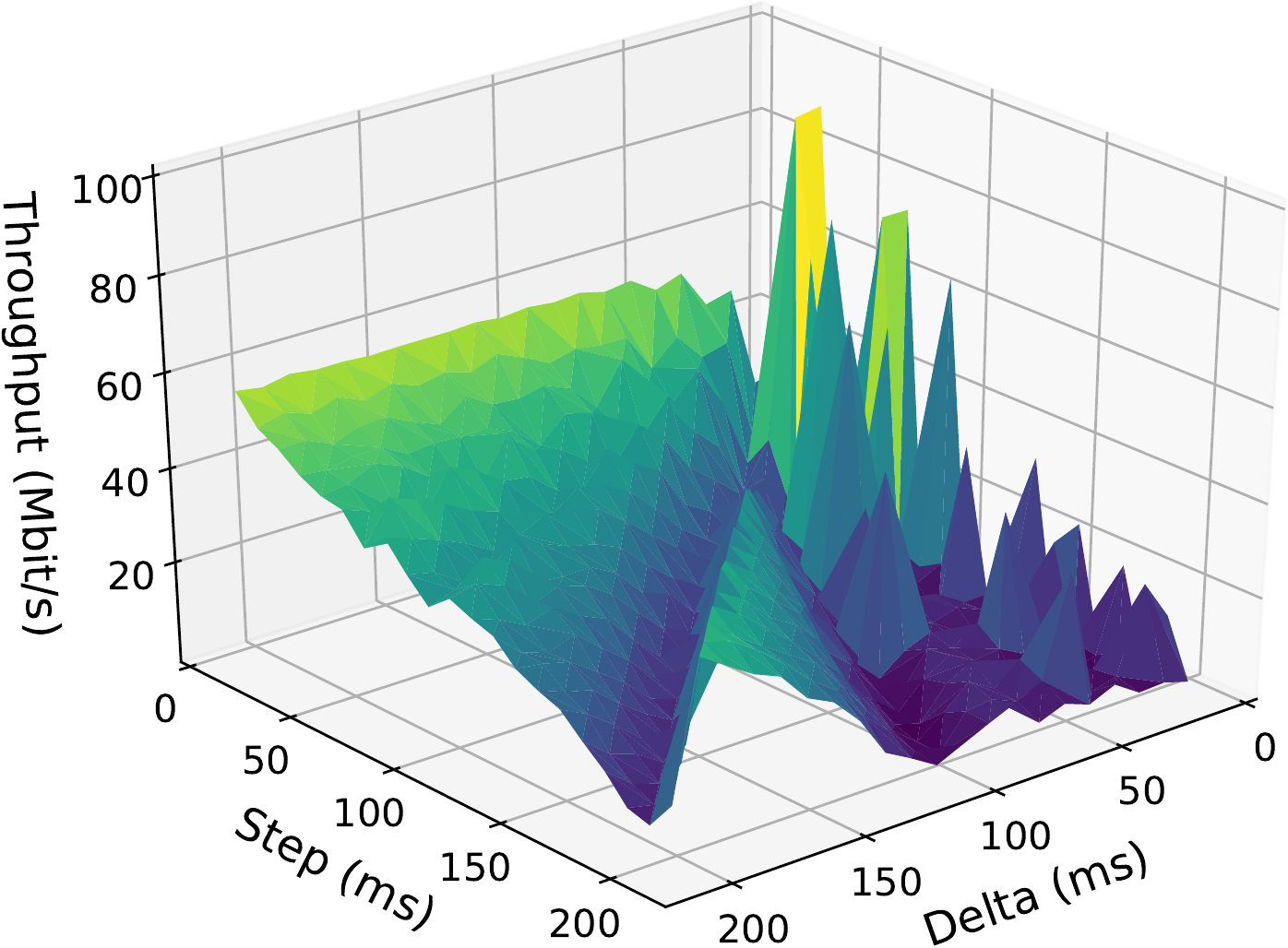} }}%
    \subfloat[\bf TCP CDG]{{\includegraphics[width=0.5\textwidth]{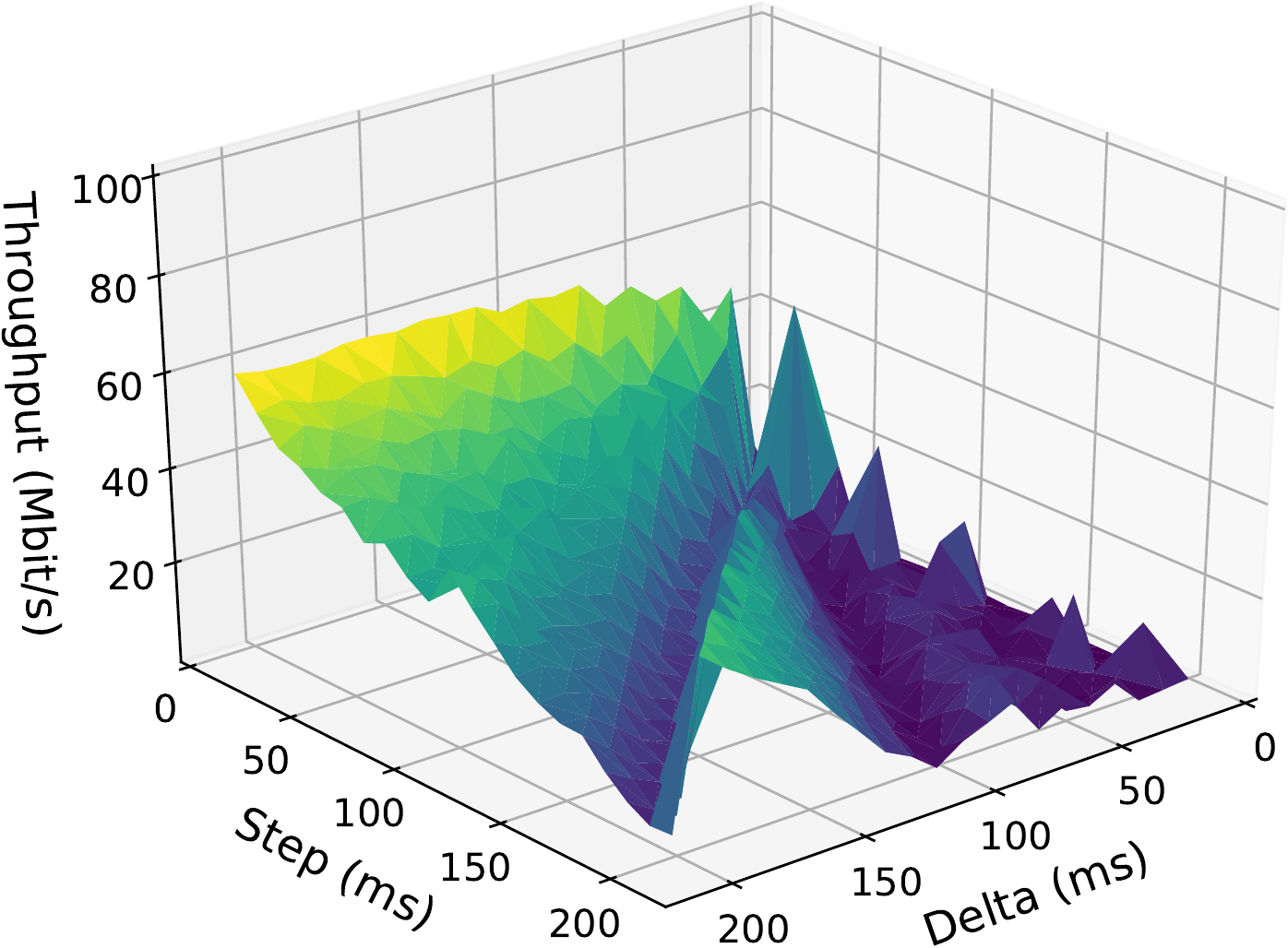} }}\\%
    \vspace{-0.45cm}
    \hspace*{-0.4cm}
\subfloat[\bf Indigo]{{\includegraphics[width=0.5\textwidth]{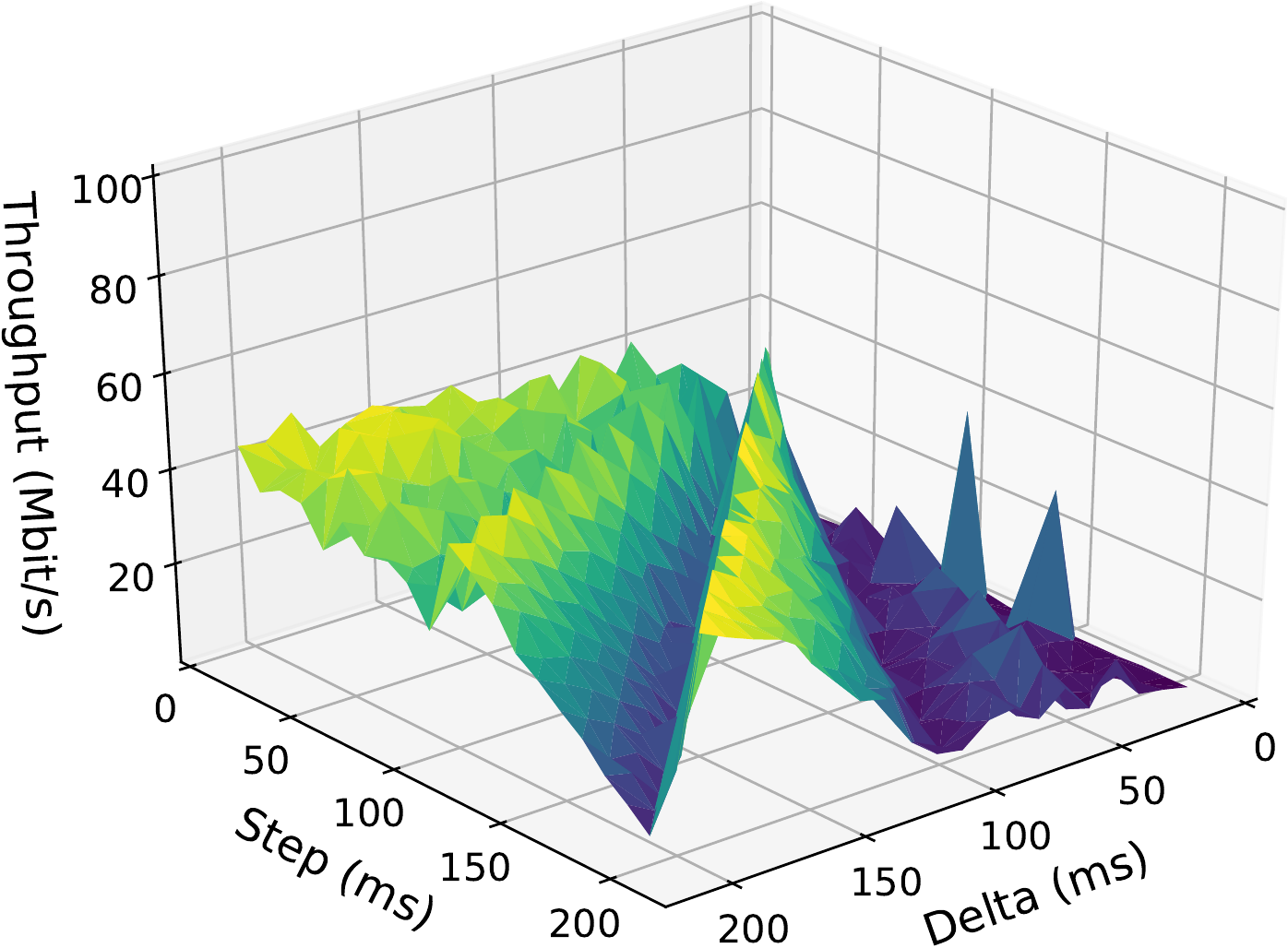} }}%
    \subfloat[\bf PCC]{{\includegraphics[width=0.5\textwidth]{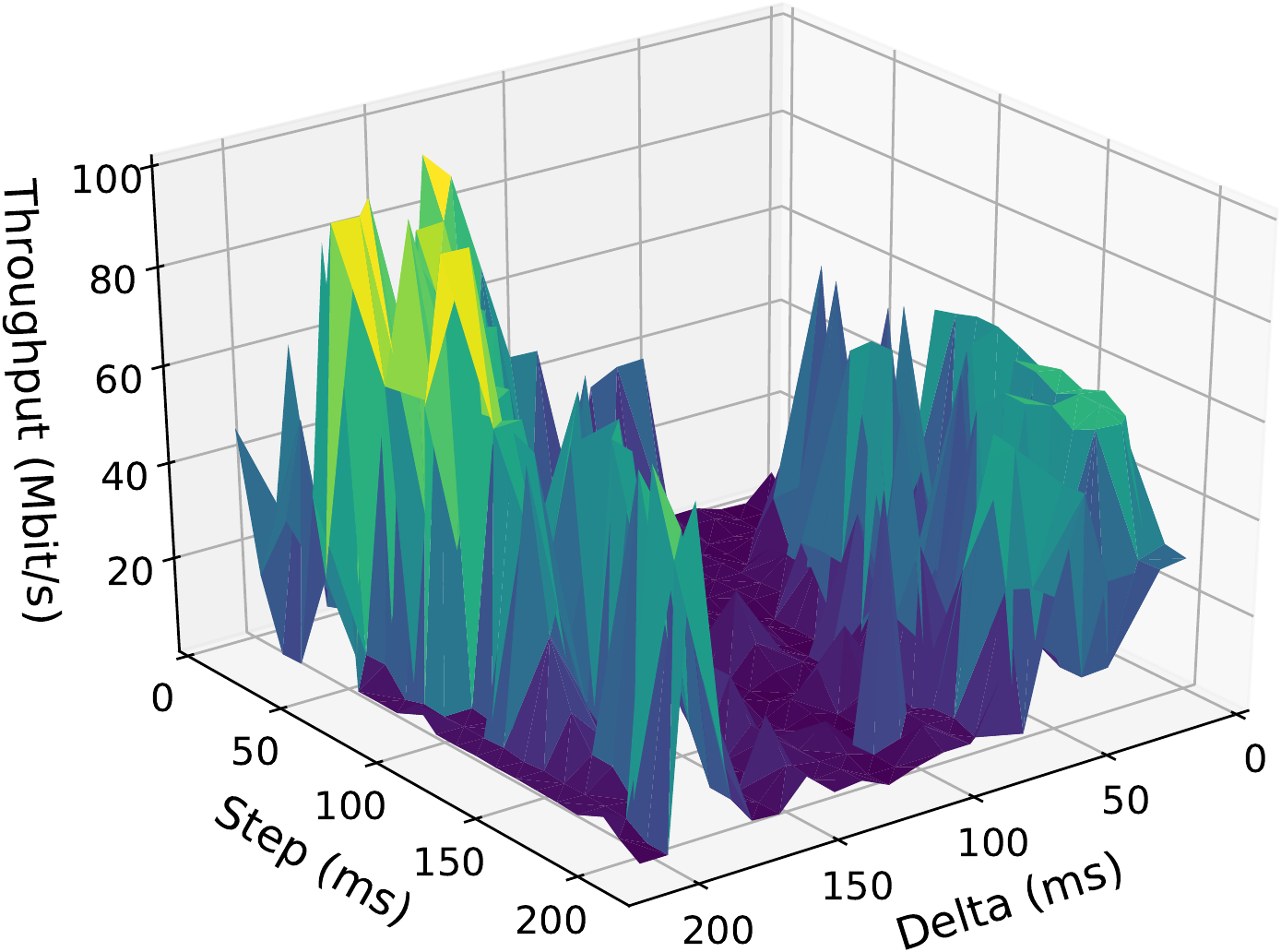} }}
    \caption{\mbox{Left side view rate plots for the schemes under the square-wave delay.}}%
    \label{fig:leftsideview}
    \end{adjustbox}
\end{figure}

\begin{figure}[h!]
\captionsetup[subfigure]{labelformat=empty}
    \begin{adjustbox}{minipage=\linewidth,totalheight=\textheight}
\vspace*{-0.9cm}
    \centering
    \subfloat[\bf TCP Cubic]{{\includegraphics[width=0.5\textwidth]{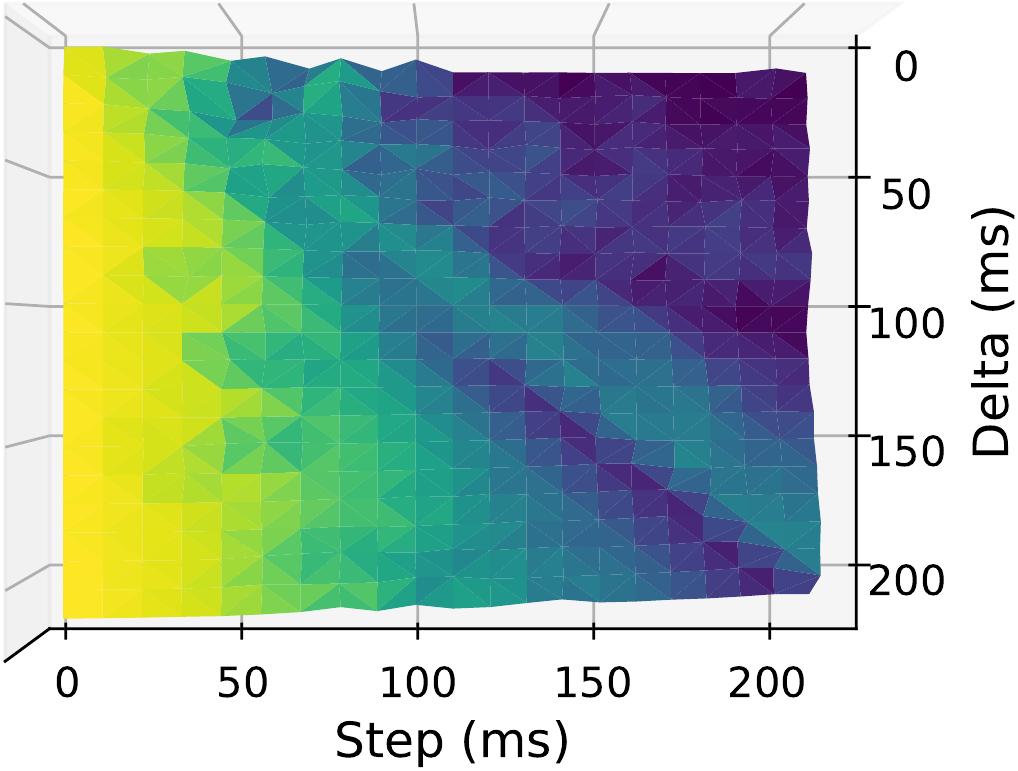} }}%
    \subfloat[\bf QUIC Cubic]{{\includegraphics[width=0.5\textwidth]{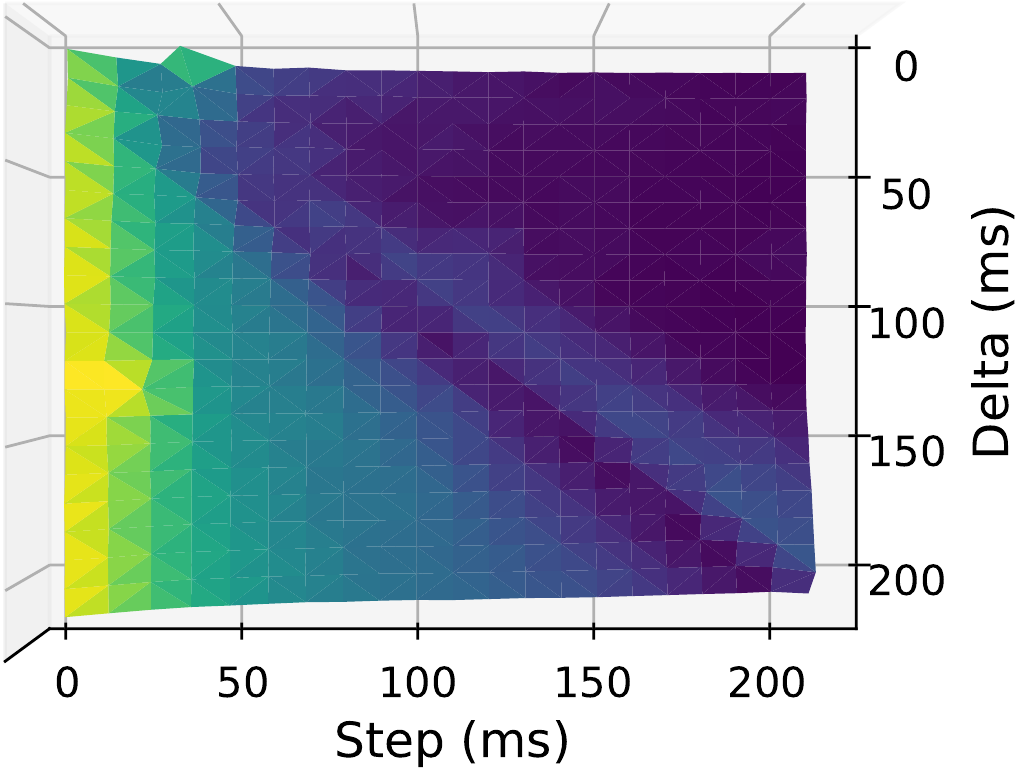} }}\\
    \vspace{-0.4cm}
    \subfloat[\bf TCP Vegas]{{\includegraphics[width=0.5\textwidth]{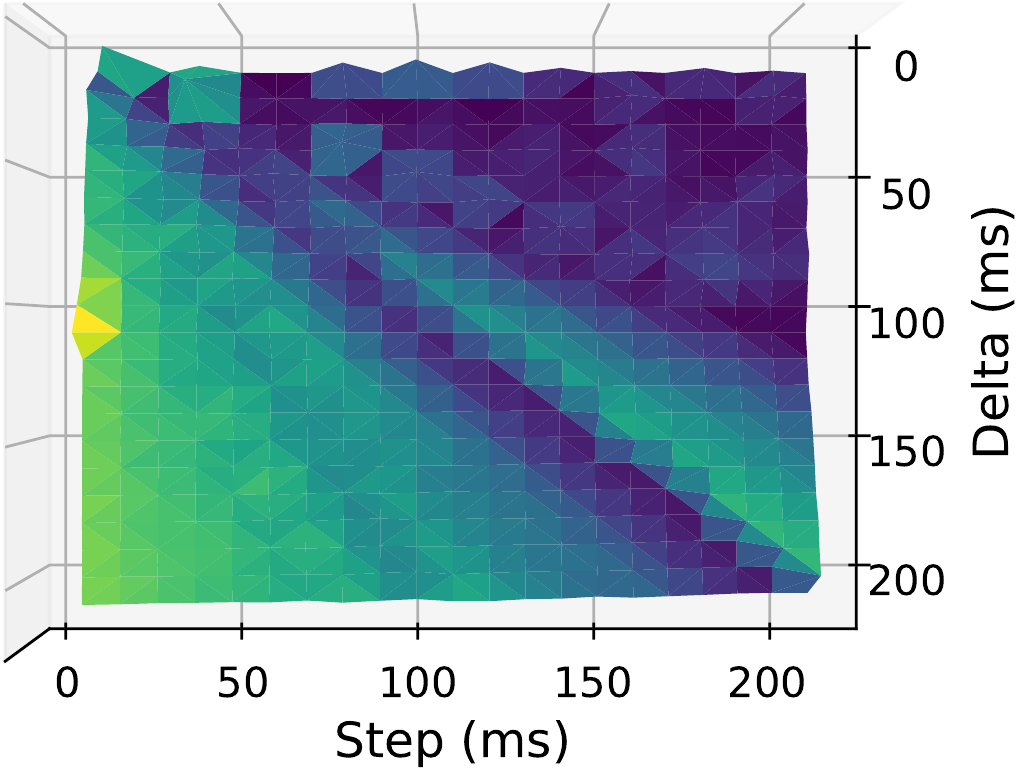} }}%
    \subfloat[\bf Copa]{{\includegraphics[width=0.5\textwidth]{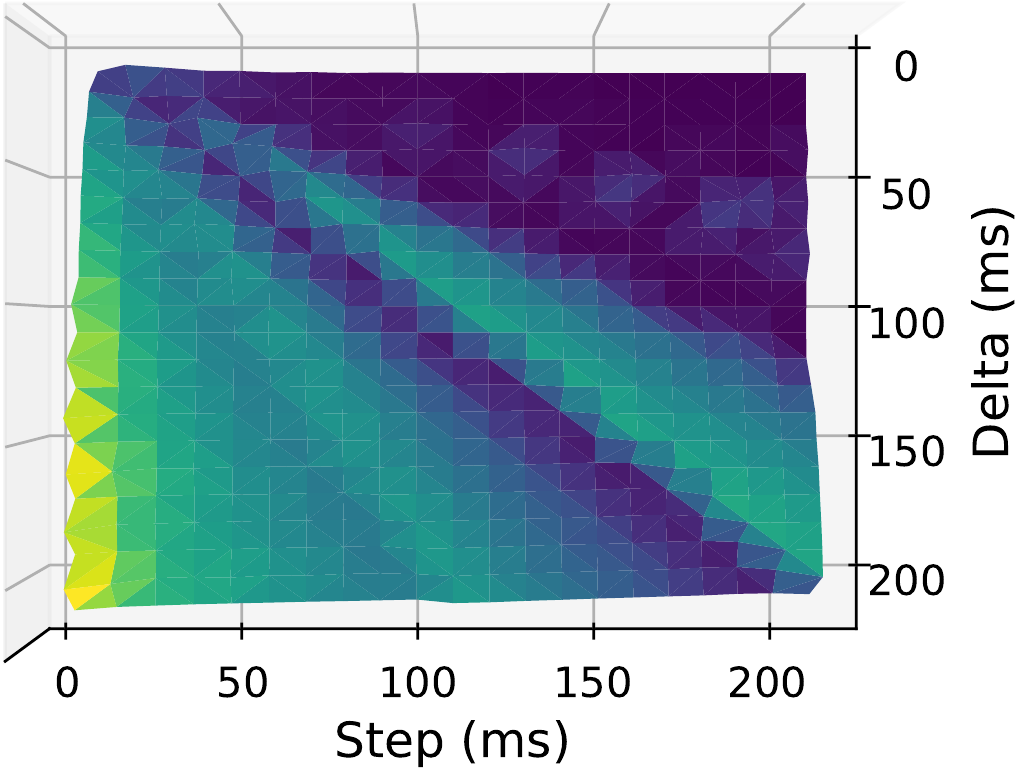} }}\\
    \vspace{-0.4cm}
    \subfloat[\bf TCP BBR]{{\includegraphics[width=0.5\textwidth]{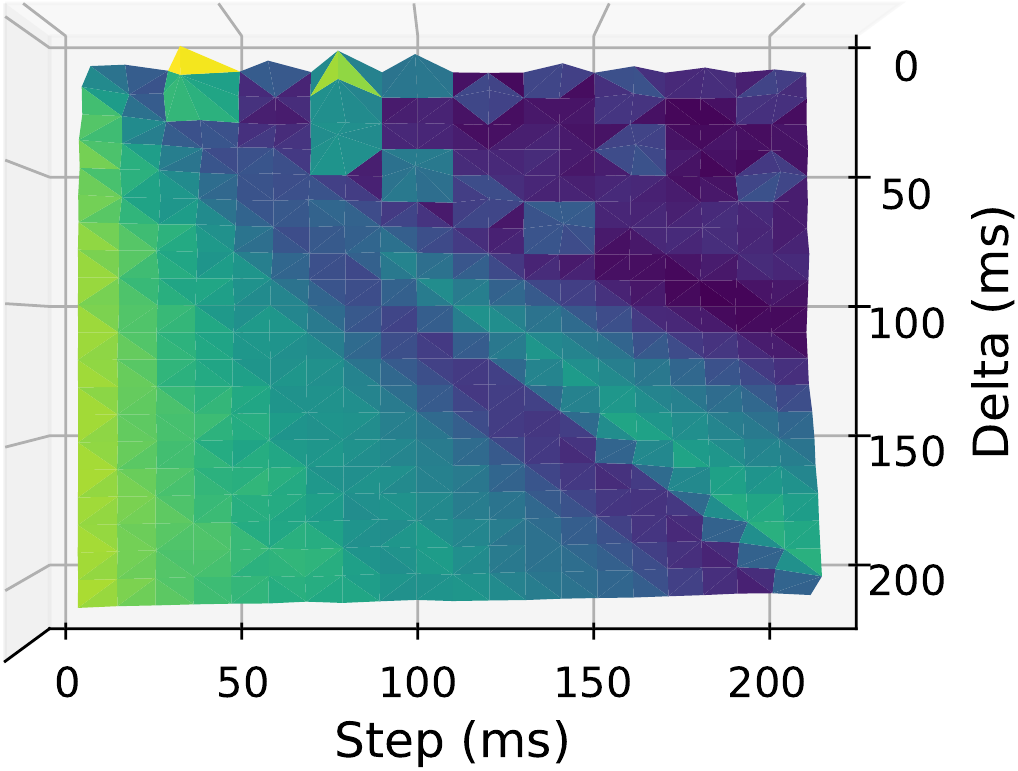} }}%
    \subfloat[\bf TCP CDG]{{\includegraphics[width=0.5\textwidth]{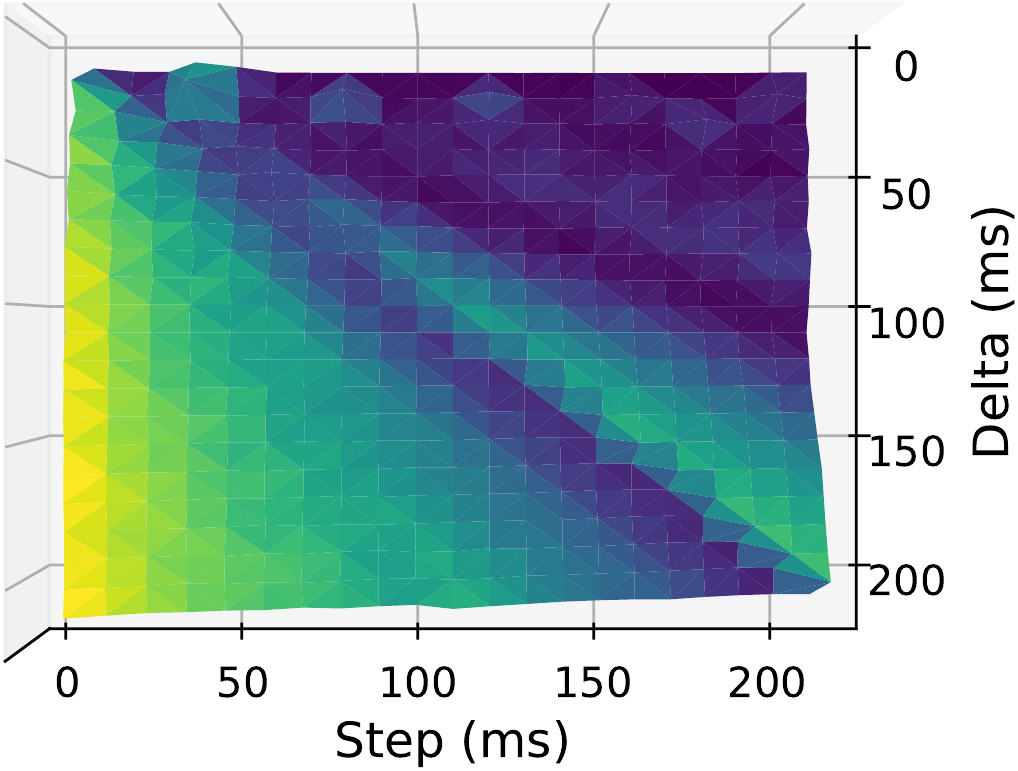} }}\\%
    \vspace{-0.4cm}
\subfloat[\bf Indigo]{{\includegraphics[width=0.5\textwidth]{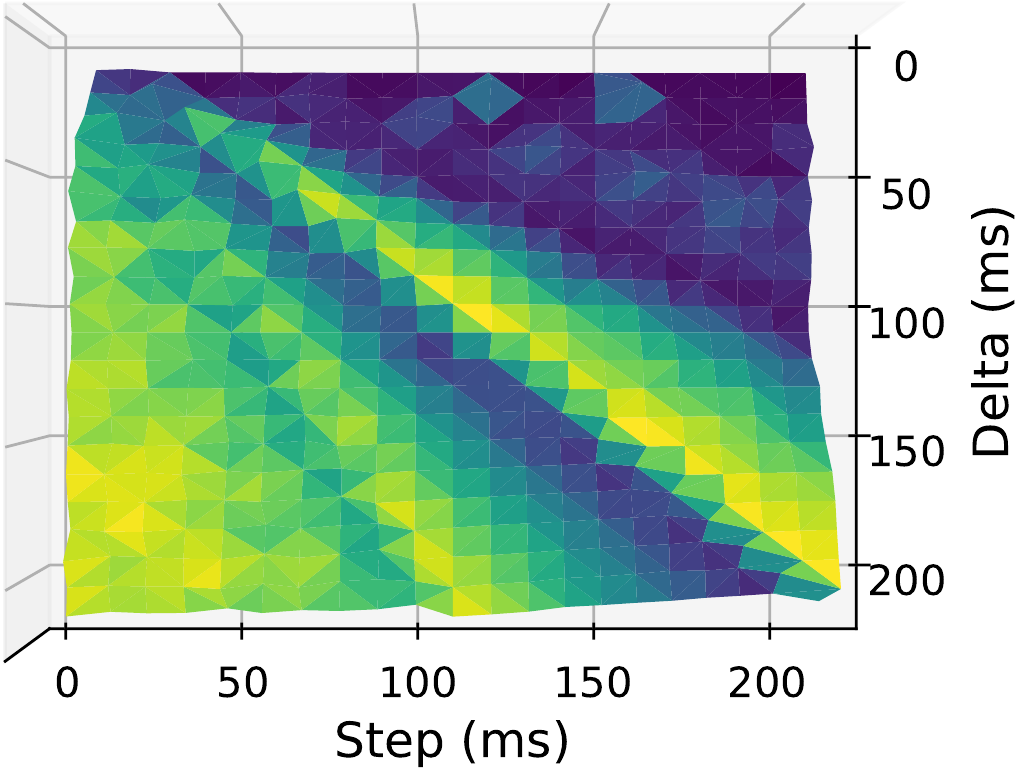} }}%
    \subfloat[\bf PCC]{{\includegraphics[width=0.5\textwidth]{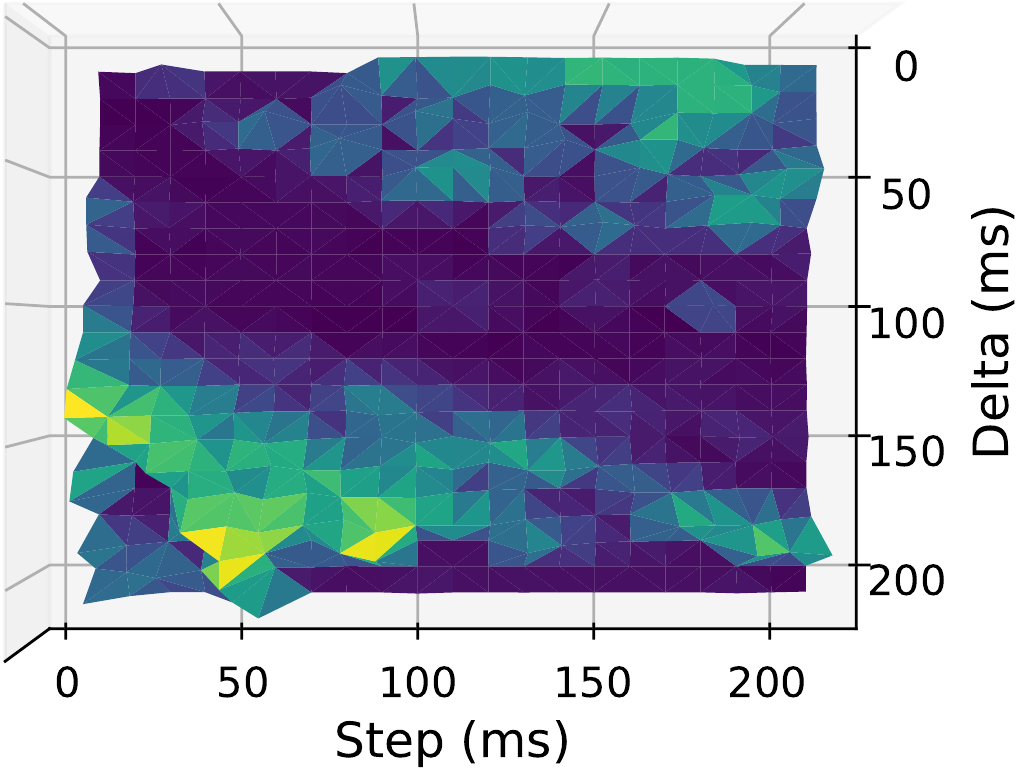} }}
    \caption{\mbox{Top view rate plots for the schemes under the square-wave delay.}}%
    \label{fig:topview}
    \end{adjustbox}
\end{figure}

Figure~\ref{fig:rightsideview} shows the right side view plots of the results.  Figure~\ref{fig:leftsideview} shows the left side view plots of the same results. Figure~\ref{fig:topview} shows the top view plots of the same results.

In general, the results observed in the plots are according to expectations: the greater the delta and the smaller the step of the square-wave delay, the bigger the resulting throughput is. It can be seen in plots from all the three points of view, but most distinctly in the top view plots in Figure~\ref{fig:topview}. 

Meanwhile, in the left side view plots of all the schemes in Figure~\ref{fig:leftsideview}, it is easy to notice a distinctive ``valley" in the throughput surface. This ``valley" corresponds to a dark-blue diagonal clearly visible in the corresponding top view plots in Figure~\ref{fig:topview}. The dark-blue diagonal stretches from the left upper corner to the right lower corner of each plot and indicates the remarkable drop in the throughput of the tested schemes when the step is close or equal to the delta.

The observation of the data, over which the plots were built, gives more details on the pattern. When the step value is approaching the delta value, the throughput starts dropping. The throughput becomes very low and this continues till the step is equal to the delta. But as soon as the step exceeds the delta, the throughput radically increases, and the increase is abrupt, without any gradual transition. For the sake of example, Tables~\ref{tab:vddata150} and~\ref{tab:vddata190} contain the throughput values of TCP Cubic, TCP BBR, and Copa for the delta 150 ms and 190 ms correspondingly. The discussed pattern is highlighted in both the tables with red color.

\FloatBarrier

\newcolumntype{Z}{>{\centering\arraybackslash}p{1cm}}
\newcolumntype{Y}{>{\centering\arraybackslash}X}
\definecolor{mylr}{HTML}{FFE6E6}

\begin{table*}[h]
\centering
\small
\caption{Chosen throughputs for Cubic, BBR, Copa for the delta 150 ms.\\The delta and step values are in ms, the throughputs are in Mbit/s.}
\begin{tabu}{|ZV{5}Y|Y|Y|Y|Y|Y|Y|Y|Y|YV{5}}
\hline
\multicolumn{1}{|c|}{\parbox[c][0.8cm]{0.9cm}{\bf delta}} & 150 & 150 & 150 & 150 & 150 & 150 & 150 & 150 & 150 & \multicolumn{1}{c|}{150}\\
\hline
\multicolumn{1}{|c|}{\parbox[c][0.8cm]{0.7cm}{\bf step}} & 90 & 100 & 110 & 120 & 130 & 140 & 150 & 160 & 170 & \multicolumn{1}{c|}{180}\\
\cline{1-1}\noalign{\vskip-1pt}\tabucline[2pt]{2-11}
\pbox[c][0.8cm]{100cm}{\bf Cubic} & 64.38 & 57.93 & 49.71 & 41.52 & \cellcolor{mylr}31.57 & \cellcolor{myr}18.89 & \cellcolor{myr}12.21  & 40.49 & 45.33 & 40.89 \\
\cline{1-1}\noalign{\vskip-1pt}\tabucline[2pt]{2-11}
\pbox[c][0.8cm]{100cm}{\bf BBR} & 30.13 & \cellcolor{mylr}27.83 & \cellcolor{mylr}24.28 & \cellcolor{myr}16.70 & \cellcolor{myr}10.66 & \cellcolor{myr}8.33 & \cellcolor{myr}15.63 & 40.22 & 36.05 & 32.20\\
\cline{1-1}\noalign{\vskip-1pt}\tabucline[2pt]{2-11}
\pbox[c][0.8cm]{100cm}{\bf Copa} & 37.98 & 31.92 & \cellcolor{mylr}26.10 & \cellcolor{mylr}21.19 & \cellcolor{myr}12.94 & \cellcolor{myr}5.26 & \cellcolor{myr}9.31 & 47.26 & 43.10 & 41.04\\
\cline{1-1}\noalign{\vskip-1pt}\tabucline[2pt]{2-11}
\end{tabu}
\label{tab:vddata150}
\end{table*}

\begin{table*}[h]
\centering
\small
\caption{Chosen throughputs for Cubic, BBR, Copa for the delta 190 ms.\\The delta and step values are in ms, the throughputs are in Mbit/s.}
\begin{tabu}{|ZV{5}Y|Y|Y|Y|Y|Y|Y|Y|Y|YV{5}}
\hline
\multicolumn{1}{|c|}{\parbox[c][0.8cm]{0.9cm}{\bf delta}} & 190 & 190 & 190 & 190 & 190 & 190 & 190 & 190 & 190 & \multicolumn{1}{c|}{190}\\
\hline
\multicolumn{1}{|c|}{\parbox[c][0.8cm]{0.7cm}{\bf step}} & 120 & 130 & 140 & 150 & 160 & 170 & 180 & 190 & 200 & \multicolumn{1}{c|}{210}\\
\cline{1-1}\noalign{\vskip-1pt}\tabucline[2pt]{2-11}
\pbox[c][0.8cm]{100cm}{\bf Cubic} & 51.17 & 42.04 & 34.68 & 37.52 & \cellcolor{mylr}30.80 & \cellcolor{mylr}23.93 & \cellcolor{myr}13.73  & \cellcolor{myr}10.36 & 46.02 & 41.30 \\
\cline{1-1}\noalign{\vskip-1pt}\tabucline[2pt]{2-11}
\pbox[c][0.8cm]{100cm}{\bf BBR} & 27.01 & \cellcolor{mylr}23.34 & \cellcolor{mylr}23.63 & \cellcolor{myr}19.94 & \cellcolor{myr}14.83 & \cellcolor{myr}10.91 & \cellcolor{myr}6.99 & \cellcolor{myr}13.59 & 40.86 & 38.25\\
\cline{1-1}\noalign{\vskip-1pt}\tabucline[2pt]{2-11}
\pbox[c][0.8cm]{100cm}{\bf Copa} & 33.98 & 28.98 & \cellcolor{mylr}23.65 & \cellcolor{mylr}20.76 & \cellcolor{myr}16.71 & \cellcolor{myr}10.37 & \cellcolor{myr}6.09 & \cellcolor{myr}10.84 & 46.41 & 44.71\\
\cline{1-1}\noalign{\vskip-1pt}\tabucline[2pt]{2-11}
\end{tabu}
\label{tab:vddata190}
\end{table*}

The thesis author checked if setting the base delay in the experiments to a small value, rather than to zero, changes the situation. The following variants were tried for the topology in Figure~\ref{fig:vd-dumbbell}: 1 ms, 3 ms or 5 ms base delay at the central\nolinebreak[4] link; \mbox{1 ms} delay at both the left and right links alongside 1ms or 3 ms base delay at the central link. None of these adjustments changed the results of the experiments considerably.

The next step was to explore how the change of the congestion window depends on the delta and step values for TCP schemes. The information on the change of the congestion window over time cannot be learned from a PCAP dump file, as it is not kept there. The current size of the congestion window of a TCP connection is maintained at the sender host, as it is the sender host that performs the end-to-end congestion control.

The thesis author solved the problem in the following way. As noted above, one thread of CoCo-Beholder testing tool is varying the delay at the central link during the whole runtime, while the other thread is starting a batch of new flows each second of the runtime. In this series of experiments, there is only one flow started at the $0^{\text{th}}$ second of the runtime. This means that this other thread has no tasks to perform during the rest of the runtime. The thesis author made the thread execute the Linux command \verb+ss -ti+~\cite{ss} at the sender virtual host in a cycle every 50 ms and save its\nolinebreak[4] output into a log file. The thesis author ensured that this temporary change to the source code of CoCo-Beholder did not influence the results and execution time.

\begin{figure}[h!]
\captionsetup[subfigure]{labelformat=empty}
    \begin{adjustbox}{minipage=\textwidth,totalheight=\textheight}
    \centering
    \includegraphics[width=\textwidth]{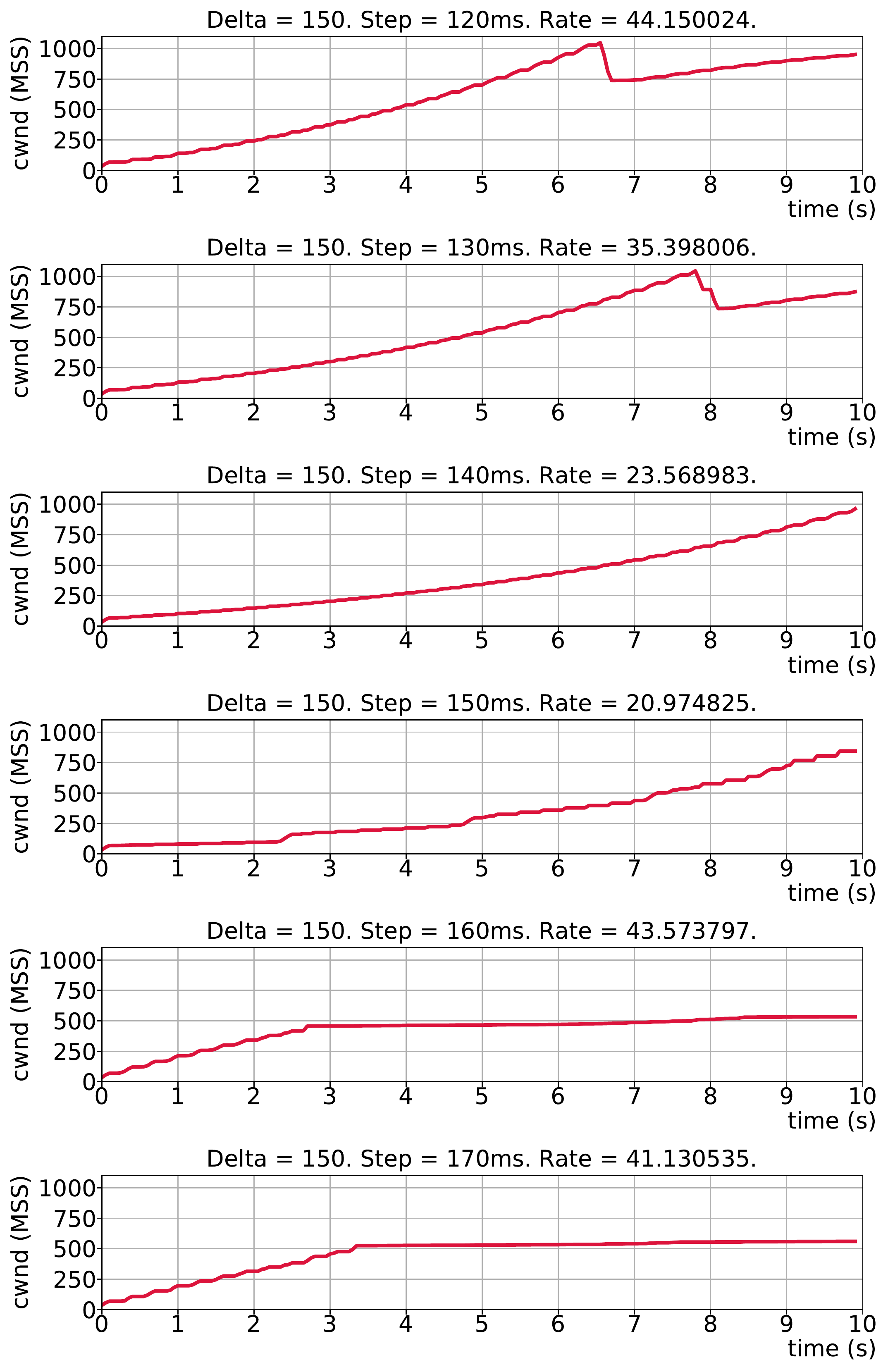}
    \caption{\mbox{Cubic's congestion window evolution under the square-wave delay.}}%
    \label{fig:cubic-cwnd}
    \end{adjustbox}
\end{figure}

\begin{figure}[h!]
\captionsetup[subfigure]{labelformat=empty}
    \begin{adjustbox}{minipage=\textwidth,totalheight=\textheight}
    \centering
    \includegraphics[width=\textwidth]{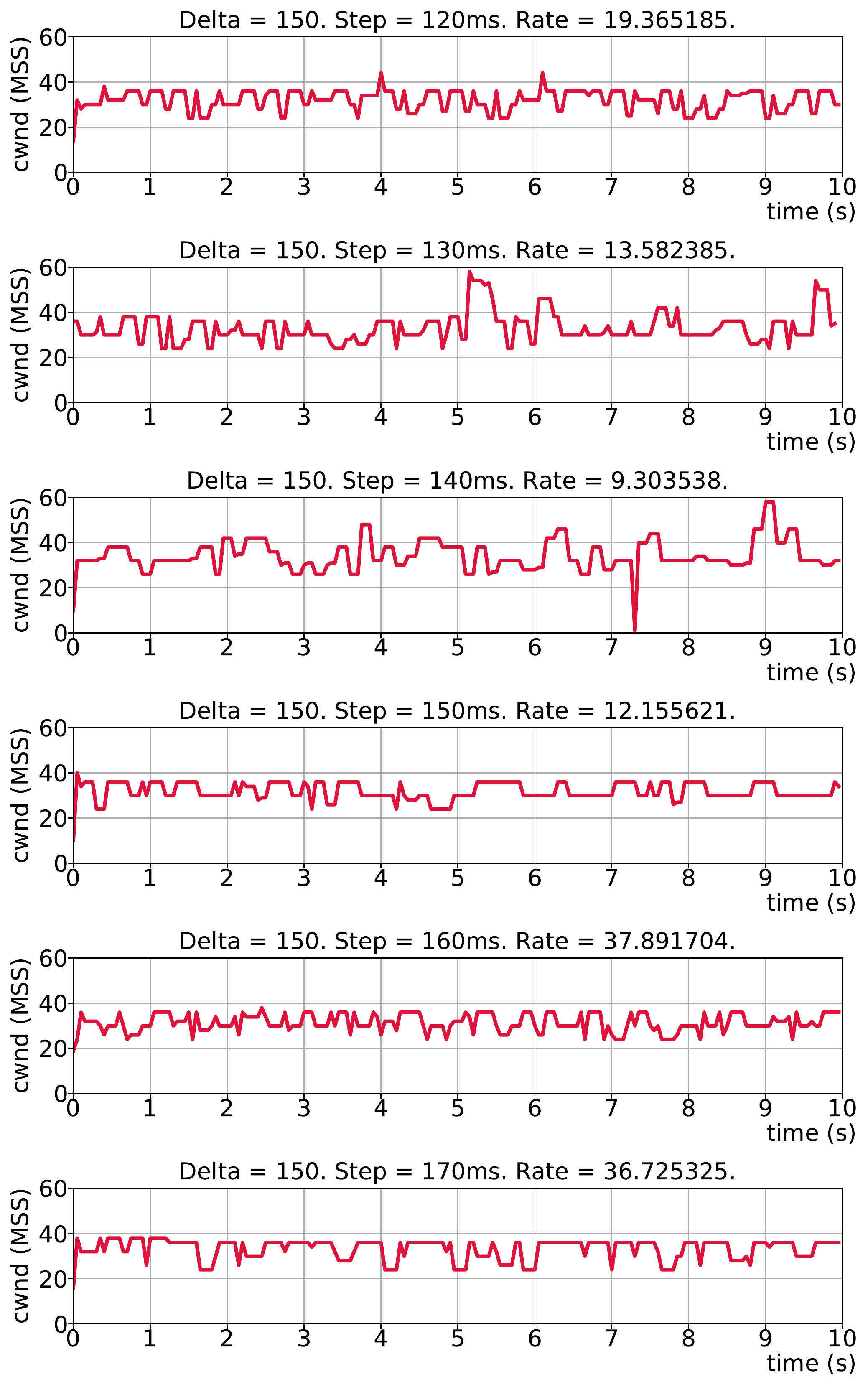}
    \caption{\mbox{BBR's congestion window evolution under the square-wave delay.}}%
    \label{fig:bbr-cwnd}
    \end{adjustbox}
\end{figure}

\FloatBarrier

The output of the \verb+ss -ti+ command provides the information on the single TCP connection (the flow of the tested TCP scheme) at the virtual host including the current size of the congestion window measured in Maximum Segment Size (MSS)~\cite{kurose}.

For TCP Cubic and TCP BBR, the modified experiments were run with the delta 150\nolinebreak[4] ms and the steps that are close or equal to the delta: 120, 130, ..., 170 ms. After that, the output log files were parsed and plotted. The resulting plots showing the evolution of the congestion window under the square-wave delay with different steps can be seen in Figure~\ref{fig:cubic-cwnd} for Cubic and in Figure~\ref{fig:bbr-cwnd} for BBR.

For Cubic, it can be observed that the congestion window expands much slower when the step is less than the delta but close to it, and the closer the step and the delta are, the slower the expansion of the congestion window is. However, once the step exceeds the delta, the congestion window is not restraint anymore and reaches the necessary size during the first three seconds of the runtime. To see these observations better in the plots, it can be convenient to choose a second at the x-axis, e.g. the $5^{\text{th}}$ second, and to watch how the corresponding congestion window size changes from the top plot (the\nolinebreak[4] step 120\nolinebreak[4] ms) to the bottom one (the step 170 ms).

BBR's congestion window does not show much change when the step is approaching the delta. The thesis author guesses that this is due to the fact that BBR is a hybrid scheme, so it tries not to fill the queues, in order to keep the latency low. BBR does not expand the congestion window that aggressively, as Cubic does when it can. Therefore, when some problem prevents Cubic from expanding the congestion window, it is striking at once, while it is not that noticeable when BBR is affected by the same problem.   

The further discussion and plots will be related to BBR only. The thesis author will analyze the RTT plots of the experiments, and this is more convenient to do for BBR, as it tries to maintain the RTT as low as possible, and so the RTT plots are neater and better correspond to the delay installed at the central link of the topology. 

The plots for Cubic, analogous to those, which will be further presented for BBR (throughput, one-way delay, and RTT plots), can be found in Appendix~\ref{appendix:AppendixE}. The reader can compare the plots for BBR and Cubic and notice that they look similar in the sense of the general shape and the periodicity of the curves. Most importantly, the RTT plots for Cubic and BBR have the common patterns, highlighted later in this section. The thesis author claims that the discussion and the conclusions, made in this section for BBR, are true for Cubic also. The thesis author suggests that they are true for all other schemes, above-mentioned in this section, experiencing the drop of the throughput when the step is approaching the delta ``from the left".\\

The three cases were considered: the step is less than the delta but close to it, the step and the delta are equal, and the step is greater than the delta but close to it. The thesis author discovered that the plots for the three cases differ but the plots in the scope of one case are similar. This is why the task was simplified to discuss here only the plots for the three narrow cases: $step==(delta-10 ms)$; $step==delta$; $step==(delta+10 ms)$. The delta considered here is 150 ms, and so the steps are 140, 150, 160 ms.

\begin{figure}[h!]
\vspace*{-0.3cm}
\begin{adjustbox}{minipage=\linewidth,totalheight=\textheight}
    \centering
    \subfloat[Step = 140 ms]{{\includegraphics[height=0.3\textheight]{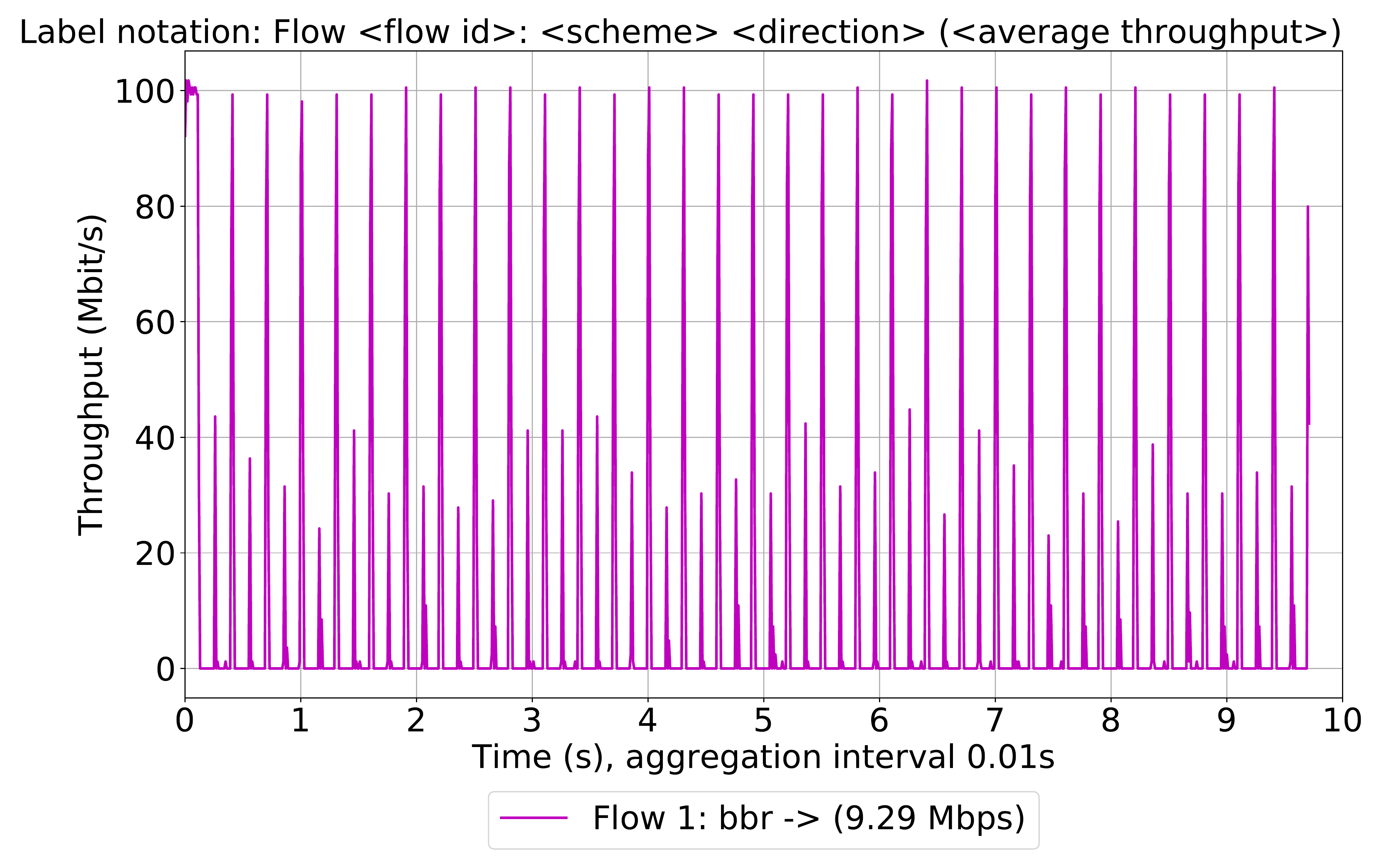} }}\\
    \vspace*{-0.3cm}
    \subfloat[Step = 150 ms]{{\includegraphics[height=0.3\textheight]{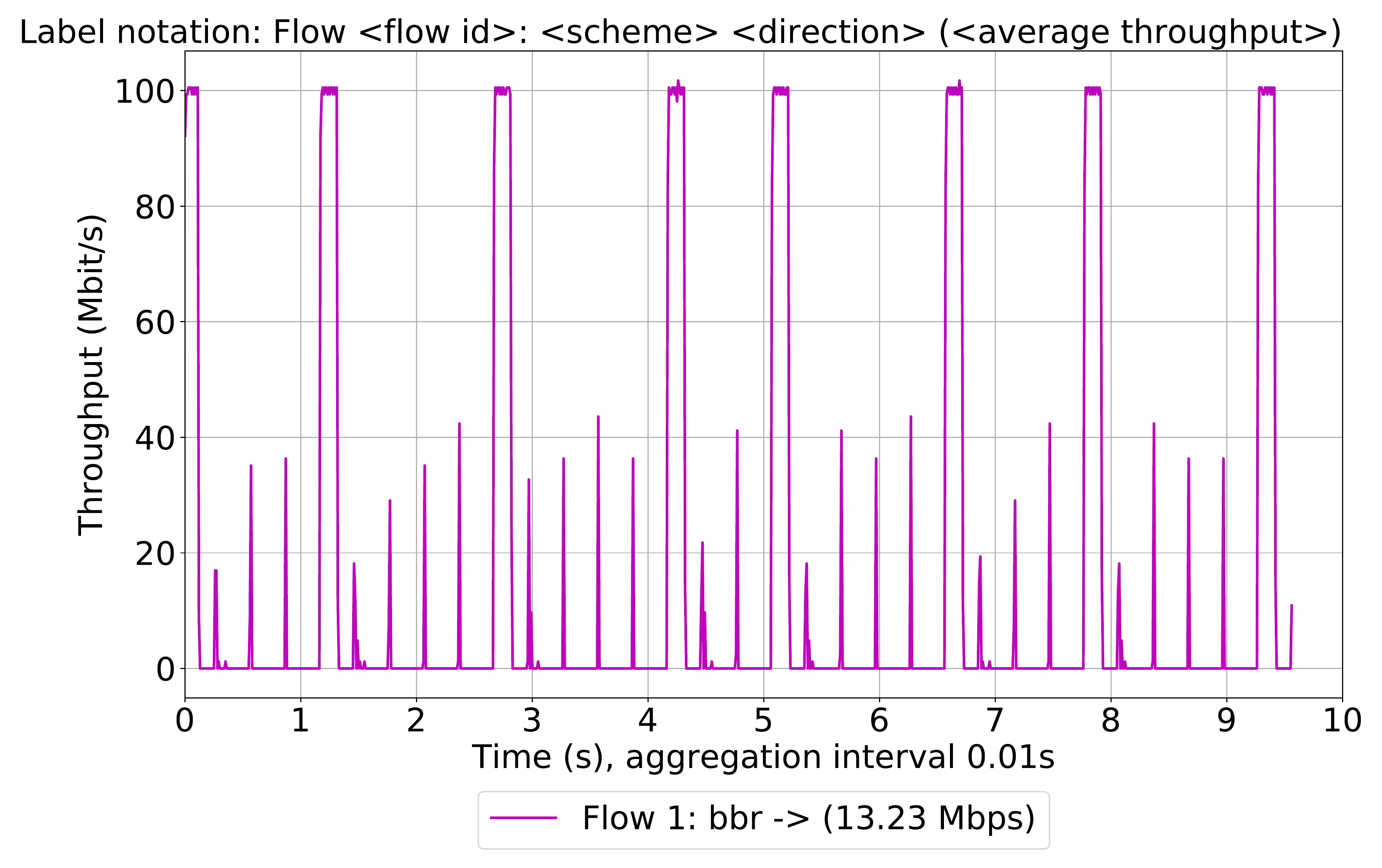} }}\\
    \vspace*{-0.3cm}
    \subfloat[Step = 160 ms]{{\includegraphics[height=0.3\textheight]{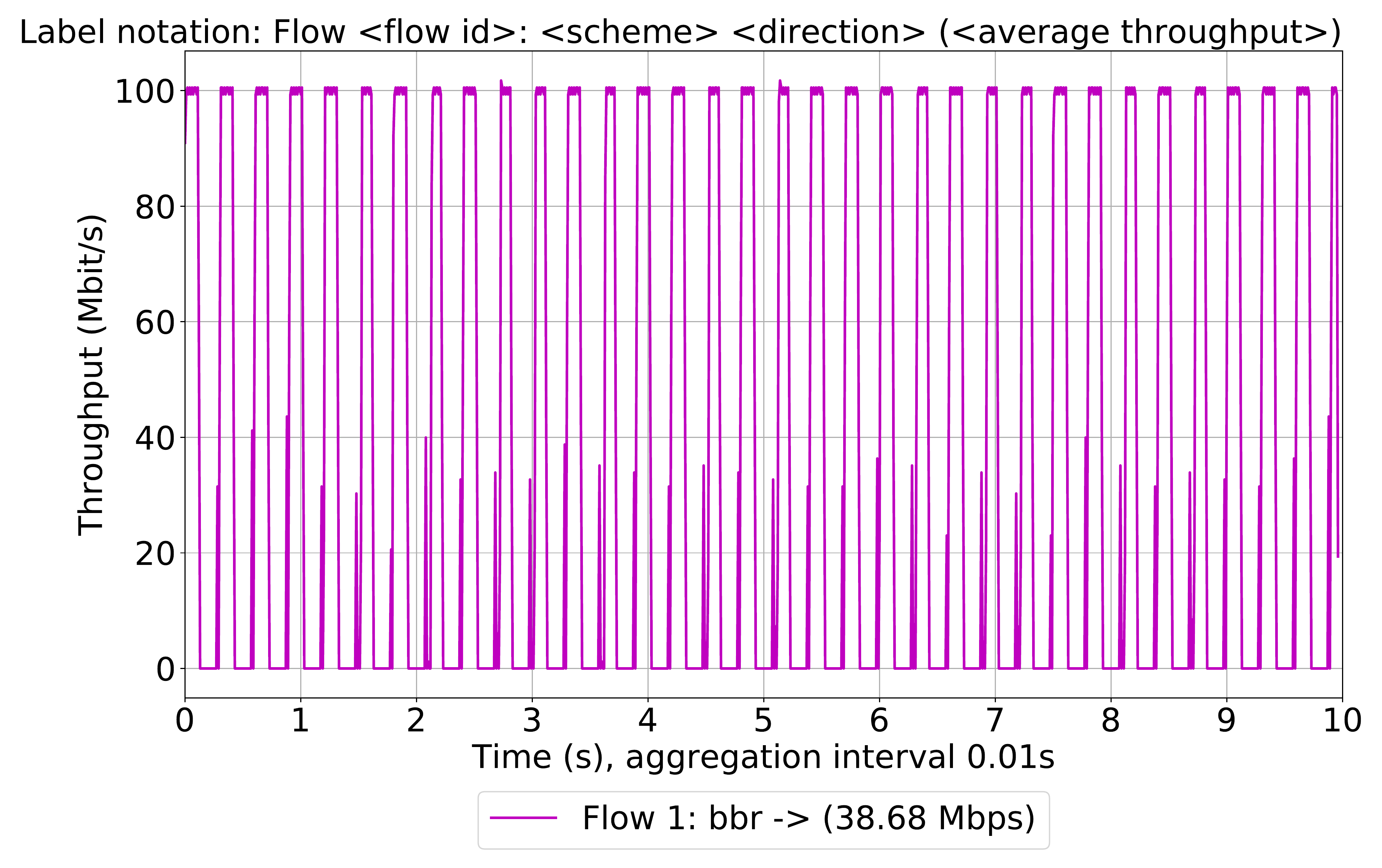} }}%
    \vspace*{-0.1cm}
    \caption{BBR's average throughput plots for the delta 150 ms.}%
    \label{fig:vd-bbr-rate}
\end{adjustbox}
\end{figure}

There are BBR's average throughput plots for the three steps in Figure~\ref{fig:vd-bbr-rate} generated by CoCo-Beholder. The top and bottom plots -- for the step 140 and 160 ms -- are periodic, and the period is twice the delta, i.e. 300 ms. In the top plot, the rate jumps for an instant up to 100 Mbit/s every 300 ms but is zero most of the time, which explains the low overall average throughput statistic. The bottom plot shows the best throughput of all the three plots: half the runtime the rate is zero and half the runtime the rate is 100 Mbit/s. The center plot, belonging to the case when the step and delta are equal, looks like a transitional one between the top and the bottom plots. 

\begin{figure}[h!]
\vspace*{-0.3cm}
\begin{adjustbox}{minipage=\linewidth,totalheight=\textheight}
    \centering
    \subfloat[Step = 140 ms]{{\includegraphics[height=0.3\textheight]{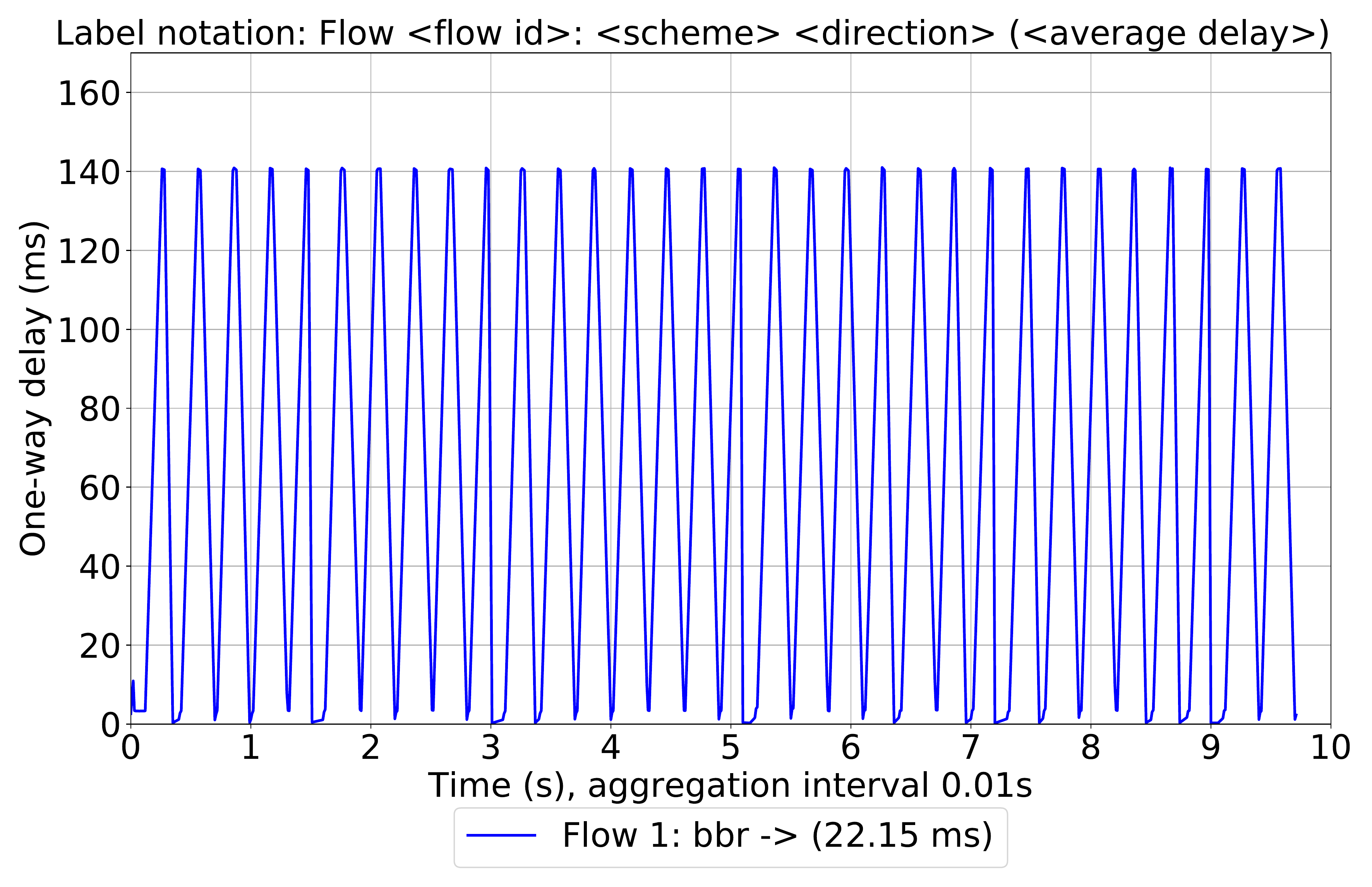} }}\\
    \vspace*{-0.3cm}
    \subfloat[Step = 150 ms]{{\includegraphics[height=0.3\textheight]{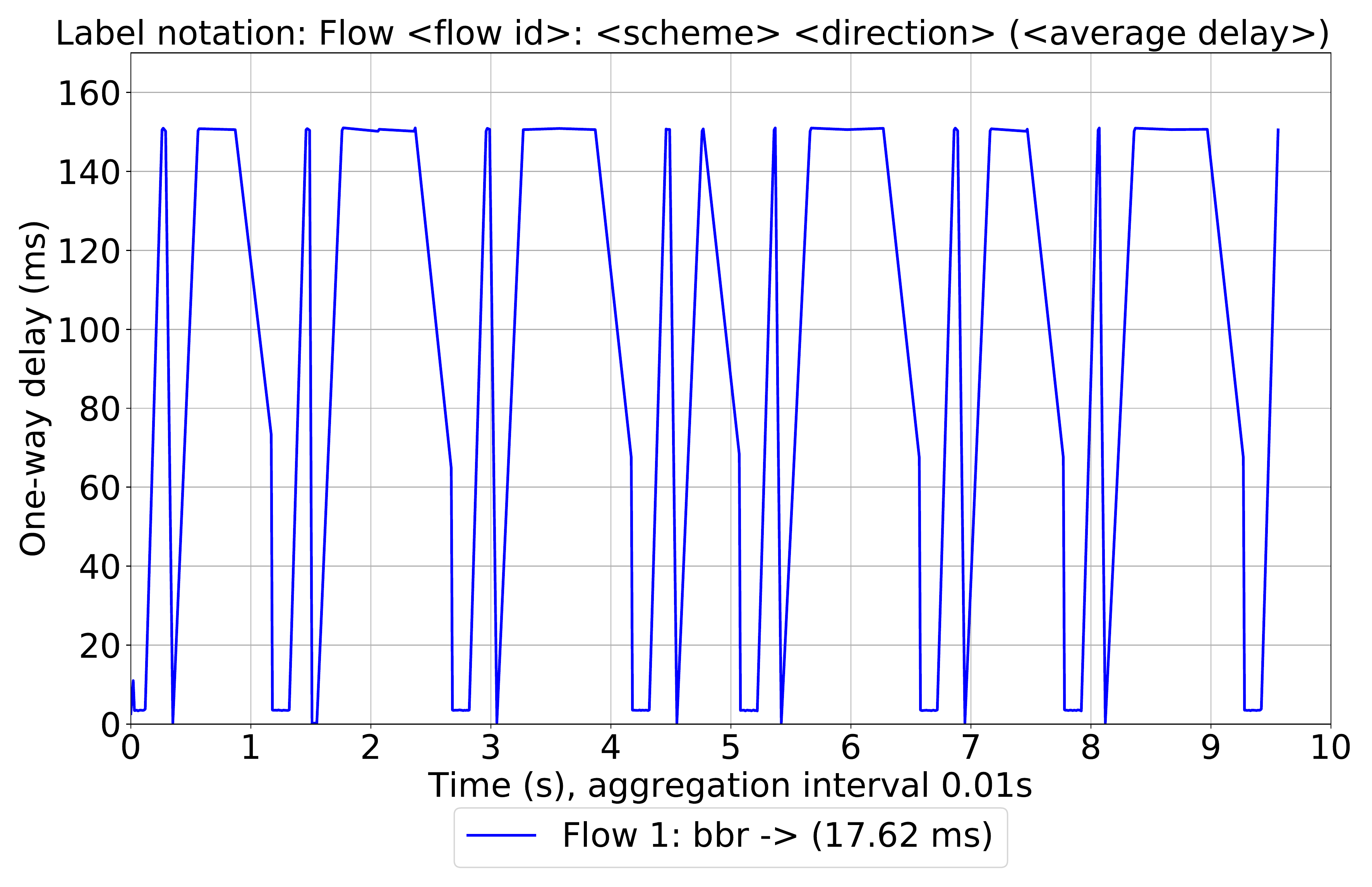} }}\\
    \vspace*{-0.3cm}
    \subfloat[Step = 160 ms]{{\includegraphics[height=0.3\textheight]{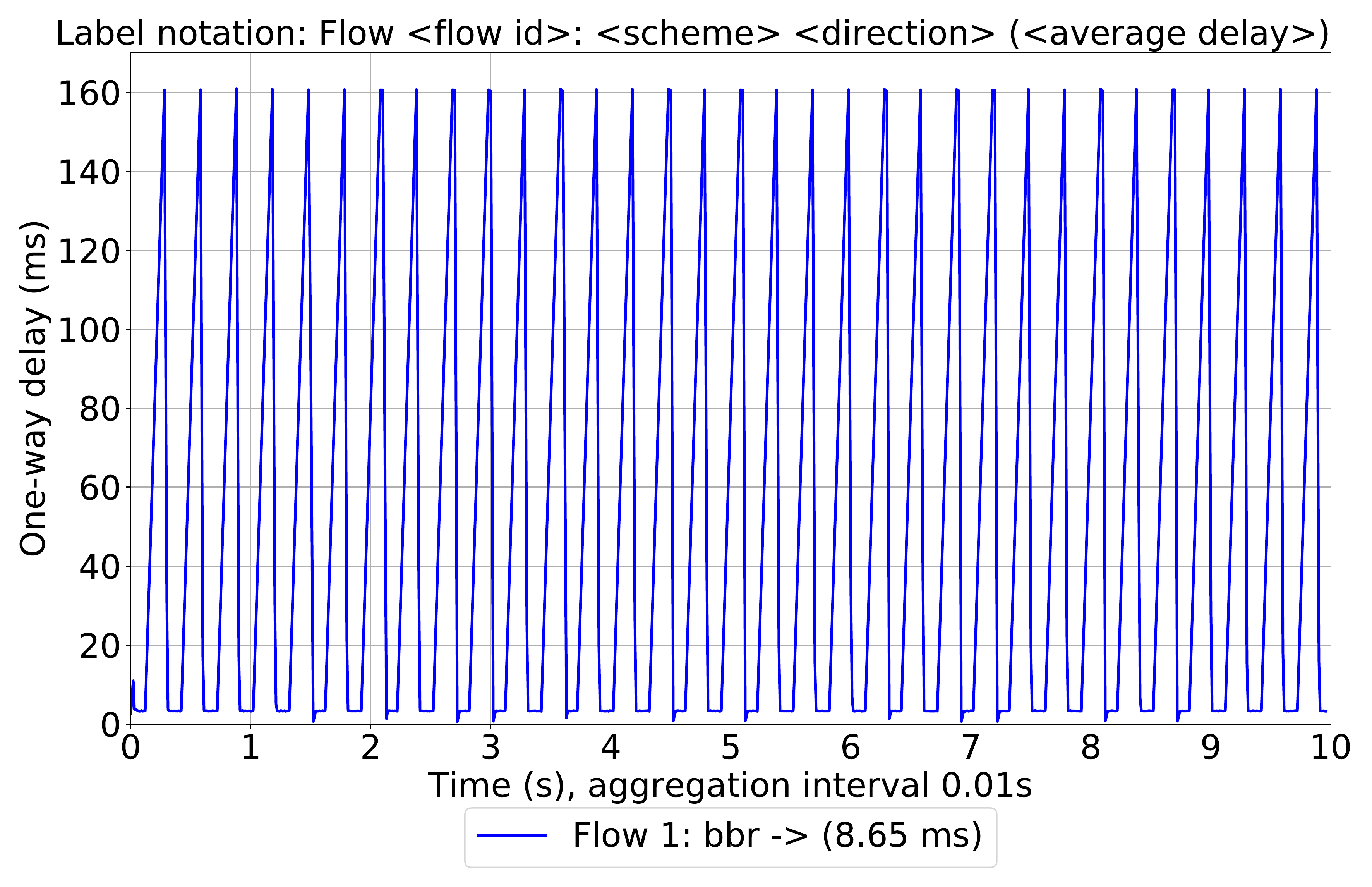} }}%
    \vspace*{-0.1cm}
    \caption{BBR's average one-way delay plots for the delta 150 ms.}%
    \label{fig:vd-bbr-avgdelay}
\end{adjustbox}
\end{figure}

\begin{figure}[h!]
\vspace*{-0.3cm}
\begin{adjustbox}{minipage=\linewidth,totalheight=\textheight}
    \centering
    \subfloat[Step = 140 ms]{{\includegraphics[height=0.3\textheight]{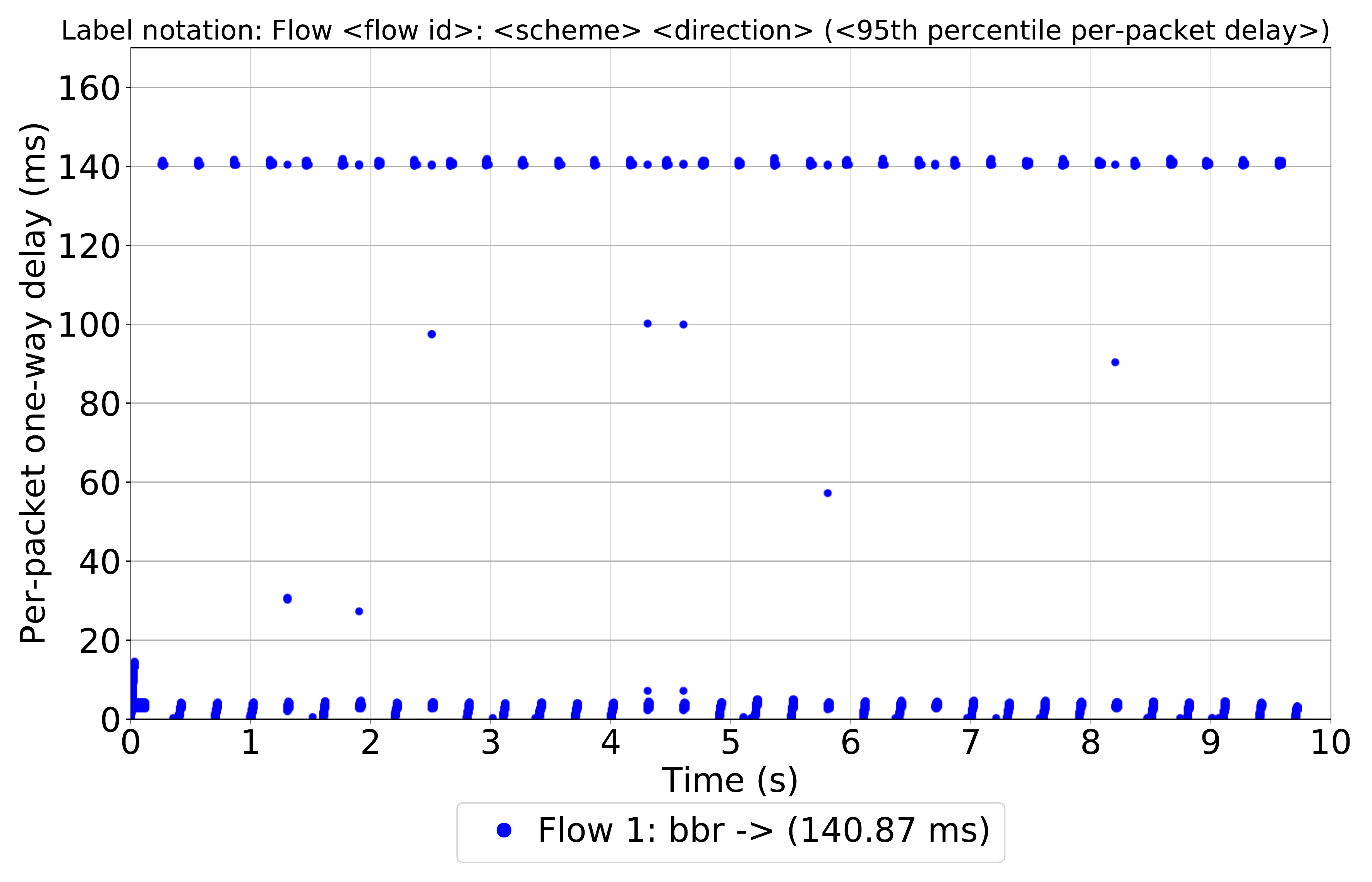} }}\\
    \vspace*{-0.3cm}
    \subfloat[Step = 150 ms]{{\includegraphics[height=0.3\textheight]{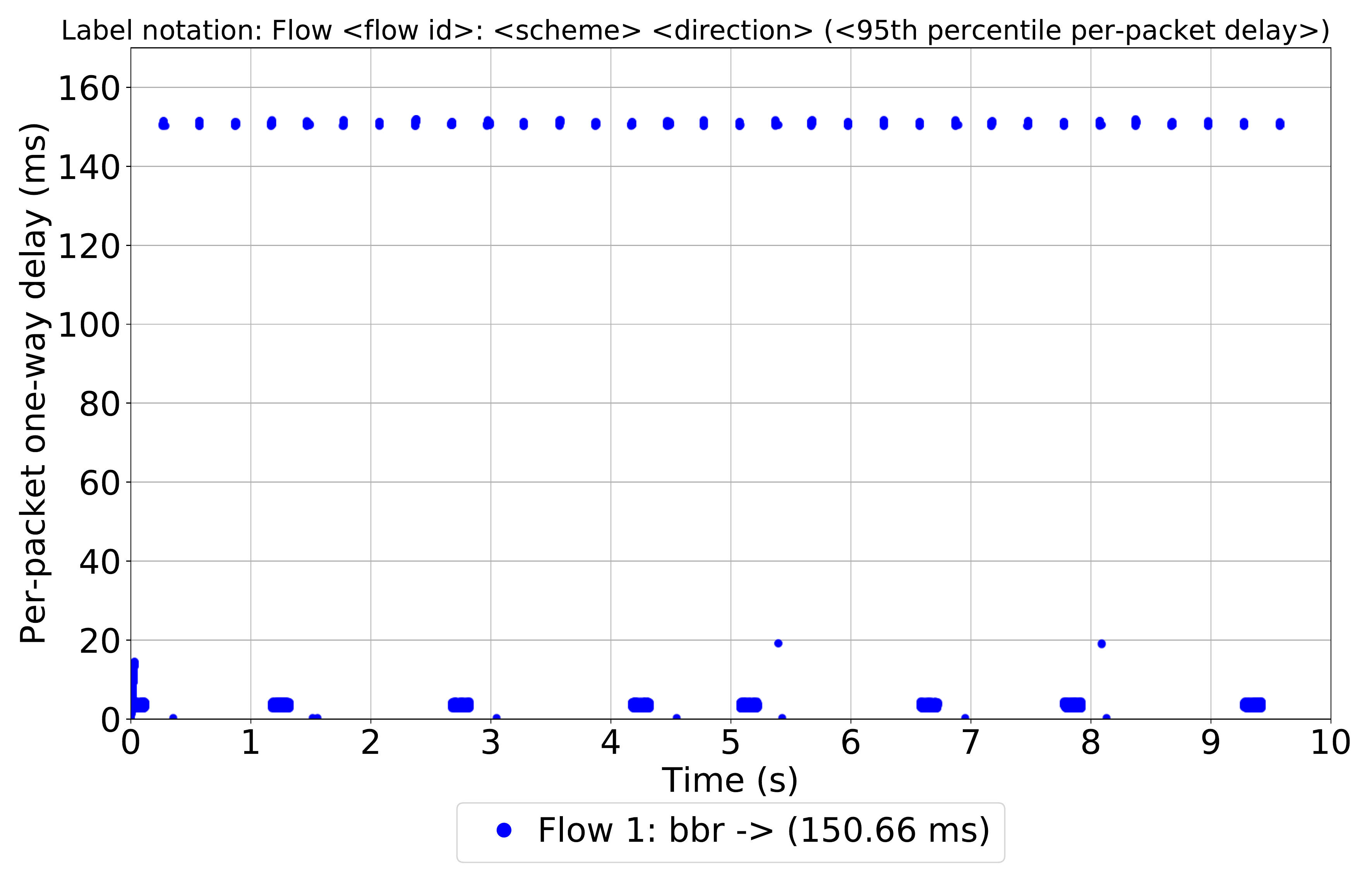} }}\\
    \vspace*{-0.3cm}
    \subfloat[Step = 160 ms]{{\includegraphics[height=0.3\textheight]{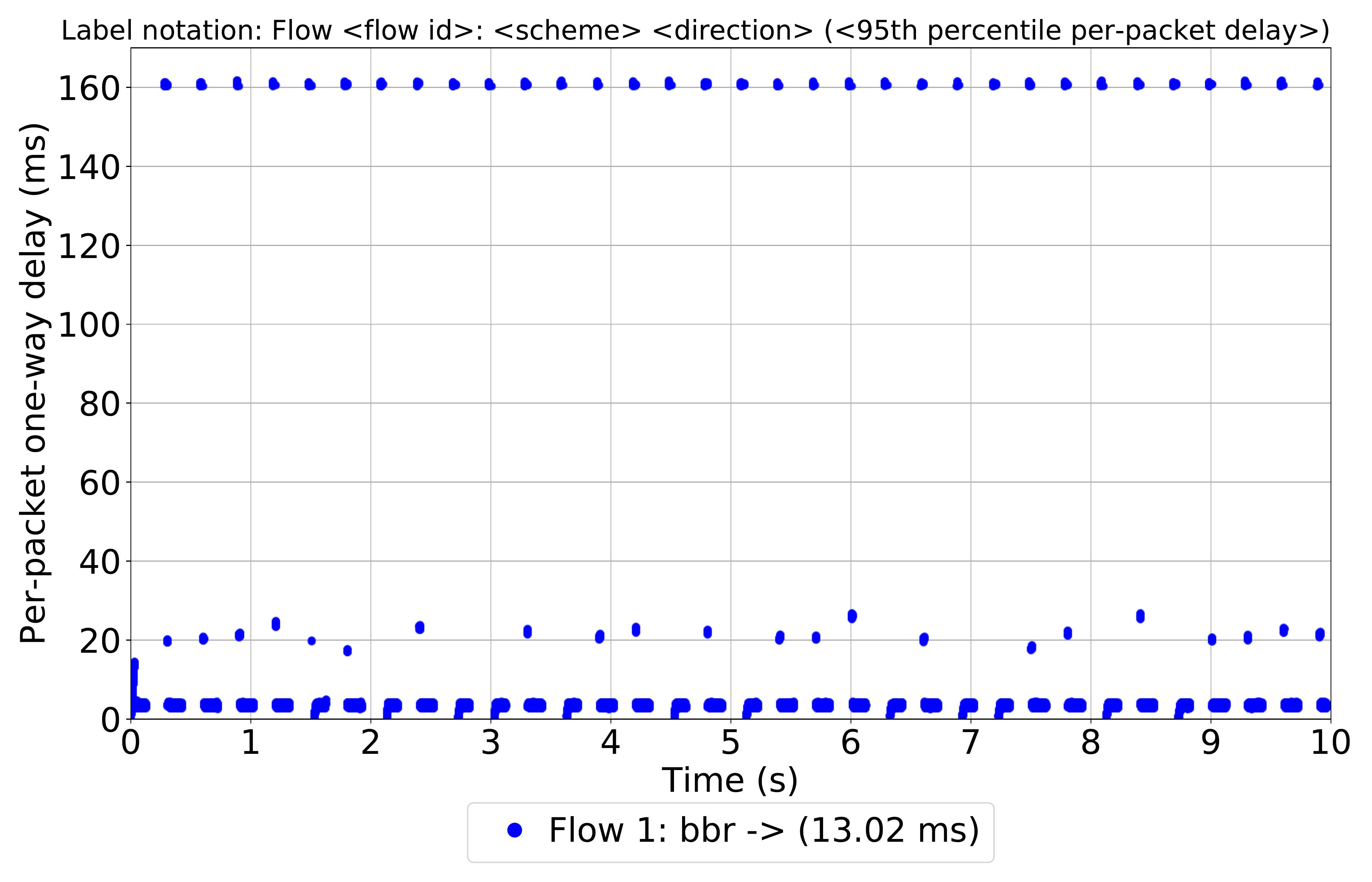} }}%
    \vspace*{-0.1cm}
    \caption{BBR's per-packet one-way delay plots for the delta 150 ms.}%
    \label{fig:vd-bbr-pptdelay}
\end{adjustbox}
\end{figure}

There are BBR's average one-way delay and per-packet one-way plots for the three steps in Figures~\ref{fig:vd-bbr-avgdelay} and~\ref{fig:vd-bbr-pptdelay} generated by CoCo-Beholder. It should be reminded for clarity that CoCo-Beholder is currently able to generate plots and statistics only for data traffic, rather than acknowledgment traffic, further referred to as ACK traffic. That is, the one-way delay plots~\ref{fig:vd-bbr-avgdelay} and~\ref{fig:vd-bbr-pptdelay} relate to the data traffic packets.

The average and per-packet one-way delay plots for each step are complementary and help to understand each other. At first glance,  the top and bottom average one-way delay plots look very similar: the curves oscillate with the period 300 ms between zero and the step value, i.e. 140 ms for the top plot and 160 ms for the bottom plot. 
However, it is easy to see that the concentration of packets that have one-way delays close to zero is much bigger in the bottom per-packet one-way delay plot than in the top per-packet one-way delay plot.
 
The latter fact is also reflected in the overall $95^\text{th}$ percentile one-way delay statistics depicted in the flow's labels in the per-packet one-way delay plots. This statistic is 140.87 ms for the step 140 ms and only 13.02 ms for the step 160 ms, which agrees with the considerably higher throughput in the latter case. 

The center one-way delay plots with the delta and step being 150 ms, both average and per-packet, again look transitional between the top and the bottom plots. That is, the time intervals when packets have one-way delays close to zero are longer but they a not periodic and there are only a few of them, while most of the runtime packets experience 150 ms one-way delays.

\FloatBarrier

\FloatBarrier

\FloatBarrier

\begin{figure}[h!]
\vspace*{-0.3cm}
\begin{adjustbox}{minipage=\linewidth,totalheight=\textheight}
    \centering
    \subfloat[Step = 140 ms]{{\includegraphics[height=0.3\textheight]{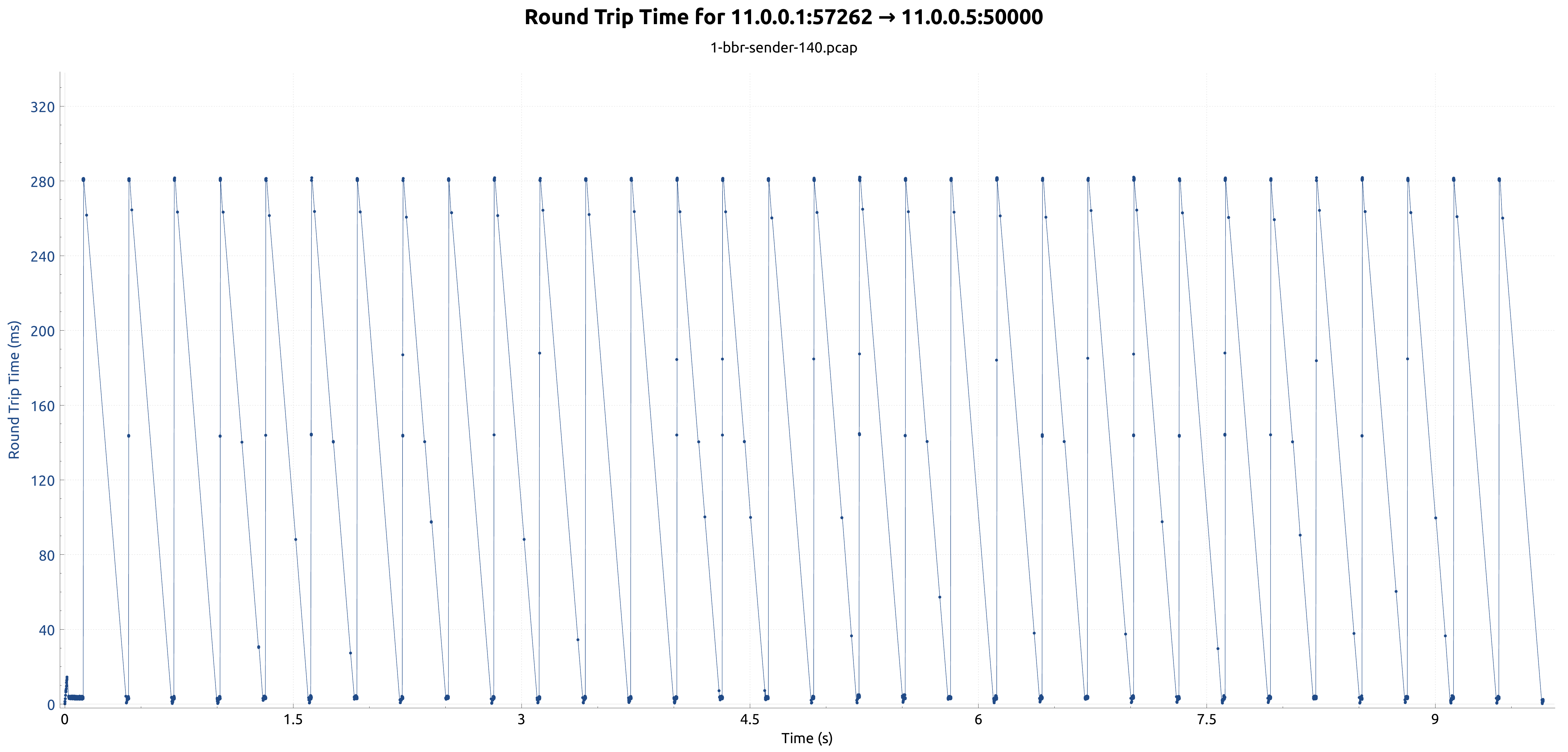} }}\\
    \vspace*{-0.2cm}
    \subfloat[Step = 150 ms]{{\includegraphics[height=0.3\textheight]{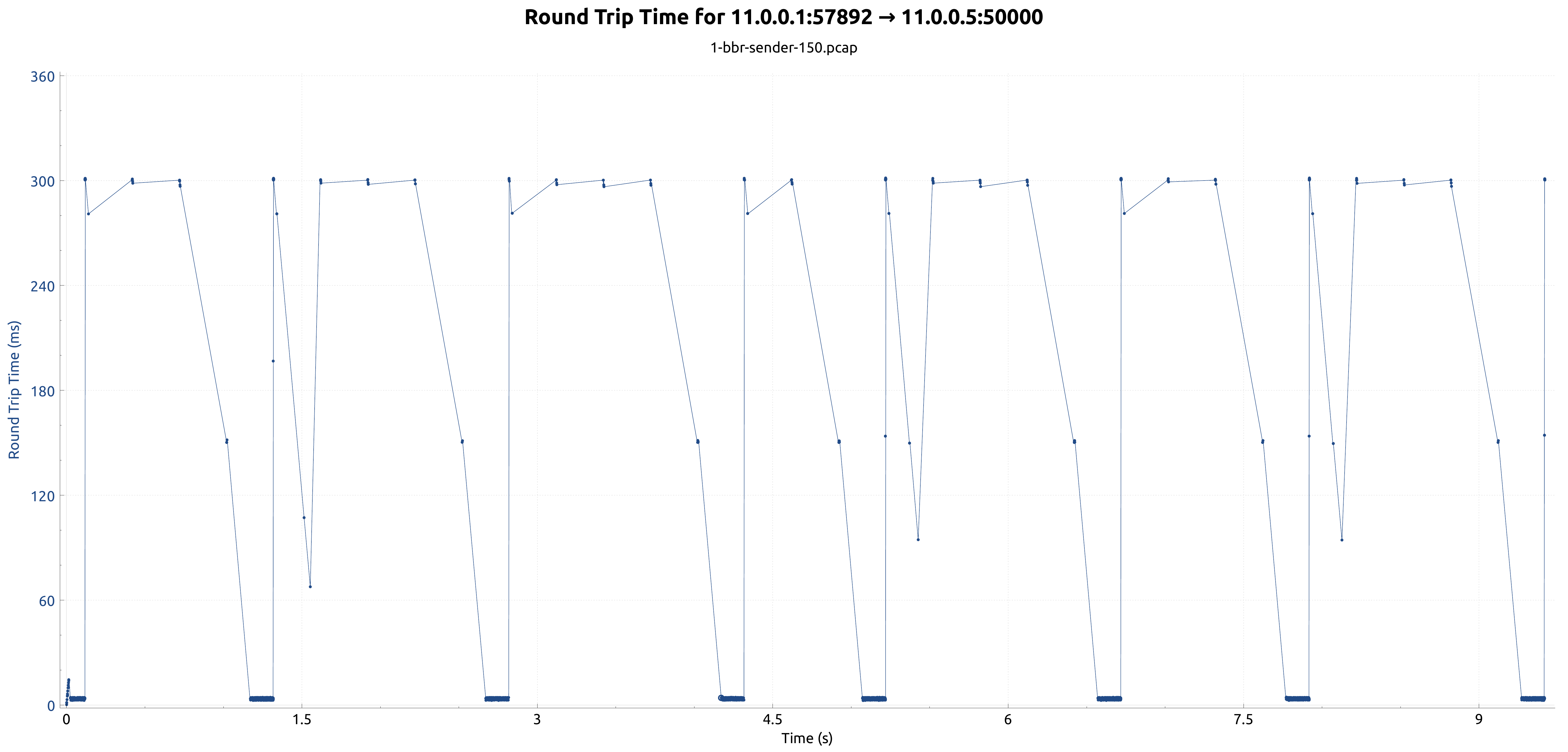} }}\\
    \vspace*{-0.2cm}
    \subfloat[Step = 160 ms]{{\includegraphics[height=0.3\textheight]{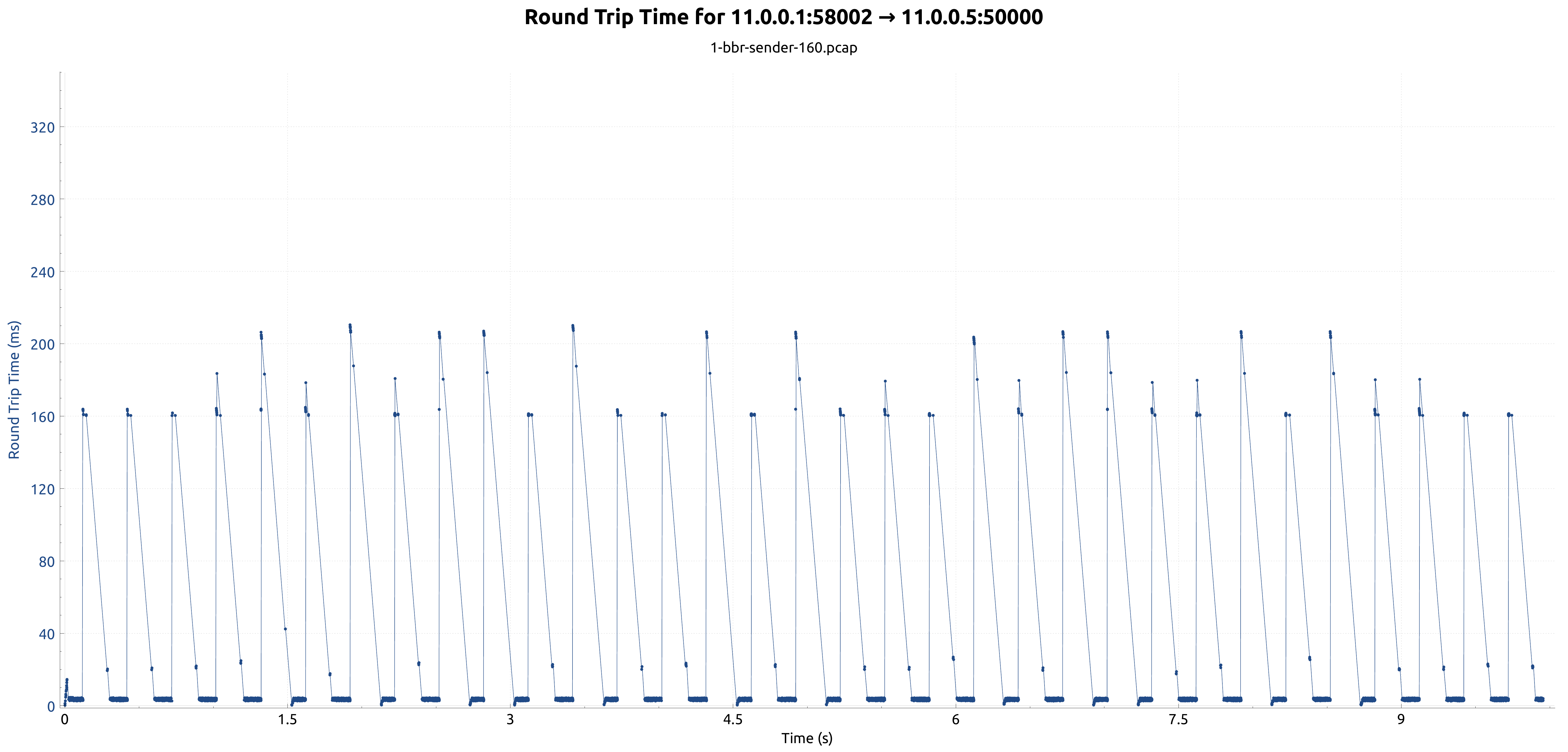} }}%
    \vspace*{-0.1cm}
    \caption{BBR's per-packet RTT plots for the delta 150 ms.}%
    \label{fig:vd-bbr-rtt}
\end{adjustbox}
\end{figure}

\FloatBarrier

In Figure~\ref{fig:vd-bbr-rtt}, there are BBR's per-packet RTT plots of data traffic for the three steps, which will compensate for the lack of the one-way delay plots of ACK traffic. These vector-graphics plots were generated using I/O Graph window of Wireshark applied to PCAP dump files recorded by CoCo-Beholder testing tool at the sender virtual host of the topology with \verb+tcpdump+, as discussed in Section~\ref{sec:testing_tool}. 

The analysis of the three RTT plots will be performed here. For the purpose, it should be reminded that the flow of the scheme is rightward. This is important because, when CoCo-Beholder testing tool changes the delay at the central link of the topology every delta time, it first changes the delay of the left interface and then of the right interface at the ends of the link. Therefore, if there was no delay at the link and then the delay is installed, the packets at the left interface get affected by the newly installed delay first, i.e. retained in the queue of the interface for the delay. This means that if the flow is rightward, it is the data packets of the flow that get affected by the delay first and the ACK packets of the flow only get affected a couple of milliseconds later, while if the flow is leftward, it is vice versa.

\begin{figure}[b!]
\begin{tikzpicture}  
\node[anchor=south west,inner sep=0] at (0,0) {\includegraphics[width=\textwidth]{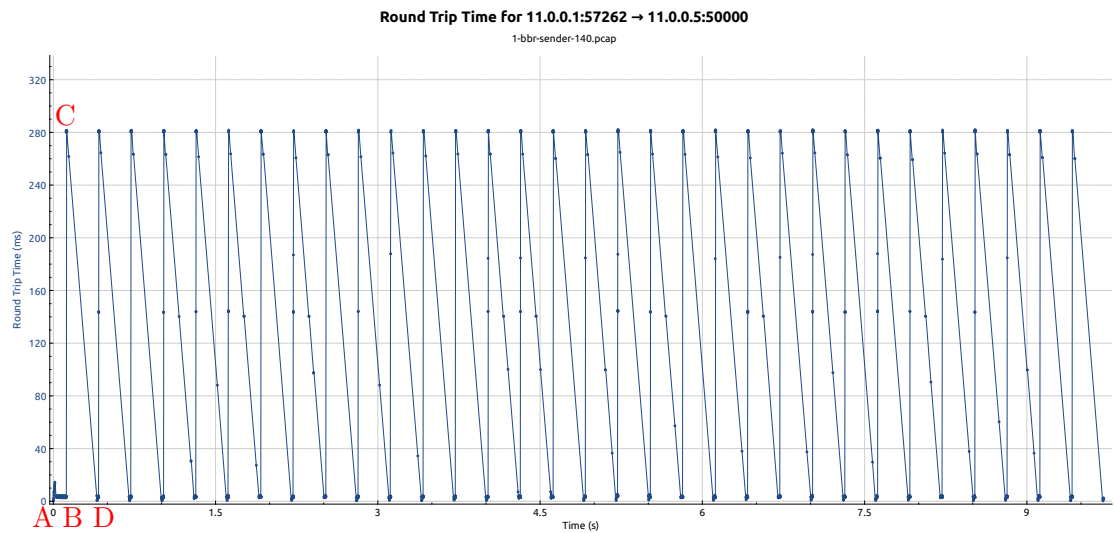}};
\node[text width=0.1cm, right, align=center, red] at (0.2,0.25) {A};
\node[text width=0.1cm, right, align=center, red] at (0.6,0.25) {B};
\node[text width=0.1cm, right, align=center, red] at (0.5,5.58) {C};
\node[text width=0.1cm, right, align=center, red] at (1.0,0.25) {D};
\end{tikzpicture}
\caption{BBR's per-packet RTT plot for the delta 150 ms and step 140 ms.}
\label{fig:vd-bbr-rtt-140}
\end{figure}

Though the RTT plots for the leftward and rightward flows look a little bit different, the fact is only relevant for a better understanding of the plots. The thesis author checked that the change of the flow's direction does not influence the resulting throughputs of a scheme in this series of experiments in any considerable way.
\begin{center}
\vspace{-0.3cm}
$\ast$~$\ast$~$\ast$
\vspace{-0.3cm}
\end{center}
The analysis is begun with BBR's RTT plot for the step 140 ms. The top plot of Figure~\ref{fig:vd-bbr-rtt} is now present individually in Figure~\ref{fig:vd-bbr-rtt-140} with some places of the plot marked with red letters.

\FloatBarrier

At point $A$, the flow traffic starts to run. No delay is installed between routers $R1$ and $R2$ in the topology shown in Figure~\ref{fig:vd-dumbbell}, so the data packets have their RTTs close to zero during the time interval $A-B$, the length of which equals the 150 ms delta.

At point $B$, the 140 ms delay is installed at the interface of router $R1$ and the interface of router $R2$ at the ends of the central link. $Sender$ sends off the last small group of packets denoted with letter $C$ before taking a pause, as ACK packets stop coming. 

This group of packets is detained in the queue of the configured interface at $R1$ for \mbox{140 ms}. When these data packets reach $Receiver$, it sends the ACK packets. However, the 150 ms delta has not yet expired. Hence, the delay is still installed at the routers, and $R2$ detains the ACK packets for 140 ms also. This is why the RTTs of the group of packets $C$ are 280 ms, as well as the length of the time interval $B-D$.

$Sender$ gets the ACK packets at $(150+280)=430$ ms since the start of the flow at point\nolinebreak[4] $A$, i.e., the length of the time interval $A-D$ is 430 ms. As the delta is 150 ms, this means that when the ACK packets come to $Sender$, there has been no delay at the central link for 130 ms and there will be no delay for 20 ms more. 

That is, though it is difficult to discern in the plot, the letter $D$ marks not a point but the 20 ms time interval, during which quite a big number of data packets is being sent by $Sender$ and acknowledged by $Receiver$ immediately, which results in the RTTs of the packets being close to zero.

To summarize, the interval $B-D$ is twice the step, and the interval denoted by $D$ is twice the delta minus twice the step. The thesis author checked that in the resulting RTT plot for another step 120 ms, which is also less than the delta but close to it, \mbox{the interval} $B-D$ is 240 ms, while $D$ lasts ($2\cdot150-240)=60$ ms indeed. 

The process in $B, C, D$ repeats with the period $2\cdot delta$ ms, and the traffic is actually sent only during $2\cdot (delta - step)$ ms out of these $2\cdot delta$ ms. In particular, for the delta \mbox{150 ms} and step 140 ms, the traffic is sent only for 20 ms out of every 300 ms. This explains the low throughput in the currently considered case when the step is approaching the delta ``from the left".

\begin{center}
\vspace{-0.3cm}
$\ast$~$\ast$~$\ast$
\vspace{-0.3cm}
\end{center}

The analysis is continued with BBR's RTT plot for the step 160 ms. The zoomed-in bottom plot of Figure~\ref{fig:vd-bbr-rtt} is now present individually in the top row of Figure~\ref{fig:vd-bbr-rtt-160} with some marks in red and violet colors.

At point $A$, the flow traffic starts to run. No delay is installed between routers $R1$ and $R2$, so the data packets have their RTTs close to zero during the time interval $A-B$, the length of which equals the 150 ms delta.

\usetikzlibrary{arrows,calc,decorations.markings,math,arrows.meta}

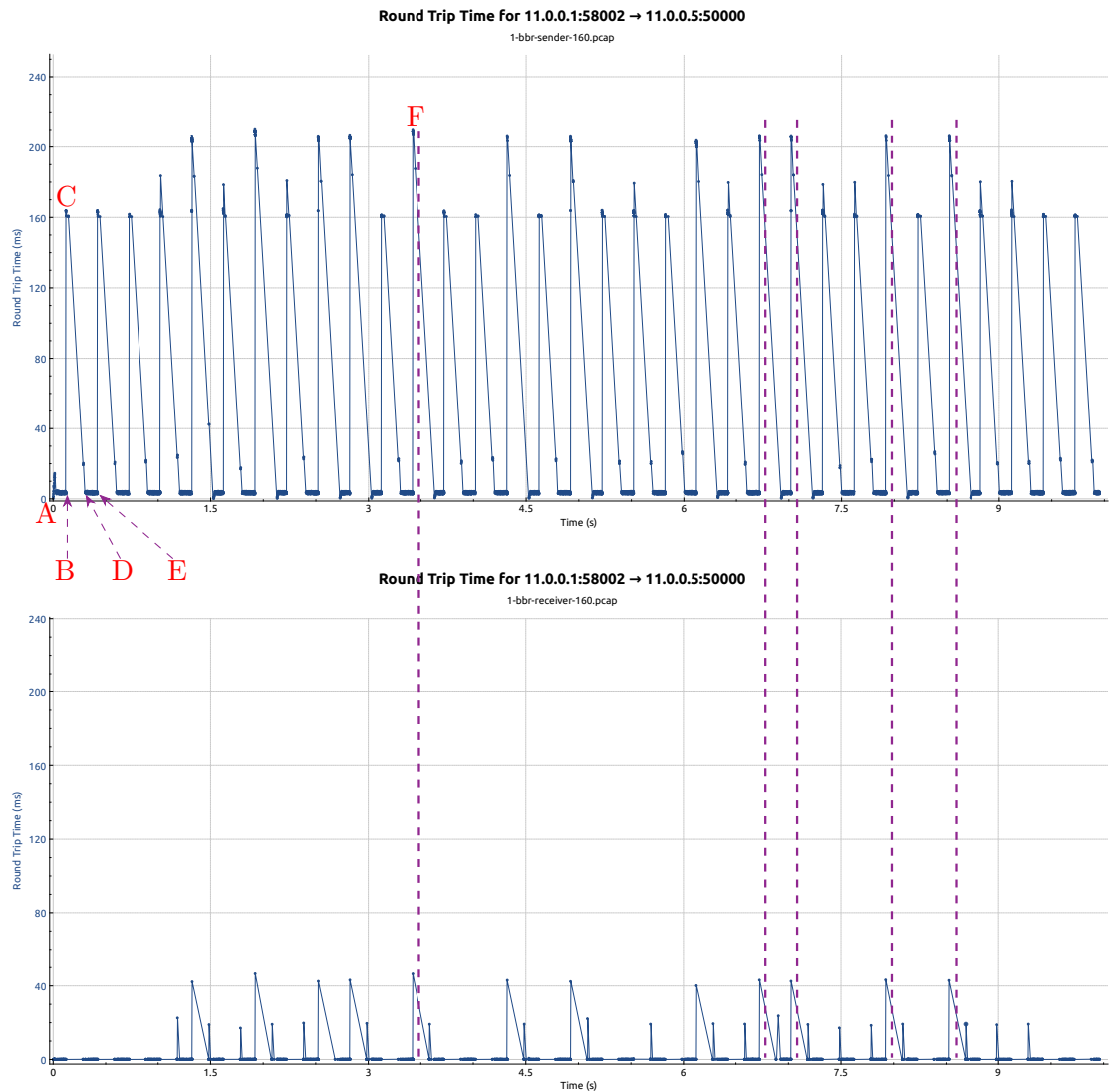
\begin{figure}[t!]
\begin{tikzpicture}  
\node[anchor=south west,inner sep=0] at (0,0) {\includegraphics[width=\textwidth]{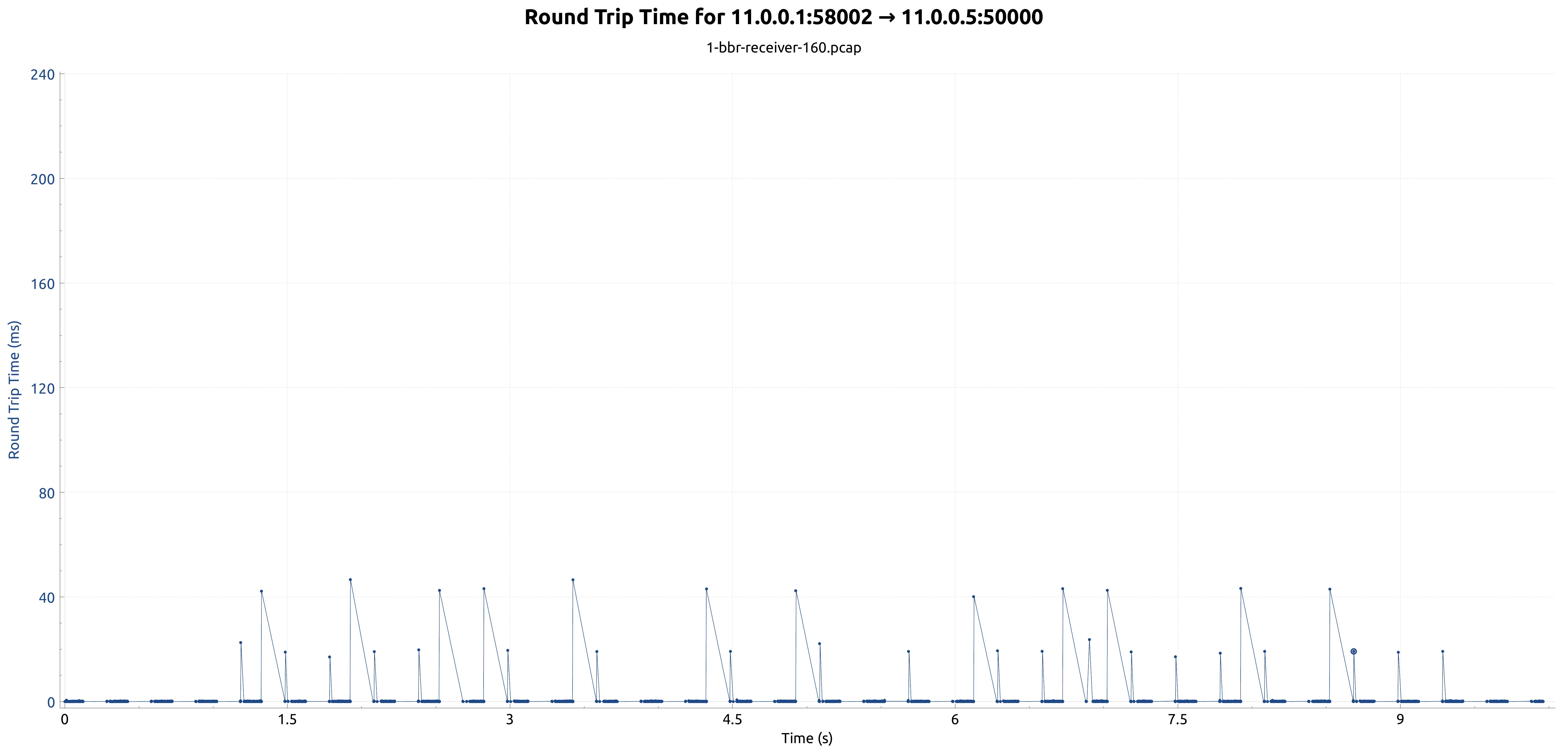}};
\node[anchor=south west,inner sep=0] at (0,7.5) {\includegraphics[width=\textwidth]{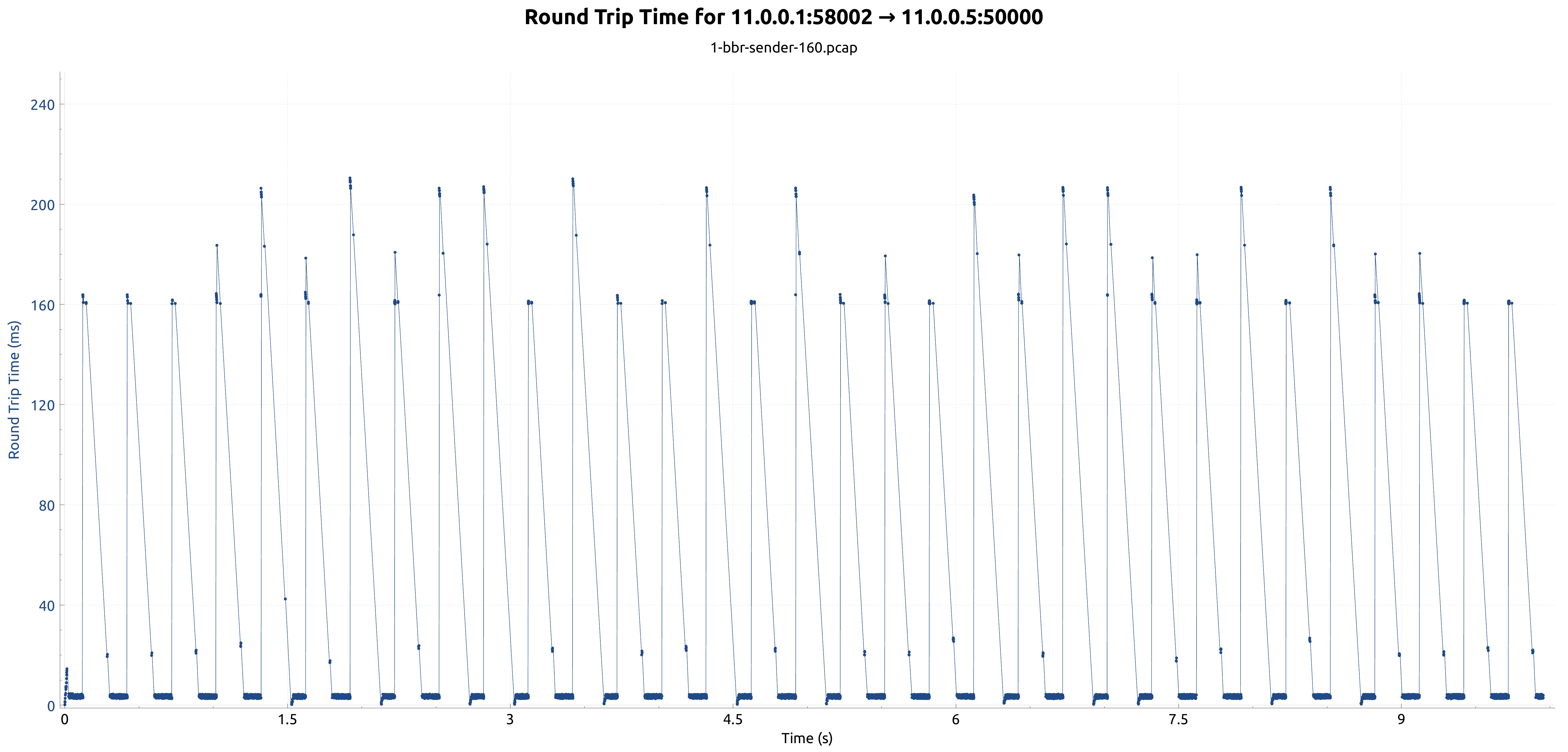}};
\draw[dashed, Plum, line width=0.3mm] (12.55, 13)--(12.55, 0.5);
\draw[dashed, Plum, line width=0.3mm] (11.7, 13)--(11.7, 0.5);
\draw[dashed, Plum, line width=0.3mm] (10.03, 13)--(10.03, 0.5);
\draw[dashed, Plum, line width=0.3mm] (10.45, 13)--(10.45, 0.5);
\draw[dashed, Plum, line width=0.3mm] (5.45, 12.85)--(5.45, 0.5);

\node[text width=0.1cm, right, align=center, red] at (0.24,7.75) {A};
\node[text width=0.1cm, right, align=center, red] at (0.5,7) {B};
\node[text width=0.1cm, right, align=center, red] at (0.51,11.97) {C};
\node[text width=0.1cm, right, align=center, red] at (1.25,7) {D};
\node[text width=0.1cm, right, align=center, red] at (1.99,7) {E};
\node[text width=0.1cm, right, align=center, red] at (5.15,13.05) {F};
\draw[-{Stealth[scale=1]}, dashed, Plum] (0.8, 7.15)--(0.8, 8);
\draw[-{Stealth[scale=1]}, dashed, Plum] (1.5, 7.15)--(1.04, 8);
\draw[-{Stealth[scale=1]}, dashed, Plum] (2.2, 7.15)--(1.22, 8.0);
\end{tikzpicture}
\caption{BBR's per-packet RTT plots for the delta 150 ms and step 160 ms over the dumps recorded at the sender (the top row) and at the receiver (the bottom row).}
\label{fig:vd-bbr-rtt-160}
\end{figure}

\FloatBarrier

At point $B$, the 160 ms delay is installed at the interface of router $R1$ and the interface of router $R2$ at the ends of the central link. $Sender$ sends off the last group of packets denoted with letter $C$ before taking a pause, as ACK packets stop coming. The group of packets $C$ lies in the time interval 15--20 ms long. However, the length of the interval is always that small and does not depend on the delta or the step (as long as the step is greater than the delta but close to it). This is why the thesis author neglects the length of the interval $C$ in the further discussion.

The group of packets $C$ is detained in the queue of the configured interface at $R1$ for 160 ms. When these data packets reach $Receiver$, it sends the ACK packets. The 150 ms delta has already expired, so there is again no delay at the central link. Hence, the ACK packets reach $Sender$ immediately. This is why the RTTs of the group of packets $C$ are 160 ms, as well as the length of the time interval $B-D$.

$Sender$ gets the ACK packets at $(150+160)=310$ ms since the start of the flow at point\nolinebreak[4] $A$, i.e., the length of the time interval $A-D$ is 310 ms. As the delta is 150 ms, this means that when the ACK packets come to $Sender$, there has been no delay at the central link for 10 ms and there will be no delay for 140 ms more. 

That is, during the 140 ms time interval $D-E$, $Sender$ sends  data packets that get immediately acknowledged by $Receiver$  and have their RTTs close to zero.

To summarize, the interval $B-D$ equals the step, and the interval $D-E$ is twice the delta minus the step. The process $B, C, D, E$ repeats with the period $2\cdot delta$ ms, and the traffic is actually sent during $(2\cdot delta - step)$ ms out of every $2\cdot delta$ ms. In particular, for the delta \mbox{150 ms} and step 160 ms, the traffic is sent for 140 ms out of every 300 ms. This explains the high throughput in the currently considered case when the step is approaching the delta ``from the right".

As denoted with letter $F$, sometimes the last group of packets, sent off by $Sender$ after the delay 160 ms is installed at the central link, experience the RTTs around 210 ms, i.e., around 50 ms higher than the installed delay 160 ms. The thesis author found the explanation of the fact when exploring the RTT plot present in the bottom row of Figure~\ref{fig:vd-bbr-rtt-160}. This RTT plot is the only one in this section that was built over the PCAP dump recorded at $Receiver$, rather than at $Sender$ (CoCo-Beholder records the traffic at all the hosts in the topology). This per-packet RTT plot indicates how soon $Receiver$ has sent the ACK packet off after receiving a data packet from $Sender$.

As marked with the long violet dashed lines in Figure~\ref{fig:vd-bbr-rtt-160}, the peaks, like one denoted with $F$ in the top plot, always have the corresponding peaks in the bottom plot, the height of which is around 40--50 ms. The thesis author explored the PCAP dumps and concluded that these peaks are due to TCP delayed acknowledgment feature. When $Receiver$ gets a data packet, it waits for another one. If the other data packet comes, $Receiver$ immediately acknowledges the pair of data packets with a single ACK packet. If for 40 ms (on Linux~\cite{ren2014survey}) the  other data packet does not come, $Receiver$ acknowledges the single data packet at last. This technique of combining a pair of acknowledgments into a single response is used to enhance the network performance~\cite{rfc1122}.

\begin{center}
\vspace{-0.3cm}
$\ast$~$\ast$~$\ast$
\vspace{-0.3cm}
\end{center}

\begin{figure}[t!]
\begin{tikzpicture}  
\node[anchor=south west,inner sep=0] at (0,0) {\includegraphics[width=\textwidth]{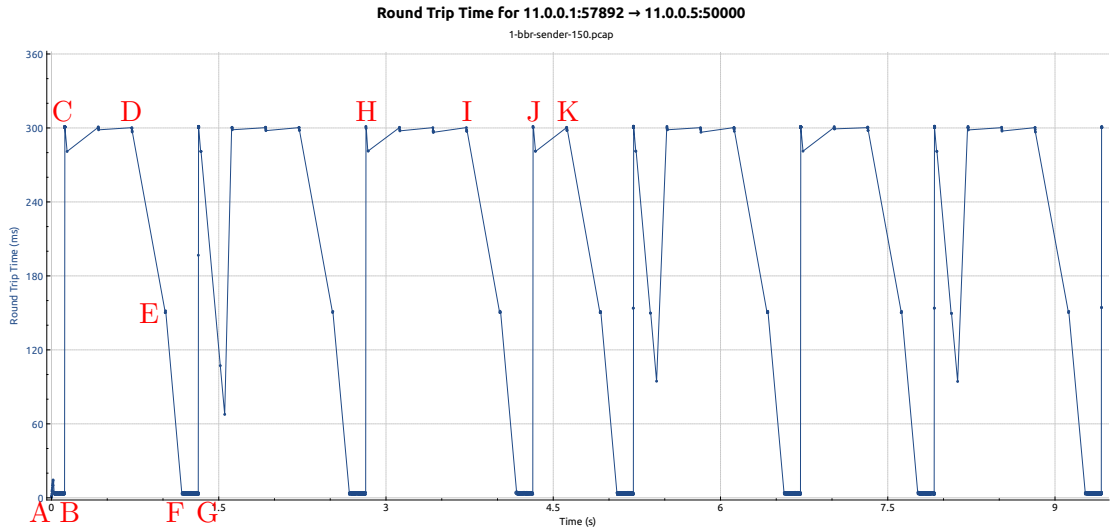}};
\node[text width=0.1cm, right, align=center, red] at (0.2,0.25) {A};
\node[text width=0.1cm, right, align=center, red] at (0.6,0.25) {B};
\node[text width=0.1cm, right, align=center, red] at (0.5,5.58) {C};
\node[text width=0.1cm, right, align=center, red] at (1.4,5.58) {D};
\node[text width=0.1cm, right, align=center, red] at (1.65,2.9) {E};
\node[text width=0.1cm, right, align=center, red] at (2.0,0.25) {F};
\node[text width=0.1cm, right, align=center, red] at (2.4,0.25) {G};
\node[text width=0.1cm, right, align=center, red] at (4.5,5.58) {H};
\node[text width=0.1cm, right, align=center, red] at (5.9,5.58) {I};
\node[text width=0.1cm, right, align=center, red] at (6.75,5.58) {J};
\node[text width=0.1cm, right, align=center, red] at (7.15,5.58) {K};
\end{tikzpicture}
\caption{BBR's per-packet RTT plot for the delta 150 ms and step 150 ms.}
\label{fig:vd-bbr-rtt-150}
\end{figure}

The analysis is concluded with BBR's RTT plot for the step and delta being 150 ms. The central plot of Figure~\ref{fig:vd-bbr-rtt} is now present individually in Figure~\ref{fig:vd-bbr-rtt-150} with some places of the plot marked with red letters.

At point $A$, the flow traffic starts to run. No delay is installed between routers $R1$ and $R2$, so the data packets have their RTTs close to zero during the time interval $A-B$, the length of which equals the 150 ms delta.

\FloatBarrier

At point $B$, the 150 ms delay is installed at the interface of router $R1$ and the interface of router $R2$ at the ends of the central link. $Sender$ sends the group of packets $C$ and the packets get delayed for 150 ms at router $R1$. When the packets reach $Receiver$, the delta 150 ms has already expired, so theoretically there should be already no delay at the central link. However, the plot shows that in practice, due to the imperfect precision of the variable delay implementation, the 150 ms is still there between $R1$ and $R2$, when $Receiver$ sends the ACK packets. Hence, the ACK packets get delayed for 150 ms at $R2$, and the RTTs of the data packets $C$ are 300 ms.

The process described in the previous paragraph can repeat a different number of times. E.g., it repeats three times for $C-D$, four times for $H-I$, and two times for $J-K$. The thesis author suggests that the exact number of times depends on the pure luck. The time interval $C-D-E$ is $3\cdot300=900$ ms.

Finally, at a certain point, a group of data packets $E$ gets lucky to reach $Receiver$ when there is already no delay at the central link -- the way it would happen always if the variable delay implementation was perfectly precise. The ACK packets for the data packets $E$ are sent off by $Receiver$ and come to $Sender$ at once. This is why the RTTs of the data packets $E$ are 150 ms, as well as the time interval $E-F$. 

When $E$ gets acknowledged, there is still no delay at the central link and there will be no delay for 150 ms more. This is why during the 150 ms interval $F-G$, $Sender$ freely sends the data packets, the RTTs of which are close to zero.

To summarize, due to the imperfect precision of the variable delay implementation, the case when the delta and step are equal demonstrates the remarkable instability. 

The synchronization problem makes the RTTs of data packets $(2\cdot delta)=(2\cdot step)$ ms. The problem may reiterate itself again and again suffocating $Sender$ completely. The cycle is broken only occasionally, and when it happens, the traffic is being eventually sent for the delta ms. This way, the resulting throughput is unpredictable and low.

If the variable delay implementation was perfectly precise, the traffic would be sent for the delta ms exactly and then would not be sent for the delta ms exactly. And this would be repeated again and again (i.e., with the period $2\cdot delta$ ms). That is, the situation would be exactly like in the previously considered case when the step is greater than the delta but close to it, and so the throughput would be very high.

\begin{center}
\vspace{-0.3cm}
$\ast$~$\ast$~$\ast$
\vspace{-0.3cm}
\end{center}

The exploration of the RTT plots allow to make the following summary:

\begin{itemize}
\item When the step is less than the delta but close to it, the traffic is sent only for $(2\cdot delta - \mathbf{2} \cdot step)$ ms, which is why the throughput is low.

\item When the step is greater than the delta but close to it, the traffic is sent for $(2\cdot delta - step)$ ms, which is why the throughput is high.

\item When the step and delta are equal, the throughput is low due to the imprecision problem, though theoretically it must be as high, as in the previous case.
\end{itemize}

These facts explain the behavior of the tested schemes under the square-wave delay and the nature of the 3D throughput plots in Figures~\ref{fig:rightsideview}--\ref{fig:topview}.

\loadchapter{conclusion}{Conclusion}

In recent years, significant effort was undertaken to make the research and evaluation of congestion control algorithms more collaborative. Pantheon of Congestion Control~\cite{pantheon} organized a collection of congestion control schemes and invites the research community to expand the collection by submitting new algorithms. The periodical live testing of all the schemes in the collection is performed, and the results are made publicly accessible.

However, Pantheon virtual network emulator, supplied together with the collection of the schemes, has very limited options for emulation. The only provided topology is point-to-point. Flows of different schemes cannot run in the topology at the same time. There is no way to have different network conditions per flow. Those are the most noticeable disadvantages. 

The thesis author attended to the problem and built CoCo-Beholder: a virtual network emulator enabling the highly customizable testing of the congestion control schemes in the collection. CoCo-Beholder provides the dumbbell topology of any size, very popular among researchers. All the links in the topology may have their individual settings: the rate, delay, and queue size. Flows of different schemes can run together in the topology, and for each flow, the user can specify its direction and when it should be started.

Written in Python, CoCo-Beholder is very human-friendly: it is easy to install and use. The emulator records the traffic of the tested schemes into regular PCAP dump files, which the user can easily explore, and enables the flexible generation of various plots and statistics for throughput, delay, and fairness.

To ensure the reliability of CoCo-Beholder, the thesis author tested 29 congestion \mbox{control} schemes in the simple point-to-point topology both with Pantheon and CoCo-Beholder. The two emulators showed similar results. CoCo-Beholder demonstrated more plausible results for a high-bandwidth link. 

\newpage

With CoCo-Beholder, the thesis author successfully reproduced the experiments from the modern paper~\cite{turkovic2019fifty}. The paper assessed the performance, intra- and inter-fairness, and intra-RTT-fairness of the three notable loss-based, delay-based, and hybrid schemes using the real hardware dumbbell testbed. The dumbbell testbed had a bottleneck rate at the central link and per-flow delays installed at the side links. The plots and statistics produced by the real testbed and CoCo-Beholder emulator are comparable. Some discrepancies in the results were observed for BBR. Understanding the reason of the BBR's different behavior can be addressed in the future work. 

As one of its features, CoCo-Beholder allows the user to have a variable delay at the central link of the emulated dumbbell topology. The thesis author explored the behavior of different congestion control schemes under the square-wave delay -- the simplest case of the variable delay. The results showed that the throughput of the schemes is dropping when the delay's step is approaching the delay's delta ``from the left" and increases again as soon as the step exceeds the delta. The case when the step and delta are equal appeared to be intractable by CoCo-Beholder due to the precision of the variable delay implementation being not perfect, though high. 

The thesis author has not encountered any research works analogous to the research on congestion control under the square-wave variable delay conducted in this thesis.

The future work may have two directions. The first one is to further improve CoCo-Beholder emulator: e.g., to add the opportunity to analyze not only data traffic but also acknowledgment traffic or more options for the topology configuration: a variable rate, alternative queue management policies, a probabilistic loss, packet reordering, etc. The second direction is to perform further experiments using CoCo-Beholder. In particular, it is possible to test different schemes under the variable delay of a more complex shape and to explore the inter-RTT-fairness of the schemes.   

The thesis author invites researchers to download and try to use CoCo-Beholder that is completely open-source. The thesis author is open for  discussion and questions.

\cleardoublepage

\appendix
\loadappendix{AppendixA}{The Help Messages of CoCo-Beholder}
\loadappendix{AppendixB}{Example of CoCo-Beholder Plots and Statistics Generation}
\loadappendix{AppendixC}{The Tested Schemes}
\loadappendix{AppendixD}{Results of the Testing CoCo-Beholder vs Pantheon}
\loadappendix{AppendixE}{Cubic Under the Square-Wave Variable Delay}

\backmatter

\label{Bibliography}
\hypersetup{urlcolor={black}}
\addtotoc{Bibliography}
\pagestyle{fancy}
\fancyhf{}
\fancyhead[LE,RO]{\thepage}
\fancyhead[RE,LO]{\emph{Bibliography}}
\bibliographystyle{unsrtnat}  
\bibliography{ms}  
\end{document}